\documentclass[twocolumn]{aastex61}
\usepackage{nicefrac}
\usepackage{txfonts}
\usepackage{amsfonts}
\usepackage{natbib}
\usepackage{epstopdf}
\usepackage{longtable}
\usepackage{booktabs}

\received{Sep 22, 2017}
\revised{Nov 4, 2017}
\accepted{Feb 8, 2018}

\submitjournal{ApJS}

\shorttitle{Herschel-PACS legacy of low-mass protostars}
\shortauthors{Karska et al.}

\begin{document}

\title{The Herschel-PACS legacy of low-mass protostars: \\
Properties of warm and hot gas and its origin in far-UV illuminated shocks}

\correspondingauthor{Agata Karska}
\email{agata.karska@umk.pl}

\author[0000-0001-8913-925X]{Agata Karska}
\affil{Centre for Astronomy, Faculty of Physics, Astronomy and Informatics,
 Nicolaus Copernicus University, Grudziadzka 5, 87-100 Torun, Poland}
\affil{Max-Planck Institut f\"{u}r Extraterrestrische Physik (MPE),
   			  Giessenbachstr. 1, D-85748 Garching, Germany}
\affil{Leiden Observatory, Leiden University, P.O. Box 9513,
          	  2300 RA Leiden, The Netherlands}

\author{Michael J. Kaufman}
\affil{Department of Physics and Astronomy, San Jose State University,
One Washington Square, San Jose, CA 95192-0106, USA}

\author{Lars E. Kristensen}
\affil{Centre for Star and Planet Formation, Niels Bohr Institute and Natural History Museum of Denmark, University of Copenhagen, 
{\O}ster Voldgade 5-7, DK-1350 Copenhagen K, Denmark}

\author{Ewine F. van Dishoeck}
\affil{Leiden Observatory, Leiden University, Niels Bohrweg 2, NL-2333 CA Leiden, The Netherlands}
\affil{Max-Planck Institut f\"{u}r Extraterrestrische Physik (MPE),
   			  Giessenbachstr. 1, D-85748 Garching, Germany}

\author{Gregory J. Herczeg}
\affil{Kavli Institute for Astronomy and Astrophysics, Peking University,
Yi He Yuan Lu 5, Haidian Qu, 100871 Beijing, People’s Republic of China}

\author{Joseph C. Mottram}
\affil{Max Planck Institute for Astronomy, Königstuhl 17, 69117 Heidelberg, Germany}

\author{\L{}ukasz Tychoniec}
\affil{Leiden Observatory, Leiden University, Niels Bohrweg 2, NL-2333 CA Leiden, The Netherlands}

\author{Johan E. Lindberg}
\affil{NASA Goddard Space Flight Center, Astrochemistry Laboratory, Mail Code 691, 8800 Greenbelt Road, Greenbelt, MD 20771, USA}

\author{Neal J. Evans II}
\affil{Department of Astronomy, The University of Texas at Austin, Austin, TX 78712, USA}

\author{Joel D. Green}
\affil{Space Telescope Science Institute, Baltimore, MD, USA}
\affil{Department of Astronomy, The University of Texas at Austin, Austin, TX 78712, USA}

\author{Yao-Lun Yang}
\affil{Department of Astronomy, The University of Texas at Austin, Austin, TX 78712, USA}

\author{Antoine Gusdorf}
\affil{LERMA, Observatoire de Paris, Ecole normale superieure, PSL Research University, CNRS, Sorbonne Universités, UPMC Univ.
Paris 06, F-75231, Paris, France}

\author{Dominika Itrich}
\affil{Centre for Astronomy, Faculty of Physics, Astronomy and Informatics,
 Nicolaus Copernicus University, Grudziadzka 5, 87-100 Torun, Poland}

\author{Natasza Si\'odmiak}
\affil{N. Copernicus Astronomical Center, Rabianska 8, 87-100 Torun, Poland}

\begin{abstract}

Recent observations from \textit{Herschel} allow the identification of important 
mechanisms responsible for the heating of gas surrounding low-mass protostars and its subsequent 
cooling in the far-infrared (FIR). Shocks are routinely invoked to reproduce some 
properties of the far-IR spectra, but standard models fail to reproduce the emission from key 
molecules, e.g. H$_2$O. Here, we present the \textit{Herschel}-PACS far-IR
 spectroscopy of 90 embedded low-mass protostars (Class 0/I). 
 The \textit{Herschel}-PACS spectral maps covering $\sim55-210$ $\mu$m with a field-of-view
 of $\sim$50'' are used to quantify the gas excitation conditions and spatial extent using rotational transitions
 of H$_{2}$O, high-$J$ CO, and OH, as well as [O I] and [C II]. 
 We confirm that a warm ($\sim$300 K) CO reservoir is ubiquitous and that a hotter component
   ($760\pm170$ K) is frequently detected around protostars. The line emission is extended beyond $\sim$1000 AU spatial scales in 40/90 objects, typically 
   in molecular tracers in Class 0 and atomic tracers in Class I objects.
   High-velocity emission ($\gtrsim90$ km s$^{-1}$) is detected 
  in only 10 sources in the [O I] line, suggesting that the bulk of [O I] arises from gas that 
  is moving slower than typical jets. Line flux ratios show an excellent agreement with models of $C$-shocks illuminated by UV photons 
 for pre-shock densities of $\sim$$10^5$ cm$^{-3}$ and UV fields 0.1-10 times the interstellar value. 
 The far-IR molecular and atomic lines are a unique diagnostic of feedback from UV emission and shocks in envelopes of deeply embedded protostars.
 
\end{abstract}


\section{Introduction} \label{sec:intro}
Complex physical processes are at play during the earliest stages of low-mass
star formation, when high accretion rates translate to significant feedback from  
a protostar on its surroundings \citep[e.g.][]{Fr14}. The launching of collimated jets  
and wide-angle winds leads to the formation of outflow cavities, and generates shock waves that compress 
and heat the envelope material to hundreds or thousands of K \citep{ND89,Ar07}. 
Ultraviolet (UV) photons produced as a result of accretion onto the protostar and created in 
shocks penetrate to large distances due to the low densities and scattering by dust in
  the outflow cavities \citep{Sp95,vK10,Vi12}. As a result,   
  the gas around low-mass protostars is heated to hundreds of K, altering the 
  chemistry and physics of the available mass reservoir and lowering the 
  efficiency of star formation \citep[e.g.][]{Of09,Dr15}.
  
Previous studies of feedback from low-mass protostars concentrated 
on the low-temperature ($T_\mathrm{kin}<100$ K) gas probed by low-$J$ ($J\leq10$) rotational transitions of CO 
at submillimeter wavelengths \citep[e.g.][]{BT99,Ar07,Ma13,Du14,Yi15}. Those transitions trace the entrained outflow gas
but show significant differences in 
spatial extent of emission and line profiles with respect to the more highly-excited lines 
\citep{Ni10,Sa14,Kr13,Kr17b}. Thus, the energetic processes seen in 
high-$J$ CO lines ($J>10$) are likely different and need to be characterized separately.

The most efficient cooling of hot and dense gas occurs in the far-infrared (IR) domain and 
thus the best tracers of the heating mechanisms are rotational lines of 
H$_2$O, high-$J$ CO, OH, and forbidden transitions of [O I] and [C II] \citep{GL78,Hol89}.
The first multi-line surveys of low-mass protostars were performed using the 
Long-Wavelength Spectrometer onboard the \textit{Infrared Space Observatory} \citep{ISO,LWS}.
The line spectra showed rich molecular emission which is spatially extended along the outflow direction 
and excited in relatively dense ($\sim10^{4}-10^{5}$ cm$^{-3}$) and warm ($\sim500$ K) gas
 \citep{Gi01,Ni02}. A combination of a slow, non-dissociative shock and a fast, dissociative shock from the  
jet was invoked to explain the observed molecular and [O I] emission, respectively. 
Yet, the low sensitivity of the instrument and the large telescope beam ($\sim80^{\prime\prime}$)
covering the entire extent of young stellar objects prevented study of the location of the far-IR emission.

Far-IR spectral maps from the Photodetector Array Camera and Spectrometer \citep[PACS;][]{Po10} on
board the \textit{Herschel} Space Observatory \citep{Pi10}\footnote{Herschel 
was an ESA space observatory with science instruments provided by European-led Principal Investigator
  consortia and with important participation from NASA.} are well-suited to revisit the 
  \textit{ISO}/LWS results due to enhanced sensitivity, along with improved spatial and spectral 
  resolution. The array of $5\times5$ elements covered a total field of view of $\sim47''$ with
  $9.4''$ pixels, corresponding to the spatial resolution scales of order $\sim1000$ AU at the typical distances to 
   nearby protostars ($\sim$200 pc).
    
Three large surveys of nearby low-mass protostars were carried out using \textit{Herschel}.
The `Water in Star forming regions with Herschel' (WISH) survey focused 
on H$_2$O and related species to characterize the physical and chemical processes in 
about 80 low-, intermediate-, and high-mass young stellar objects from pre-stellar cores to disks
 \citep{WISH} using mostly high-resolution spectra from 
the Heterodyne Instrument for the Far-Infrared \citep[HIFI,][]{dG10}. The PACS spectra centered at protostar
position were obtained 
for 18/29 deeply-embedded low-mass protostars, typically at selected transitions \citep{Ka13}. 
The `Dust, Ice, and Gas in Time' (DIGIT) open time program obtained complementary, full PACS spectra 
toward 30 Class 0/I protostars to quantify the dust and gas evolution in the far-IR \citep{Gr13,Gr16}, 
including 8 overlap sources with the WISH survey. 
Finally, the \lq William Herschel Line Legacy' (WILL) open time survey (\citealt{Mo17}) obtained PACS and HIFI
spectra toward about 50 additional low-mass protostars 
from the recent {\em Spitzer} and \textit{Herschel} Gould Belt imaging surveys
\citep[e.g.][]{Ev09,An10,Du15}. The WILL survey balances the samples of the WISH and DIGIT surveys  
both in the evolutionary stage (Class 0/I) and luminosities of protostars, thus 
ensuring that the combined sample of these three surveys is more representative
 of the general population of low-mass protostars \citep[see ][for details]{Mo17}.

Additional surveys focused on populations of protostars in specific clusters, located 
at similar distances. As part of the `\textit{Herschel} Orion Protostar Survey', 
CO emission covering the full PACS range was characterized for 21 protostars 
in Orion \citep{Ma12,Ma16}. Far-IR emission maps for the eight, youngest protostars identified in this survey
 \citep[the PACS Bright Red Sources, PBRSs]{Amy13}, were analyzed by \citet{To16}. The
`Herschel Study of Star Formation Feedback on Cloud Scales' (PI: H. Arce) obtained  
PACS and HIFI maps of the entire NGC1333 region in Perseus in the selected atomic ([O I], [C II]) 
and molecular lines (e.g. CO 10-9, H$_2$O 557 GHz), respectively \citep{Di16}. 
\cite{RM16} provide a catalog of H$_2$O and [O I] lines observed with PACS for 362 young stellar 
objects from Class 0 to Class III.

The above programs have proven H$_2$O as an important tracer of energetic physical processes 
in the earliest stages of star formation. Line profiles of H$_2$O are complex, but dominated 
by gas moving at several tens of km s$^{-1}$ \citep{Kr12,Mo14,Mo17}. The spatial extent of H$_2$O 
resembles the emission from H$_2$ and follows the outflow direction \citep{Ni10,Sa12}. 
The far-IR gas cooling budget is dominated by H$_2$O and high-$J$ CO lines, following 
the predictions from models of non-dissociative, $C$ shocks \citep{Ka13}. Thus, H$_2$O emission is 
closely linked to the outflow shocks and its excitation is in agreement with stationary 
(1D plane-parallel) shock models \citep[e.g. Paris-Durham code, ][]{FP15}. 
Yet, these models cannot reproduce the chemistry: 
low abundances of H$_2$O and O$_2$ and low
H$_2$O / CO and H$_2$O / OH flux ratios \citep{MK15,Ka14b}. UV photons can photodissociate
H$_2$O and reconcile the models and observations, but the models of shocks irradiated by 
UV have not been available for detailed comparisons until now.

In this paper, we analyze together the far-IR PACS spectra of a large and uniform sample of low-mass YSOs 
obtained as part of the WISH, DIGIT, and WILL surveys, and compare them to the stationary shock 
models from \citet{FP15}, models of photodissociation regions from \citet{Ka99}, and UV irradiated
$C$-shock models from \citet{MK15} to address the following questions: What is the typical 
far-IR spectrum of a low-mass protostar? Is the emission in various far-IR atomic and molecular 
species linked? What is the spatial extent of the line emission? What are the 
typical rotational temperatures of CO, H$_2$O, and OH, and the corresponding physical conditions 
of the gas? What is the gas cooling budget in the far-IR? What is the origin of the far-IR emission?
In particular, what kinds of shocks and densities of pre-shock material are involved? What is the 
strength of UV radiation and does it affect the chemistry of low-mass protostars? Finally, 
what is the evolution of the far-IR line properties during the Class 0/I phases and the corresponding 
changes in the physical processes (shocks, UV radiation)?
 
The paper is organized as follows. \S 2 describes our sample selection, observations,
and data reduction. \S 3 compiles results on detection statistics, spatial extent 
of line emission and resolved profiles. \S 4 shows analysis of molecular excitation 
and far-infrared line cooling. Comparisons of absolute line 
emission of atomic lines to shocks and photodissociation regions models
are presented in \S 5, along with comparisons to UV-illuminated shock models.
\S 6 discusses the results obtained in previous sections and  
\S 7 presents the summary and conclusions. 

\section{Observations}
\subsection{Sample selection}
The low-mass embedded protostars analyzed here were initially observed as part of the `Water In Star forming 
regions with Herschel' \citep[WISH,][]{WISH} and `Dust, Ice, and Gas In Time' \citep[DIGIT,][]{Gr13} surveys,
which comprised 18 and 29 protostars targeted with PACS, respectively, including 8 overlap sources. 
This sample was subsequently expanded by the `William Herschel Line Legacy' survey (WILL, \citealt{Mo17}),
where a further 49 sources were observed, including 37 Class 0/I objects. Three additional sources were 
 located in the PACS spectral maps of the primary targets, increasing the sample to 
90 sources in total (for details, see Table 1).
\begin{figure}[h]
\hspace{-1.5cm}
\includegraphics[angle=-270,height=8cm]{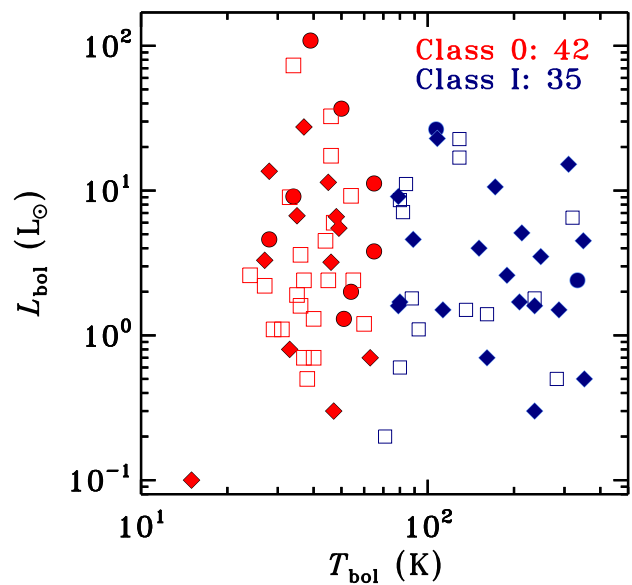}
\caption{\label{lboltbol} Distribution of bolometric luminosities and temperatures for 
the Class 0 (in red) and Class I (in blue) protostars. Different symbols show 
objects observed as part of the WISH survey (filled circles), the DIGIT survey (filled 
diamonds), and the WILL survey (open squares). Sources observed as part of both 
WISH and DIGIT surveys are shown only once, as filled diamonds.}
\end{figure}

Most of the sources are located in the Perseus (30 sources), the Aquila Rift complex (7 sources in W40, 6 sources in Aquila, 
 and 2 sources in Serpens South), and Taurus (12 sources).
 The remaining sources are from Ophiuchus (9 sources), Corona Australis (5 sources), Serpens Main (3 sources), and other molecular 
 clouds. All protostars are located at distances $\lesssim450$ pc \citep[for distance references 
see][]{Kr12,Gr13,Mo17}.

The selection procedures for the WISH and DIGIT sources are discussed in detail in \citet{WISH} and 
\citet{Gr13} with a general rule that all of them are well-known and extensively studied protostars. The WILL sources
were selected based on unbiased mid-IR and millimeter continuum observations with \textit{Spitzer} and various ground-based telescopes
\citep{Mo17}.

The sources are divided into classes based on the shapes of their spectral energy distributions (SEDs), 
constructed using flux densities obtained with \textit{Herschel}-PACS (see \citealt{Ka13} for WISH, 
\citealt{Gr16} and \citealt{Ma16} for DIGIT, \citealt{Mo17} for WILL). Sources with bolometric 
temperatures $T_\mathrm{bol}\leq70$ K are classified as Class 0, and those with 
$70 \mathrm{~K} < T_\mathrm{bol} < 650$ K are classified as Class I. However, 12 sources 
in the original WILL sample were re-classified as either Class II or pre-stellar, based on the absence of an 
entrained molecular outflow as identified by broad line wings in CO 3-2 maps and in HIFI H$_2$O and CO 10-9 spectra, 
as well as on the morphology and intensity of HCO$^+$ 4-3 and C$^{18}$O 3-2 emission (see \citealt{Mo17} and the description of the method 
in \citealt{vK09d} and \citealt{Ca16}). Additionally, in 5 out of those 12 sources and in one Class 0 source, 
\cite{Mo17} detected a narrow, bright CO 10-9 emission indicative of photodissociation
 regions (PDRs) that are not associated with young stellar objects. Excluding pre-stellar, Class II and PDRs, the final sample consists of 42 Class 0 and 35 Class I 
 sources, covering a broad range of bolometric luminosities, $L_\mathrm{bol}$ (see Figure \ref{lboltbol}).
The comparison with the global Spitzer Gould Belt sample shows that the sample here is representative 
 for the Class 0 and young Class I sources \citep{Mo17}.
 
We note that in the three sources (ID numbers 4, 82, 85), the separation of protostars is large
 enough to decompose the continuum emission from PACS spectral maps into distinct objects (see \citealt{Lee13} 
  for L1448 MM and \citealt{Li14} for sources in Corona Australis). The rest of the sources are treated as 
single objects, even though some of them are close multiples \citep[e.g.][]{VANDAM}. The impact of this 
treatment depends on the evolutionary stage and relative brightness of the respective components. \citet{Mu16}
 distinguish three cases: (i) the combined SED is simply doubled when the two components have 
similar SEDs; (ii) the SED appears odd, double-peaked when the two components are non-coeval;
(iii) the brightest component dominates the combined SED, when the other one is noticeable dimmer and 
younger. Similar cases are expected to apply for the emission lines. However, without 
a strong indication that the sources are co-eval, the calculated luminosities cannot be properly corrected 
\citep{Mu16}, and the sources are therefore treated as single sources in the following.

\subsection{Observations and data reduction}
Single footprint spectral maps of all sources were obtained with the PACS instrument onboard \textit{Herschel}. 
Each map consists of 25 spatial pixels (\textit{spaxels}) 
of $9\farcs4\times9\farcs4$ arranged in a 5$\times$5 array with a total field of view of 
$\sim47''\times47''$. Each spaxel contains a (sub-)spectrum observed in the first (red) or second (blue) order,
 within the wavelength ranges of 102-210 $\mu$m and 51-105 $\mu$m, respectively. 
 Due to flux calibration issues at the extreme ends of the spectra, the ranges from $\sim$55-100 $\mu$m and 104-190$\mu$m are used 
 in the analysis. The spectral resolving power increases with wavelength from about 1000 to 2000
 (corresponding to velocity resolutions of \mbox{$\sim$140 to 320 km s$^{-1}$}) in the first 
 order and from about 1500 to 3000 (\mbox{$\sim$100 to 210 km s$^{-1}$}) 
 in the second order. 
 
 Two main observing schemes were used: line spectroscopy mode for the WISH and WILL sources; 
 and range spectroscopy mode for the DIGIT sources and four sources from the WISH survey (Serpens SMM1 and NGC1333 IRAS 4A, 4B, and 2A). 
 The line spectroscopy mode yields 
 observations of small spectral regions ($\Delta\lambda\sim$0.5-2 $\mu$m) around selected lines and is particularly 
 suited for deep integrations. The range spectroscopy mode provides the full spectrum 
 from $\sim50$ to 210 $\mu$m but the spectral sampling within a resolution element is about 
 3-4 times coarser than in the line spectroscopy mode.
 For both schemes, the chopping / nodding observing mode from the source 
was used to subtract the background emission within 6$^{\prime}$.

Data reduction for both observing modes was performed with the Herschel Interactive Processing 
 Environment \citep[HIPE,][]{Ot10} version 13. The flux was normalized to the telescopic
background and calibrated using observations of Neptune. Spectral flatfielding within HIPE was
used to increase the signal-to-noise ratio (for details, see \citealt{He12}, \citealt{Gr13}, and \citealt{Sturm13}).
The overall flux uncertainty is about 20\% from cross-comparisons of sources in common within our programs. 

A 1D spectrum is obtained for each source by summing a custom number of spaxels chosen after investigation 
of the 2D spectral maps \citep{Ka13}, using the technique applied to the \lq\lq CDF'' (COPS-DIGIT-FOOSH protostar)
 archive \citep{Gr16}. This archive is freely available as a User Provided Data Product in the
  Herschel Science Archive\footnote{https://www.cosmos.esa.int/web/herschel/user-provided-data-products}. Most notably, the 2016 update in PACS spectroscopy includes a correction for
   pointing and jitter offsets during observations.

For sources with extended line emission, the co-addition of spaxels with 
detected emission increases the S / N, smooths the continuum, and enables correction for significant differences in 
beam sizes over the wide spectral range covered by PACS. 
 For sources with point-like 
emission in all lines, only the central spaxel spectrum is used, but the line fluxes are 
multiplied by the wavelength-dependent instrumental correction factors ($\sim1.4$ at 70 $\mu$m 
and $\sim2.3$ at 180 $\mu$m, see PACS Observer's
Manual\footnote{http://herschel.esac.esa.int/Docs/PACS/html/pacs\_om.html}).  

Since the lines are spectrally unresolved (except [O I], see \S 3.3), line fluxes are 
calculated by fitting Gaussians to the final 1D spectrum. Single Gaussians are used 
for well-isolated lines and double or triple Gaussians for closeby lines, including blends. 
The line width of the Gaussians is fixed to the instrumental value for unresolved lines, 
except the [O I] line at 63 $\mu$m which in several sources shows high-velocity wings \citep{vK10,Ka13,RM16}.
In this case, integration and / or broad Gaussian fitting are applied. All Gaussian fits 
were visually inspected to avoid possible confusion. 

\startlongtable
\begin{deluxetable*}{lllcrrrrcllll} 
\tablecaption{Luminosities and Bolometric Temperatures of Embedded Protostars \label{cat_paper4} }
\tablewidth{0pt}
\tablehead{
\colhead{ID}           &
\colhead{RA (J2000)}   &
\colhead{Dec (J2000)}  & 
\colhead{Cloud}        & 
\colhead{$D$\tablenotemark{a}}          &
\colhead{$L_\mathrm{bol}$}  &
\colhead{$\frac{L_\mathrm{submm}}{L_\mathrm{bol}}$}  &
\colhead{$T_\mathrm{bol}$}  &
\colhead{Class}  &
\colhead{Other names} & 
\colhead{Survey\tablenotemark{b}}  \\    
\colhead{~}   &        
\colhead{({h}\phn{m}\phn{s})}   &
\colhead{(\phn{\arcdeg}~\phn{\arcmin}~\phn{\arcsec})} &
\colhead{~}                                          &               
\colhead{(pc)}   &            
\colhead{($L_{\odot}$)}  & 
\colhead{(\%)}    & 
\colhead{(K)}  & 
\colhead{~} &
\colhead{~} &
\colhead{~} 
}
\startdata
1 & 03 25 22.33 & $+$30 45 14.0 & Per & 235  &  4.5 &  2.7 &  44 & 0 & L1448 IRS2, Per-emb 22, PER 01 & WL\\
2 & 03 25 36.48 & $+$30 45 22.3 & Per & 235  &  9.2 &  1.7 &  54 & 0 & L1448 IRS3/N(A), Per-emb 33, PER 02 & WL\\
3 & 03 25 38.82 & $+$30 44 06.3 & Per & 235  &  5.5 &  0.4 &  49 & 0 & L1448 MMS/C(N), Per-emb 26 & D\\
4 & 03 25 39.10 & $+$30 43 58.0 & Per & 235  &  1.7 &  2.1 &  80 & I & L1448 C(S), Per-emb 42  & D \\
5 & 03 26 37.46 & $+$30 15 28.0 & Per & 235  &  1.2 &  4.2 & 60  & 0 & I03235$+$3004, Per-emb 25, PER 04 & WL\\
6 & 03 27 39.09 & $+$30 13 03.0 & Per & 235  &  6.6 &  1.3 &  48 & I & L1455-IRS1, I03245$+$3002,  Per-emb 17  & D\\
7 & 03 28 00.40 & $+$30 08 01.3 & Per & 235  &  0.3 &  6.3 & 236 & I & L1455-IRS3, I03249$+$2957, Per-emb 46  & D\\
8 & 03 28 37.09 & $+$31 13 30.7 & Per & 235  & 11.1 &  0.6 &  84 & I & NGC1333 I1, Per-emb 35, PER 05 & WL \\
9 & 03 28 55.56 & $+$31 14 36.6 & Per & 235  & 36.8 &  0.5 &  50 & 0 & NGC1333 I2A, Per-emb 27 & WH \\
10 & 03 28 57.36 & $+$31 14 15.7 & Per & 235 &  7.1 & \nodata & 82 & I & NGC1333 I2B, Per-emb 36, PER 06 & WL\\
11 & 03 29 00.52 & $+$31 12 00.7 & Per & 235 &  0.7 &  3.9 &  37 & 0 & HRF 65, Per-emb 3, PER 07 & WL\\
12 & 03 29 01.57 & $+$31 20 20.7 & Per & 235 & 16.9 &  1.3 & 129 & I & HH 12, Per-emb 54, PER 08 & WL\\
13 & 03 29 07.76 & $+$31 21 57.2 & Per & 235 & 22.7 & \nodata & 129 & I & I03260$+$3111(W), Per-emb 50, PER 09 & WL\\
14 & 03 29 10.50 & $+$31 13 31.0 & Per & 235 &  9.1 &  3.0 &   34 & 0 & NGC1333 I4A, Per-emb 12 & WH\\
15 & 03 29 10.68 & $+$31 18 20.5 & Per & 235 &  6.0 &  2.2 &   47 & 0 & HRF 46, Per-emb 21, PER 10 & WL\\
16 & 03 29 12.04 & $+$31 13 01.5 & Per & 235 &  4.6 &  4.0 &   28 & 0 & NGC1333 I4B, Per-emb 13 & WH\\
17 & 03 29 13.52 & $+$31 13 58.0 & Per & 235 &  1.1 &  8.7 &   31 & 0 & NGC1333 I4C, Per-emb 14, PER 12 & WL\\
18 & 03 29 51.82 & $+$31 39 06.1 & Per & 235  & 0.7 &  5.0 &   40 & 0 & I03267$+$3128, Per-emb 9, PER 13  & WL\\
19 & 03 30 15.12 & $+$30 23 49.2 & Per & 235 &  1.8 &  1.6 &   88 & I & I03271$+$3013, Per-emb 34, PER 14 & WL\\
20 & 03 31 20.96 & $+$30 45 30.2 & Per & 235 &  1.6 &  5.8 &   36 & 0 & I03282$+$3035, Per-emb 5, PER 15 & WL\\
21 & 03 32 17.95 & $+$30 49 47.6 & Per & 235 &  1.1 & 13.3 &   29 & 0 & I03292$+$3039, Per-emb 2, PER 16  & WL\\
22 & 03 33 12.85 & $+$31 21 24.1 & Per & 235 &  4.5 &  0.5 &  349 & I & I03301$+$3111, Bolo76, Per-emb 64 & D\\
23 & 03 33 14.40 & $+$31 07 10.9 & Per & 235 &  0.2 & \nodata& 71 & I & B1 SMM3, Per-emb 6, PER 17 & WL\\
24 & 03 33 16.45 & $+$31 06 52.5 & Per & 235 &  0.5 & \nodata& 38 & 0 & B1 d, Per-emb 10, PER 18 & WL\\
25 & 03 33 16.66 & $+$31 07 55.2 & Per & 235 &  1.5 & 0.4 &   113 & I & B1 a, I03301$+$3057, Per-emb 40 & D\\ 
26 & 03 33 17.85 & $+$31 09 32.0 & Per & 235 &  3.2 & 0.2  & 46 & 0 & B1 c, Per-emb 29 & D \\
27 & 03 33 27.28 & $+$31 07 10.2 & Per & 235 &  1.1 & 1.7  & 93 & I & B1 SMM11, Per-emb 30, PER 19 & WL\\
28 & 03 43 56.53 & $+$32 00 52.9 & Per & 235 &  2.2 & 6.3  & 27 & 0 & HH 211 MMS, Per-emb 1, PER 20 & WL\\
29 & 03 43 56.85 & $+$32 03 04.6 & Per & 235 &  1.9 & 3.8  & 35 & 0 & IC348 MMS/SW, Per-emb 11, PER 21 & WL\\
30 & 03 44 43.94 & $+$32 01 36.1 & Per & 235 &  2.4 & 3.4  & 45 & 0 & IC348 a, Per-emb 8, PER 22 & WL\\
31 & 04 04 42.9 & $+$26 18 56.3 & Tau & 140 &  3.5  &  0.7  &  248 & I & L1489 & D,WH \\
32 & 04 19 58.4 & $+$27 09 57.0 & Tau & 140 &  1.5  &  3.3  &  136 & I & I04169$+$2702, TAU 01 & WL\\
33 & 04 21 11.4 & $+$27 01 09.0 & Tau & 140 &  0.5  &  0.8  &  282 & I & I04181$+$2654A, TAU 02 & WL \\
34 & 04 21 56.9 & $+$15 29 45.9 & Tau & 140 &  0.1  &  2.0  &  15  & 0 & IRAM 04191$+$1522 & D\\
35 & 04 22 00.6 & $+$26 57 32.0 & Tau & 140 &  0.4  &  0.2  &  196 & II & FS Tau B, TAU 03 & WL\\	
36 & 04 27 02.6 & $+$26 05 30.0 & Tau & 140 &  1.4  &  1.5  &  161 & I & DG Tau B, TAU 04 & WL \\
37 & 04 27 57.3 & $+$26 19 18.0 & Tau & 140 &  0.6  &  2.7  &   80 & I & I04248$+$2612 AB, TAU 06 & WL \\
38 & 04 29 30.0 & $+$24 39 55.0 & Tau & 140 &  0.6  &  0.2  &  169 & II & I04264$+$2433, TAU 07 & WL \\
39 & 04 31 34.1 & $+$18 08 04.9 & Tau & 140 & 22.9  &  0.7  &  108 & I & L1551 IRS5 & D\\
40 & 04 35 35.3 & $+$24 08 19.0 & Tau & 140 &  1.0  &  1.7  &  82  & II & I04325$+$2402 A, TAU 09 & WL\\
41 & 04 39 53.9 & $+$26 03 09.8 & Tau & 140 &  1.6  &  3.1  &   79 & I & L1527, I04368$+$2557 & D,WH \\
42 & 04 39 13.9 & $+$25 53 20.6 & Tau & 140 &  4.0  &  0.5  &  151 & I & TMR 1, I04361$+$2547 AB & D,WH \\
43 & 04 39 35.0 & $+$25 41 45.5 & Tau & 140 &  2.6  &  0.8  &  189 & I & TMC 1A, I04365$+$2535 & D,WH \\
44 & 04 41 12.7 & $+$25 46 35.9 & Tau & 140 &  0.7  &  3.0  &  161 & I & TMC 1, I04381$+$2540 & D,WH \\
45 & 08 25 43.9 & $-$51 00 36.0 & Core & 450 & 26.5 &  1.5 &  107 & I & HH 46 & WH \\
46 & 11 06 47.0 & $-$77 22 32.4 & Cha &  178 &  2.0 &  0.1 &   54 & 0 & Ced110 IRS4 & WH \\
47 & 11 09 28.51 & $-$76 33 28.4 & Cha & 150 &  1.6 & \nodata& 189 & II & ISO-ChaI 192, CaINa2, CHA 01 & WL \\
48 & 12 01 36.3 & $-$65 08 53.0 & Core & 200 & 11.4 &  2.5 &  45 & 0 & BHR71 & D,WH \\
49 & 12 53 17.23 & $-$77 07 10.7 & Cha & 178 & 28.3 &  0.2 & 605 & II & DK Cha, I12496-7650 & D \\	
50 & 12 59 06.58 & $-$77 07 39.9  & Cha & 178 & 1.8 &  0.6  & 236   & I  & ISO-ChaII 28, CHA 02 & WL \\
51 & 15 43 01.29 & $-$34 09 15.4 & Lup & 130 &  1.3 &  1.5 &  51    & 0  & I15398-3359 & WH \\
52 & 16 26 21.48 & $-$24 23 04.2 & Oph & 125 & 10.6 &  0.2  & 172 & I & GSS30 IRS1, Elias 21, Oph-emb 8  & D \\
53 & 16 26 25.80 & $-$24 24 28.8 & Oph & 125 &  3.3 &  4.3 &  27 & 0 & VLA 1623, Oph-emb 3  & D \\
54 & 16 26 44.2 & $-$24 34 48.4 & Oph & 125 &   1.6 &  1.8  & 236 & I & WL 12  & D \\
55 & 16 26 59.1 & $-$24 35 03.3 & Oph & 125 &   4.3 & \nodata & 69 & II+PDR? & WL 22, ISO-Oph 90, OPH 01  & WL \\
56 & 16 27 09.36 & $-$24 37 18.4 & Oph & 125 & 15.2 &  0.2    & 310 & I & Elias 29, WL 15, Oph-emb 16  & D \\
57 & 16 27 28.1 & $-$24 39 33.4 & Oph & 125 &  5.1 & \nodata  & 213  & I & IRS 44, Oph-emb 13  & D \\
58 & 16 27 29.4 & $-$24 39 16.1 & Oph & 125 &  0.5 & \nodata  & 352  & I & IRS 46 & D \\
59 & 16 31 35.76 & $-$24 01 29.2 & Oph & 125 &  1.5 &  3.0   & 287  & I & IRS 63, Oph-emb 17 & D \\
60 & 16 32 00.96 & $-$24 56 42.7 & Oph & 125 &  8.6 &  0.1 &  80 & I & Oph-emb 10, OPH 02  & WL \\
61 & 16 34 29.3 & $-$15 47 01.4 & Core & 125 &  2.4 &  0.5 & 333  & I  &  RNO 91 & WH \\
62 & 16 46 58.27 & $-$09 35 19.8 & Sco & 125 &  0.5  &  0.6  & 201  & II & L260 SMM1, SCO 01  & WL \\
63 & 18 17 29.9 & $-$04 39 39.5 & Core & 200 & 11.1 &  0.5  &  56 & 0 & L 483 MM & WH \\
64 & 18 29 03.82 & $-$01 39 01.5 & Aqu & 436 &  2.6  &  11.8  &  24 & 0 & Aqu-MM2, AQU 01 & WL \\
65 & 18 29 08.60 & $-$01 30 42.8 & Aqu & 436 &  9.0  &   7.8  &  33 & 0 & Aqu-MM4, I18265-0132,AQU 02 & WL \\  
66 & 18 29 49.56 & $+$01 15 21.9 & Ser & 429 &  108.7 &  1.5 &  39 & 0 & Ser-emb 6, Ser SMM1, FIRS1 & WH\\
67 & 18 29 56.7 & $+$01 13 17.2 & Ser & 429 &  13.6 & 2.5 &   28 & 0 & Ser SMM4 & D,WH\\
68 & 18 29 59.3 & $+$01 14 01.7 & Ser & 429 &  27.5 & 0.3 &   37 & 0 & Ser SMM3 & D,WH\\
69 & 18 29 37.70 & $-$01 50 57.8 & SerS & 436 & 17.4 & 3.9    &   46 & 0 & SerpS-MM1, SERS 01 & WL \\
70 & 18 30 04.13 & $-$02 03 02.1 & SerS & 436 & 73.2 & 4.6    &   34 & 0 & SerpS-MM18, SERS 02 & WL \\	
71 & 18 30 25.10 & $-$01 54 13.4 & Aqu & 436 &  3.5 & 5.3 & 246 & II & Aqu-MM6, I18278-0156, AQU 03 & WL \\  
72 & 18 30 28.63 & $-$01 56 47.7 & Aqu & 436 &  6.5 & 4.5 & 320 & I  & Aqu-MM7, I18278-0158, AQU 04 & WL\\ 
73 & 18 30 29.03 & $-$01 56 05.4 & Aqu & 436 &  2.4 & 9.2 & 37  & 0 & Aqu-MM10, AQU 05 & WL \\ 
74 & 18 30 49.94 & $-$01 56 06.1 & Aqu & 436 &  1.3 & 8.2 & 40  & 0 & Aqu-MM14, AQU 06 & WL \\ 
75 & 18 31 09.42 & $-$02 06 24.5 & W40 & 436 & 13.3 & 7.4 & 40 & 0+PDR & W40-MM3, W40 01 & WL \\
76 & 18 31 10.36 & $-$02 03 50.4 & W40 & 436 & 32.6 & 3.7 & 46 & 0 & W40-MM5, W40 02 & WL  \\
77 & 18 31 46.54 & $-$02 04 22.5 & W40 & 436 &  8.3 & 20.6 & 15 & PS?+PDR & W40-MM26, W40 03 & WL  \\
78 & 18 31 46.78 & $-$02 02 19.9 & W40 & 436 &  6.1 & 9.4 & 16  & PS?+PDR & W40-MM27, W40 04 & WL   \\
79 & 18 31 47.90 & $-$02 01 37.2 & W40 & 436 &  5.9 & 27.3 & 14 & PS?+PDR & W40-MM28, W40 05 & WL  \\
80 & 18 31 57.24 & $-$02 00 27.7 & W40 & 436 &  4.1 & 2.2 &  33 & PS?+PDR & W40-MM34, W40 06 & WL \\
81 & 18 32 13.36 & $-$01 57 29.6 & W40 & 436 &  3.6 & 3.3 &  36 & 0 & W40-MM36, W40 07 & WL  \\
82 & 19 01 48.03 & $-$36 57 22.2 & CrA & 130  &  1.7 &  0.8  & 209 & I & RCrA IRS 5A & D \\
83 & 19 01 48.47 & $-$36 57 14.9 & CrA & 130  &  0.7 &  1.9  &  63 & 0 & RCrA IRS 5N & D \\
84 & 19 01 55.33 & $-$36 57 22.4 & CrA & 130  &  9.1 &  1.0  &  79 & I & RCrA IRS 7A ($+$ SMM 1C) & D \\
85 & 19 01 56.42 & $-$36 57 28.3 & CrA & 130  &  4.6 &  2.1  &  89 & I & RCrA IRS 7B & D \\
86 & 19 02 58.67 & $-$37 07 35.9  & CrA & 130 &  2.4 &  2.2 &   55 & 0 & CrA-44, IRAS 32c, CRA01 & WL \\
87 & 19 17 53.7 & $+$19 12 20.0 & Core & 300 &   3.8 &  2.0  &  65  & 0 & L 723 MM & WH \\
88 & 19 37 00.9 & $+$07 34 09.6 & Core & 106 &   0.8 &  5.0  &  33 & 0 & B335 & D \\
89 & 20 39 06.3 & $+$68 02 16.0 & Core & 325 &   6.7 &  3.3  &  35 & 0 & L1157 & D \\
90 & 21 24 07.5 & $+$49 59 09.0 & Core & 200 &   0.3 & 11.1  & 47 & 0 & L1014 & D \\
\enddata
\tablenotetext{a}{Distances come from \cite{WISH} for the WISH sources, \cite{Gr13} for the DIGIT sources, 
and \cite{Mo17} for the WILL sources with the exception of sources in Serpens, which have been
 updated to use the latest distance by \cite{OL17}. PS refers to possible pre-stellar cores and PDR to photodissociation regions.}
\tablenotetext{b}{Survey names refer to: D - DIGIT, WH - WISH, and WL - WILL programs on \textit{Herschel}.}
\tablecomments{Numbered Per-emb and Oph-emb names come from \cite{En09}. Aqu, SerpS, and W40 numbered names are from
\citet{Ma11}. Chamaeleon names come from \cite{Sp13} and \cite{Wi12}. The final entries are other names used by the WILL program and as such also in the \textit{Herschel} archive.}
\end{deluxetable*}
On the other hand, the analysis of spatial extent of line and continuum emission 
required spaxel-by-spaxel information about the fluxes (see Sec. 3.2). For that purpose, 
we used the CDF archive \citep{Gr16}, where the line fitting 
process was automated and performed for the WISH and DIGIT sources\footnote{\url{http://www.cosmos.esa.int/web/herschel/user-provided-dataproducts}}. The main steps of the process included
taking the lines from a pre-compiled database, establishing the threshold for detection, 
and then generating tables of line flux, width, centroid, and uncertainties for every 
detected line, along with an upper limit to the flux for every undetected line. 
After producing a line detection database, the integrity of the line fits was tested
 to better characterize the signal-to-noise ratio, and decouple any blended lines. 
We performed a similar automatic line fitting for the WILL sources.
\begin{figure*}[t]
\begin{center}
\includegraphics[angle=90,height=11cm]{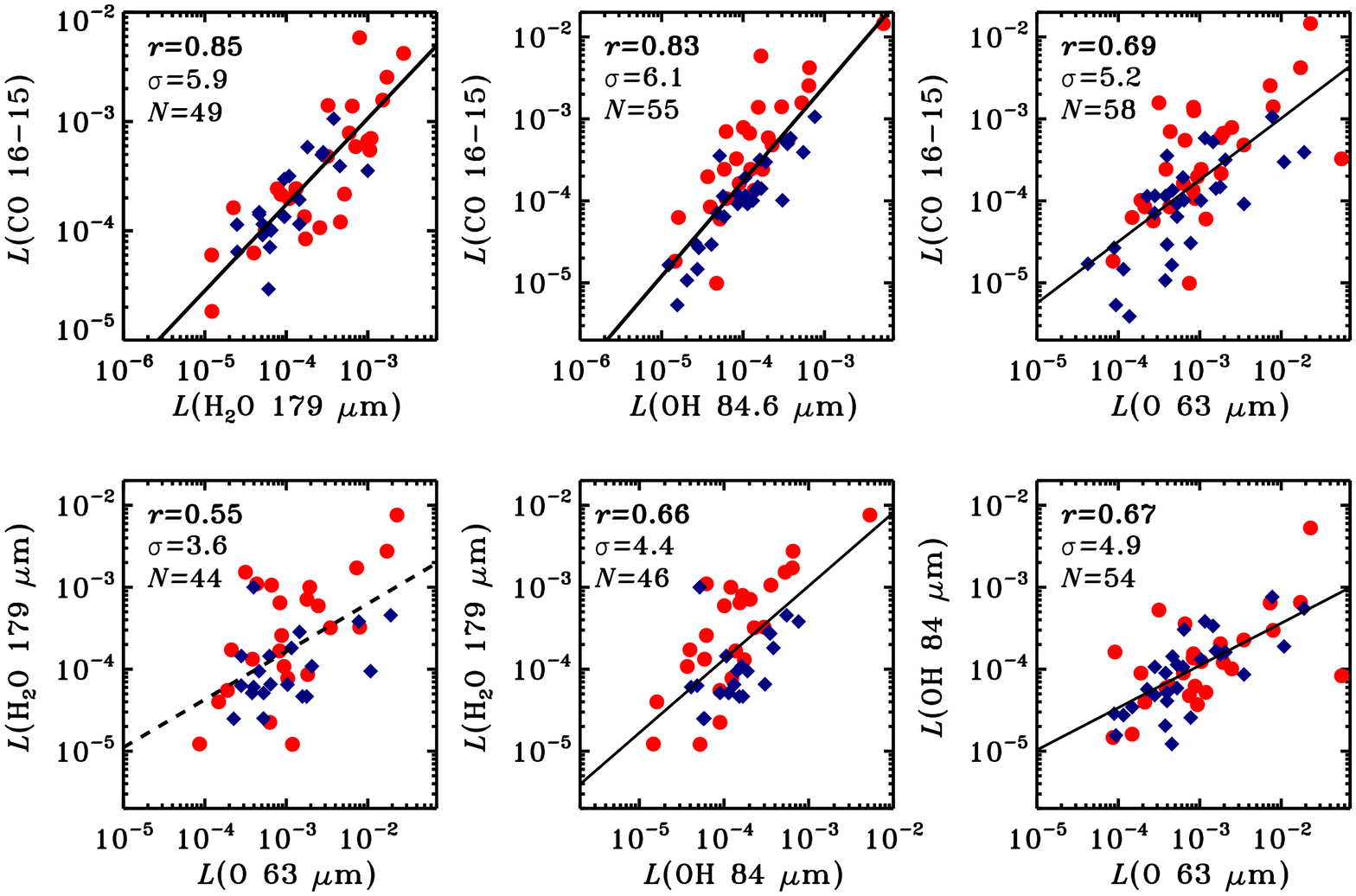}
\caption{\label{new_corr} \textit{Top:} Correlations between the line luminosities in units 
of $L_\odot$ of 
the CO 16-15 line and the H$_2$O 2$_{12}$-1$_{01}$ line at 179 $\mu$m, 
the OH line at 84.6 $\mu$m, and the [O I] 63 $\mu$m line (from left to right). 
 \textit{Bottom:} Correlations between the line luminosities of 
the H$_2$O line at 179 $\mu$m and the [O I] 63 $\mu$m line, and the OH line at 84.6 $\mu$m,
as well as between the [O I] 63 $\mu$m line and the OH line at 84.6 $\mu$m. 
Class 0 sources are shown as red circles and Class I sources as blue diamonds. Solid lines show the best 
power-law fits obtained with a least-squares method. The dashed line shows the weakest correlation, 
with $\sigma<4$. Correlation coefficients ($r$), 
significance of the correlations ($\sigma$), and the number of sources with line 
		detections ($N$) are shown on the plots.} 
\end{center}
\end{figure*}
\begin{figure*}[!tb]
\begin{center}
\includegraphics[angle=0,height=17cm]{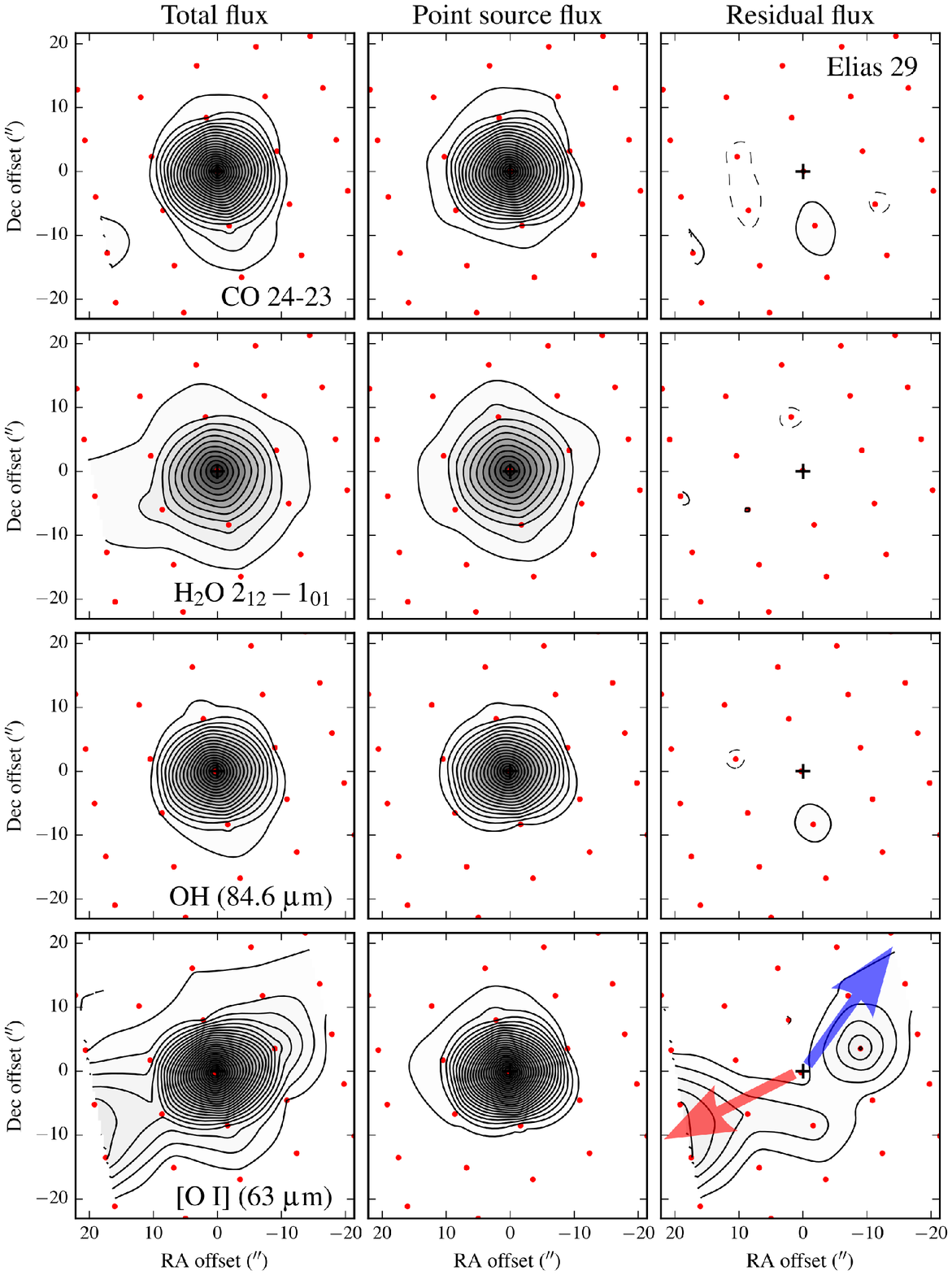}

\caption{\label{extent} Spatial extent of line emission in an example source Elias 29. The panels show
(from top to bottom) the line emission in the CO 24-23 line at 108.7 $\mu$m, the H$_2$O 2$_{12}$-1$_{01}$ 
line at 179.5 $\mu$m, the OH 84.6 $\mu$m line, and the [O I] line at 63 $\mu$m. 
The left column shows the observed line emission in 3 $\sigma$ contours, the central column 
shows the point-source model using the simulated PSF of \textit{Herschel}, and the right column shows 
the residual between the observations and the models. The dashed contours represent negative values. 
The arrows indicate the directions of the red and blue outflow lobes in CO 3-2 from \cite{Yi15}.}
\end{center}
\end{figure*}

We note that the line spectra for protostars analyzed here have been presented in previous survey papers: 
in Figs. 1 and 2 of \citealt{Ka13} (the WISH 
survey), Figs. 11-13 of \citealt{Gr13} (L1014, L1551 IRS5, and Elias 29), and in Fig. 4 of 
\citealt{Mo17} (the WILL survey). Mini-surveys or individual sources have 
been analyzed by \citealt{vK10} (HH 46), \citealt{vK10b} (DK Cha), \citealt{He12} (NGC1333 I4B),
 \citealt{Li14} (CrA sources), \citealt{Lee13} (L1448-MM), \citealt{Lee14} (Taurus sources),
 \citealt{Je15} (GSS30 IRS1), and \citealt{YL17} (BHR 71). Any differences between the sources physical parameters ($L_\mathrm{bol}$, $T_\mathrm{bol}$) 
 and line fluxes between those papers and the current work stem from the adopted distances to the 
 sources and re-reduction of the spectra with the 
 newer version of HIPE.
\section{Results}
\subsection{Line detections}
PACS spectra of low-mass protostars show exclusively rotational lines of CO, H$_2$O, OH, and 
forbidden transitions of [O I] and [C II]. CO, H$_2$O and OH lines are seen in emission, consistent 
with velocity-resolved profiles obtained with HIFI, which are dominated by emission 
components and only very narrow absorptions from the envelope 
\citep{Mo17}. The exception in PACS data is the H$_2$O 2$_{12}$-1$_{01}$ line at 179 $\mu$m seen in absorption in a few fields 
with multiple outflows, likely because significant extent of the emission and possible off position contamination. 

The [O I] lines are seen in emission in the majority of cases, except where the off position is 
contaminated due to extended emission from photodissociation regions (PDRs) associated with the cloud surface.
The [C II] line at 158 $\mu$m is even more sensitive to the cloud emission and 
is often also detected in the nod positions, leading to the appearance of negative emission
\citep{Be16}. In total, the CO 16-15 line targeted in the WISH, 
DIGIT, and WILL programs, is detected in 64 out of 77 sources (Class II, PS, and PDRs excluded), the OH doublet  
$^{2}\Pi_{\nicefrac{3}{2}}$ $J=\nicefrac{7}{2}-\nicefrac{5}{2}$ at 84 $\mu$m in
57 sources, and the H$_2$O 2$_{12}$-1$_{01}$ 
line in 49 sources. Atomic emission associated with a YSO is detected in at least\footnote{We exclude 
the cases where a significant contamination by the off-position or the other outflows 
are seen.} 65 sources in the 
[O I] 63 $\mu$m line and in 10 sources in the [C II] line.

In addition to the CO 16-15 ($E_\mathrm{up}=752$ K) and H$_2$O 2$_{12}$-1$_{01}$ ($E_\mathrm{up}=114$ K) lines,
higher-excitation transitions of these molecules are also detected. CO transitions with upper energies above 2400 K ($J_\mathrm{up}>$29) are 
detected in 35 objects, with transitions up to $J=49-48$ detected in the most remarkable Class 0 source, NGC1333 IRAS4B \citep[see][]{He12}.
The H$_2$O 8$_{18}$-7$_{07}$ line at 63 $\mu$m ($E_\mathrm{up}\sim1000$ K) is seen in 25 objects, where -- in all cases -- the highly-excited 
($J_\mathrm{up}\gtrsim 30$) CO emission is also present. 

Self-absorption may have a small effect on the H$_2$O 2$_{12}$-1$_{01}$ fluxes \citep{Mo14}, but 
not on the higher-excitation H$_2$O lines, CO, and OH lines \citep{Wa11,Kr17b}. 
The OH lines 
primarily trace the outflow and the self-absorption is not expected to be significant \citep{Wa11},
but the ground-state lines at 119 $\mu$m have not been accessible to HIFI. The [O I]
line at 63 $\mu$m shows some self-absorption toward more massive sources \citep{Le15}, but
the same is not expected for the low-mass low-luminosity sources \citep{Kr17a}. Thus, 
self-absorption is not likely to play an important role in the analysis presented here.

\begin{deluxetable*}{llccccccccccccccc} 
\tabletypesize{\footnotesize}
\tablecaption{\label{ext} Sources with extended line and continuum emission}
\tablehead{
\colhead{ID}           &
\colhead{Source}   &
\multicolumn{3}{c}{H$_2$O 179 $\mu$m} & 
\multicolumn{3}{c}{CO 186 $\mu$m} &
\multicolumn{3}{c}{OH 84 $\mu$m} &
\multicolumn{3}{c}{[O I] 63 $\mu$m}  &
Remarks \\  
\colhead{~} &
\colhead{~}  &
\colhead{Line} &
\colhead{Cont} &
\colhead{Align} &   
\colhead{Line} &
\colhead{Cont} &
\colhead{Align} &   
\colhead{Line} &
\colhead{Cont} &
\colhead{Align} &   
\colhead{Line} &
\colhead{Cont} &
\colhead{Align} &   
\colhead{~} 
}
\startdata
1 & L1448 IRS2\tablenotemark{a} &    -- & X & n & -- & X & n &  -- & X & n & X & -- & n & \\
3  & \textbf{L1448 C(N)}  & ? & X & n &    X & X & n &   -- & -- & n &    X & -- & n & multiple sources \\
4  & L1448 C(S)		& ? & X & n &   -- & X & n &   -- & -- & n &    X & -- & n & \citet{Lee13} \\
6  & \textbf{L1455 IRS1}     & X	& X	& n &	 X & X & n &   -- & -- & n &    X & -- & n & 		\\
12 & HH12\tablenotemark{a} 			& X & X & n &    X & X & n &   -- & X & n &    X & X & n &  \\ 
14 & NGC1333 I4A	& X & X & n &    X & X & n &   -- & X & y &    -- & -- & y & mispointing \\
16 & NGC1333 I4B	& X & X & n &	 X & X & n &    X & X & n &    -- & ? & n & \citet{He12} \\
25 & \textbf{B1 a}		& -- & X & n &   X & X & n &   -- & -- & -- &  X  & -- & n &    \\
26 & B1 c 			& X & X & n &	 X & X & n &   -- & -- & -- &  -- & -- & -- &  \\ 
27 & B1 SMM11\tablenotemark{a} 		& -- & -- & -- &  -- & -- & -- & -- & -- & -- & X & -- & n & \\
28 & HH211 MMS\tablenotemark{a}		& X & X & n &  -- & X & n & -- & -- & -- & X & -- & n & \\
29 & IC348 MMS\tablenotemark{a}		& X & X & n & -- & X & n & -- & -- & -- & X & -- & n & \\
30 & IC348 a\tablenotemark{a}		& -- & X & n & -- & X & n & -- & -- & -- & X & -- & n & \\
31 & L1489 		    & -- & X & n & 	-- & X & n &    -- & X & n &  X & -- & n & mispointing \\
32 & I04169\tablenotemark{a} 		& -- & X & n & -- & X & n & -- & -- & y & X & -- & n & \\
33 & I04181\tablenotemark{a}			& -- & -- & y & -- & -- & y & -- & -- & y & X & -- & n & \\	
34 & IRAM04191      & -- & -- & y & -- & -- & y & -- & -- & y & 	X & -- & n & \\	
36 & DG Tau B\tablenotemark{a}		& -- & -- & y & -- & -- & y & -- & -- & y & 	X & -- & n & \\	
37 & I04248\tablenotemark{a} 		& ? & X & n & ? & X & n & ? & -- & ? & X & -- & n & \\
38 & I04264\tablenotemark{a} 		& -- & -- & y & -- & -- & y & -- & -- & y & X & -- & n & \\
39 & L1551 IRS5     & ? & X & ? &    X & X & y &   -- & -- & -- &  X & X & y & mispointing \\
41 & L1527 			& X & X & y &    X & X & y &   -- & X & n &  X & X & y & mispointing \\
42 & TMR1			& X & X & y & 	 X & X & y &    X & X & y &  X & X & y & mispointing \\
43 & TMC1A 			& -- & X & n &   -- & X & n & -- & X & n & X & X & n & mispointing \\
44 & \textbf{TMC1} 			& -- & X & n &   X & X & n &   -- & -- & -- & X & -- & n &  weak CO \\
45 & \textbf{HH 46}  		& -- & X & n &    X & X & n &    -- & X & n &    X & -- & n & \citet{vK10} \\
49 & DK Cha			& X & X & y &	 X & X & y &	X & X & y &	   X  & X & y & mispointing \\
51 & \textbf{I15398} 		& X & X & n &    X & X & n &   -- & X & y &    X  & X & n & mispointing \\ 
52 & GSS30 IRS1     & X & X & y & 	 X & X & y &   X  & X & y &    X  & X & y &  multiple sources \\
53 & \textbf{VLA1623} 		& X & X & n &    X & X & n &   -- & X & n &    X & X & n &  multiple sources \\
56 & \textbf{Elias29}		& -- & X & n &   X & X & n &   -- & -- & -- &  X  & X & n &  small CO resid.  \\
61 & RNO 91			& ? & X & ? &    -- & X & n &  -- & X & n &    X & X & y &  cont/line overlap \\
63 & L483			& X & X & n &    X & X & y &   -- & X & n &	   X  & X & y &  cont/line overlap \\
64 &\textbf{Aqu-MM2}\tablenotemark{a}  	& X & X & n &    X & X & n &   ? & ? & ? &  X  & ? & n & \\
66 & \textbf{Ser SMM1} 		& X & X & n &  X & X & n & X & X & n & X & X & n & \citet{Di14}\\
67 & \textbf{Ser SMM3}		& X & X & n &  X & X & n & X & --  & n  & X & --  & n  & contamin. by SerSMM6\\		
68 & \textbf{Ser SMM4} 		& X & X & n &  X & X & n & X & --  & n  & X & --  & n  & \citet{Di13}\\	
86 & CrA-44\tablenotemark{a} 		& -- & X & n &  -- & X & n & -- & X & n & X & X & n & \\
88 & B335   		& -- & X & n &  -- & X & n &   -- & -- & -- &  X  & -- & n &     \\
89 & \textbf{L1157}			& X & X & n &   X & X & n &    -- & -- & -- &  X  & -- & n & weak CO	\\
\enddata
\tablenotetext{a}{For these WILL sources, the CO 24-23 or CO 21-20 lines are used instead of the CO 14-13 line at 186 $\mu$m.}
\tablecomments{X notes the cases where emission is extended, whereas y/n refers to whether the 
emission in line and in continuum is aligned (y) or not (n). Non-detections of extended emission 
is marked with '--'. Question mark (?) is used when the emission is not detected on the map. 
Sources with detected extended emission in both CO 14-13 and [O I] 63 $\mu$m are
in boldface.}
\end{deluxetable*}

The detections and line fluxes of various species are related to each other (see Figure \ref{new_corr}). 
There is a significant correlation, at $\sim 6\sigma$, between the flux in the CO 16-15 line and in the H$_2$O 2$_{12}$-1$_{01}$
line, and the CO 16-15 line and OH 84.6 $\mu$m line. A weaker, yet significant correlation is obtained between the 
CO and [O I] line, OH and [O I] line, and H$_2$O and OH lines ($\sim 4-5\sigma$). The H$_2$O and [O I] lines 
correlate at the lowest, 3.6 $\sigma$ level. 

Class 0 and Class I sources show similar distributions in Figure \ref{new_corr},
except that the OH lines are brighter for the Class I sources with 
respect to CO and H$_2$O lines (see CO - OH and H$_2$O - OH plots). 
The only outlier is L1448 C(S) containing
ice features that indicate a dense envelope \citep{Lee13}. Similar correlation strengths 
between the OH and [O I] lines in Class 0/I sources suggest that the fraction of OH associated
 with the component traced by the [O I] line increases for more evolved sources. 
 The different origin of the [O I] emission is further supported by its
 strong correlation with $L_\mathrm{bol}$, whereas molecular tracers correlate stronger with
 the envelope mass \citep{Mo17}.
 
In summary, rich molecular line emission is seen in $\sim70$\% of the targeted sources, allowing 
a statistical analysis of the largest sample of protostars so far. In \S 4,
multiple lines of molecular species are used to constrain the excitation of molecules 
and to calculate the cooling budget of the gas in the far-infrared.

\subsection{Spatial extent of line emission}
Fully-sampled maps of far-infrared line emission exist for a handful of Class 0/I protostars in a few lines and show extended emission along
the outflow direction \citep{Ni10,Ni15,He12}. While these observations clearly associate the emission in H$_2$O and [O I] with jets
and outflows, statistical properties of the far-IR emitting gas could not be established. Thus, the single footprint
maps from PACS (FOV$\sim47\times47$'', see Section 2) provide a unique dataset to link the emission in various species and to
test whether the extended far-IR emission is indeed common among protostars.

In order to study the prevalence and shape of any extended line emission, we remove the point-source emission
associated with the continuum peaks, i.e. the position(s) of the protostar(s). For that reason, we first calculate the line (or
continuum) emission from the point sources located on the map and then subtract them from the entire PACS map after convolution
with the simulated point spread function (PSF) of \textit{Herschel} \citep[the POMAC method, see Section 3.1 in ][]{Li14}. The
residual emission is thus not associated with any known point source, and likely originates in extended structures. Because the
PACS maps are sparse, the method relies on pre-defined source coordinates (e.g. from infrared photometry or submillimeter
interferometric observations) and is sensitive to pointing errors. Comparison of the residual line and continuum emission is
thus useful to double-check whether the pattern of emission differs and truly indicates the line emission that is extended
beyond the continuum peaks.

 Figure \ref{extent} illustrates the above procedure for the case of Elias 29. The observed line emission (left column) appears to be extended in all
 lines, but that is due to the compact emission from the point source being enlarged and distorted in shape by the non-circular
 PSF of \textit{Herschel} (central column). Subtraction of the simulated emission yields negligible residuals in the CO, H$_2$O,
 and OH lines. The emission is extended only in the [O I] line, where the residual emission shows two peaks corresponding to the
 blue and red outflow lobes (right column, see \citealt{Yi15} for outflow directions). The continuum emission is point-like in the vicinity of the considered lines (not
 shown here).

 We applied a similar procedure for all sources in our sample and list the sources with extended line emission in 
 at least one species in Table \ref{ext}. There are 37 sources with the extended [O I] emission at 63 $\mu$m, 
 23 with extended CO emission (typically 14-13 at 186 $\mu$m, see Table 2), 19 with extended H$_2$O emission in the 2$_{12}$-1$_{01}$ line at 179 $\mu$m, and 8 with extended OH 
 emission at 84.6 $\mu$m. These statistics include the sources where the line emission is clearly 
 spatially offset from 
 the continuum and the line emission is likely linked to the outflows \citep[see Fig.A.2. in ][]{Mo17}. The remaining sources, where line and continuum 
 emission is well-aligned, may be caused by an off-center location of the source.
 Thus, the residual extended emission might be a result of imperfect subtraction of the PSF from the maps \textbf{for this small subset of sources (with ID 39-42, 49, 52,63)}. 
 
 Patterns of emission are often similar in certain species. Out of 40 sources in Table \ref{ext},
 both CO and H$_2$O emission are extended in 17 objects and are not extended in 12 sources. 
 20 out of 37 sources with extended [O I] emission also show extended CO emission.
 The exceptions are, for example, NGC1333 I4A and I4B, which 
 show very weak and compact [O I] emission but clearly extended emission in the molecular species.  
 Conversely, there are 13 sources where [O I] is extended, but the CO emission is compact \textbf{(Figure \ref{extent})}. 
 Many of these sources are located in Taurus and form a uniform group with compact molecular 
 emission and prominent [O I] emission associated with jets \citep{Po12}.
 
 The OH emission is typically compact,
 apart from a few sources with very bright extended emission in H$_2$O and CO (e.g. NGC1333 I4B).  
 As noted in \citet{Ka13}, where a subsample of the sources was analyzed, the OH emission 
 does not resemble the emission in other molecular species. Here, only a few sources show compact [O I] and OH 
 emission and, at the same time, extended H$_2$O and CO (e.g. NGC1333 I4A and B1 c). 
 
 Extended emission in at least one species (atomic or molecular) is detected in 19 Class 0, 19 Class I, and 2 
 Class II sources (see Table 1 and 2). Thus, there is no clear indication that the 
 evolutionary stage strongly influences the extent of the observed emission in general.
 However, among the sources which show extended emission both in H$_2$O and CO, 11 
 out of 17 are Class 0 objects (65\%). At the same time, only 5 out of 13 sources 
 with the extended emission seen merely in [O I] are Class 0s (38\%). 
 
 In conclusion, extended emission is detected in 40 out of 90 sources (44\%).
 A similar fraction of the sources with extended and compact emission (50\%) was seen in the WISH survey alone \citep{Ka13}.
  However, only 28\% of sources show extended emission in molecular species i.e., excluding [O I]. 
 The line emission is more often extended in molecular species for the less evolved 
 sources, and in atomic species for the more evolved ones. This is consistent with 
 the idea put forward by \citealt{Ni15} that the jet becomes more atomic over time.

\begin{figure*}[tb]
\hspace{-1cm}

\includegraphics[angle=0,height=9cm]{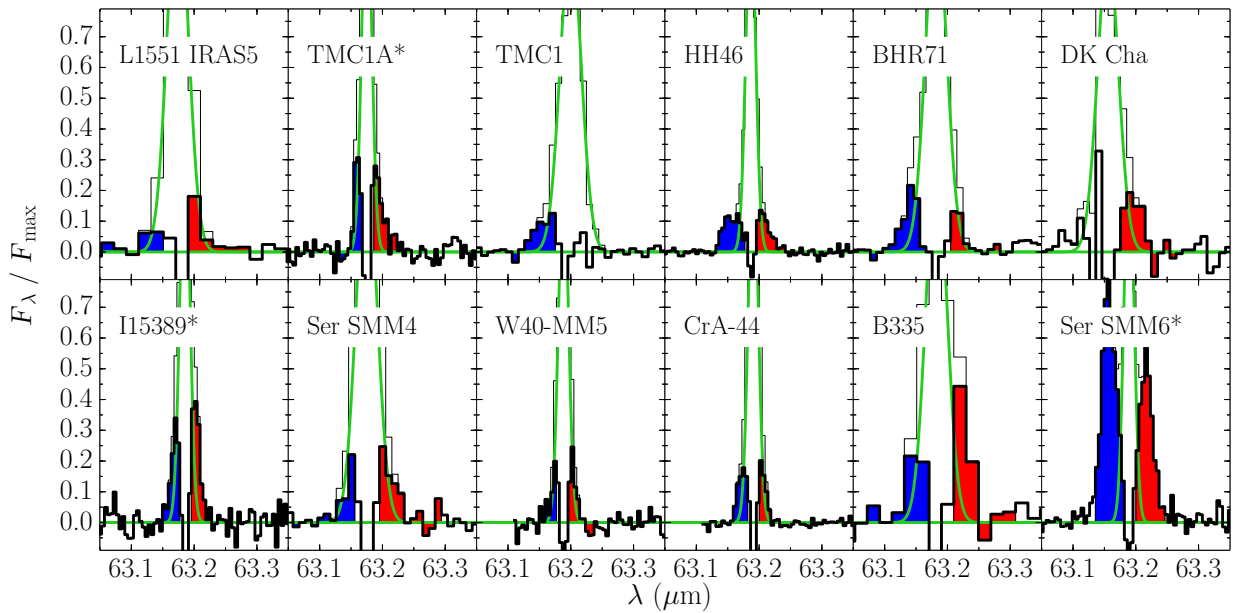}

\caption{\label{highvel} Line profiles of the [O I] 63 $\mu$m for the sources where 
the high-velocity emission is detected. The spectra are taken at 
the central position of the continuum peak except the sources marked with 
a star, where a few spaxels were summed due to mispointing.
The sampling of the spectra are different for the line scans and full scans. Blue and 
red areas show the line wing emission beyond typically $\pm90$ km s$^{-1}$ resulting from the subtraction of the Gaussian profile 
(in green) from the line profile.}
\end{figure*}

\subsection{High-velocity emission in [O I]}
The [O I] 63 $\mu$m line often includes emission from a fast jet, which is detectable
 if the velocity exceeds $\sim$90 km s$^{-1}$, the approximate spectral resolution of PACS at this wavelength. 
 The high-velocity emission is produced by an atomic jet that is embedded in the molecular emission
 \citep{vK10,Ni15}.

Figure \ref{highvel} shows the [O I] profiles for the sources where the high-velocity
 line wings are detected. The spectra are extracted from the single spaxel where the 
 continuum peaks to avoid 
 the effects of instrumental line shifts between spaxels that can be introduced by the instrument.
 The residuals are calculated by subtracting the Gaussian profile fitted with the instrumental line width 
 and the line center left as a free parameter. The line wing emission is reported for the 
 sources where the integrated emission in the wing exceeds the 3 $\sigma$ flux uncertainty.
  
 The most remarkable [O I] line wings are detected at the position of Serpens SMM6 -- a 
 source which was not specifically targeted by the surveys, but is located at the edge of the PACS map of Serpens SMM3. 
 The strong, blue-shifted wing in HH 46 reported in \citet{vK10} and \citet{Ka13} traces the 
 well-known outflow extending beyond the dense core. L1551-IRS5 shows extended 
 [O I] emission in the PACS maps \citep[see][]{Lee14} consistent with the detected line wings. 
 Similar results are obtained for the fully-sampled [O I] maps in \cite{Ni15} and \cite{Di16}.

In comparison to HIFI line profiles, only BHR71 and HH46 show the high-velocity wings in
both [O I] and H$_2$O $1_{10}-1_{01}$ \citep{Kr12,Gr13,Mo17}. The remaining sources show 
narrow emission in the H$_2$O $1_{10}-1_{01}$ transitions, suggesting that the [O I] line 
does not trace the same kinematic or physical structures as H$_2$O \citep[see also][]{Ni15,Mo17}. 
However, further observations with the German REceiver for Astronomy at Terahertz Frequencies \citep[GREAT][]{Hey12}
 heterodyne instrument on the Stratospheric Observatory for Infrared Astronomy are needed to fully resolve the [O I] line profile.

In summary, 12 sources show high-velocity wings in the [O I] 63 $\mu$m line, including
Ser SMM6 which was additionally detected on the map of Ser SMM3.
All of those protostars also show extended [O I] emission on the PACS maps (see \S 3.2),
suggesting that they drive fast, atomic jets. The presence of high-velocity emission 
does not correlate with the evolutionary stage, since the detections are equally split 
between the Class 0 and Class I sources. The lack of high-velocity emission in the
majority of sources indicates that most of the [O I] emission is emitted at 
velocities below $\sim90$ km s$^{-1}$.

\begin{figure*}[!tb]
  \begin{minipage}[t]{.3\textwidth}
  \begin{center}
       \includegraphics[angle=90,height=5.2cm]{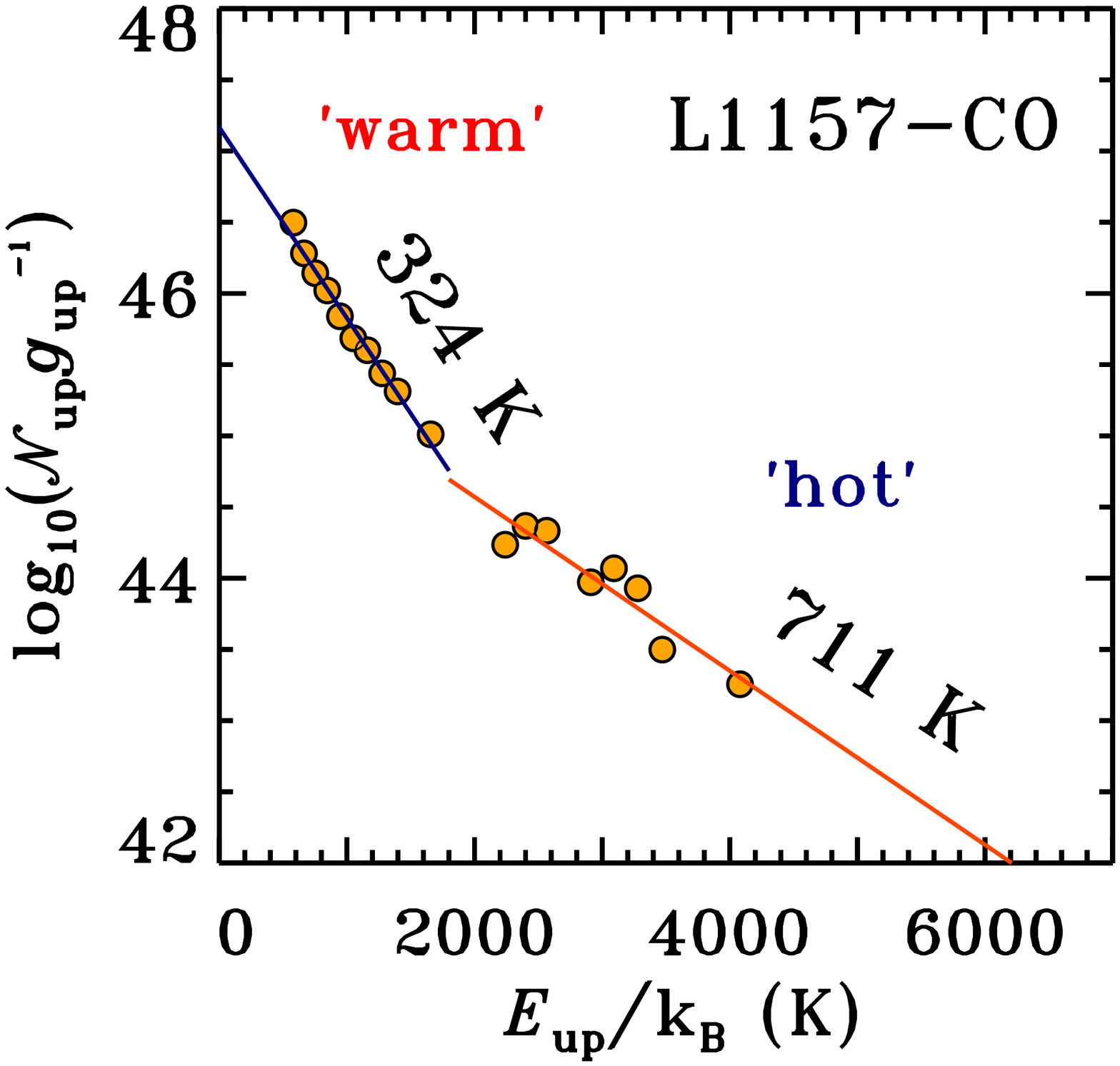}   
  \end{center}
  \end{minipage}
  \hfill
  \begin{minipage}[t]{.3\textwidth}
      \begin{center}
    \includegraphics[angle=90,height=5.2cm]{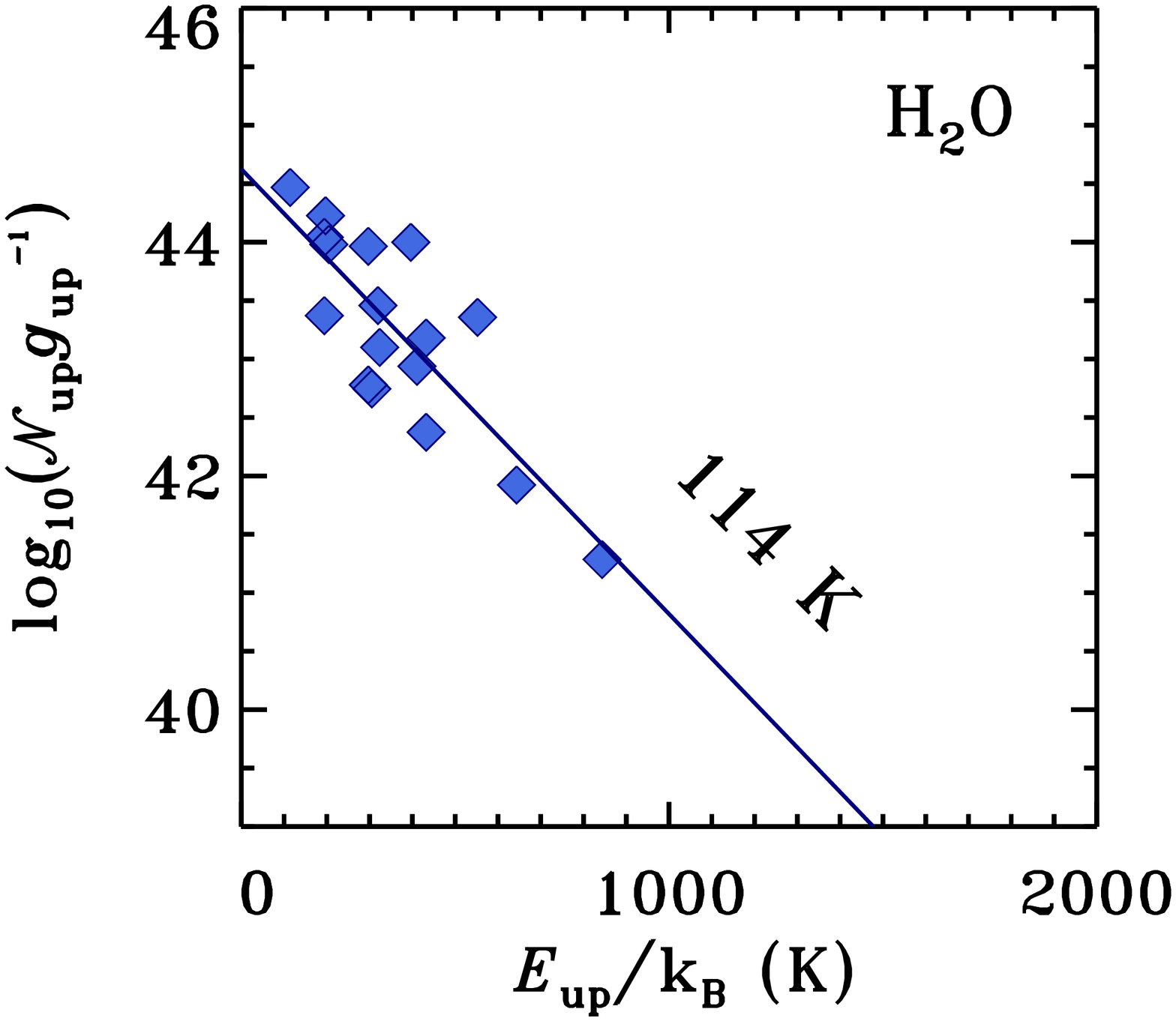} 
      \end{center}
  \end{minipage}
    \hfill
   \begin{minipage}[t]{.3\textwidth}
      \begin{center}
    \includegraphics[angle=90,height=5.2cm]{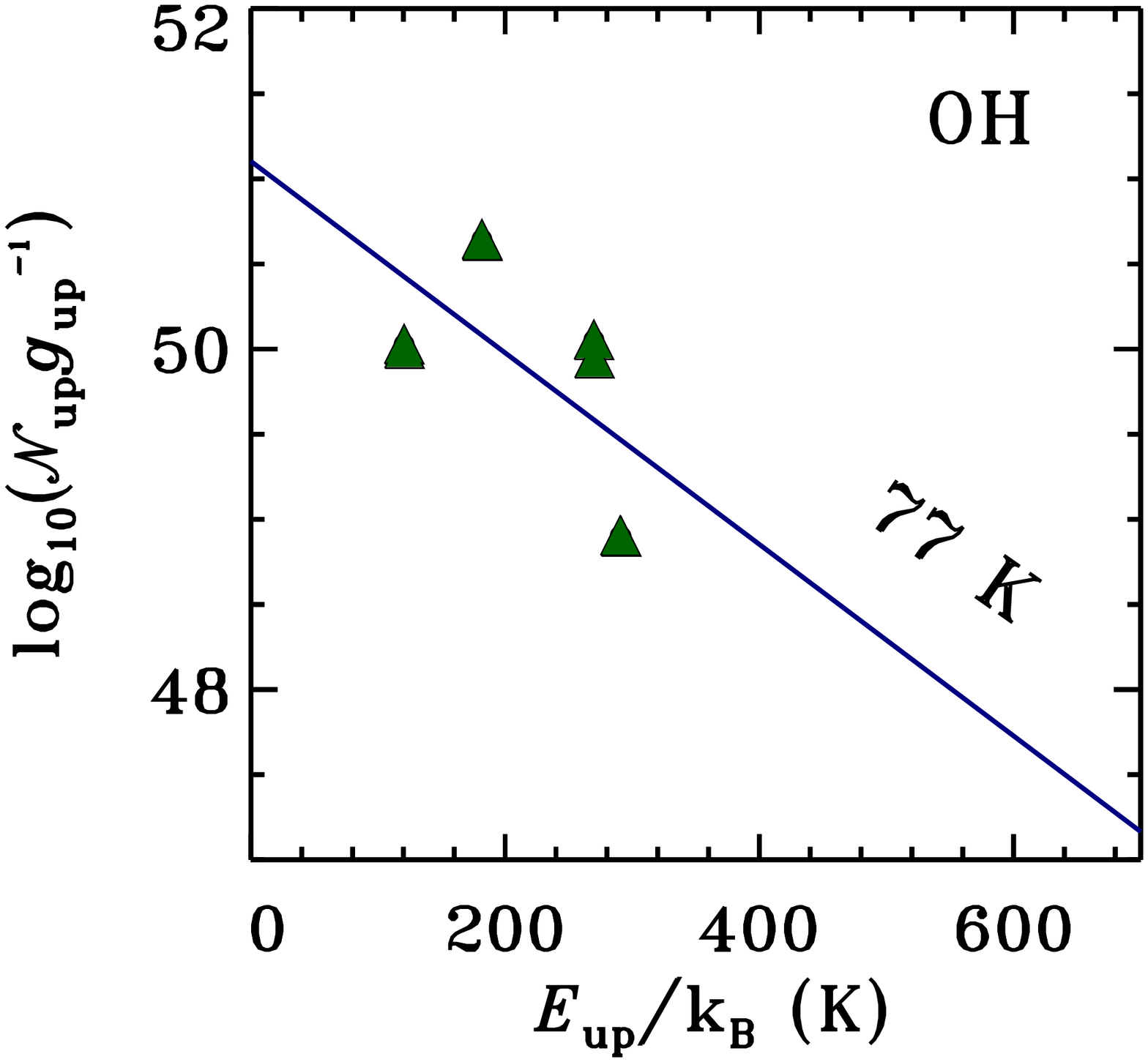} 
      \end{center}
  \end{minipage}
      \hfill
        \caption{\label{molexc} Rotational diagrams of CO, H$_2$O, and OH for example 
        protostar L1157. The base-10 logarithm of the number of emitting molecules from a level $u$,
  $\mathcal{N}_\mathrm{u}$, divided by the degeneracy of the level,
  $g_\mathrm{u}$, is shown as a function of energy of the upper level
  in kelvins, $E_\mathrm{up}$. Blue and red solid lines show linear fits to the data and the
    corresponding rotational temperatures.}
\end{figure*}
\begin{figure}[!tb]
\hspace{-1cm}

\includegraphics[angle=90,height=6cm]{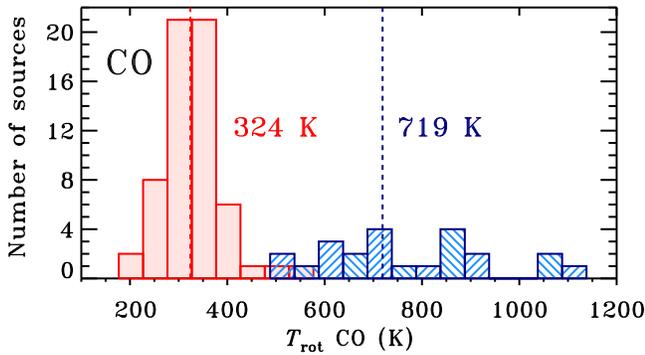}

\caption{\label{hist} Histograms of the CO rotational temperatures for the sources 
with full spectra from $\sim60-190$ $\mu$m. The pink color shows the distributions 
of temperatures calculated using the CO $J_\mathrm{up}=14-25$ lines ($E_\mathrm{up}=580-1800$ K)
and the light-blue color the CO $J_\mathrm{up}\geq26$ ($E_\mathrm{up}=580-1800$ K) lines. Median 
values for those two components are drawn with the dashed lines.}
\end{figure}
\begin{figure}[!tb]
\hspace{-1cm}

\includegraphics[angle=90,height=6cm]{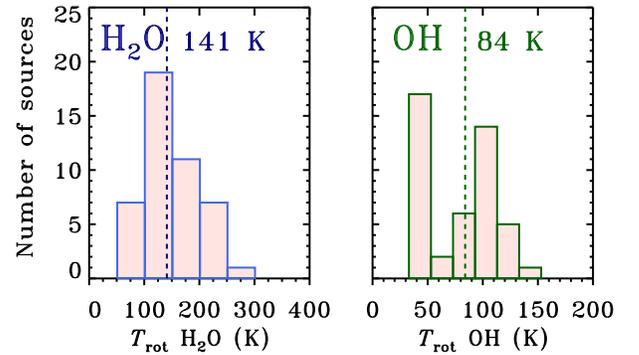}

\caption{\label{hist2} Histograms of the H$_2$O (left) and OH (right) rotational temperatures for the sources 
with full spectra from $\sim60-190$ $\mu$m. Median values for those two temperatures are drawn with the dashed lines.}
\end{figure}
\section{Analysis}

\subsection{Molecular excitation}
The detection of multiple transitions of a given molecule enables the excitation conditions,
 and ultimately the physical conditions, of the gas to be determined \citep[see Sec. 3.1.,][]{Ka13}. 
Here, we present rotational diagrams of H$_2$O, CO, and OH and determine the statistics of 
rotational temperatures ($T_\mathrm{rot}$) and numbers of emitting molecules ($\mathcal{N}$) for the 
entire sample. The diagrams are used in Section 4.2 to calculate the total line emission in each species.

Figure \ref{molexc} shows example rotational diagrams obtained for the central region of L1157, a well-known
low-mass protostar driving a large scale outflow \citep{Ba01,Be12,Le12}. The observations of CO are 
well described by two independent linear fits, which correspond to rotational temperatures of $\sim320$ K and $\sim710$ K,
referred to as the \lq warm' and \lq hot' components \citep{Ka13}. 
The break in the diagram is located 
at CO $J_\mathrm{up}\sim25$, with upper level energies of $\sim1800$ K at $\lambda\sim$100 $\mu$m,
 where the spectrum
 is not properly calibrated, thus the fluxes of CO 25-24 and 26-25 lines at 104 and 100 $\mu$m 
are not readily available. The two temperature components are associated with different velocity 
components of the line profile of CO 16-15 in velocity-resolved observations \citep{Kr17b}.
The H$_2$O and OH lines,
on the other hand, show large scatter in rotational diagrams and only a single 
temperature component is fitted to the data (Figure \ref{molexc}), corresponding to temperatures of $\sim110$ K and $\sim80$ K, respectively. 
The scatter is due to high critical densities of H$_2$O and OH lines, which are subthermally excited and 
often optically thick \citep{He12,Ma12,Wa13,Mo14}.

The rotational temperatures obtained for L1157 are representative of the low-mass protostars in general
(see diagrams for the remaining sources in the Appendix: Figures \ref{cowill}-\ref{ohwill2} - for sources with line scans and 
Figure \ref{dig1} for sources with the full spectra). Figure \ref{hist} 
shows histograms of rotational temperatures of the CO warm and hot components calculated for 
sources where the full PACS spectrum was obtained (see also Table \ref{exc_table}). The distribution 
shows a clear peak for the CO warm ($J\sim14-24$) with a median at $324$ K (mean of 325$\pm$62 K). The hot component ($J\gtrsim24$) is less commonly detected and 
shows a much broader range of values, from about 600 K to 1100 K, with the median at $719$ K (mean of 764$\pm$174 K). The 
median rotational temperatures of H$_2$O and OH are $141$ K and $84$ K, respectively, 
and are typical for the majority of the sources (see Figure \ref{hist2}). The high-temperature 
tail of the H$_2$O temperature distribution indicates sources with additional hot
emission in the H$_2$O rotational diagrams, similar to CO. Decomposition of the H$_2$O diagrams for nine 
sources with the largest number of detected hot water lines ($E_\mathrm{up}\gtrsim600$ K), 
yields a median value of \lq hot' H$_2$O rotational temperature of $410$ K. The residual temperature of the 
 \lq warm' H$_2$O is $300$ K. However, since H$_2$O and OH are subthermally excited \citep{Mo14}, rotational temperatures do not reflect the 
true temperatures of the gas.

The above results are consistent with the similar measurements from smaller 
surveys of low-mass protostars, where the prevalence of a 300 K component in CO rotational diagrams 
was identified (\citealt{Ma12}, \citealt{Ka13}, \citealt{Gr13}, \citealt{Li14}, \citealt{Lee14}).
The extended sample presented here allows us also to derive typical rotational temperatures 
of H$_2$O ($130$ K) and OH ($84$ K) and to isolate additional hot components in the 
H$_2$O and CO diagrams. The rotational temperatures of the hot components show a broad range 
of values and trace the most energetic processes in young stellar objects. The diagrams for all sources
will be used in \S 4.2.1 to calculate the total line emission in each molecular species and 
thus to determine the cooling rate of hot and warm gas in low-mass protostars. 

\newpage
\startlongtable
\begin{deluxetable*}{llccccccccc} 
\tabletypesize{\scriptsize}
\tablecaption{\label{exc_table}Rotational temperatures and numbers of emitting molecules of CO, H$_2$O, and OH}
\tablehead{
\colhead{ID}           &
\colhead{Source}   &
\multicolumn{2}{c}{CO $J_\mathrm{up}=14-26$} &
\multicolumn{2}{c}{CO $J_\mathrm{up}\geq27$} &
\multicolumn{2}{c}{H$_2$O} &
\multicolumn{2}{c}{OH} &
\colhead{Mode}  \\ \cline{3-4}  \cline{5-6} \cline{7-8}  \cline{9-10}
\colhead{}   &        
\colhead{}   &
\colhead{$T_\mathrm{rot}$(K)} &
\colhead{$\mathrm{log}_\mathrm{10}\mathcal{N}$}                                          &               
\colhead{$T_\mathrm{rot}$(K)}   &            
\colhead{$\mathrm{log}_\mathrm{10}\mathcal{N}$}  & 
\colhead{$T_\mathrm{rot}$(K)}    & 
\colhead{$\mathrm{log}_\mathrm{10}\mathcal{N}$}    & 
\colhead{$T_\mathrm{rot}$(K)}  & 
\colhead{$\mathrm{log}_\mathrm{10}\mathcal{N}$} &
\colhead{F/L} 
}
\startdata
1 & L1448 IRS2 & 319(23) & 49.1(0.1) & 492(10) & 48.5(0.1) & 143(18) & 45.8(0.2) & 33(6) & 54.2(0.6) & L  \\
2 & L1448 IRS3 & 281(16) & 49.7(0.1) & 678(66) & 48.3(0.2) & 141(19) & 46.0(0.2) & 39(12) & 53.9(0.9) & L \\
3 & L1448 C(N) & 351(14) & 49.1(0.1) & 894(75) & 48.2(0.1) & 152(18) &  46.3(0.2) &  123(17) & 52.2(0.2) & F \\
4 & L1448 C(S)  & 294(16) & 49.0(0.1) & -- & -- &  84(11) & 46.3(0.3) & 94(18) & 51.7(0.3) & F \\		
5 & I03235 & -- & -- & -- & -- & -- & -- & -- & -- & L \\
6 & L1455 IRS1	&	334(31) &	48.3(0.1) & -- &	-- & 124(30) & 45.6(0.3) & 37(15) &	53.2(1.2) & F \\
7 & L1455 IRS3	&	177(38)	&   48.4(0.4) & -- &	-- & -- & -- & -- & -- & F \\
8 & NGC1333 I1 & 387(24) & 48.4(0.1) & -- & -- &  140(27) & 45.6(0.3) & -- & -- & L \\
9 & NGC1333 I2A &   334(93)	&	48.9(0.3) &	-- &	-- & 127(62) & 46.5(0.4) &	-- & -- & F \\
10 & NGC1333 I2B & 357(11) & 48.5(0.0) & -- & -- & 154(20) & 45.1(0.2) & 38(6) & 53.5(0.4) & L \\
11 & HRF 65 & -- & -- & -- & -- & -- & -- & -- & -- & L \\
12 & PER08 & 303(31) & 48.9(0.2) & -- & -- & 210(28) & 45.5(0.2) & 36(6) & 53.8(0.5) & L \\
13 & PER09 & 313(30) & 49.1(0.2) & 699(212) & 48.3(0.5) & 175(12) & 45.9(0.1) & 35(5) & 54.2(0.5) & L \\
14 & NGC1333 I4A & 260(9) & 49.5(0.1) & 633(130) & 48.1(0.4) & 105(18) & 46.4(0.3) & 150(214) & 51.6(0.9) & FL \\
15 & PER10 & 342(51) & 48.8(0.2) & -- & -- & -- & -- & 40(14) & 53.6(1.0)& L  \\
16 & NGC1333 I4B & 343(9) & 49.6(0.1) & 896(42) & 48.6(0.1) & 183(15) & 46.5(0.1) &  84(14) & 52.8(0.3) & F \\
17 & NGC1333 I4C & -- & -- & -- & -- & -- & -- & -- & -- & L \\
18 & I03267 & -- & -- & -- & -- & -- & -- & -- & -- & L \\
19 & I03271 & 379(51) & 48.5(0.2) & 572( 85) & 48.1(0.3) & 166(19) & 45.5(0.2) & 35(6) & 53.8(0.5)& L  \\
20 & I03282 & 318(10) & 48.4(0.1) & -- & -- & -- & --  & -- & -- & L \\
21 & I03292 & -- & -- & -- & -- & -- & -- & -- & -- & L \\
22 & IRAS03301	&   355(25)	&   48.4(0.1) & 1089(370) &	47.3(0.4)  	& 176(22) & 45.4(0.2) & 109(20) & 52.1(0.3) & F \\
23 & B1 SMM3 & -- & -- & -- & -- & -- & -- & -- & -- & L \\
24 & B1 d & 390(78) & 48.6(0.3) & -- & -- & -- & -- & -- & -- & L \\
25 & B1 a	&	293(10) & 48.8(0.1) &	-- &	-- & 211(54) & 45.3(0.3) & 103(20) & 51.8(0.3) & F \\
26 & B1 c	&	228(15) & 48.8(0.1)	& 	-- &  -- & 51(21) & 46.3(0.7) & --	& -- & F \\
27 & B1 SMM11 & 322(14) & 48.6(0.1) & -- & -- & 127(29) & 45.3(0.3) & 35(8) & 53.9(0.7) & L \\
28 & HH 211 MMS & 273(25) & 49.4(0.2) & -- & -- & 71(15) & 46.5(0.4) & 41(10) & 53.3(0.6) & L \\
29 & IC348 MMS & 411(146) & 48.3(0.5) & -- & -- & -- & -- & 35(10) & 53.5(0.9) & L \\
30 & IC348 a & 259(19) & 49.0(0.2) & -- & -- & 96(18) & 45.4(0.3) & 34(4) & 53.7(0.4) & L \\
31 & L1489 & 347(13) & 48.3(0.1) & 867(108) & 47.2(0.2) & 191(22) & 45.1(0.2) &	102(18) & 51.5(0.3) & F \\
32 & I04169 & -- & -- & -- & -- & -- & -- & -- & -- & L \\
33 & I04181 & 446(86) & 47.4(0.2) & -- & -- &  -- & -- & 38(6) & 52.7(0.4) & L \\
34 & IRAM04191	& 242(16) & 47.9(0.1) & -- & --	 & 130(32) & 44.8(0.2) & 83(31) & 50.4(0.4) & F \\
35 & FS Tau B & -- & -- & -- & -- & -- & -- & -- & -- & L \\
36 & DG Tau B & -- & -- & -- & -- & -- & -- & -- & -- & L \\
37 & I04248   & -- & -- & -- & -- & -- & -- & -- & -- & L \\
38 & I04264 & 434(74) & 47.7(0.2) & -- & -- & 182(22) & 44.6(0.2) & 34(3) & 53.2(0.3) & L \\
39 & L1551 IRS5	& 562(95) & 48.4(0.1) & 751(171) & 47.7(0.4) & --  & --  &	39(11) & 53.3(0.8) & F \\				
40 & I04325 & 290(43) & 47.9(0.3) & -- & -- & 310(202) & 44.4(0.4) & -- & -- & L \\
41 & L1527 & 380(55) & 47.9(0.2) & -- & -- & 185(40) & 45.2(0.2) &	84(51) & 51.7(0.7) & F \\
42 & TMR1 & 314(11) & 48.5(0.1) &  -- & -- & 193(31) & 45.4(0.2) & 108(24) & 51.8(0.3) & F \\
43 & TMC1A & 269(53) & 48.3(0.3) &  -- & -- &  -- & -- &  -- & -- & F \\
44 & TMC1 & 338(8)	& 48.3(0.1)	& 856(174) & 47.2(0.3)	& 148(24) & 44.9(0.2) &	103(19) & 51.5(0.3) & F \\
45 & HH 46 & 372(60) & 49.1(0.2) & -- & -- & 68(9) & 46.4(0.2) & 100(60) & 52.4(0.6) & L \\
46 & Ced110 IRS4 & 484(88) & 47.4(0.2) & -- & -- & 106(44) & 44.9(0.5) & 128(139) & 51.1(0.8) & L \\
47 & CHA01 & 278(30) & 48.3(0.2) & -- & -- & 155(82) & 44.7(0.4) & 34(7) & 53.3(0.7) & L \\
48 & BHR71 & 353(13) & 49.3(0.1) & 1072(85)	& 47.9(0.1)	& 125(23) & 45.8(0.3) & 87(17) & 52.1(0.4) & F \\
49 & DK Cha & 393(21) & 49.0(0.1) & 866(338) & 48.2(0.5) & -- & -- & 67(12) & 52.7(0.4) & F \\
50 & CHA02 & -- & -- & -- & -- & -- & -- & -- & -- & L \\
51 & I15398	& 291(8) & 48.9(0.1) & -- & -- & 78(82) & 45.6(1.1) & 79(45) & 51.7(0.7) & L \\
52 & GSS30 IRS1	& 342(16) & 49.2(0.1) & 865(84)	 & 48.0(0.2) & 245(22) & 46.0(0.1) &	99(18) & 52.2(0.3) & F \\
53 & VLA 1623	& 286(14) &	48.9(0.1) & 1074(406) & 47.1(0.4) & 165(40) & 45.3(0.3) &	71(16) & 51.9(0.4) & F \\
54 & WL12	&	328(9)  & 48.5(0.1)	& 690(58) & 47.4(0.2)	& 202(25) & 45.0(0.2) &	94(17) & 51.5(0.3) & F \\
55 & WL22   & -- & -- & -- & -- & -- & -- & -- & -- & L \\
56 & Elias29 &  347(10) & 49.2(0.1)	& 638(10) & 48.4(0.1)	& 288(33) & 46.0(0.1) & 112(27) & 52.1(0.3) & F \\
57 & IRS44 & 385(32) & 47.0(0.1) & -- & --	& -- & -- & -- & -- & F \\
58 & IRS46	&	310(35)	& 47.8(0.2)	& -- & --	& -- & -- & -- & -- & F \\
59 & IRS63	&	364(45)	& 47.2(0.2)	& -- & --	& -- & -- & -- & -- & F \\
60 & OPH02 & 358(10) & 48.4(0.1) & 488(143) & 48.1(0.7) & 165(19) & 45.2(0.2) & 37(8) & 53.5(0.6) & L \\
61 & RNO91 & 231(30) & 47.9(0.2) & -- & -- & -- & -- & 107(104) & 51.0(0.9) & L \\
62 & L260 SMM1 & 423(63) & 47.6(0.2) &  -- & -- & 185(22) & 44.5(0.2) & 36(3) & 53.0(0.3) & L \\
63 & L483 & 371(32) & 48.6(0.1) & 606(96) & 48.0(0.3) & 146(17) & 45.7(0.2) & 109(84) & 51.9(0.7) & L \\
64 & Aqu-MM2 & 284(40) & 49.1(0.3) &  -- & -- & -- & -- & -- & -- & L \\
65 & Aqu-MM4 & 278(46) & 49.2(0.3) &   -- & -- & -- & -- & -- & -- & L \\
66 & Ser SMM1 & 349(20) & 50.6(0.1) & 606(32) & 49.7(0.1) &  103(12) & 47.3(0.2) & 83(13) & 53.6(0.3) & F \\
67 & Ser SMM4 & 269(13) & 50.2(0.1) & -- & --	& 76(28) & 47.0(0.5) &  67(49) & 53.0(1.0) & F \\
68 & Ser SMM3 & 298(13) & 50.0(0.1) & -- & --	& 123(27) & 46.8(0.2) &  -- & -- & F \\
69 & SerpS-MM1 & -- & -- & -- & -- & -- & -- & -- & -- & L \\
70 & SerpS-MM18 & 204(34) & 50.1(0.5) &   -- & -- & 123(19) & 46.0(0.3) & -- & -- & L \\
71 & Aqu-MM6 & -- & -- & -- & -- & -- & -- & -- & -- & L \\
72 & Aqu-MM7 & -- & -- & -- & -- & -- & -- & -- & -- & L \\
73 & Aqu-MM10 & -- & -- & -- & -- & -- & -- & -- & -- & L \\
74 & Aqu-MM14 & -- & -- & -- & -- & -- & -- & -- & -- & L \\
75 & W40-MM3 & -- & -- & -- & -- & -- & -- & -- & -- & L \\
76 & W40-MM5 & 339(54) & 48.4(0.3) &   -- & -- & -- & -- & -- & -- & L \\
77 & W40-MM26 & -- & -- & -- & -- & -- & -- & -- & -- & L \\
78 & W40-MM27 & -- & -- & -- & -- & -- & -- & -- & -- & L \\
79 & W40-MM28 & -- & -- & -- & -- & -- & -- & -- & -- & L \\
80 & W40-MM34 & -- & -- & -- & -- & -- & -- & -- & -- & L \\
81 & W40-MM36 & -- & -- & -- & -- & -- & -- & -- & -- & L \\
82 & RCrA IRS5A	& 298(9) & 48.5(0.1) & -- & -- & 219(30) & 45.1(0.2) &	113(17) & 51.9(0.3) & F \\
83 & RCrA IRS5N	& 299(75) & 47.8(0.4) & --	& -- & -- & -- & -- & -- & F \\
84 & RCrA IRS7A	& 307(10) & 49.5(0.1) & 719(55)	& 48.4(0.2) & 218(24) & 46.0(0.2) &	114(17) & 52.9(0.3) & F \\
85 & RCrA IRS7B	& 281(8) & 49.1(0.1) & 837(141)	& 47.5(0.3) & 208(28) &	45.3(0.2) & 119(17) & 52.0(0.3) & F \\
86 & CrA-44 & 343(15) & 48.2(0.1) &   -- & -- & 117(49) & 44.6(0.5) & 37(5) & 53.1(0.4) & L \\
87 & L723 & 354(31) & 48.7(0.1) & -- & -- & 105(49) & 45.3(0.5) & 50(17) & 52.6(0.8) & L \\
88 & B335 & 283(11)	& 48.4(0.1)	& 840(411) & 47.0(0.6)	& 140(40) & 44.7(0.4) & 101(63) & 51.1(0.6) & F \\
89 & L1157	& 324(13) &	49.2(0.1) &	711(106) & 48.2(0.3) & 114(20) & 46.3(0.3) &  77(40) & 52.5(0.7) & F \\
90 & L1014	& --	& --	& --	& --	& -- & -- & -- & -- & F \\
\enddata
\tablecomments{Mode refers to the line (L) and full (F) spectroscopy with PACS. The lack of 
detection of the components on rotational diagram is marked with '--'. Uncertainties are shown in parentheses.}
\end{deluxetable*}
\subsection{Molecular and atomic cooling}
The fractions of gas cooling emitted in different molecular and atomic species can help to constrain 
 the physical processes that heat the gas. For example, a high ratio of molecular  
 to atomic line emission is predicted by continuous C-type shock models \citep{KN96,FP10}, whereas 
 the increase of atomic emission as the protostar evolves is explained by increasing
 UV fields \citep{Ni02,Vi12} or increased shock velocity \citep{ND89}. Below, we calculate 
 the cooling budget of the gas for the full sample of sources, which is larger than 
 any previous study by a factor of three and evenly distributed over the Class 0 and I stages.

 \subsubsection{Calculations}
 The procedure to calculate line luminosities in each species depends on whether the full 
 spectrum in the 55-210 $\mu$m was observed (the full spectroscopy mode) or only the selected 
 lines were targeted (the line scan mode). 
 
 The procedure for the full scan mode is the following. 
 The total luminosity in the atomic lines is calculated by summing the fluxes of the two 
 [O I] lines in the PACS spectra, at 63 and 145 $\mu$m, and using the distances listed in Table \ref{cat_paper4}.
  The [C II] line flux is not included because only 10 sources show reliable detections (see \S3.1).
  The line luminosity of CO is obtained by the addition of fluxes of the detected lines 
  and extrapolated fluxes of lines located around 100 $\mu$m or blended with other species 
  (e.g. CO 23-22 at 113 $\mu$m and CO 31-30 at 84 $\mu$m). The latter are obtained 
  using the linear fits to CO diagrams if available (see \S 4.1). When 
  reliable fits to the rotational diagrams are not possible, typically in the \lq hot' 
  component, the line fluxes of the detected lines are summed and the fluxes 
  of the lower-$J$ lines are determined using the fit to the \lq warm' component. 
  The lines 
  included in the interpolation are only those with upper energies lower than the 
  energy of the highest detected level. 
  The line luminosities of H$_2$O and OH lines are 
  calculated by the addition of detected lines. Fluxes of lines blended with CO transitions are 
  determined by subtracting the CO fluxes obtained from the fit to the CO diagrams. No 
  further extrapolation is implemented.
  
  In order to calculate the line luminosities for the sources observed in line spectroscopy mode,
  we take advantage of the results obtained for the sources with full PACS spectra. The mean 
  value of the ratio of [O I] 63 $\mu$m to [O I] 145 $\mu$m is 11.6$\pm$4.6 for the 
  sources observed in the WISH and DIGIT surveys where both lines were targeted. The value 
  is consistent with the ratio of 10.5 obtained for the WISH survey alone \citep{Ka13}. 
  We therefore multiply the fluxes of the [O I] 63 $\mu$m by a factor of $\sim1.086$ ($\pm$0.034) to calculate
   the total line luminosities of [O I] for the sources from the WILL survey. 

\startlongtable
\begin{deluxetable}{lrrrrrrrrrrrrr} 
\tabletypesize{\scriptsize}
\tablecaption{\label{cool_table} Total line luminosities of H$_2$O, CO, OH, and [O I]}
\tablehead{
\colhead{ID}           &
\colhead{H$_2$O} &
\colhead{CO warm}  & 
\colhead{CO hot}  &
\colhead{OH}  &
\colhead{[O I] 63} &
\colhead{[O I] tot}  &
\colhead{Rem.}     \\ \cline{2-7}
\colhead{~}   &        
\multicolumn{6}{c}{($10^{-3}$ $L_{\odot}$) } &
\colhead{~} 
}
\startdata
1  &  3.29  &  4.17  &  0.79  &  1.83  &  2.61  &  2.84 & c\\
2  &  5.05  &  11.03  &  1.25  &  2.33  &  6.86  &  7.45 & c\\
3  &  12.62  &  5.37  &  3.00  &  3.21  &  0.66  &  0.69 &  \\
4  &  4.04  &  2.70  &  --  &  0.59  &  0.40  &  0.44 & h \\
5  &  --  &  --  &  --  &  --  &  0.39  &  0.43 & c \\
6  &  1.10  &  1.34  &  --  &  0.30  &  0.19  &  0.22 & h \\
7  &  --  &  0.10  &  --  &  0.05  &  0.15  &  0.16 &  \\
8  &  1.83  &  1.20  &  0.11  &  0.34  &  0.28  &  0.31 & c,lh \\
9  &  2.50  &  2.10  &  --  &  0.21  &  --  &  -- &  \\
10  &  0.75  &  1.39  &  -- &  1.04  &  1.31  &  1.42 & c \\
11  &  --  &  --  &  --  &  --  &  0.19  &  0.20 & c \\
12  &  2.38  &  2.41  &  0.10  &  1.36  &  10.82 & 11.75 & c,lh \\
13  &  5.87  &  4.18  &  1.25  &  2.59  &  --  &  -- & c,oc \\
14	&  5.95  &  6.07  &  0.68  &  0.47  &  0.43 & 0.43 & no \\
15  &  --  &  2.14  &  0.08  &  1.40  &  -- & -- & c,lh,oc \\
16  &  25.57 &  14.92 & 7.72 & 4.72 & 0.32  &  0.30 & \\
17  &  --  &  --  &  --  &  --  &  0.12  &  0.13 & c \\
18  &  0.02  &  0.01  &  --  &  --  &  --  &  -- & lw,lh \\
19  &  2.06  &  1.32  &  0.52  &  1.07  &  0.37  &  0.40 & c \\
20  &  --  &  0.77  &   --  &  0.09  &  0.21  &  0.23 & c \\
21  &  --  &  0.05  &   --  &  --   &  0.23  &  0.25 & c \\
22  &  2.03  &  0.94  &  0.31  &  1.82  &  0.64  &  0.75 &  \\
23  &  0.00  &  0.01  &  --  &  --  &  0.14  &  0.15 & c,lc \\
24  &  0.24  &  2.09  &  --  &  0.03  &  1.83  &  1.99 & c,lh,loh \\
25  &  1.31  &  1.65  &  --  &  0.92  &  0.62  &  0.75 &  \\
26  &  1.12  &  0.91  &  --  &  --  &  --  &  -- &  \\
27  &  0.52  &  1.32  &  --  &  1.15  &  1.78  &  1.93 & c \\
28  &  6.44  &  4.99  &  --  &  0.99  &  1.94  &  2.11 & c \\
29  &  0.00  &  1.01  &  0.03  &  0.49  &  0.88  &  0.95 & c,lh \\
30  &  0.74  &  1.75  &  --    &  0.76  &  1.19  &  1.29 & c \\
31  &  0.88  &  0.79  &  0.20  &  0.39  &  0.28  &  0.30 &  \\
32  &  0.02  &  0.05  &  --  &  0.08  &  0.66  &  0.72 & c,lm \\
33  &  0.05  &  0.16  &  --  &  0.18  &  0.12  &  0.13 & c,lw \\
34  &  0.07  &  0.12  &  --  &  0.04  &  0.09  &  0.10 &  \\
35  &  0.03  &  0.02  &  --  &  0.03  &  0.06  &  0.06 & c,lm \\
36  &  --  &  0.03  &  0.01  &  0.05  &  0.47  &  0.51 & c,lm \\
37  &  --  &  --  &  --  &  --  &  0.10  &  0.11 & c \\
38  &  0.21  &  0.28  &  0.01  &  0.30  &  0.02  &  0.02 & c,lh \\
39  &  0.14  &  2.06  &  0.48  &  0.28  &  3.49  &  3.67 &  \\
40  &  0.11  &  0.20  &  0.01  &  0.04  &  0.49  &  0.53 & c,lh,loh \\
41  &  0.28  &  0.44  &  --  &  0.36  &  0.40  &  0.43 & lh \\
42  &  1.39  &  0.97  &  --  &  0.91  &  0.38  &  0.43 &  \\
43  &  0.00  &  0.41  &  --  &  0.06  &  1.18  &  1.18 & no \\
44  &  0.31  &  0.65  &  0.17  &  0.40  &  0.53  &  0.57 &  \\
45  &  2.11  &  4.90  & 0.23 & 3.69 & 19.12 & 20.30 &  \\
46  &  0.19  &  0.17  &  --  & 0.26 &  0.75 & 0.77 & \\
47  &  0.26  &  0.50  &  0.02  &  0.24  &  0.52  &  0.57 & c,lh \\
48  &  2.43  &  9.01  & 3.11  &  1.35  &  2.47 & 2.61 &  \\
49  &  0.76  &  4.93  &  1.61  &  1.10  &  3.46  &  3.59 &  \\
50  &  0.06  &  0.04  &  --  &  0.09  &  0.13  &  0.14 & c,lm \\
51  & 0.73 & 1.82 & -- & 0.50 & 0.94 & 0.99 & \\
52  &  9.08  &  5.51  &  2.01  &  1.92  &  1.46  &  1.55 &  \\
53  &  1.00  &  2.10  &  0.21  &  1.00  &  0.38  &  0.44 &  \\
54  &  0.64  &  1.13  &  0.20  &  0.35  &  0.23  &  0.29 &  \\
55  &  0.02  &  0.02  &  --  &  --  &  2.95  &  3.20 & c,lm \\
56  &  8.57  &  6.06  &  1.56  &  2.30  &  1.16  &  1.27 &  \\
57  &  --  &  0.02  &  --  &  --  &  0.14  &  0.15 &  \\
58  &  0.08  &  0.16  &  --  &  --  &  0.04  &  0.04 & no \\
59  &  0.02  &  0.05  &  --  &  0.03  &  0.09  &  0.10 &  \\
60  &  0.81  &  0.92  &  0.26  &  0.77  &  0.53  &  0.57 & c \\
61  &  0.02 &  0.11 & -- & 0.10 & 0.38 & 0.40 & \\
62  &  0.17  &  0.22  &  0.01  &  0.23  &  0.28  &  0.31 & c,lh \\
63  &  2.05 & 1.63 & 0.47 & 1.14 & 0.45 & 0.49 & \\
64  &  1.13  &  3.31  &  --  &  0.00  &  0.30  &  0.33 & c,lw \\
65  &  0.52  &  3.52  &  0.06  &  0.11  &  0.27  &  0.30 & c,lm \\
66  &  66.07  &  142.98  &  29.26  &  31.69  &  22.86  &  25.71  &  \\
67  &  10.82  &  35.62  &  8.31  &  4.82  &  17.24  &  18.52 &  \\
68  &  8.63  &  24.25  &  1.24  &  1.94  &  7.32  &  7.84 &  \\
69  &  --  &  --  &  --  &  --  &  --  & 2.814 & 3.056\\
70  &  3.78  &  11.39  &  0.10  &  0.12  &  --  &  -- & oc \\
71  &  --  &  --  &  --  &  --  &  --  &  -- &  \\
72  &  --  &  --  &  --  &  --  &  --  &  -- &  \\
73  &  0.10  &  0.12  &  --  &  --  &  --  &  -- &  \\
74  &  --  &  0.04  &  --  &  0.05  &  --  &  -- & c,lc,loh \\
75  &  0.05  &  0.10  &  --  &  --  &  --  &  -- & lm,oc \\
76  &  0.06  &  1.01  &  --  &  0.10  &  55.02  &  59.75 & c,lm \\
77  &   --   &   --   &   --   &   --   &   --   &   --  & oc \\
78  &   --   &   --   &   --   &   --   &   --   &   --  & oc \\
79  &   --   &   --   &   --   &   --   &   --   &   --  & oc \\
80  &   --   &   --   &   --   &   --   &   --   &   --  & oc \\
81  &  --  &  -- &  --  & --  &  0.55	& 0.60 & c \\
82  &  0.82  &  0.82  &  0.13  &  0.69  &  19.61  &  20.86 &  \\
83  &  0.08  &  0.16  &  --  &  0.04  &  8.09  &  8.82 &  \\
84  &  8.23  &  8.83  &  2.39  &  5.84  &  147.35  &  157.64 &  \\
85  &  1.43  &  2.72  &  0.44  &  0.97  &  39.14  &  43.39 &  \\
86  &  0.16  &  0.57  &  0.02  &  0.35  &  0.95  &  1.03 & c,lh \\
87  &  0.29  &  1.85 & 0.15 & 0.27  & 0.63  & 0.69  &  \\
88  &  0.18  &  0.57  &  0.09  &  0.12  &  0.15  &  0.16 &  \\
89  &  3.92  &  5.65  &  1.53  &  1.73  &  1.79  &  1.92 &  \\
90  &  --  &  --  &  --  &  0.04  &  0.09  &  0.09 & no \\
\enddata
\tablecomments{'CO warm' (w) and 'CO hot' (h) in the header refer to the CO lines with $J_\mathrm{up}=14-26$
and $J_\mathrm{up}\geq27$, respectively. The '[O I] tot' is the sum of the line luminosities of
the [O I] lines at 63 and 145 $\mu$m. Remarks mean: c - values derived from line scans and 
corrected for the missing lines; no - non-detection of the [O I] line at 145 $\mu$m; oc - contamination of the 
[O I] emission on the map; lw, lc, lh, loh, lm - lower limits of the H$_2$O, CO w, CO h,  
OH, and all molecules, respectively; h - line luminosities of CO h are added to the CO w.
Non-detections are shown as '--'. }
\end{deluxetable}
\begin{figure*}[!tb]
  \begin{minipage}[t]{.5\textwidth}
  \begin{center}
       \includegraphics[angle=90,height=7cm]{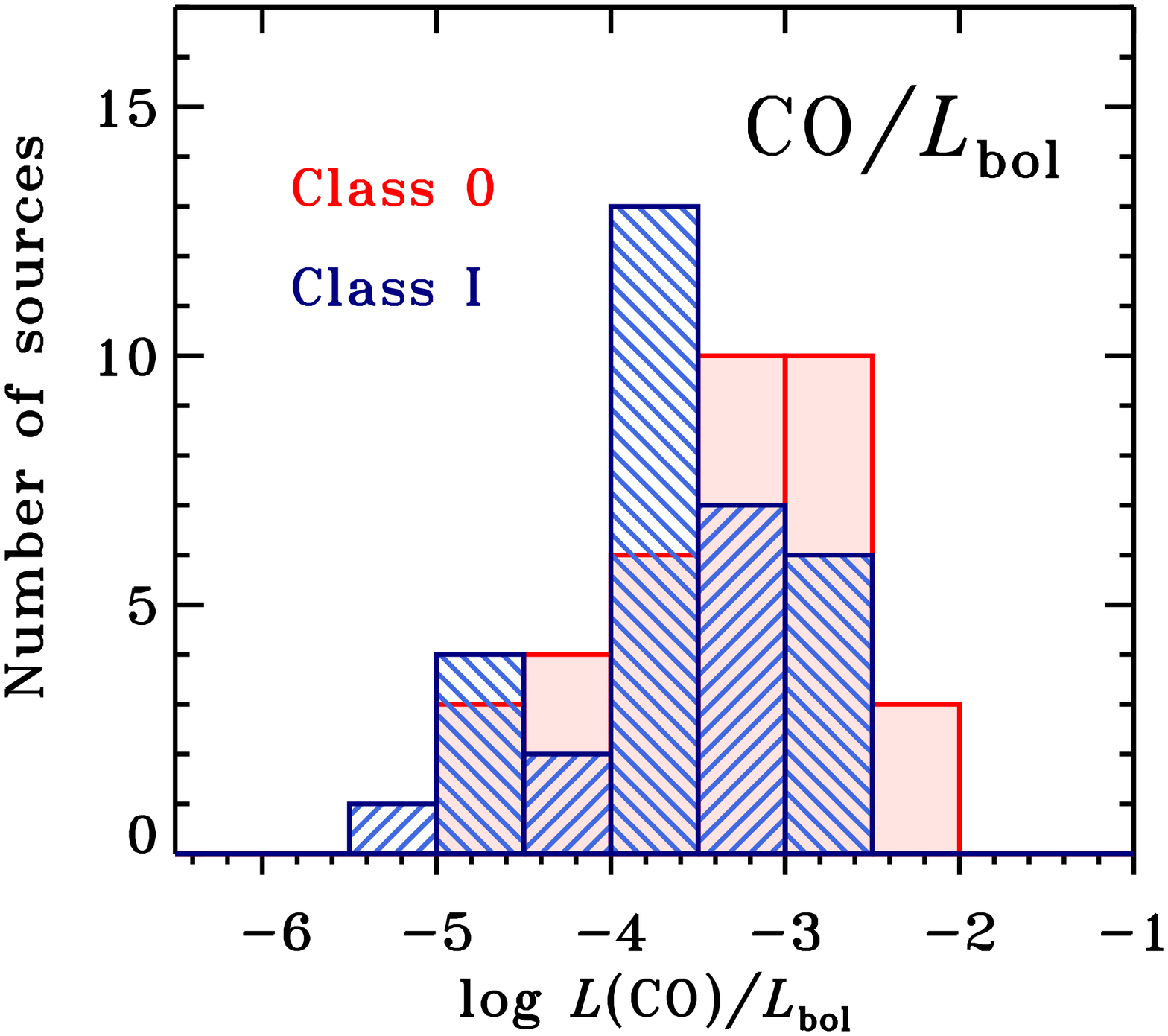}   
       \includegraphics[angle=90,height=7cm]{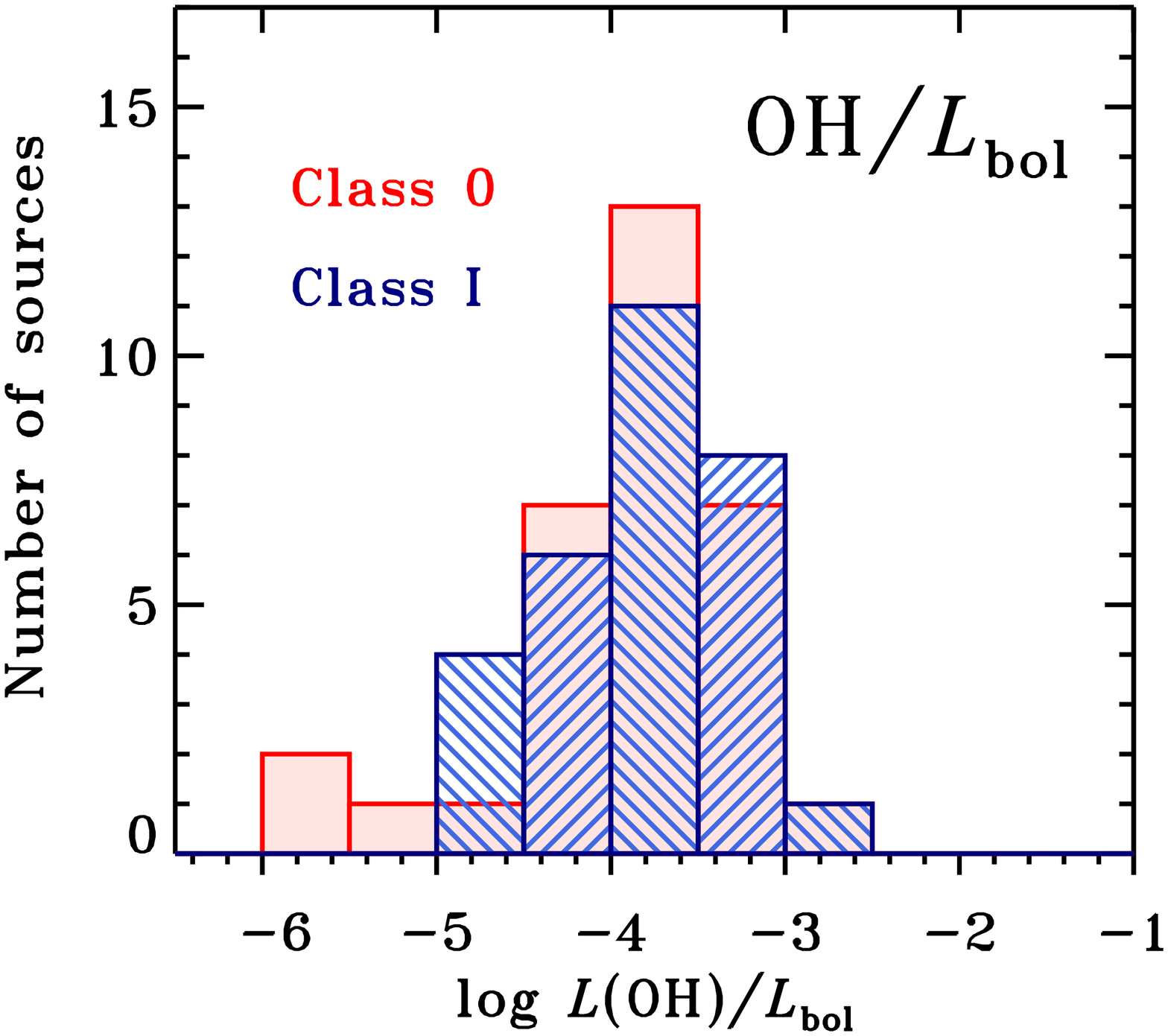}   
  \end{center}
  \end{minipage}
  \hfill
  \begin{minipage}[t]{.5\textwidth}
      \begin{center}
    \includegraphics[angle=90,height=7cm]{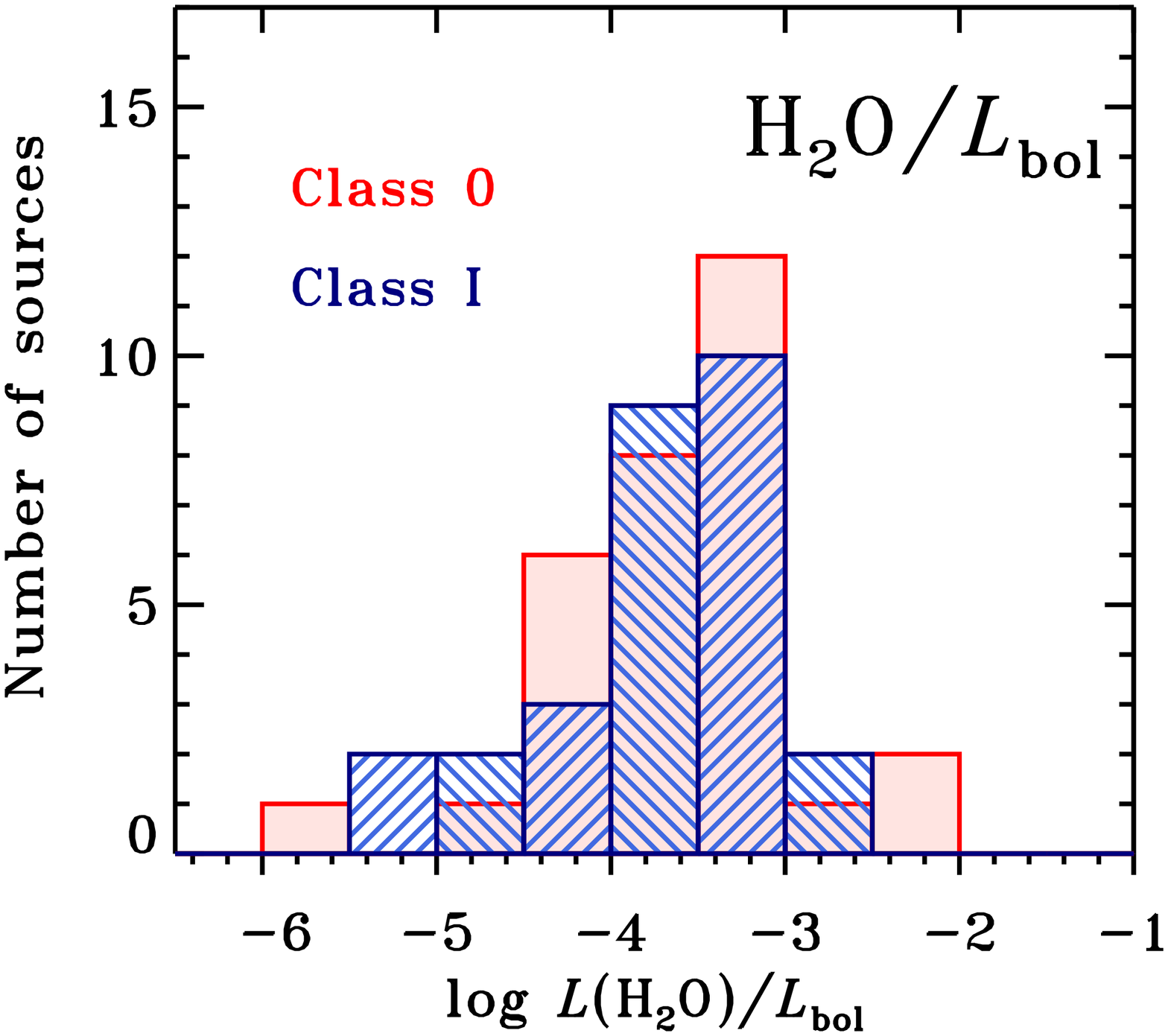} 
    \includegraphics[angle=90,height=7cm]{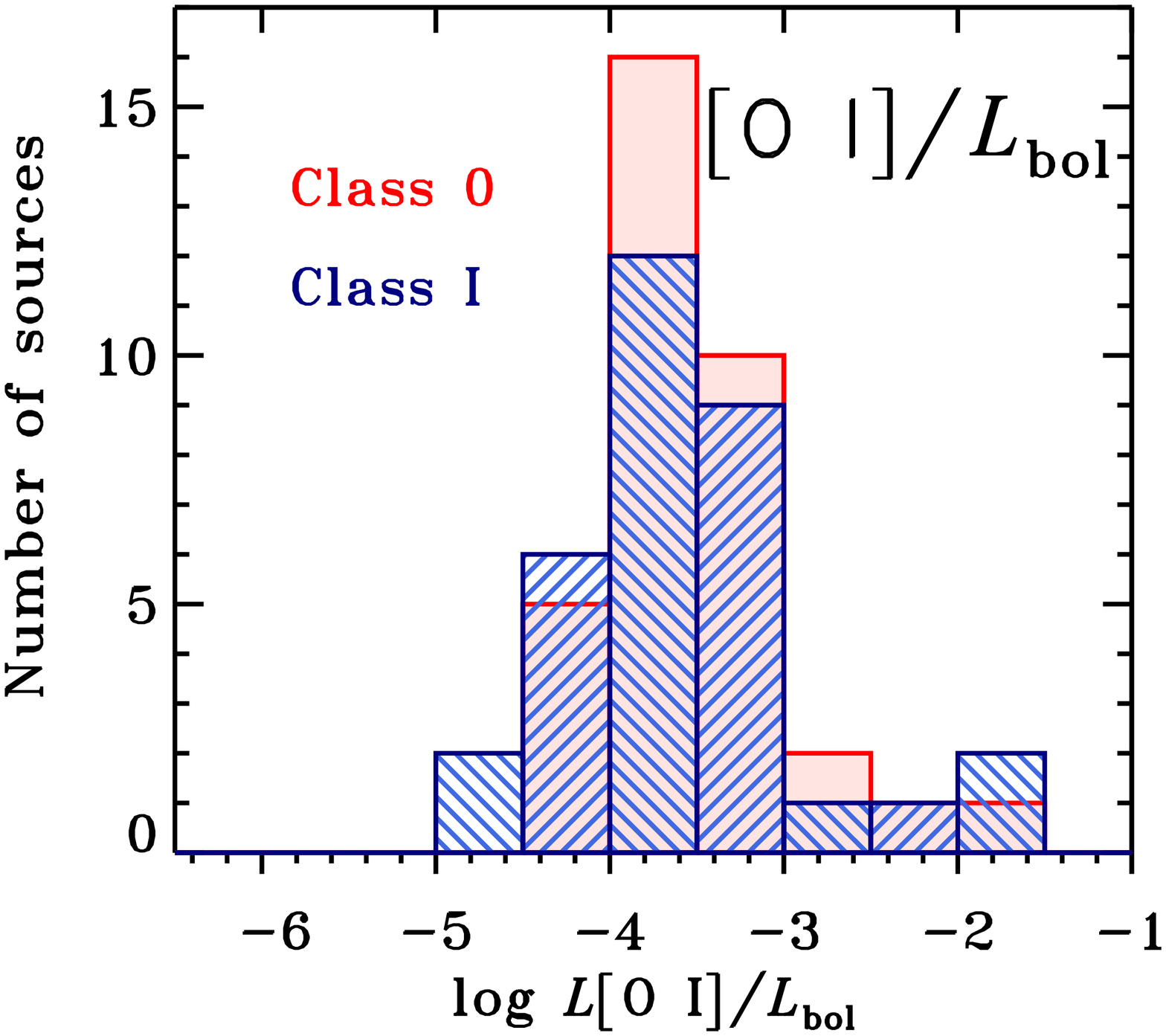} 
      \end{center}
  \end{minipage}
    \hfill
        \caption{\label{hist_new} Histograms of the CO (top left), H$_2$O (top right), 
        OH (bottom left), and [O I] (bottom right) cooling over bolometric luminosities for 
        all sources with at least one detection in a given species except Class II sources 
        and sources possibly contaminated by PDR emission (see Table 1 for details). 
        The red color shows the distributions for Class 0 sources and blue for Class I sources.}
\end{figure*}

  The line luminosities of CO are obtained using linear fits to the rotational diagrams,
  in the same manner as for the full scan observations. In order to determine the H$_2$O 
  and OH line luminosities, we first used the full PACS spectra to calculate the respective
  line luminosities if the objects were observed in the line scan mode.
  The ratio of the 
  observed-to-calculated luminosities using limited number of lines is 3.0$\pm$1.4  for H$_2$O and 2.1$\pm$0.6 for OH in the WILL 
  settings, and 2.1$\pm$0.7  for H$_2$O and 1.4$\pm$0.4 for OH in the WISH line scan settings.
 	We adopt those values to 
  calculate the total line luminosities only for the sources where the number of detected 
  lines allows us to fit the linear component of the rotational diagrams (see Table \ref{exc_table}). 
  For the remaining sources, we refrain from using this correction and list only the lower limits
  for the H$_2$O and OH luminosities. 
  
  The full list of derived total line luminosities is presented in Table \ref{cool_table}. We 
 refrain from listing the CO line luminosities with $J_\mathrm{u}<14$, which are inaccessible to PACS. 
  The APEX/CHAMP$^+$ and HIFI survey of \citet{Yi13} shows that those lower-$J$ CO lines trace
  a physical component that is intrinsically different from the one seen in the PACS CO lines, 
  namely the entrained outflow gas with temperatures of $T\lesssim 100$ K. Recent analysis of full
  CO ladders with PACS and SPIRE confirm that the temperatures and distribution of line emission 
  vary significantly for CO lines above and below $J_\mathrm{u}\sim14$ \citep[][Yang et al. in prep.]{YL17}. 
  The inclusion of those $J_\mathrm{u}<14$ CO
  line fluxes could therefore add some confusion to the interpretation of the far-infrared lines.
    
 \subsubsection{Absolute line luminosities}
\begin{figure*}[!tb]
  \begin{minipage}[t]{.2\textwidth}
       \includegraphics[angle=90,height=4cm]{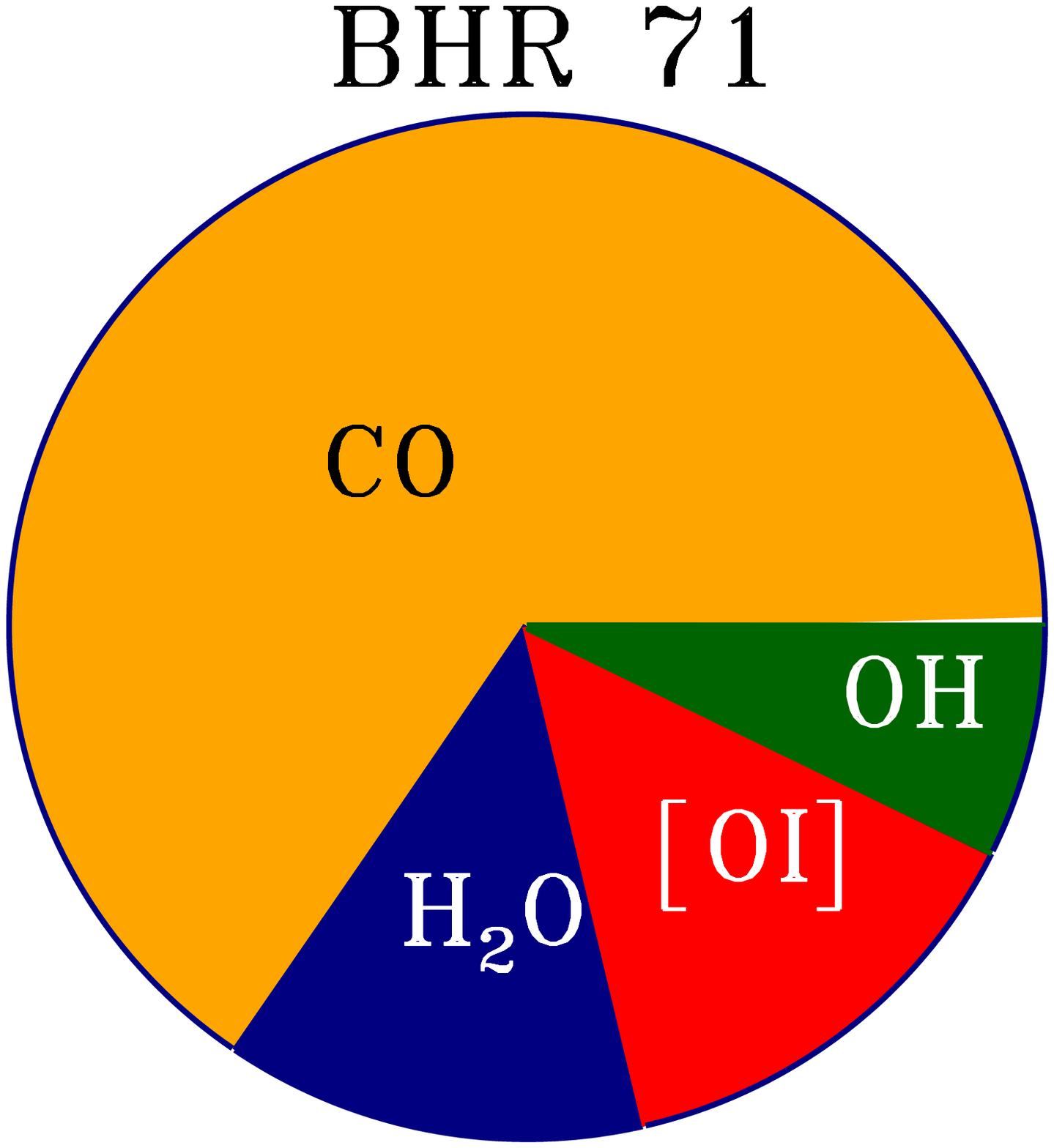}   
       \includegraphics[angle=90,height=4cm]{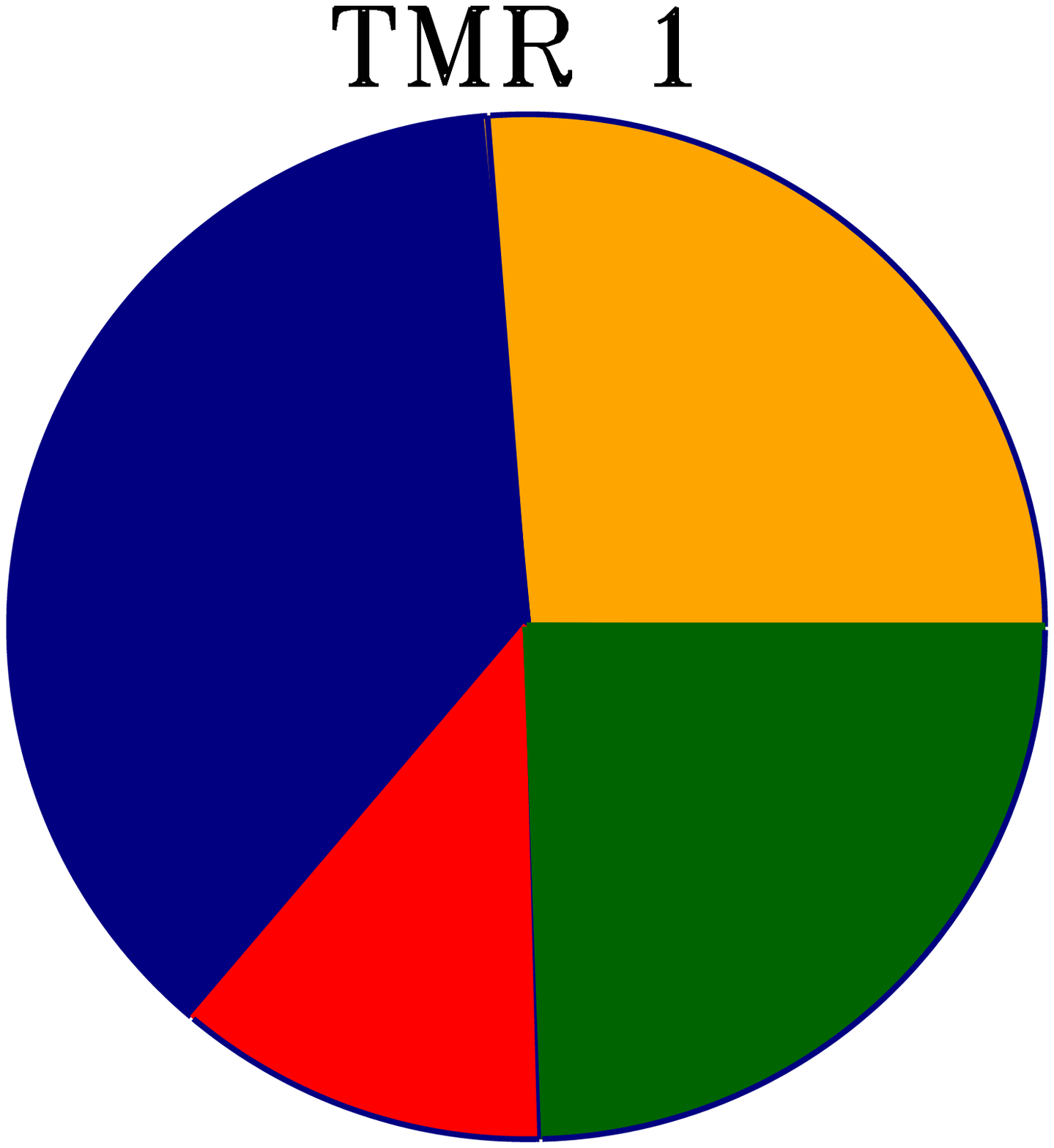}   
  \end{minipage}
  \hfill
  \begin{minipage}[t]{.24\textwidth}
     \includegraphics[angle=90,height=4cm]{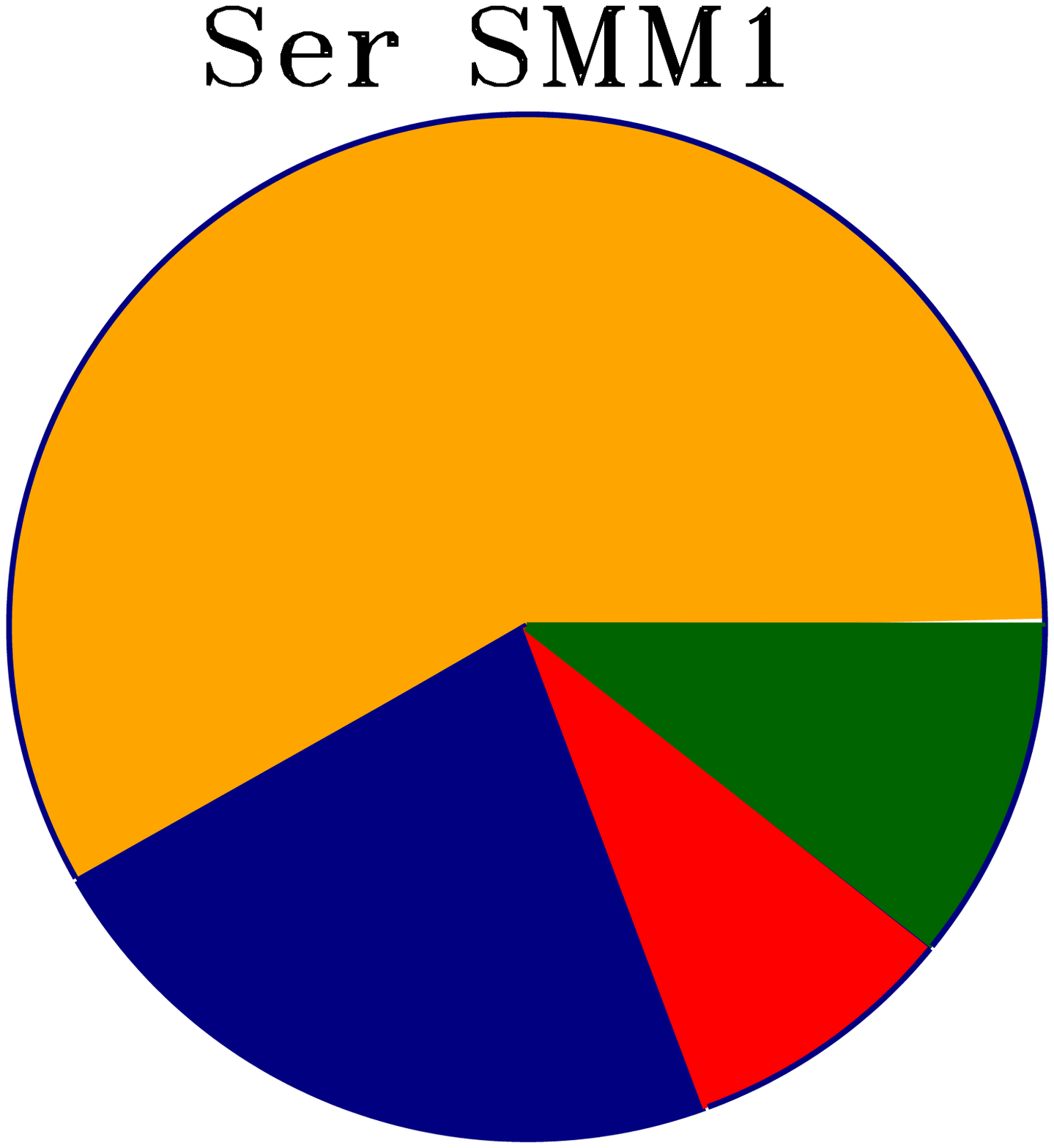} 
    \includegraphics[angle=90,height=4cm]{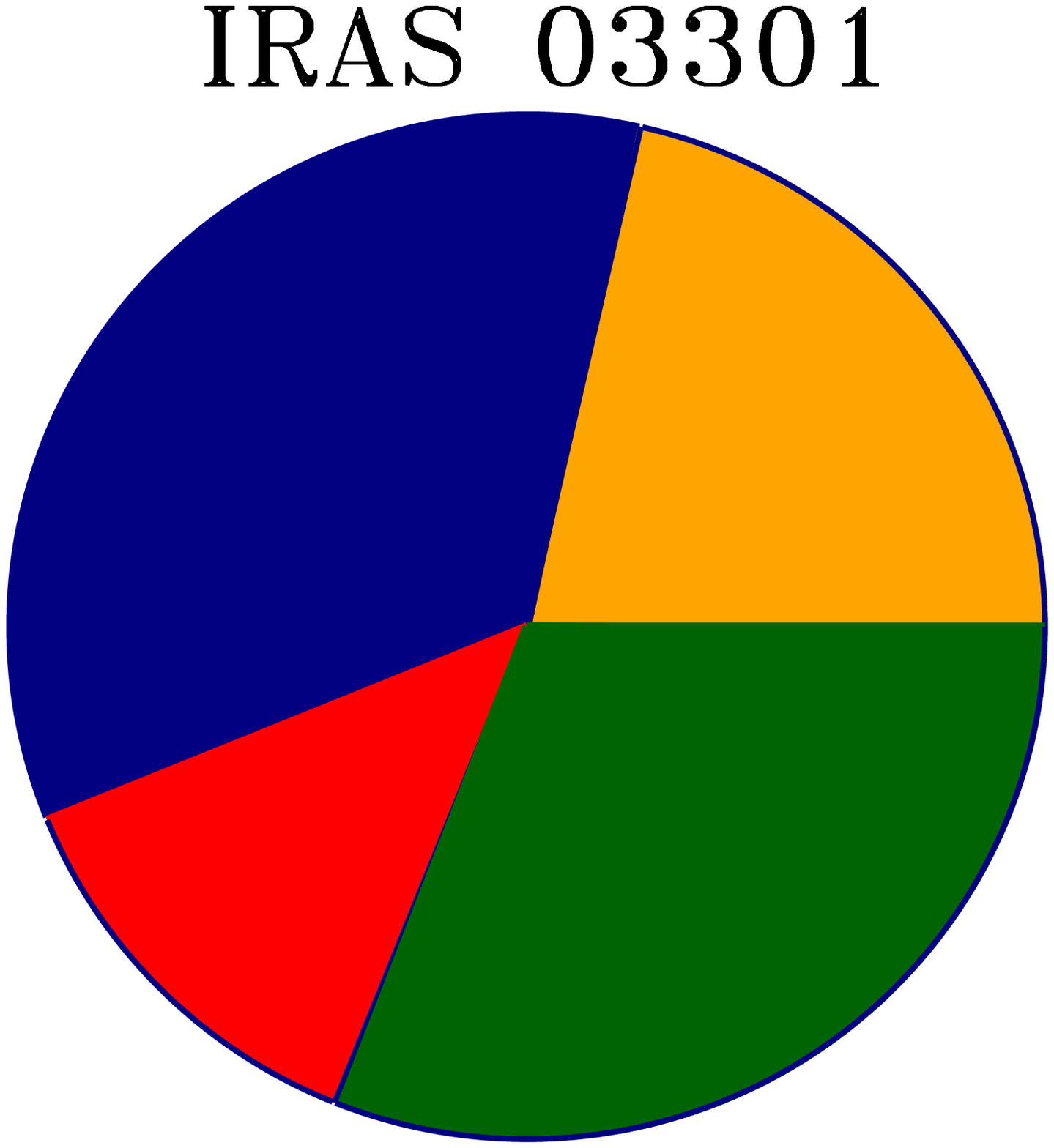} 
  \end{minipage}
   \hfill
  \begin{minipage}[t]{.24\textwidth}
     \includegraphics[angle=90,height=4cm]{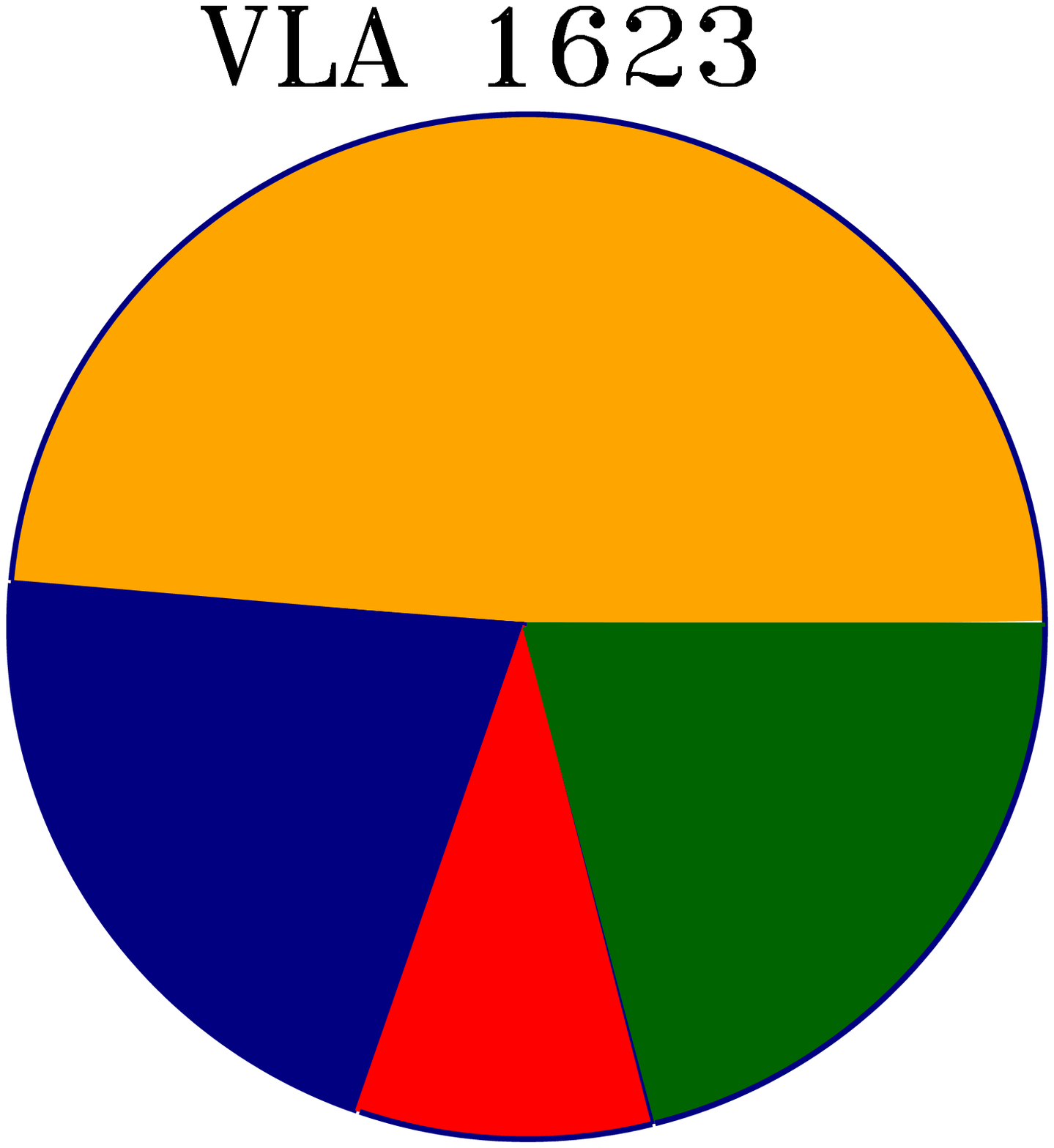} 
    \includegraphics[angle=90,height=4cm]{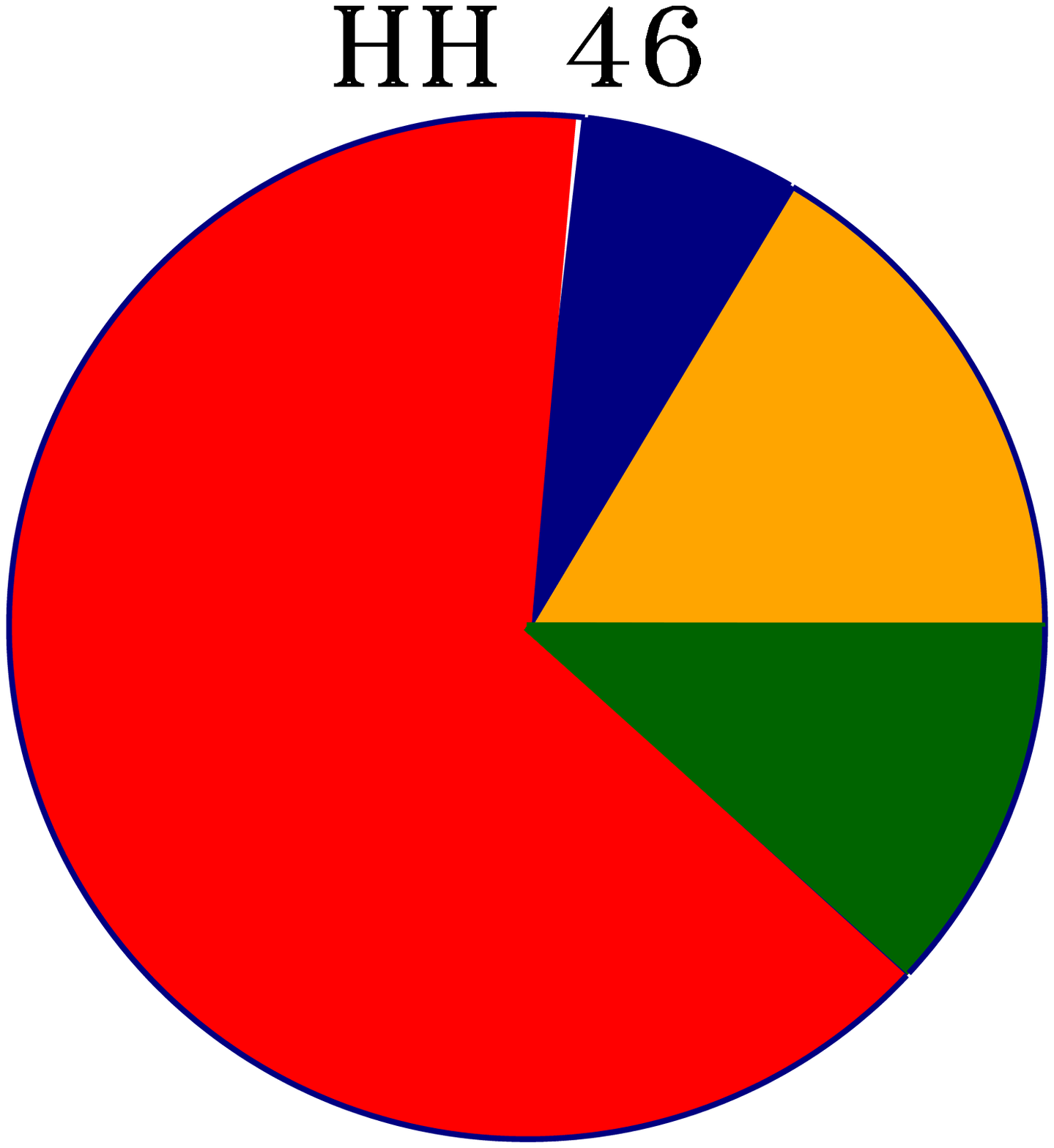} 
  \end{minipage}
 \hfill
  \begin{minipage}[t]{.24\textwidth}
     \includegraphics[angle=90,height=4cm]{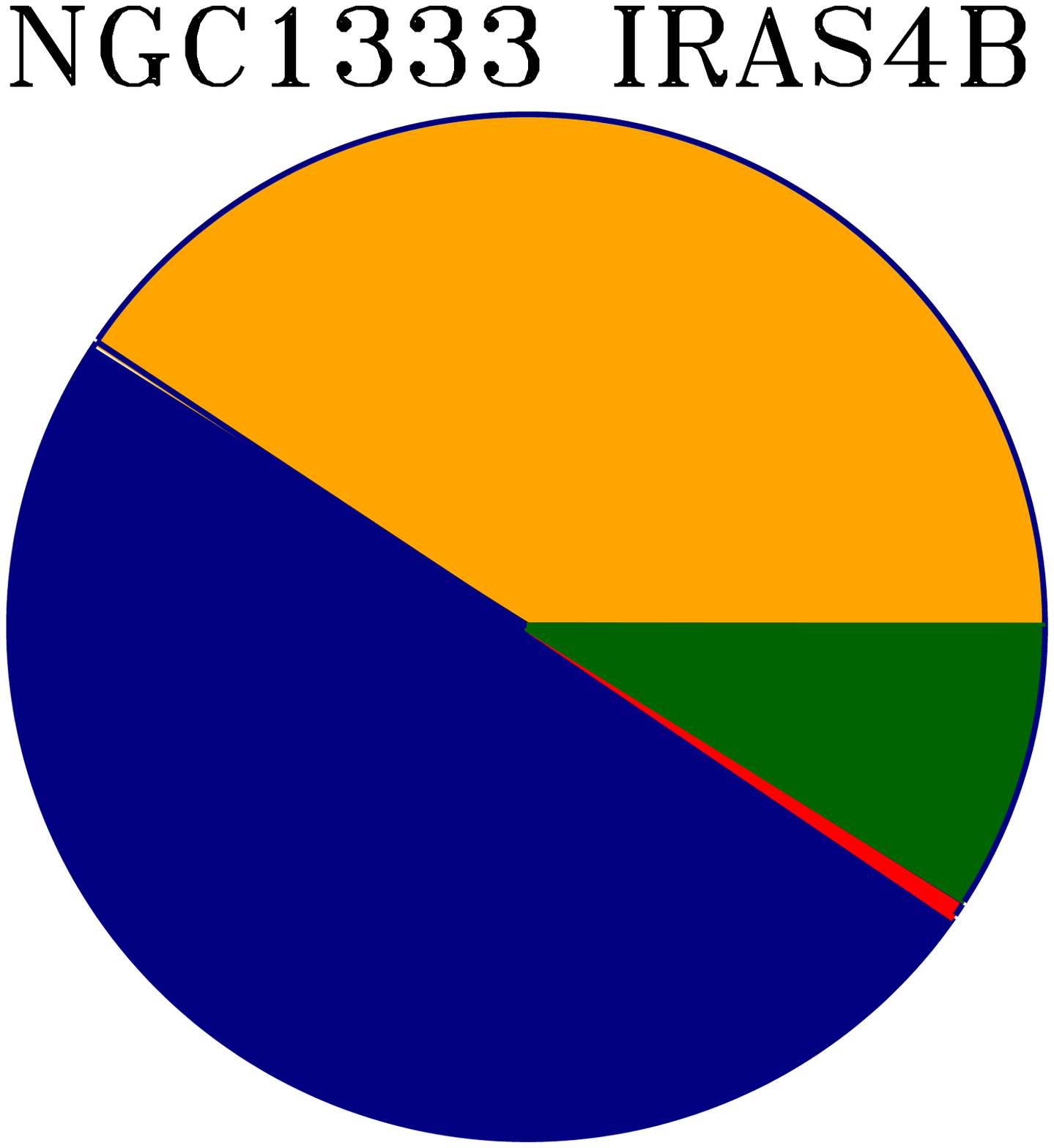} 
    \includegraphics[angle=90,height=4cm]{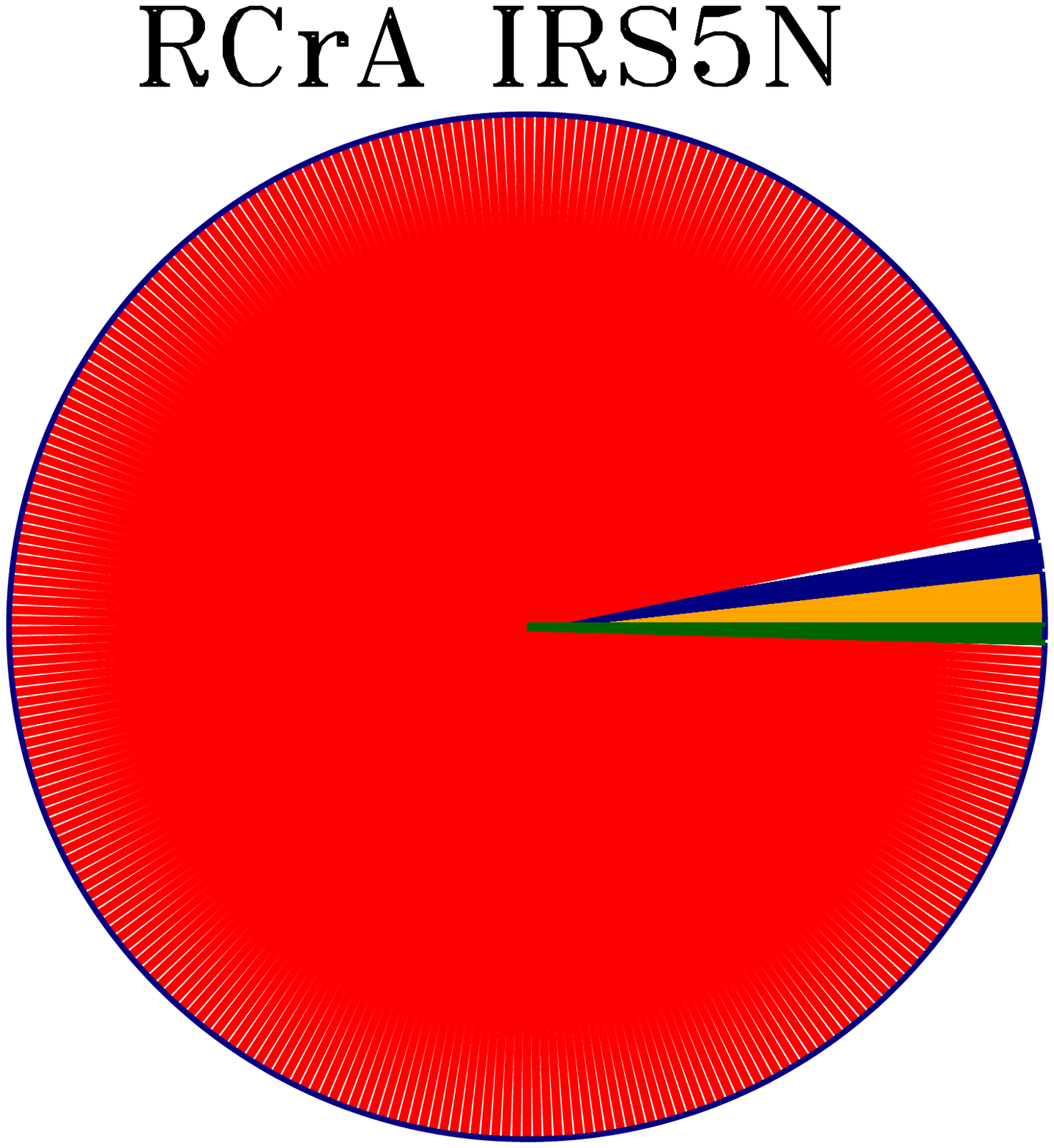} 
  \end{minipage}
    \hfill
        \caption{\label{cool_indiv} Fractions of gas cooling contributed by CO (orange), H$_2$O (blue),
         OH (green), and [O I] (red), to the total far-infrared line cooling in eight example sources.}
\end{figure*}

The distributions of the line to bolometric luminosity ratio for various species across the sample 
of Class 0 and I sources is illustrated in Figure \ref{hist_new} (see also Figure \ref{hist_new_abs} for absolute fluxes).
 The largest median line luminosity divided by 
 $L_\mathrm{bol}$ is that of CO ($3.3\cdot10^{-4}$), followed by H$_2$O ($3.0\cdot10^{-4}$), [O I] ($2.4\cdot10^{-4}$), and
  OH ($1.5\cdot10^{-4}$). The distributions vary depending on the considered species.
  
For CO, the total line luminosities divided by $L_\mathrm{bol}$ are significantly larger for 
Class 0 than for Class I sources (see Figure \ref{hist_new}); the median for the former is $6.2\cdot10^{-4}$ 
versus $2.7\cdot10^{-4}$ for Class I. Similar trends are not immediately seen in the rest 
of the molecular species. Both H$_2$O and OH show almost identical distributions 
for Class 0 and I sources, with median line luminosities over $L_\mathrm{bol}$ that are 
almost equal ($3.1\cdot10^{-4}$ vs. $2.7\cdot10^{-4}$ for H$_2$O, and $1.4\cdot10^{-4}$ vs. $1.5\cdot10^{-4}$ for OH).
The [O I] line luminosities over $L_\mathrm{bol}$ are also similar between the two classes, 
and show a clear peak concentrated in the 10$^{-4}$-10$^{-3}$ bins.

   These results are qualitatively consistent with the far-IR line luminosities from the WISH 
   survey alone \citep{Ka13}. The cooling by CO was found to be dominant, with the median value for Class 0 of 
   $\sim 5.7\cdot 10^{-3}$ $L_\mathrm{\odot}$ and for Class I of $\sim 0.6\cdot 10^{-3}$ $L_\mathrm{\odot}$.
   Similarly, H$_2$O was the second most important coolant at $\sim 5.3\cdot 10^{-3}$ $L_\mathrm{\odot}$ 
   for Class 0 and $\sim 0.5\cdot 10^{-3}$ $L_\mathrm{\odot}$ for Class I. Median line luminosities for the 
   entire sample were $\sim 2.0\cdot 10^{-3}$ for both CO and H$_2$O, and the median bolometric luminosity was 3.8 $L_\odot$.
   
   It is noteworthy that the CO and H$_2$O absolute luminosities obtained 
   in the WISH survey for the Class 0 sources are a factor of 3-5 higher than the median values for the 
   entire sample probed here. This illustrates the bias toward the brightest outflow sources targeted in 
   WISH, which is reduced by the addition of the DIGIT and WILL samples. In contrast, the CO and H$_2$O 
   luminosities for Class I analysed here are a factor of 2 higher than for the Class I sources in the WISH sample, reducing the 
   differences due to the evolutionary stage.
   
   The cooling in [O I] and OH is consistent to within a factor of 2 between the combined and WISH-only samples. 
   The ratios of the [O I] luminosities for Class 0 and I sources are about
   1.4 in the extended sample and 2.2 in the WISH survey, illustrating that the differences in the [O I] cooling with evolutionary stage 
   are less significant than previously observed \citep[also with respect to the ISO results, see ][]{Ni02,Mo17}.
    The trends in the line luminosities from Class 0 to Class I are further discussed in \S 6.1.
   
\subsubsection{Contributions of various species to the total far-infrared gas cooling}
Low-mass protostars show a wide range of relative contributions from H$_2$O, CO, OH, and [O I] 
to the total far-infrared line luminosities, as illustrated in Figure \ref{cool_indiv} by eight example 
sources. The highest fraction
of cooling due to CO ($\sim$65 \%) is seen in BHR 71, where prominent emission from high-$J$ CO lines
 is detected and accompanied by much lower, comparable amounts of cooling in H$_2$O and \mbox{[O I]} ($\lesssim$15 \%), 
 and to smaller extent OH ($\sim$7 \%). A considerable fraction of gas cooling in CO is also detected in NGC1333 
 IRAS4B ($\sim$40 \%), but in this source the emission from a large number of H$_2$O lines dominates 
 the cooling budget ($\sim$50 \%), in the presence of weak [O I] emission \citep{He12}. This source 
 most resembles the results found for PACS Bright 
 Red Sources \citep{To16}. The highest relative line emission of OH is 
 seen in IRAS 03301 ($\sim$30 \%), with a comparable fraction contributed by H$_2$O and to a smaller 
 extent CO and [O I]. Protostars in the Corona Australis region, e.g. R CrA IRS 5N, show the dominant far-IR 
 cooling via [O I] ($\sim$97 \%), likely due to the presence of increased UV fields from a nearby 
  Herbig Be star \citep{Li14}.

The fractions of cooling contributed by various species do not depend on the source 
parameters, i.e. its bolometric temperature and luminosity (see Figures \ref{cool2} and \ref{cool4} in the Appendix). The spread in 
the line luminosities divided by the total far-infrared line luminosities (FIRL) is 
substantial for all species, but clearly the largest for [O I]. We find, however, that the highest 
ratios of $L_\mathrm{CO}$ / FIRL are found for Class 0 sources, whereas 
the highest ratios of $L_\mathrm{OH}$ / FIRL are seen in Class I sources (see also Section 6.3).

A better quantity to describe the far-infrared budget is the ratio of line luminosities 
of various species. As illustrated by Figure \ref{new_corr}, the correlations 
between molecular species are very strong ($\sim 6 \sigma$), and only somewhat weaker 
between the molecules and [O I] ($\sim 4-5 \sigma$). As expected, the 
line ratios of those species cover a narrow range of values for the majority of the 
protostars (see Figure \ref{cool_ratio} in the Appendix). Similar characteristics 
have been seen in the sample of protostars in Perseus. 
This uniformity implies that the entire sample can be treated as one, a consequence which 
will be utilized below.
\section{Comparisons to models of shocks and photodissociation regions}
\begin{figure*}[t]
\begin{center}
\vspace{-4cm}
\includegraphics[angle=90,height=10cm]{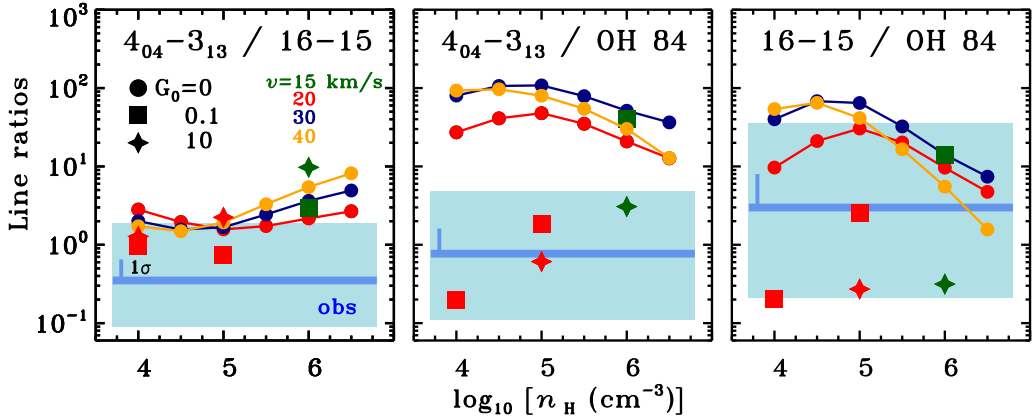}
\caption{\label{obsrat} Ratios of line fluxes in units of erg cm$^{-2}$ s$^{-1}$ comparing different species as a
function of logarithm of density of the pre-shock gas, $n_{H}$. Filled symbols
show models of fully shielded $C$ shocks from \citealt{KN96} (circles), 
$C$ shocks irradiated by UV fields of $G_\mathrm{0}=0.1$ (squares) and $G_\mathrm{0}=10$ (stars) from
\citet{MK15} and Kaufman (in prep.). Models are calculated for pre-shock densities of 
$\varv=15$ km s$^{-1}$ (green), $\varv=20$ km s$^{-1}$ (red), $\varv=30$ km s$^{-1}$ 
(orange), and $\varv=40$ km s$^{-1}$ (navy blue). Not shown on the 
central and right figures are the model ratios for pre-shock density log$_{10}n=4$ and 
$G_\mathrm{0}=10$, which have values $\sim2\times10^{-2}$.
The observations are shown as light blue boxes with the horizontal blue line indicating 
the median values of ratios for all sources and the vertical line - the standard
deviation.}
\end{center}
\end{figure*}
Far-infrared line emission can be compared to shock models in order 
to derive the physical conditions of the medium and the properties of shock waves. 
A critical part of such comparisons is ensuring that the relevant species are emitted from the same location(s)
in young stellar objects.

\citet{Ka13} analyzed the patterns of spatially-resolved far-IR line emission toward 18 protostars from the WISH survey 
and concluded that (i) the emission follows the outflow directions as defined from CO 6-5 
\citep{Yi12,Yi15}; (ii) high$-J$ CO emission ($J_\mathrm{u}\ge14$) seen in PACS is well-aligned 
with the emission in H$_2$O, but differs with respect to [O I]. For the same protostars, 
\citet{Sa12} and \citet{Kr17b} show that the velocity-resolved profiles of CO 16-15 and H$_2$O
$1_{10}-1_{01}$ are almost identical, which strengthens the conclusion that the two species are connected.

Assuming the same origin of high-$J$ CO and H$_2$O, \citet{Ka14b} compared the observed line ratios of molecular species 
obtained for 22 protostars in Perseus (WILL survey) to the available stationary C- and J-type shock models
in dark clouds fully-shielded from UV. In this way, the uncertainty in the distance and any large-scale environmental effects 
were minimised. The observed 
line ratios of various same-molecule pairs of H$_2$O, CO, and OH lines cover a narrow range  
and agree with the models of $C$-type stationary shocks propagating at (pre-shock) densities of 10$^4$-10$^5$ cm$^{-3}$ 
(but not with the $J$-type models, see top panel of Figure 8 in \citealt{Ka14b}). 
However, the line ratios of H$_2$O / CO, H$_2$O / OH, and CO / OH lines are overestimated by 
stationary C- and J-shock models by 1-2 orders of magnitude with respect to observations. 
In order to reconcile these models and observations, \citet{Ka14b} proposed that some 
H$_2$O is photodissociated by FUV photons and thus its abundance is lowered with respect
to OH and CO. 

A number of low-mass sources show velocity-resolved emission from light (ionized) hydrides,
 e.g., CH, CH$^+$, OH$^+$ \citep{Kr13,Be16}. The velocity structure is such that this
  emission originates in shocks. The observations were compared to models of fully dissociative
   shocks \citep{ND89}: the molecular emission is consistent with the molecules
    reforming in the post-shock zone. The velocities required to dissociate the gas
     ($\gtrsim$ 50 km s$^{-1}$) are inconsistent with the observed velocities ($\sim$ 10 km s$^{-1}$);
      instead, dissociation must come from elsewhere. \citet{Kr13} proposed that the gas is dissociated
       by UV photons from the nearby accreting protostar, and that the observed signatures present
        direct evidence of UV irradiated shocks.
        
Further evidence comes from observations of H$_2$O and high-$J$ CO. When the H$_2$O abundance
 is inferred from comparison to high-$J$ CO 16--15 emission \citep{Kr17b}, a total abundance of 
 $\sim$ 10$^{-6}$ is inferred. In warm shocked gas where $T > 300$ K, all oxygen is expected
  to be driven into H$_2$O \cite[e.g.][]{Be98}, and the total H$_2$O abundance reaches
   $\sim$ 10$^{-4}$. The observed decrease of two orders of magnitude implies that not all
    oxygen is driven into H$_2$O. From a simple chemistry calculation, \citet{Kr17b} showed that the
     abundance is consistent with a moderate UV field dissociating the H$_2$O ($G_0 \sim 10^{-1}-1$),
      thus again demonstrating that UV irradiation likely dominates the chemistry whereas the
       shocks dominate the excitation.

While other solutions cannot be excluded, these would have to explain the multi-species line ratios,
 light (ionized) hydrides, and low absolute water abundance. One specific example of such a potential
  solution is non-stationary C-type shocks \citep{FP99,Le04a,Le04b}. \citet{FP13} found satisfying non-stationary
  fits to PACS observations of CO, H$_2$O, and OH in one protostar, NGC1333 IRAS 4B. However, the predictions
   of these models for the emission from ionised hydrides have not been investigated yet.

In the following sections, we measure the line ratios of H$_2$O / CO, H$_2$O / OH, and CO / OH lines
for the entire sample presented here, and compare them with state-of-the-art models of stationary $C$-type
 shocks illuminated by UV irradiation (Sect. 5.1). These models predict that a significant fraction of molecules are photodissociated, which increases
 the abundances of atomic and ionic species, particularly O and C$^+$ and light ionized hydrides
observed with HIFI \citep{Kr13}. This in turn changes the
  emission spectrum of these atomic species. To test this hypothesis, we first update 
   the observed molecular line ratios for the full sample, introducing the UV-irradiated shocks (\S5.1), and comparing the atomic/ionic
    emission to stationary shock models (\S 5.2), then PDR models (\S 5.3) and finally irradiated
     shock models (\S 5.4). The reason for this approach is to isolate the effects of shocks
      and UV irradiation (PDR) separately, before combining them.
\subsection{Molecular line ratios vs. models of UV-irradiated shocks}
UV photons have an important impact on the chemical and thermal structure of a shock \citep{Les13,MK15}. 
The increase in the ionization fraction results in a tighter coupling between the ionized and neutral fluids
and thus decreased coupling length. The decreased coupling length effectively means
  that the mechanical energy is deposited more rapidly into the gas, leading to 
  higher peak temperatures for a given shock velocity than in fully-shielded shocks. 
In addition, the molecular and atomic abundances depend strongly on the strength of the radiation
 field and the gas density. As illustrated in Figure 1 of \citealt{MK15}, the pre-shock abundance of 
 gas-phase atomic oxygen decreases as the density increases, because more oxygen is frozen out 
 onto the dust grains when shielded from UV photons. At the same time, the oxygen abundance scales  
 directly with the strength of the UV field, which photodissociates H$_2$O. At pre-shock densities
  of 10$^5$ cm$^{-3}$, the atomic oxygen abundance is $10^{-6}$ to $10^{-4}$ for UV fields $G_\mathrm{0}$ of 0.1 to 10. 
 H$_2$O photodissociation also takes place in the post-shock gas, thus efficiently 
  lowering its column-averaged abundance.

 The physical conditions in these irradiated shocks depend very strongly on the charge balance,
 and as such it is crucial to keep accurate track of it through the shock. This is done in the
  shock models of \citet{MK15}, where both the gas and grain charge balances are calculated at each 
  step of the models, and the model results show that the shocks are stable.

Figure \ref{obsrat} show the observed line ratios of selected CO, H$_{2}$O, and OH lines
for the full sample of protostars and the results from stationary C-type shock models \citep{KN96}.
There is a general trend for these shock models to overestimate the observed emission ratios, 
most clearly seen for the H$_2$O $4_{04}-3_{13}$ / OH 84 $\mu$m line ratio.
The results obtained for the extended sample of sources presented here are consistent
with the narrow ranges obtained for the sample of 22 protostars from Perseus \citep{Ka14b}.
Fully-shielded $C-$type shock models from \citet{KN96} still overestimate the ratios 
when compared to the new observations. The same ratios calculated using shock models 
illuminated by UV radiation show a better agreement with the 
observations, for a range of shock parameters \citep[][Kaufman in prep.]{MK15}. 

The largest difference between the UV and non-UV $C-$shocks is seen in the 
ratio of the H$_{2}$O $4_{04}$-$3_{13}$ and OH 84.6 $\mu$m lines, which is reduced in the new models by
two orders of magnitude at pre-shock densities of 10$^4$ and 10$^5$ cm$^{-3}$. 
The ratio of the H$_{2}$O $4_{04}$-$3_{13}$ and CO 16-15 lines is best reproduced by UV-irradiated
$C$-shocks at pre-shock densities of 10$^4$-10$^5$ cm$^{-3}$, but the differences with respect to the 
models without UV are small. Similar conclusions hold also for 
other pairs of H$_{2}$O/CO, H$_{2}$O/OH, and CO/OH transitions (see Figure \ref{obsrat2}),
 as well as for the ratios of total line luminosities in H$_{2}$O / OH etc. (see Sec. 6.1.).
Based on these line ratios and the observed chemical signatures \citep{Kr13}, a scenario 
where emission arises in UV-irradiated stationary shocks seems very likely.

\subsection{Atomic line fluxes vs. stationary shock models}
\begin{figure}[!tb]
\begin{center}
\includegraphics[angle=0,height=10cm]{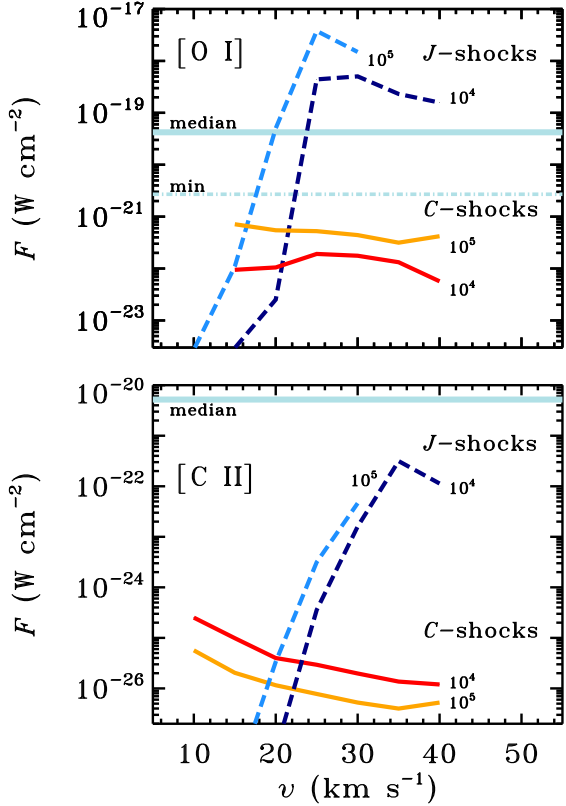}
\vspace{0.5cm}
\caption{\label{absO} Absolute fluxes of the [O I] line at 63 $\mu$m (top) and the [C II] line at 158 $\mu$m
(bottom) predicted by $C$- and $J$-shock models (full lines) of \citet{FP15} and compared with 
observations (horizontal lines). Median observed fluxes 
of [O I] shown in the top panel are calculated only for those 
sources where the emission is associated with the source and not large-scale cloud emission. 
The minimum detected flux of [O I] is shown also for reference.
In case of [C II], in the rare cases that the line is actually detected and spatially 
associated with the source, the fluxes are of the order of magnitude shown on the figure.}
\end{center}
\end{figure}
\begin{figure}[!tb]
\begin{center}
\includegraphics[angle=90,height=6.5cm]{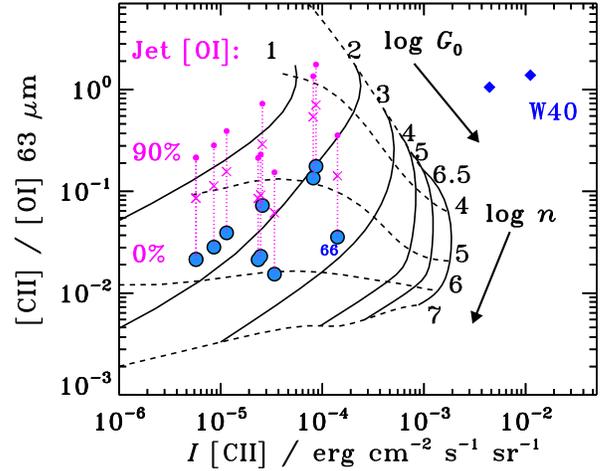}
\caption{\label{pdr} Observed line ratio of the [C II] line at 158 $\mu$m and 
the [O I] line at 63 $\mu$m as function of the [C II] line intensity 
(blue circles) and models of photodissociation regions from 
\citet{Ka99}. Models are shown for densities from $10^{3}$ to $10^{7}$ cm$^{-3}$ (in 
dashed lines) and for incident UV from 10$^{1}$ to 10$^{6.5}$ G$_\mathrm{0}$ (in solid lines). 
Small magenta dots and crosses indicate line ratios assuming that 90\% and 75\% of [O I] flux 
comes from the jet, respectively. Two cloud-related PDR regions in the W40 cloud are shown with blue diamonds. 
}
\end{center}
\end{figure}
Before comparing the observed atomic/ionic emission from O and C$^+$ to irradiated shock models, 
it is instructive to first compare to stationary shock models without external UV radiation.
 To do so, we use the line fluxes from $C$ and $J$-type shock models as presented in \citet{FP15}.

The model intensities in units of erg cm$^{-2}$ s$^{-1}$ sr$^{-1}$ are translated to 
 W cm$^{-2}$ assuming that the emission entirely 
fills one spaxel in the PACS maps (1 sr $\sim 2.1$ $10^{-9}$ $\pi$). A size of one spaxel is chosen 
despite the fact that PACS maps in the [O I]
and [C II] lines often show emission extending to more than one spaxel \citep[this work and][]{Ka13},
 because the actual emitting region observed by each spaxel is likely only a fraction of the spaxel size 
(i.e. the \lq filling factor' is below unity). A possible underestimate of a factor of a few of the emitting region size does not 
affect the conclusions below given the orders of magnitude variations probed by the models and observations.

Comparison of absolute observed and model line fluxes of the [O I] 63 $\mu$m and 
[C II] 158 $\mu$m lines are shown in Figure \ref{absO} (top panel). The median observed [O I] flux,
calculated for sources with emission spatially associated with YSOs, 
 equals $\sim 4\cdot 10^{-20}$ W cm$^{-2}$ 
and falls between the predictions for the $C-$ and $J-$type shock models for a wide range of 
shock velocities. The $C-$type shock model fluxes are insensitive to the shock velocities 
and equal $\sim$10$^{-21}$ W cm$^{-2}$ for a pre-shock density of 
$10^5$ cm$^{-3}$ and $\sim$10$^{-22}$ W cm$^{-2}$ for a pre-shock density of 
$10^4$ cm$^{-3}$, and are clearly inconsistent with the data.

The $J-$type shock model fluxes, on the other hand, fit even the brightest observed [O I] lines.
A sharp increase from about 10$^{-22}$ W cm$^{-2}$ to 10$^{-18}$ W cm$^{-2}$ occurs when the shock 
becomes dissociative, at velocities of about 20 km s$^{-1}$. Depending on the size of the 
emitting region, likely lower than the full spaxel area, the observations of all 
absolute line fluxes of [O I] can be reproduced with $J-$type shocks.

The observed fluxes and limits of the [C II] line at 158 $\mu$m greatly exceed the model predictions 
for both $C-$ and $J-$type shock models (Figure \ref{absO}, bottom panel). Whenever the line is
 detected and associated with 
a YSO, the fluxes exceed 10$^{-21}$ W cm$^{-2}$, 4 orders of magnitude above the model 
$C-$ shock fluxes and a factor of a few above the peak [C II] flux from the 
$J-$ shocks, at velocities of about 30 km s$^{-1}$. In the majority of the sources, however, the [C II] line is 
not detected and only an upper limit can be used for comparisons with the models. For those sources, 
the observed limit is consistent with both $J-$ and $C-$type shocks.

\subsection{Atomic line fluxes vs. models of photodissociation regions}
As shown in the previous section, stationary C-type shock models do not reproduce [O I]
 and [C II] emission (when detected) by several orders of magnitude.
Because [C II] is a well-known PDR tracer, it is instructive to check whether PDR models could 
explain the observed fluxes. 

Producing [C II] emission in sufficient quantities requires hard UV photons with $\lambda$ $<1100$ \AA, likely 
originating in the vicinity of the protostar, in accretion flows at the accretion shock boundary layer 
and / or fast bow-shock \citep[e.g.][]{Sp95,vK09}. Models of photodissociation regions \citep[PDRs,][]{TH85} 
can be used to constrain the UV field and densities using the absolute intensities of atomic and ionic lines 
as well as molecules. 

In our case, where most of the molecular emission and a part of the [O I] 
emission likely originates in shocks, comparisons to the absolute line intensities of [C II] 
are the most reliable measure of FUV alone. To better constrain the range of possible parameters, 
the ratio of [C II] and [O I] can be used, taking into account that a fraction of 
the [O I] emission may arise in from dissociative shocks associated with the jets. This ratio rapidly 
decreases with density, controlled mainly by [O I] with its high critical density of about 
 $\sim 5\cdot 10^5$ cm$^{-3}$, two orders of magnitude higher than for [C II] \citep{Ka99}.

Figure \ref{pdr} compares the observed [C II] / [O I] versus [C II]
intensities for the subset of sources detected in [CII] with the PDR model predictions from \citet{Ka99}. 
The [C II] intensities 
are calculated assuming a size of one spaxel and shown only for sources 
where the [C II] emission is spatially associated with YSOs. The range of observed 
[C II] intensities is $10^{-5}$-$10^{-4}$ erg cm$^{-2}$ s$^{-1}$ sr$^{-1}$,
whereas [C II] / [O I] ranges from $10^{-2}$-$10^{0}$. The corresponding 
model densities range from $10^{5}$ to $10^{6}$ cm$^{-3}$. The majority 
of the sources show a very similar incident UV field, about 10$^{2}$ G$_\mathrm{0}$ on scales of $\sim$1000 AU,
 with the exception of a more massive source, Serpens SMM1 (\#66), which is in the 10$^{3}$ G$_\mathrm{0}$ regime.
 Independent high-angular resolution observations of free-free emission on scales of $<$1000 AU illustrate the presence of a
  strong UV field toward this source \citep{Hu16}. 

Possible contributions from shocks to the [O I] intensity 
would decrease both the densities and the UV field matching the observations (see Figure \ref{pdr}). 
For example, if 90\% of [O I] flux comes from the shock, corresponding to an increase in the observed
[C II] / [O I] by a factor of 10, the best fit densities are in the range 
$10^{4}$-$10^{5}$ cm$^{-3}$ and UV field of $\sim10^{1}$-10$^{2}$ G$_\mathrm{0}$. 
In fact, the spectrally resolved line profile of [O I] in a single outflow position
in NGC1333 I4A, observed with SOFIA-GREAT by \citet{Kr17a}, resembles the 
high-velocity H$_2$O and CO 16-15 profiles associated with the shocks observed with \textit{Herschel} by \citet{San14}. Similarly, 
[C II] profiles from HIFI obtained for a subsample of WISH sources reveal that the origin of emission may be shocks rather than a PDR \citep{Kr13,Be16}.
These factors suggest that the actual position of the PDR-only emission in these sources
 lies above and to the left of the observed fluxes, and so the densities and G$_0$ values should be
  considered as upper limits.

Finally, we note that the diagnostic diagram from Figure \ref{pdr} can be used to distinguish
between PDRs associated with YSOs and clouds. For comparison with the local PDRs, two sources 
observed within a strong PDR in the W40 cloud (W40-MM26 and W40-MM27) are shown. Both of them are located 
in the low-density and high-UV regime that is very distinct from the other observed YSOs.
\begin{figure}[!tb]
\begin{center}  
\includegraphics[angle=0,height=10cm]{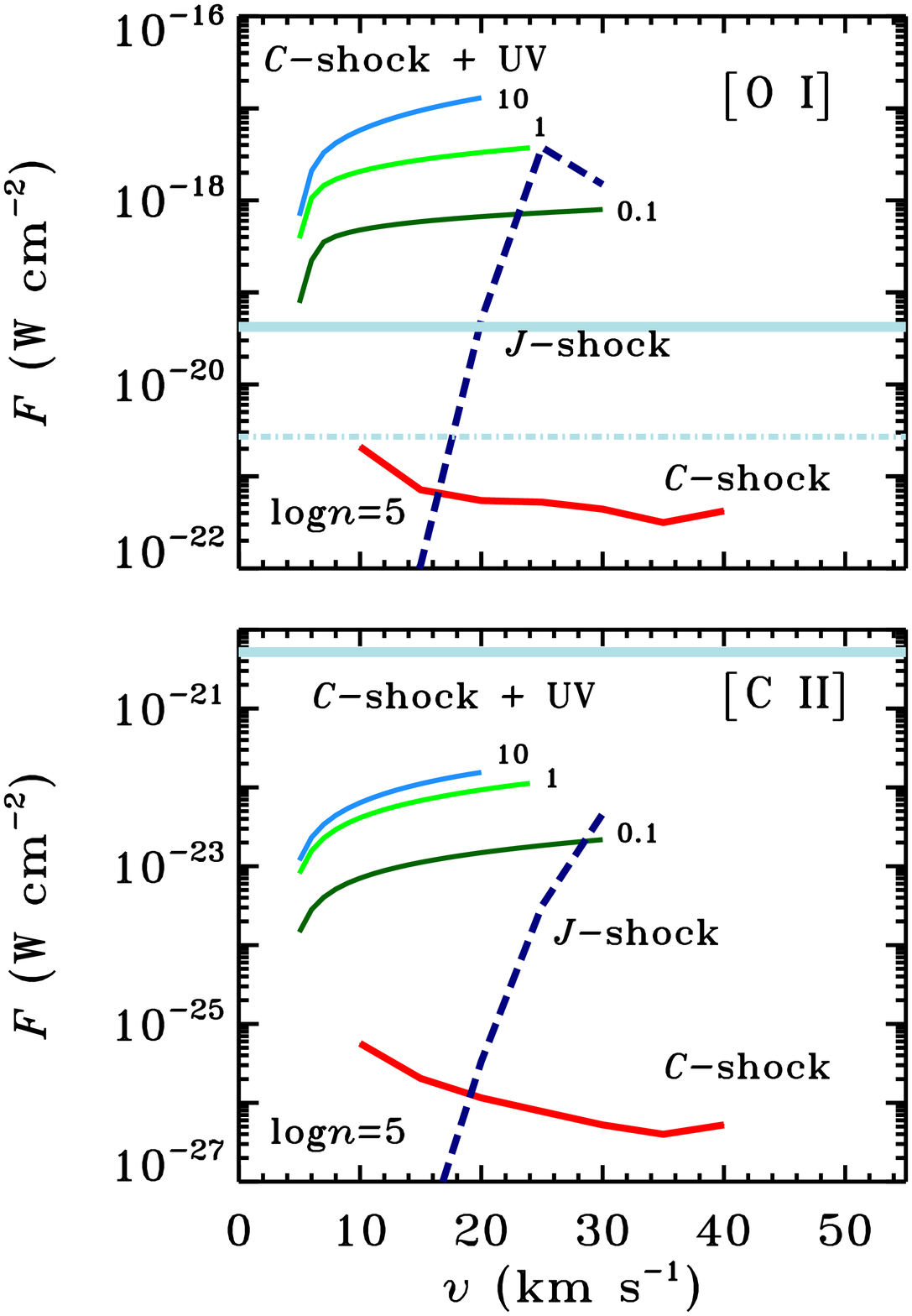}
\vspace{0.5cm}
\caption{\label{abs1} Absolute fluxes of the [O I] line at 63 $\mu$m (top) and the [C II] line at 158 $\mu$m
(bottom) predicted by $C$- (full, red) and $J$-shock (dashed) models of \citet{FP15} and UV-illuminated $C$-shock models \citep[][Kaufman in prep.]{MK15}
for $G_\mathrm{0}$ = 0.1, 1, and 10 (full lines, green to blue). All models assume a pre-shock density of $10^5$ cm$^{-3}$. 
The observed fluxes shown as horizontal lines are the same as in Figure \ref{absO}.}
\end{center}
\end{figure}
\subsection{Atomic line fluxes vs. models of UV-irradiated shocks}
Velocity-resolved spectra clearly suggests that [O I] and [C II] emission is associated with shocks 
\citep[e.g.][]{Be16,Kr17a}. 
 The two previous sections (5.2 and 5.3) suggest that where stationary shocks fail is where PDR
  models succeed, i.e. in reproducing the atomic/ionic emission. This, in turn, suggests that for
   these species, shocks set the dynamics and PDRs set the chemistry and excitation.

The presence of UV photons increases
the ionization fraction in the pre-shock gas, leading to gas that is hotter and more compressed
for a given shock velocity than without FUV photons \citep{MK15}. In addition, the pre-shock gas-phase abundances
are altered and depend strongly on the strength of the radiation field and the gas density. 
For low values of the field, more O-bearing species are locked in ices, while for high fields 
they are desorbed into the gas phase. FUV radiation can also photodissociate H$_2$O in the post-shock
gas, and thus lower the column-averaged abundance of H$_2$O and increase the O abundance 
\citep[see ][for an example calculation]{Kr17b}.

The abundances of atomic species are therefore expected to be significantly higher in UV 
illuminated shocks with respect to models of shielded shocks. Figure \ref{abs1}  
shows that the fluxes of the [O I] and [C II] scale directly with the assumed UV fields 
and indeed are a few orders of magnitude above fully-shielded $C-$shock models \citep{FP15}.
The critical density of [O I] at 63 $\mu$m is $\sim 5\times 10^{5}$ cm$^{-3}$, and so, 
in the models assuming pre-shock densities of $n=10^{5}$ cm$^{-3}$, the absolute line flux quickly
 increases even at low pre-shock velocities ($\varv\sim5-10$ km s$^{-1}$).
 At higher shock velocities, the line flux changes are small, up to a factor of 2 between
 10 and 20 km s$^{-1}$. The maximum shock velocities, above which the shock becomes dissociative,
  are lower than in the fully-shielded 
 $C-$shocks and equal about 30 km s$^{-1}$ for $G_\mathrm{0}=0.1$ and 20 km s$^{-1}$ for $G_\mathrm{0}=10$ 
 (all values for pre-shock densities of $10^{5}$ cm$^{-3}$). Similar trends are seen for the 
 [C II] line, except that the initial rise is shallower due to the lower critical density 
 of the line ($\sim 3\times 10^{3}$ cm$^{-3}$).

The model line fluxes predicted by UV illuminated $C-$shocks agree well 
with the detected [O I] line luminosities and non-detections of [C II] emission for the majority of the sources
(Figure \ref{absO}). The maximum 
value of the [O I] 63 $\mu$m flux observed with PACS is $\sim1.5\cdot 10^{-18}$ W cm$^{-2}$, 
setting the upper limit on UV fields of $G_\mathrm{0}=1$. The median [O I] flux is about two 
orders of magnitude lower, suggesting that the \lq filling factor' of the UV illuminated 
shocks is below unity, similar to the conclusions drawn from the HIFI H$_2$O line energetics 
\citep{Mo14}.
\section{Discussion}
\subsection{Origin of far-IR emission in FUV-irradiated $C$-shocks}
Far-IR line emission observed with \textit{Herschel} toward low-mass protostars is characterized 
by: (i) a spatially extended pattern along the outflow direction, in cases where emission 
is resolved, but not spatially coincident with the low-$J$ CO entrained gas (see Sec. 3.2, Karska et al. 2013, Nisini et al. 2015); (ii) broad molecular line profiles with 
velocities up to $\sim$50 km s$^{-1}$ (\citealt{Kr12,Kr17b}, \citealt{IreneCO}, \citealt{Mo14,Mo17}); (iii)
a universal CO rotational temperature of $\sim$300 K, independent of protostellar luminosity 
(see Sec. 4.1, \citealt{Ma12}, Karska et al. 2013, \citealt{Gr13}, \citealt{Ma15}, \citealt{MJ17}); 
 (iv) an overall cooling budget in far-IR lines similar to predictions of stationary C type shocks 
 (see Sect. 4.2, \citealt{Ka13}, \citealt{Lee14}). Altogether, these characteristics 
 suggest that the bulk of emission arises either in outflow cavity shocks \citep[e.g.][]{Mo14}
 or in the molecular disk wind \citep{Pa12,Yv16}.

Comparisons of specific line fluxes with shock models offer an opportunity to further determine
the pre- and post-shock density of the gas, as well as 
the shock parameters themselves (its velocity and type, dependent on the ionization fraction and the 
presence of a magnetic field). However, many different shock models reproduce the observed
 line fluxes of certain far-IR species for individual targets, resulting in a broad 
range of the derived shock parameters (e.g. \citealt{Sa13}, \citealt{Di13}, \citealt{Be12,Bu14,Lee14}). 
A more homogeneous comparison of $\sim$20 protostars in Perseus 
with stationary $C$- and $J$- type shock models from \citet{KN96} and \citet{FP10} illustrated that in fact
none of those models can reproduce line ratios of different species (e.g. H$_2$O / CO, H$_2$O / OH,
 and CO / OH, \citealt{Ka14b}). Additionally, stationary shock models alone cannot reproduce the observed CO ladders 
of Class 0 / I protostars \citep{vK10,Vi12}.

In Sec. 5, we showed the model predictions of far-IR spectra for shocks propagating at 
pre-shock densities of 10$^3$-10$^5$ cm$^{-3}$ and irradiated by UV fields of G$_0$ equal 0.1, 1 and 10 
\citep[][Kaufman in prep.]{MK15}. Line ratios such as H$_2$O $4_{04}-3_{13}$ / OH 84 $\mu$m 
generally decrease in the presence of UV photons that photodissociate H$_2$O into OH and O.
Stationary $C$-type shocks also clearly cannot account for the emission in O and C$^{+}$ (Fig. 11), 
but separating the emission from the shock and from the possible PDR is non-trivial with 
unresolved line profiles. The order of magnitude estimates of UV fields with PDR 
models require assumptions about the contribution of the jet in the [O I] emission (Fig. 12).
\begin{figure}[t]
\hspace{-0.5cm}
\includegraphics[angle=90,height=9cm]{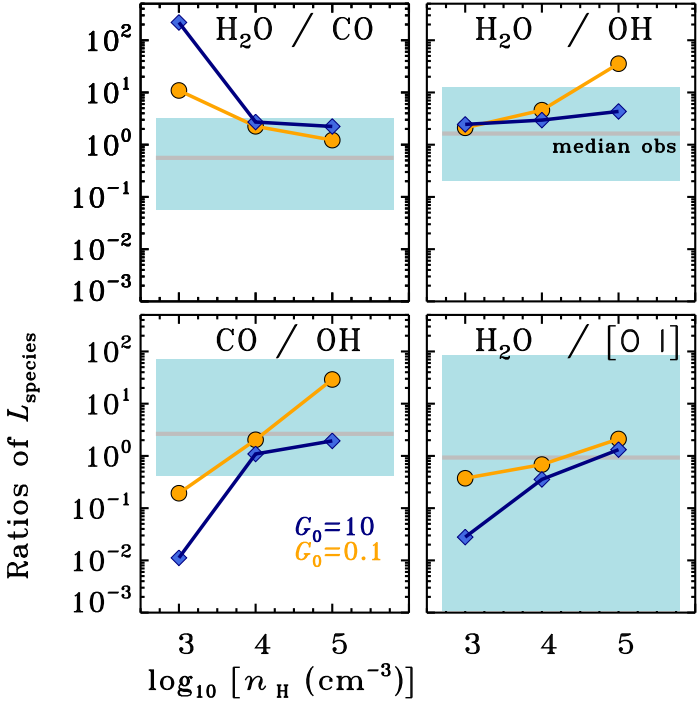}
\vspace{0.5cm}
\caption{\label{fuv_summary} Ratios of total line luminosities in various far-infrared species 
predicted by UV illuminated $C-$type shock models (solid lines) and calculated from the observations (light blue boxes) 
as a function of pre-shock densities. All models correspond to shock velocities $\varv$ equal 20 km s$^{-1}$ and UV fields parametrized by $G_\mathrm{0}$ equal  
10 (in blue) and 0.1 (in orange). 
Observed minimum and maximum values of line ratios are calculated excluding 
lower limits. The median values from observations are shown as horizontal lines.} 
\end{figure}

The $C$ shock models illuminated by FUV are in agreement with \textit{selected} line ratios presented
 in \citet{Ka14b}. A comparison involving \textit{all} observed transitions  
is shown in Figure \ref{fuv_summary}. The total far-infrared line luminosities in H$_2$O, CO, and OH were calculated
in Sec. 4.2, and their sum is indicated as $L_\mathrm{FIRL}$. The total line luminosities 
from the models are calculated using the transitions accessible with PACS: the H$_2$O lines located in the 
56 - 94.7 $\mu$m and 108.0 - 180.5 $\mu$m ranges, the CO lines with $J_\mathrm{up}=14-25$ and 
$26-49$, and 8 OH doublets (at 65, 71, 79, 84, 119, and 163 $\mu$m). 
The observed ratios of H$_2$O / CO, H$_2$O / OH, CO / OH, and H$_2$O / [O I] show a 
Gaussian distribution and therefore are representative for the entire sample (Figure B.8). 

The trends in the model line ratios result from the amount of oxygen available in the gas phase and the 
length scale over which the temperature and abundances are elevated behind the shock. The abundance 
of oxygen scales almost linearly with the logarithm of $G_0$ / $n$ \citep[][Kaufman in prep.]{MK15}, and thus, for stronger  
UV fields and for a given pre-shock density, the abundance of H$_2$O is larger (see the trends 
in the H$_2$O / CO ratio). Additionally, the length scale is inversely proportional to the 
square root of pre-shock density ($L\sim\frac{1}{\sqrt{n}}$) at the assumed $A_\mathrm{V}=10$, so for larger pre-shock densities, 
the shock-averaged column of H$_2$O decreases \citep[see Fig. 2 in ][]{MK15}. 

Since H$_2$O is efficiently formed at $T>230$ K by hydrogenation of OH \citep{Be98}, the peak abundance 
of H$_2$O behind the shock corresponds to the drop in the OH abundance. Further out from the 
shock, H$_2$O is photodissociated by the UV photons, and the higher $G_0$, the lower 
H$_2$O / OH ratio (see Fig. \ref{fuv_summary}). The increase in OH abundances with increasing 
UV field also explains the decrease of the CO / OH ratio for $G_{0}=10$. The trends in H$_2$O / [O I] 
resembles that of H$_2$O / OH, except that at densities as low as 10$^3$ cm$^{-3}$ both
H$_2$O and OH are efficiently photodissociated, and the ratio of the H$_2$O / [O I] 
for high $G_0$ drops to $\sim10^{-2}$. The balance between O, OH, and H$_2$O 
depends also on the H/H$_2$ ratio and temperature, which has to be considered.

The ratios of total line luminosities calculated from the models and from the observations 
show a good agreement with the observed range of pre-shock densities, 10$^4$-10$^5$ cm$^{-3}$. The H$_2$O / OH 
and H$_2$O / [O I] ratios are also explained by the pre-shock densities of 10$^3$ cm$^{-3}$, 
but CO is not well-reproduced at these lower densities. The UV fields considered here ($G_{0}$
 of 0.1 and 10) and illuminating $C$ shocks produce model line ratios in the same 
 range as the observed ratios. 

\subsection{Impact of UV irradiation on the CO ladder}

The shape of the CO ladder in the far-IR is universal across Class 0 and I protostars (Sec. 4.1, e.g. \citealt{Ma12,Ka13,Gr13,Li14}) as well 
as across several orders of magnitudes in luminosity \citep{Ka14,Ma15,Mas15}. In the energy range covered by $J_\mathrm{up}$=14-24, the CO rotational diagram
 typically consists of a single linear component, corresponding to the temperature of $320$ K with 
 a standard deviation $<20$\% (the so-called \lq warm' component). Out of 90 sources from our
  sample, 59 protostars show the \lq warm' component. Out of those protostars, 21 sources (more than 1/3) show an additional
   linear component corresponding to the temperature of $\sim720$ K with a standard deviation $<20$\%
  (the so-called \lq hot' component). 

Several interpretations have been proposed to explain the shape of the CO ladder in low-mass protostars. 
\citet{Vi12} modeled the shape of the PACS CO ladder in three objects with a combination of passively heated envelope, 
UV-heated cavity walls and small-scale $C$-type shocks. The bulk of emission was 
attributed to a PDR and shocks, with a tentative trend of increasing role of UV photons 
for more evolved sources. Indeed, more recent models confirm that the 300 K component in CO ladders
can be well-reproduced even by pure UV heating of the outflow cavities \citep{Lee14b,Lee15}. 
Observationally, the presence of UV photons is supported 
by extended emission in $^{13}$CO $J$=6-5, which delineates the outflow cavity walls in nearby 
protostars \citep{Yi12,Yi15}. Yet, the CO transitions in a broad range of $J$ levels cannot be explained 
by UV heating alone.

\citet{Ne12} reproduced the curved shape of the rotational diagram using a single component fit, 
assuming a subthermal excitation of CO (densities of $\sim$10$^3$ cm$^{-3}$ and temperatures above 1000 K). 
Apart from the the sources with the highest-$J$ CO transitions, successful fits of this kind were 
obtained for most of the protostars observed in the HOPS and DIGIT key programs \citep{Ma12,Gr13}. 
An alternative approach considers the ladder as the combination of two distinct physical components,
with densities above $\sim$10$^4$-10$^5$ cm$^{-3}$ \citep[e.g.][]{Ka13}. \citet{FP13} are associating 
those two components with $C$ (\lq warm' component) and $J$ (\lq hot' component) shocks, while 
\citet{Ne14} explain them as a combination of shocks with different velocities. A comprehensive 
analysis of the high-$J$ CO line profiles from HIFI associates two distinct velocity components with 
two excitation components, corresponding to the warm and hot PACS components \citep{Kr17b}.

\begin{figure}[!tb]
\hspace{-1cm}
\includegraphics[angle=90,height=8.5cm]{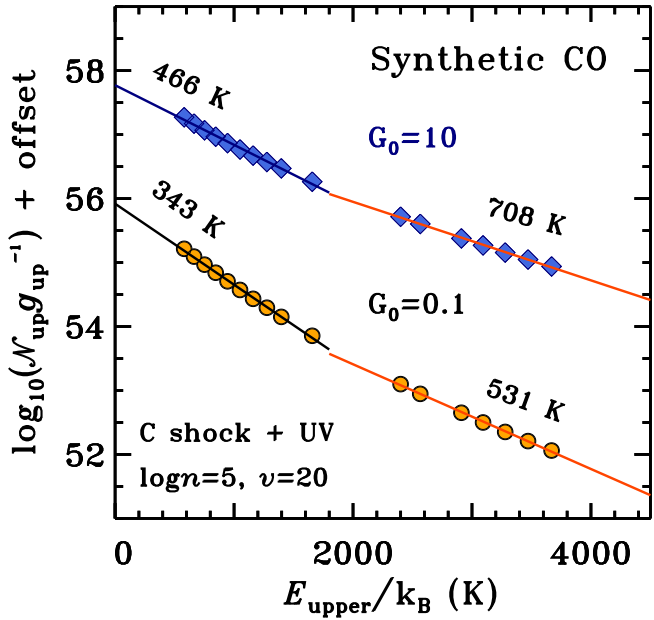}
\caption{\label{comodel} Synthetic CO ladders calculated using C shock models irradiated 
by UV photons \citep[][Kaufman in prep.]{MK15}. The models are calculated for a 
pre-shock density of $\sim$10$^5$ cm$^{-3}$ and a shock velocity of $\varv=20$ km s$^{-1}$ 
and two values of UV fields: $G_0$=0.1 (yellow) and 10 (blue).} 
\end{figure}

As shown in previous section, the observed far-IR line fluxes are mostly produced in $C$ type shocks, 
likely ubiquitous in the regions of outflow-envelope interactions. At the same time, there is a strong
spatial and kinematic association of high$-J$ CO emission with H$_2$O \citep{Kr12,Kr17b,Yi13,Mo14,Mo17}. 
The required pre-shock densities are of order $10^5-10^6$ cm$^{-3}$, translating to densities 
$>10^6$ cm$^{-3}$ adopting a compression factor of $\gtrsim10$ in these $C$ shocks \citep[see Sec. 5.3 in][]{Ka13}.

The only feasible alternative 
to $C$-type shocks are molecular disk winds, where excitation of H$_2$O and high-$J$ CO
is caused by similar processes as in molecular shocks \citep{Pa12,Yv16}. 
The wind models match the velocity-resolved H$_2$O 557 GHz and
higher-excited line profiles very well, but predictions for high-$J$ CO lines such as
presented here are not yet available. 

Fully-shielded, stationary $C$ shock models cannot reproduce at once the entire CO ladder  
\citep{vK10,Vi12}. It is therefore interesting to explore the effect of UV irradiation of shocks
on the shape of those CO ladders. Figure \ref{comodel} compares two C shock models calculated for UV fields
$G_0$ equal of 0.1 and 10 \citep[][Kaufman in prep.]{MK15}, for a single 
pre-shock density of $\sim$10$^5$ cm$^{-3}$ and a shock velocity of $\varv=20$ km s$^{-1}$. 
The datapoints correspond to the transitions located in the PACS range up to $J_\mathrm{up}=36$ (at 72.8 $\mu$m).

Two linear fits to the diagrams give rotational temperatures of $\sim$340 K and $\sim530$ K 
for the $G_0$ of 0.1 and temperatures of $\sim470$ K and $\sim710$ K for the $G_0$ of 10. 
Clearly, in the presence of very low UV fields ($G_0$=0.1), the CO rotational temperature of the \lq warm' 
component is in good agreement with observations (see Fig. 6). The temperature 
of the \lq hot' component is within the observed ranges as well. Similarly, the model 
with $G_0$ of 10 shows the \lq hot' component characterized by a higher, but still consistent 
with observations, temperature of about $\sim710$ K. The \lq warm' component temperature here 
is at the higher end of the observations (Fig. 6). A full grid of models is required
 to assess whether a single model can account for the entire CO ladder. 

\begin{deluxetable}{lccccccccc}[b!]
\tablecaption{Evolution of far-IR line luminosities \label{lum} }
\tablecolumns{5}
\tablehead{
\colhead{~} & \multicolumn{2}{c}{Class 0} & \multicolumn{2}{c}{Class I} \\ \cline{2-3}  \cline{4-5}
\colhead{~} & \colhead{ mean} & \colhead{median} & \colhead{mean} & \colhead{median} 
}
\startdata
$\log_{10} (L_\mathrm{CO}$/$L_\mathrm{bol}$) & -3.37 & -3.21 & -3.66 & -3.56 \\
$\log_{10} (L_\mathrm{CO}$/$L_\mathrm{FIRL}$) & -0.47 & -0.35 & -0.66 & -0.53 \\
&  &   &  &  \\
$\log_{10} (L_{\mathrm{H}_2\mathrm{O}}$/$L_\mathrm{bol}$) & -3.60  & -3.50 & -3.73 & -3.57 \\
$\log_{10} (L_{\mathrm{H}_2\mathrm{O}}$/$L_\mathrm{FIRL}$) & -0.83 & -0.71 & -0.85 &  -0.75 \\
&  &   &  &  \\
$\log_{10} (L_\mathrm{OH}$/$L_\mathrm{bol}$) & -3.97 & -3.86 & -3.85 & -3.82 \\
$\log_{10} (L_\mathrm{OH}$/$L_\mathrm{FIRL}$) & -1.18 & -0.98 & -0.90 & -0.81 \\
&  &   &  &  \\
$\log_{10} (L_\mathrm{[OI]}$/$L_\mathrm{bol}$) & -3.53 &  -3.55 & -3.52 & -3.65 \\
$\log_{10} (L_\mathrm{[OI]}$/$L_\mathrm{FIRL}$) & -0.81 & -0.75 & -0.60 & -0.55 \\
&  &   &  &  \\
$\log_{10} (L_{\mathrm{H}_2\mathrm{O}}$/$L_\mathrm{CO}$) & -0.30 & -0.33 & -0.19 & -0.13 \\
$\log_{10} (L_{\mathrm{H}_2\mathrm{O}}$/$L_\mathrm{OH}$) & 0.42 &  0.32 & 0.07 & 0.07\\
$\log_{10} (L_\mathrm{CO}$/$L_\mathrm{OH}$) & 0.68 & 0.68 & 0.26 & 0.23 \\
$\log_{10} (L_{\mathrm{H}_2\mathrm{O}}$/$L_\mathrm{[OI]}$) & -0.003 & 0.04 & -0.23 & -0.24 \\
\enddata
\end{deluxetable}

Observationally, the high-temperature component at CO ladders is detected in 21 sources
(Sec. 4.1.) and can be associated with a specific Gaussian component in the velocity-resolved 
line profiles from HIFI \citep{Kr17b}. The fraction of emission 
in this particular profile component compared to the rest of the emission, associated with 
the bulk outflow, is $\sim$20\% (hot component) and $\sim$80\% (warm component) of the total line emission. Thus, in agreement with the fractions 
 contributed by the \lq warm' and \lq hot' components seen on CO rotational diagrams. 
 The absence of the hot component in the majority of the sources is due to low S/N rather 
 than the presence or lack of the UV radiation.

In summary, preliminary models of $C$-shocks irradiated by FUV qualitatively explain the decrease
in the amount of H$_2$O relative to e.g. CO and OH, and perhaps the excitation of the highest-$J$ CO lines. 
The average values of UV fields that match our observations are of the order of 0.1-10 times 
the interstellar radiation field. The order-of-magnitude
 estimates of UV fields presented here are in agreement 
with a recent survey of ionized hydrides, where UV fields determined in low-mass 
protostars are 2-400 times the interstellar value \citep{Be16}. 

The next step is to calculate a full grid of shock models to verify that the
conclusions reached here, based on just a few model calculations, are
robust. This will further assist in demonstrating that the \lq warm'
component is reproduced by a large range of shock conditions.

\subsection{Evolution of the far-infrared line emission}

The changes in physical processes and conditions as the protostar
evolves from Class 0 to Class I should be reflected in the cooling by various atomic and 
molecular species and the total far-IR gas cooling budget. 

As demonstrated in Figure \ref{hist_new} (\S 4.2.2), the distributions of line luminosities 
over $L_\mathrm{bol}$ for each far-IR species are similar between Class 0 and Class I protostars 
except for CO. The overall line 
cooling is dominated by CO and H$_2$O, followed by [O I] and OH, consistent with previous
 surveys \citep{Ka13}. The spread in line luminosities -- both absolute and divided by $L_\mathrm{bol}$ --
is significant for both Class 0 and I protostars (see Table \ref{lum}). In order to track the evolutionary 
 changes, we therefore normalize the line cooling in each species with the total 
 line luminosity in the far-IR, defined by \citealt{Ni02} as $L_\mathrm{FIRL}=L_\mathrm{CO}+L_\mathrm{[O I]}+L_{\mathrm{H}_2\mathrm{O}}+L_\mathrm{OH}$.
 As a result, the spread of values within each class of objects decreases (see Table \ref{lum}).
\begin{figure}[!tb]
\hspace{-2cm}

\includegraphics[angle=90,height=7cm]{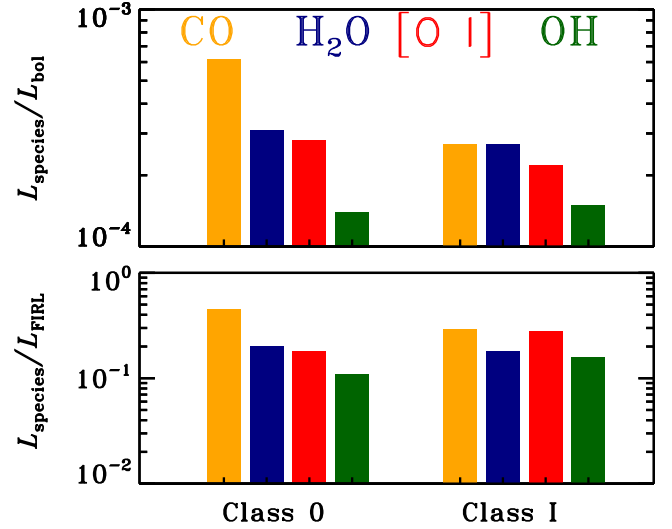}

\caption{\label{firevol} Median line cooling in CO, H$_2$O, [O I], and OH 
over bolometric luminosities (top) and total far-IR line cooling (bottom) for Class 0 and Class I sources.}
\end{figure}

Figure \ref{firevol} illustrates how line luminosities of CO, H$_2$O, [OI], and OH 
change between the Class 0 and I stages. The dominant cooling in the CO lines 
in Class 0 protostars decreases by a factor of $\sim$2.3. On the other hand, the line cooling 
in H$_2$O, [O I], and OH do not change significantly. This is confirmed by  
two-sample Kolmogorov-Smirnov (K-S) tests comparing the ratios of line luminosities over 
$L_\mathrm{bol}$ for Class 0 and I sources. Only the $L\mathrm{(CO)}/L_\mathrm{bol}$ ratio shows
 a statistically significant difference between these evolutionary stages 
 (6\% chance of being drawn from the same distribution). Thus, the less evolved 
sources emit more copiously in CO, which - unlike H$_2$O - is not easily destroyed by photodissociation \citep{Kr17b}.

\begin{figure}[h]
\hspace{-1cm}
\includegraphics[angle=90,height=7.5cm]{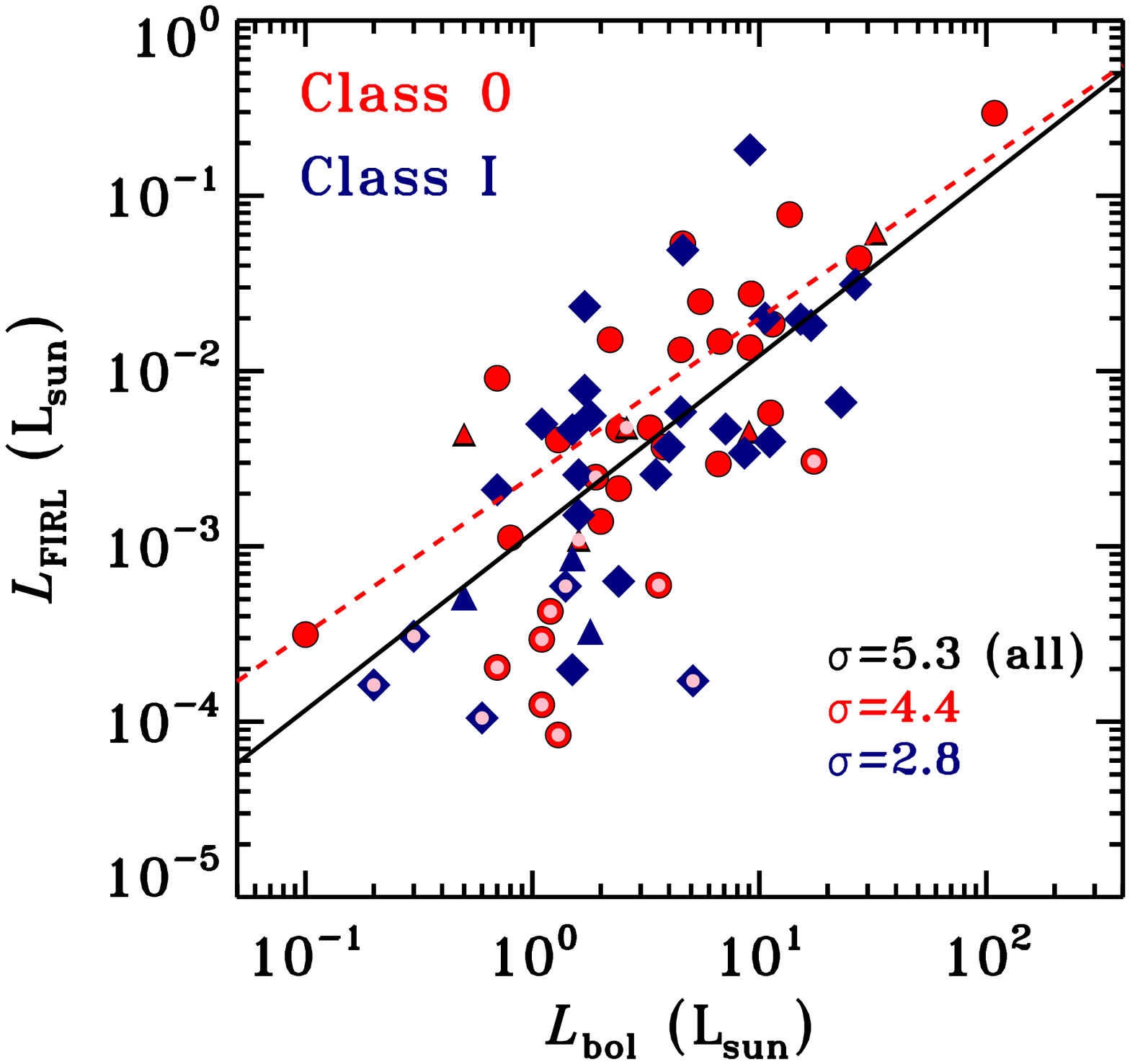}

\hspace{-1cm}
\includegraphics[angle=90,height=7.5cm]{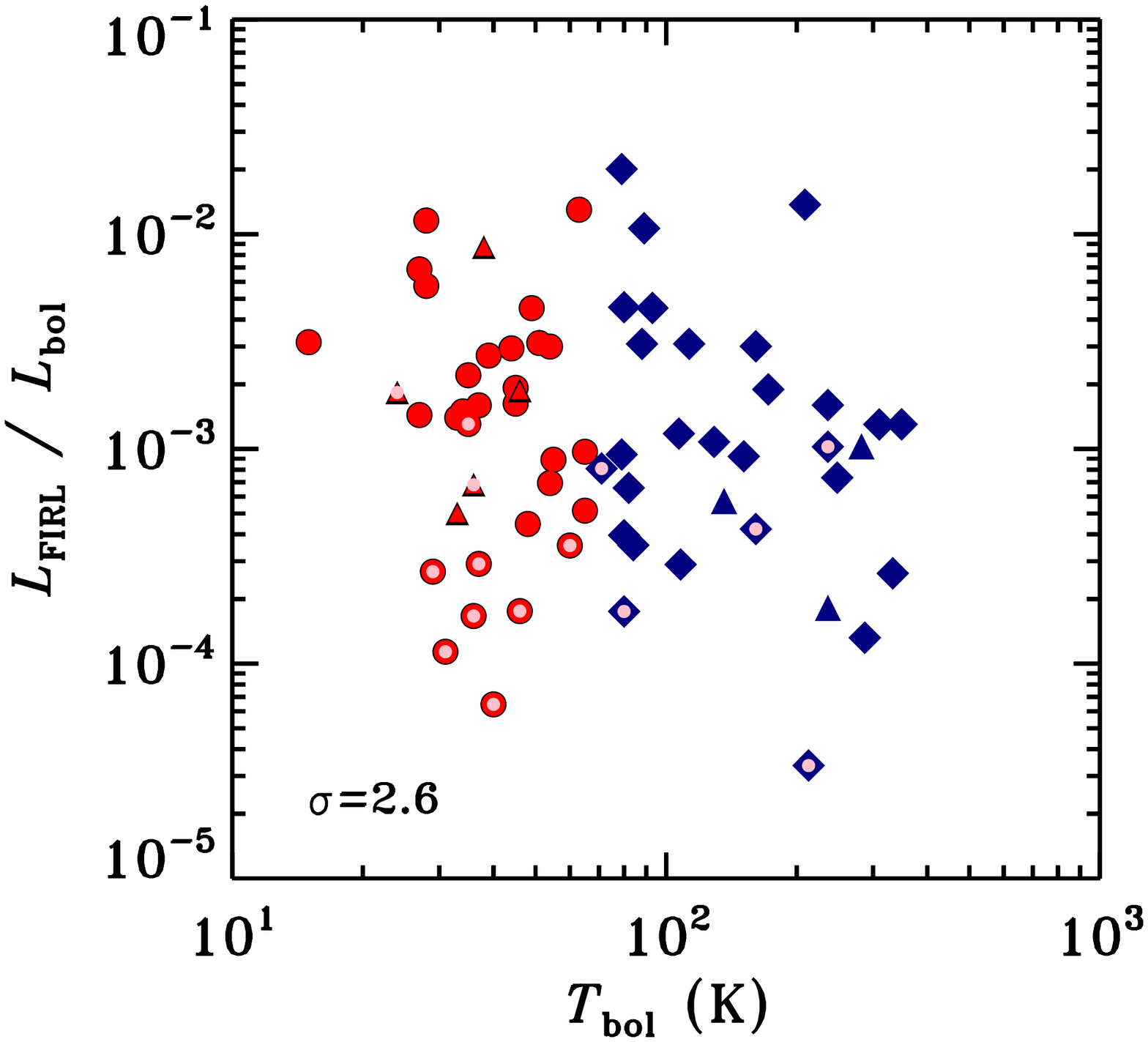}
\caption{\label{firlbol} \textit{(Top)} Total far-infrared line luminosity ($L_\mathrm{FIRL}$) versus bolometric 
luminosity for Class 0 (red symbols) and Class I sources (blue symbols), respectively. 
Triangles show the sources where lower limits of $L_\mathrm{FIRL}$ are calculated. Pink dots 
denote sources where the line cooling in at least one species is 0. The black solid line shows a fit 
to the entire sample and the red dashed line shows a fit to the Class 0 sources with at least one 
detection in each species and with lower limits excluded. \textit{(Bottom)} The same as above, but $L_\mathrm{FIRL}$/$L_\mathrm{bol}$ is plotted versus the bolometric 
temperature.} 
\end{figure}

The redistribution of cooling in molecular and atomic channels does not strongly affect 
the total far-IR line cooling ($L_\mathrm{FIRL}$), which decreases only by a factor of 1.2, 
from $4.5\times10^{-3}$ L$_{\odot}$ in Class 0 to $3.7\times10^{-3}$ L$_{\odot}$ in Class I. 
$L_\mathrm{FIRL}$ shows no significant correlation with $T_\mathrm{bol}$, but  
clearly correlates with $L_\mathrm{bol}$ (Figure \ref{firlbol}), consistent with previous studies
\citep{Ni02,Ka13}. The difference between Class 0 and I objects is confirmed by K-S tests 
with probability of less than 1\% that the $L_\mathrm{FIRL}$ are drawn from the same distribution.
Removal of the $L_\mathrm{FIRL}$-$L_\mathrm{bol}$ dependence, i.e. the 
quantity $L_\mathrm{FIRL}$/$L_\mathrm{bol}$, would allow us to isolate a very weak relation with the
 bolometric temperature, but the K-S test fails due to the large spread of values.
 This large spread of $L_\mathrm{FIRL}$/$L_\mathrm{bol}$ values on 
 top of the decrease in total far-IR cooling with time shows that the processes responsible 
 for the line cooling likely depend on object-to-object characteristics e.g. local 
 environment and the amount of UV radiation illuminating the shocked gas. 

An almost linear relation between 
the total far-IR cooling and bolometric luminosity ($L_\mathrm{FIRL}\sim10^{-2}$ $L_\mathrm{bol}$) 
was first reported based on ISO data and interpreted as a consequence of a wind/jet interactions 
with the surroundings, driven by protostellar accretion and subsequent ejection \citep{Gi01,Ni02}. 
In fact, $L_\mathrm{FIRL}$ can be used as a direct measure of mechanical luminosity 
deposited by the jet/outflow \citep{Ma09}. For $\varv=15$ km s$^{-1}$ shocks propagating 
into the ambient density of $10^{^5}$ cm$^{-3}$ and illuminated by UV fields $G_{0}=1$ (\S 5), 
the mass loss rates corresponding to the observed line cooling can be calculated 
following Equation 2 from \citealt{Ma09}:
\begin{equation}
\frac{1}{2} \dot M \varv ^2 = (1 - f_m) \frac{1}{f_x} L_\mathrm{FIRL}.
\end{equation}
Adopting $(1-f_m)\sim 0.75$ and $f_x\sim0.25$, where $f_m$ is the fraction of shock mechanical energy translated into excitation, 
and $f_x$ is the fraction of cooling due to far-IR transitions of CO, H$_2$O,
[O I], and OH for the best-fitting shock model (the remaining cooling contributed by H$_2$), we find mass-loss rates of, $\dot M $ 
$\sim5\times10^{-7}$ M$_\odot$ yr$^{-1}$ for Class 0 protostars and $\dot M \sim4\times10^{-7}$ M$_\odot$ yr$^{-1}$ 
for Class I protostars. 

Mass loss rates of the order of $10^{-7}$ M$_\odot$ yr$^{-1}$ are consistent with 
the rates calculated using the fully-sampled maps of outflows from Class 0 protostars obtained in the 
[O I] line at 63 $\mu$m \citep{Ni15,Mo17}. Using such maps, the [O I] jet tangential velocity and its length projected
 onto the plane of the sky can be determined, but similar observations are not available 
 for a statistical sample of protostars. In fact, the sources mapped as part of this project 
 are well-known for their bright outflows and the derived mass loss rates from the [O I] 
 line are likely at the higher end of the expected values. Thus, the total far-IR 
 line cooling coupled with shock models seems to be a valid method of determining the 
 mass loss rates for protostars and more reliable than the [O I] alone.
 
  \citet{Mo17} found that the ratio of mass in the entrained outflow to envelope mass 
  vary between $\sim$0.1 and 10\%, peaking at about 1\% for both Class 0 and I protostars in the WILL sample.
  In this context, similar values of the mass loss rates found here are generally in 
  agreement with their argument in favor of an approximately constant entrainment efficiency per unit length 
  during Class 0/I. The relatively low values of mass loss rate of $10^{-7}$ M$_\odot$ yr$^{-1}$ suggest that the 
  the core-to-star formation efficiency is likely higher than 50\%.
   
\section{Summary and conclusions}

\textit{Herschel} / PACS observations in the surroundings of low-mass protostars
 characterize the cooling of the gas at the earliest stages of star formation. 
 A combination of atomic and molecular tracers provides hints about the underlying 
 heating processes that produce the hot gas observed here. The bulk of molecular emission 
 traces non-dissociative shocks occurring in the outflows. 
  UV photons influence the chemistry in those shocks, altering 
 the abundances of key cooling species. The main conclusions, addressing each of the 
 questions raised in the introduction, are the following:
 
 \begin{enumerate}
 
 \item Rich far-IR line emission of H$_2$O, CO, OH and [O I] is detected in $\sim70$\% 
 of the targeted sources, allowing 
 the statistical and unbiased analysis of such emission in the largest sample of protostars so far. 
 
 \item Lines of H$_2$O and CO $J_\mathrm{up}\geq14$ are strongly correlated and a 
 weaker relation is found between the molecular species and [O I]. 
 
 \item 56\% of sources show compact atomic and molecular emission on $10^3$ AU scales.
 The remaining sources show extended emission on at least $10^4$ AU scales  
 covered within our survey, mostly in [O I]. 25 sources (28\%) show clear extended emission in molecular 
 tracers. Among 17 sources which show extended emission in both CO and H$_2$O lines, 11 are 
 Class 0 objects. At the same time, only 5 out of 13 sources with extended emission in [O I]
  are Class 0 objects. The emission is extended along the 
   direction of the outflow seen in the CO 3-2 maps of the regions, but not emitting from the same gas \citep{Mo17}. 
   
\item In 10 sources for which the [O I] shows 
 extended emission on $10^4$ AU scales, a high-velocity ($\varv\geq 90$ km s$^{-1}$) 
 component emerges in the line profiles. Thus, [O I] is closely related to the 
 atomic jet, as suggested by previous studies \citep[e.g.][]{Ni15}. 
 However, the bulk of the [OI] emission is spectrally unresolved.
 
 \item Analysis of the molecular excitation using rotational diagrams gives a median  
 $T_\mathrm{rot}\sim$320 K for CO ($E_\mathrm{up}\leq$1800 K), $T_\mathrm{rot}\sim$130 K for H$_2$O ($E_\mathrm{up}\leq$700 K), and 
 $T_\mathrm{rot}\sim$85 K for OH. Additionally, emission from hot CO and H$_2$O is detected for 24\% and 10\% of sources, respectively, 
 covering a broad range of rotational temperatures with a median at $\sim720$ K for CO and at $\sim410$ K for H$_2$O.
 Densities $\gtrsim 10^4$ cm$^{-3}$ and kinetic temperatures $\gtrsim 300$ K are required 
 to explain the excitation.
 
 \item The far-IR line cooling budget in Class 0 sources is dominated by CO ($\sim$40\%), 
 followed by H$_2$O and [O I] ($\gtrsim20$\%). In Class I sources, the cooling is dominated 
 by [O I] ($\sim40$\%), and CO and H$_2$O ($\gtrsim20$\%). The fractions of cooling contributed
  by various species do not depend strongly on bolometric luminosity. 

\item Line flux ratios of different molecules (e.g., H$_2$O/OH, H$_2$O/CO) suggest that the bulk
 of emission does not arise in standard $C$-type shocks. These ratios, along with observations
  of light hydrides \citep{Kr13,Be16} and low H$_2$O absolute abundances
   \citep{Kr17b} instead point to emission originating in UV-irradiated shocks.

\item Models of UV-irradiated shocks successfully reproduce molecular line ratios for pre-shock
 densities of $\sim10^5$ cm$^{-3}$ and $G_0$ of $\sim0.1-10$. More detailed modeling is required to match emission from
  individual sources to specific shock models.

 \item Comparisons of the atomic line fluxes with stationary shock models show that 
 the observations cannot be reproduced with non-dissociative $C$-shocks. While the 
 [O I] fluxes could be mitigated with an additional $J$-shock component, the [CII] 
 line detected in 10 sources requires UV irradiation. Comparisons to models of
  photodissociation regions allow us to estimate the strength of the UV fields $G_\mathrm{0}\sim10$,
 assuming that 90\% of the [O I] emission arises in the dissociative jet shocks.
 
 \item New models of $C$-shocks illuminated by UV can reproduce the absolute 
 line fluxes of [O I] and non-detections of [C II]. 

 \item Far-IR line cooling does not change significantly from the Class 0 to Class I stage. 
As a proxy of the mechanical energy deposited by the jet/outflow, it allows the calculation of 
mass loss rates of $10^{-7}$ M$_\odot$ yr$^{-1}$. These relatively low values suggest that  
  the core-to-star formation efficiency is likely higher than 50\%. The nearly constant 
  value obtained for Class 0 and I sources implies that the accretion/ejection and associated
  feedback mechanisms are comparable across the protostellar evolution.

 \end{enumerate}
   
\acknowledgments
\noindent \textbf{Acknowledgments}\\
We wish to thank the entire WISH, DIGIT, and WILL teams for 
stimulating discussions and the common effort in analysing and interpreting 
the \textit{Herschel} observations.
AK acknowledges support from the Foundation for Polish Science (FNP) and the Polish 
National Science Center grants 2013/11/N/ST9/00400 and 2016/21/D/ST9/01098.
Sub-millimeter astronomy in Copenhagen is supported by the European Research Council (ERC)
 under the European Union’s Horizon 2020 research and innovation programme (grant agreement No 646908)
 through ERC Consolidator Grant “S4F”. Herschel was an ESA space observatory with science instruments provided
by the European-led Principal Investigator consortia and with important
participation from NASA. Astrochemistry in Leiden is supported by the Netherlands Research
School for Astronomy (NOVA), by a Royal Netherlands Academy of Arts
and Sciences (KNAW) professor prize, by a Spinoza grant and grant
614.001.008 from the Netherlands Organisation for Scientific Research
(NWO). JCM acknowledges support from the European Research Council under the 
European Community’s Horizon 2020 framework program (2014-2020) via the 
ERC Consolidator grant `From Cloud to Star Formation (CSF)’ (project number 648505).
 Support for this work, part of the Herschel Open Time Key Project Program,
 was provided by NASA through an award issued by the Jet Propulsion Laboratory, 
 California Institute of Technology. NS acknowledges the support from the Polish National Science Center
grant 2014/15/B/ST9/02111. Research conducted within the scope of the HECOLS
  International Associated Laboratory, supported in part by the Polish NCN grant DEC-2013/08/M/ST9/00664.
\bibliographystyle{apj}
\bibliography{biblio14}

\appendix
\renewcommand\thefigure{\thesection.\arabic{figure}}    
\setcounter{figure}{0}    
\section{Rotational diagrams}
\begin{figure*}[!tb]
\begin{center}
\includegraphics[angle=0,height=16cm]{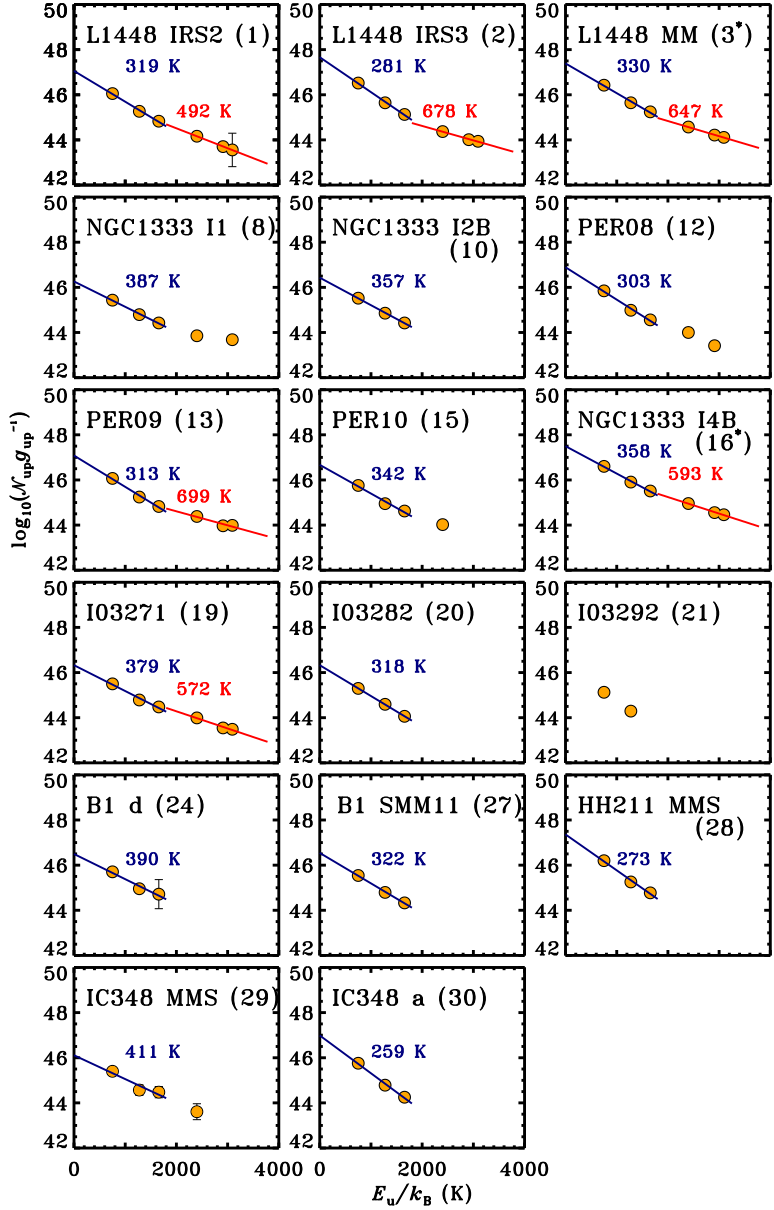}
\caption{\label{cowill} Rotational diagrams of CO for Perseus sources observed as part 
of the WILL program. Blue and red solid lines show linear fits to the data and the
    corresponding rotational temperatures, when at least 3 lines are detected in a given 
    component. 
The numbers in brackets correspond to the object identification numbers (see Table 1). 
L1448 MM and NGC1333 I4B are shown here but not listed separately in the Table \ref{exc_table} because 
existing full scans from DIGIT and WISH programs for those sources provide better estimates 
of rotational temperatures.}
\end{center}
\end{figure*}

\begin{figure*}[!tb]
\begin{center}
\includegraphics[angle=0,height=16cm]{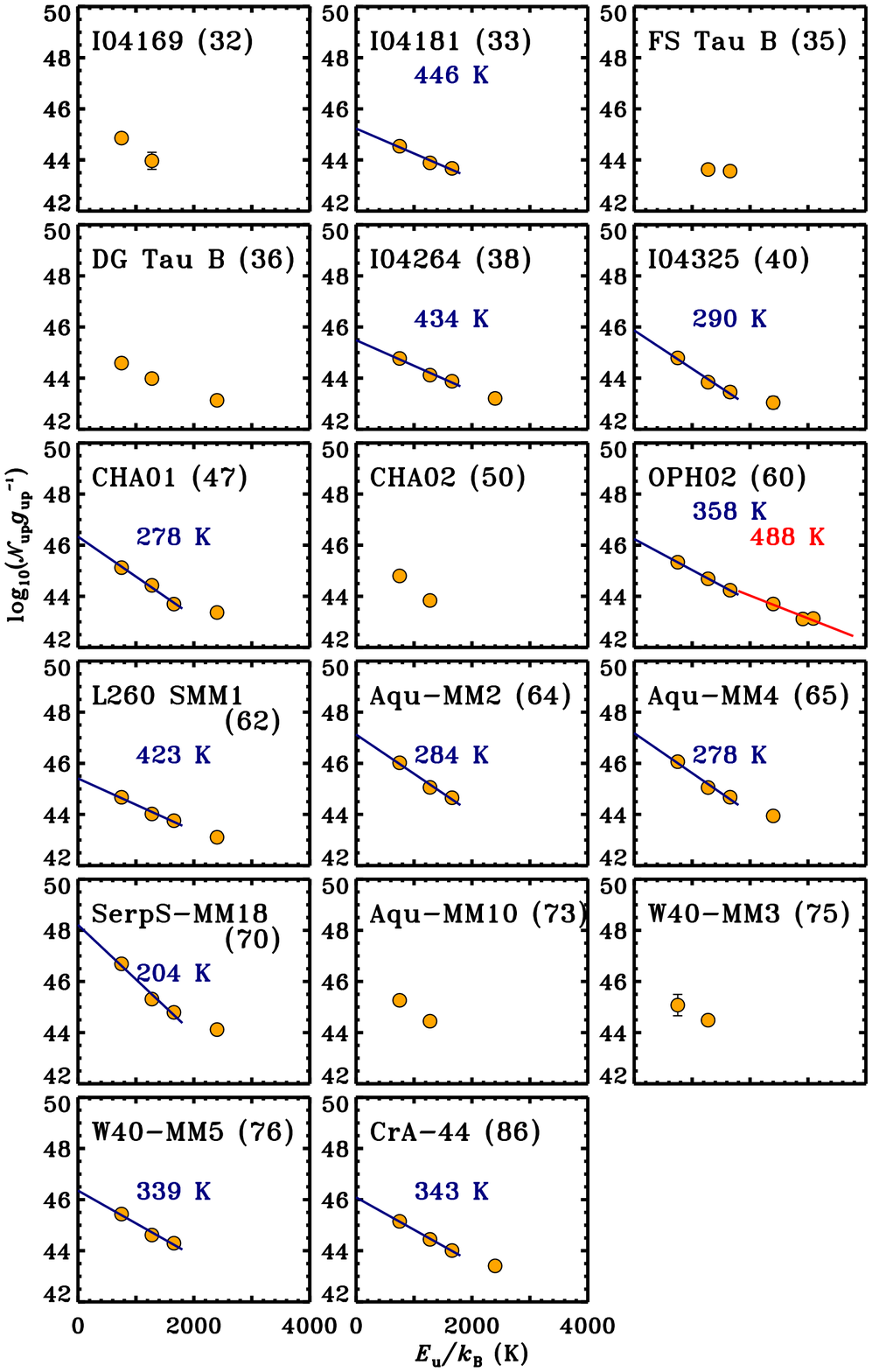}
\caption{\label{cowill2} The same as Figure \ref{cowill}, but for the remaining sources 
from the WILL program.}
\end{center}
\end{figure*}

\begin{figure*}
\begin{center}
\includegraphics[angle=0,height=16cm]{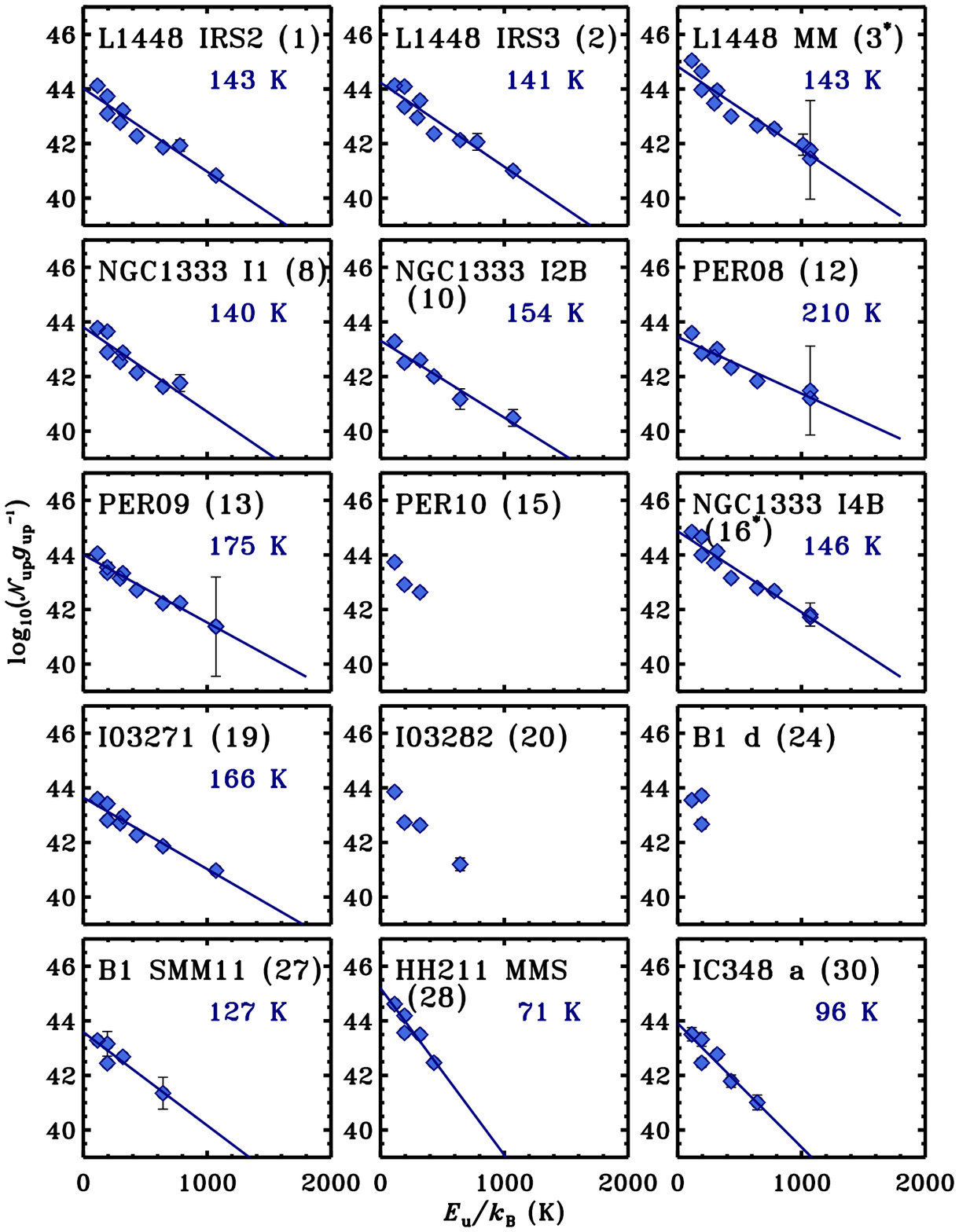}
\vspace{-2.5cm}

\caption{\label{h2owill} Rotational diagrams of H$_2$O for Perseus sources observed as part 
of the WILL program. Blue solid lines show linear fits to the data and the
    corresponding rotational temperatures, when at least 4 lines are detected. }
\end{center}
\end{figure*}

\begin{figure*}
\begin{center}
\includegraphics[angle=0,height=16cm]{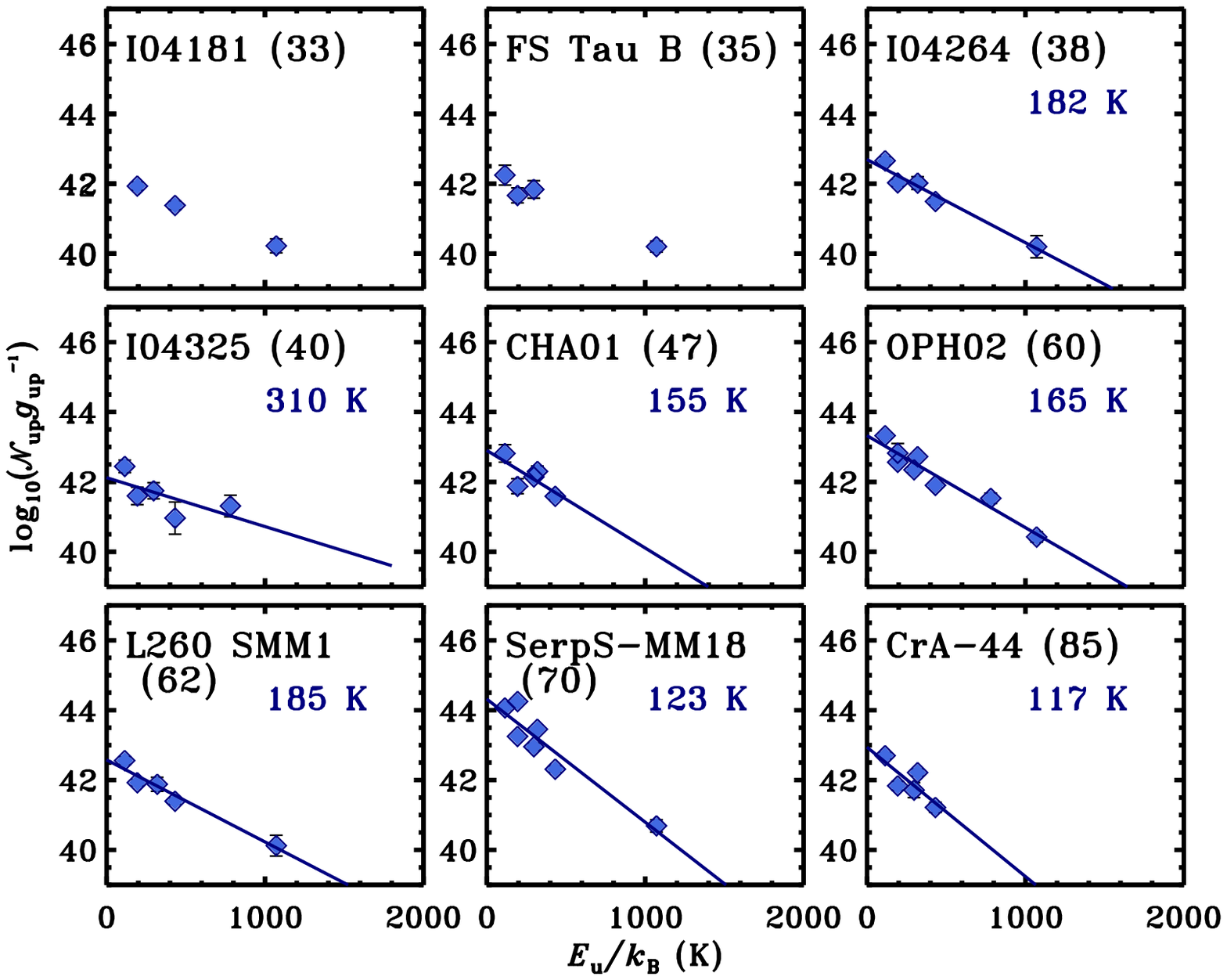}
\vspace{-0.4cm}
\vspace{-7.5cm}

\caption{\label{h2owill2} The same as Figure \ref{h2owill}, but for the remaining sources 
from the WILL program.}
\end{center}
\end{figure*}

Figures \ref{cowill} and \ref{cowill2} show CO rotational diagrams for the WILL 
sources. Separate linear fits are done to the transitions located in the \lq warm' and \lq hot' 
components, the same as used in \citet{Ka13} and \citet{Gr13}. Figures \ref{h2owill} and 
\ref{h2owill2} show H$_2$O rotational diagrams with only one linear fit to all observed lines. 
Table \ref{exc_table} summarizes the results for all sources.

Figures \ref{dig1} - \ref{dig2} show CO, H$_2$O, and OH rotational diagrams for the DIGIT 
sources. Separate linear fits are done at CO diagrams to the transitions located in the \lq warm' and \lq hot' 
components and single component fits to the H$_2$O and OH diagrams.

\begin{figure*}
\begin{center}
\includegraphics[angle=0,height=16cm]{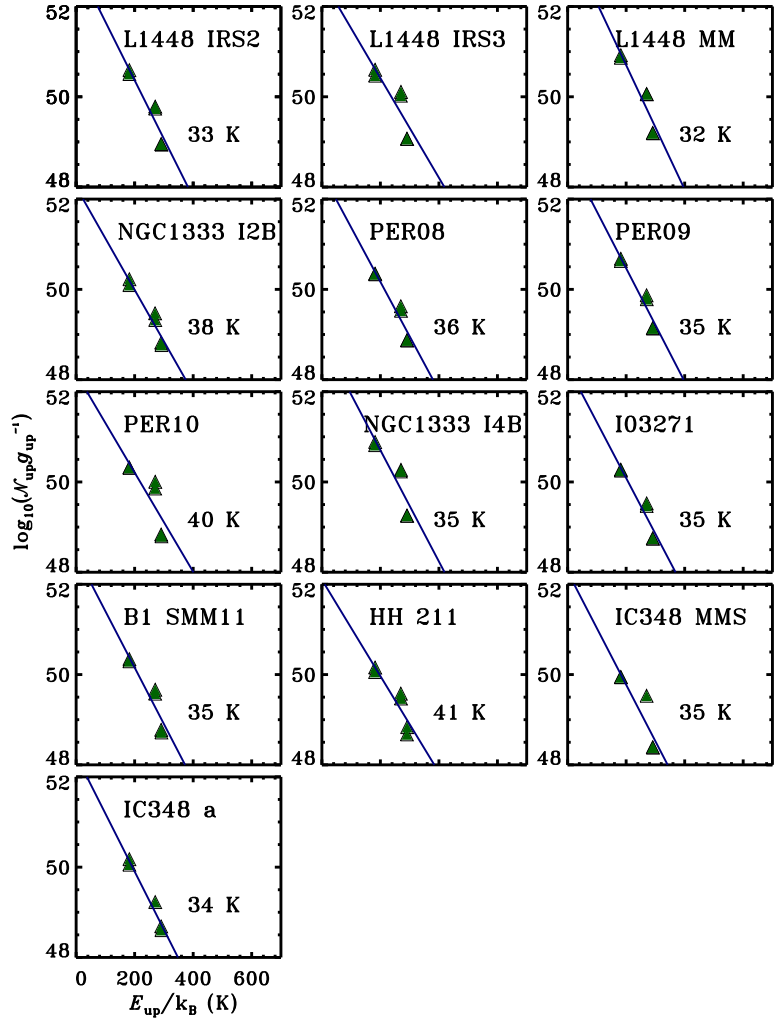}
\vspace{-2.5cm}

\caption{\label{ohwill} Rotational diagrams of OH for Perseus sources observed as part 
of the WILL program. Blue solid lines show linear fits to the data and the
    corresponding rotational temperatures, when at least 3 doublets are detected. }
\end{center}
\end{figure*}
\begin{figure*}
\begin{center}
\includegraphics[angle=0,height=16cm]{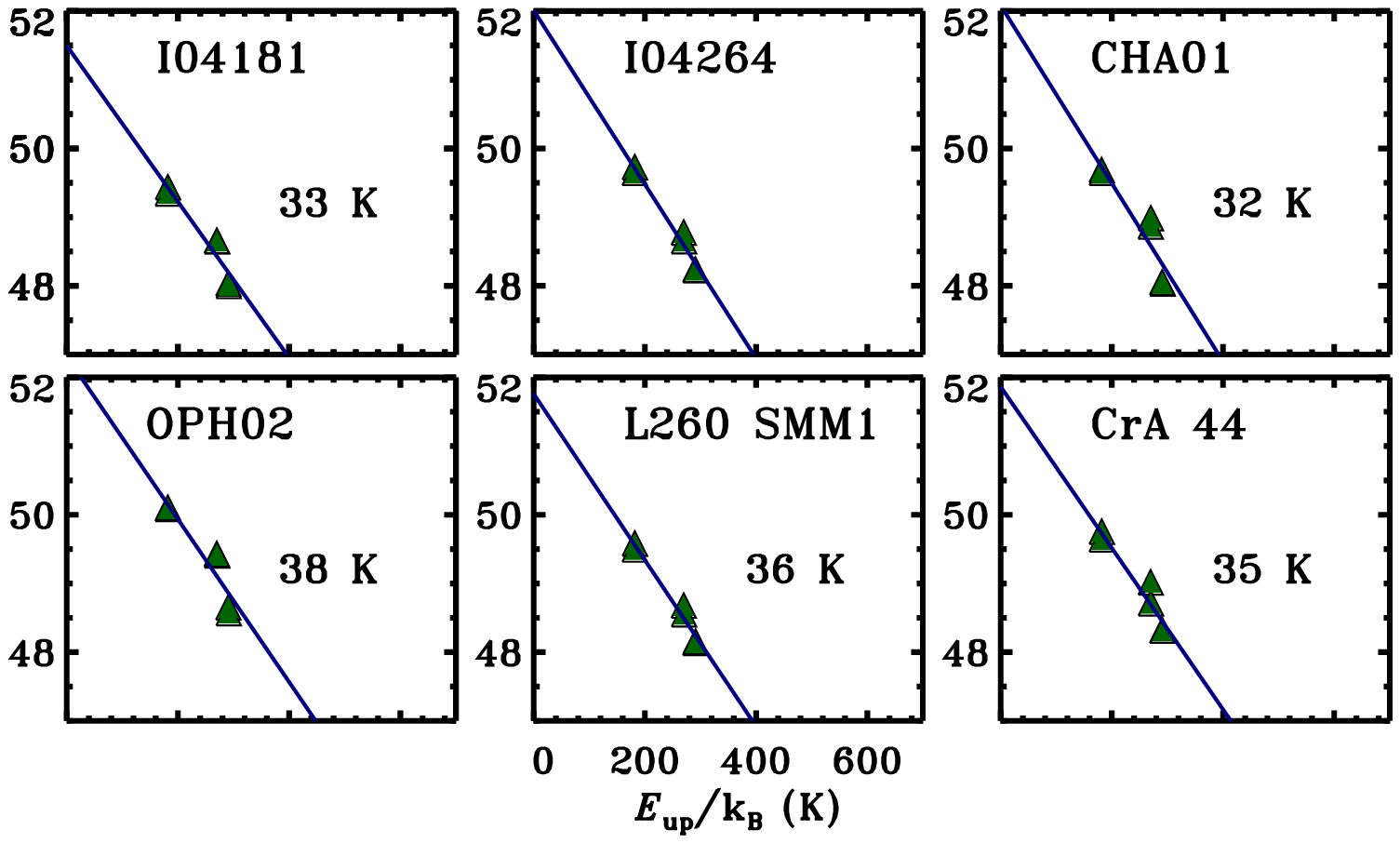}
\vspace{-0.4cm}
\vspace{-7.5cm}

\caption{\label{ohwill2} The same as Figure \ref{ohwill}, but for the remaining sources 
from the WILL program.}
\end{center}
\end{figure*}
\begin{figure*}[!tb]
  \begin{minipage}[t]{.3\textwidth}
  \begin{center}
       \includegraphics[angle=90,height=4.8cm]{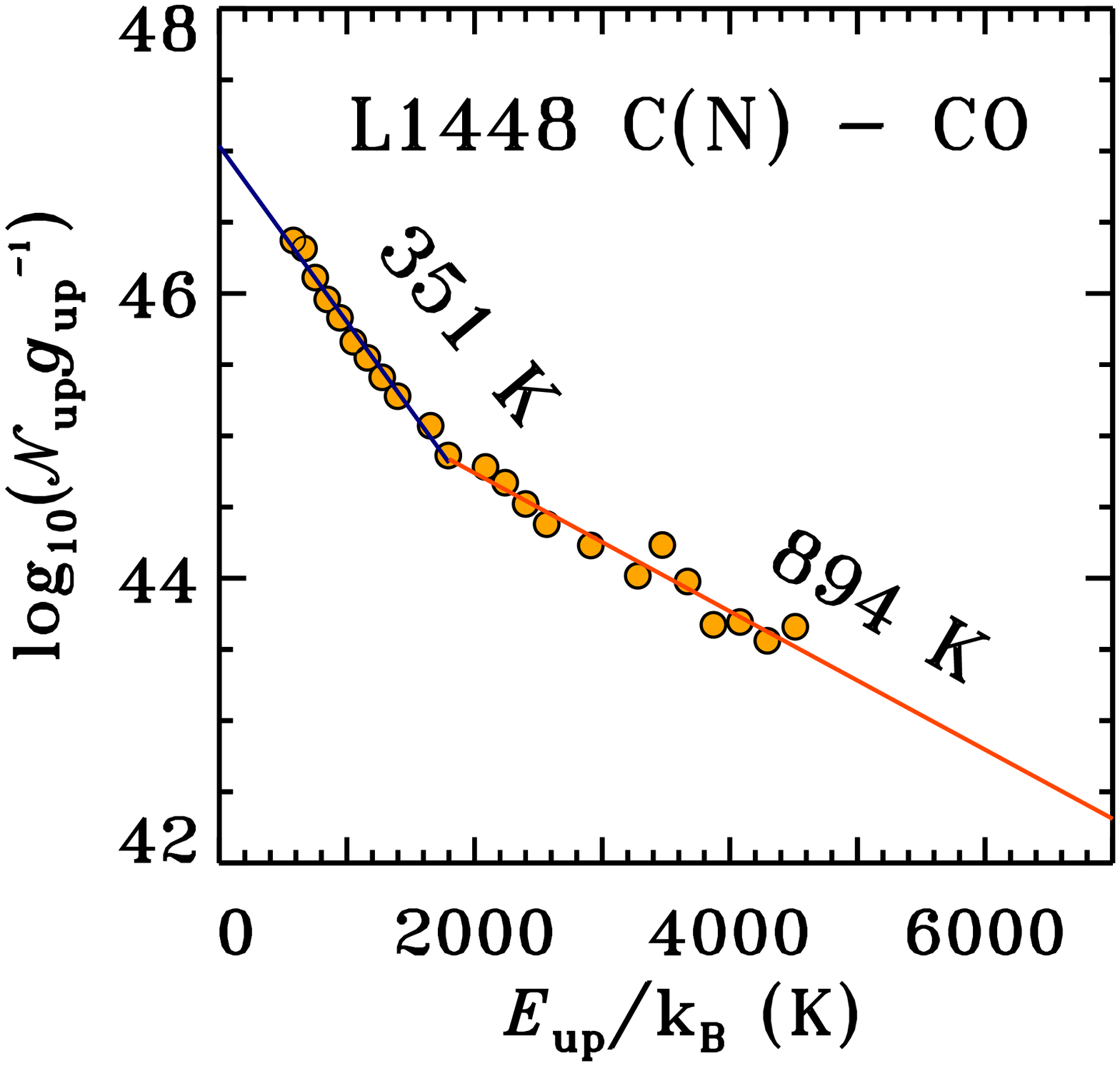} 
       \includegraphics[angle=90,height=4.8cm]{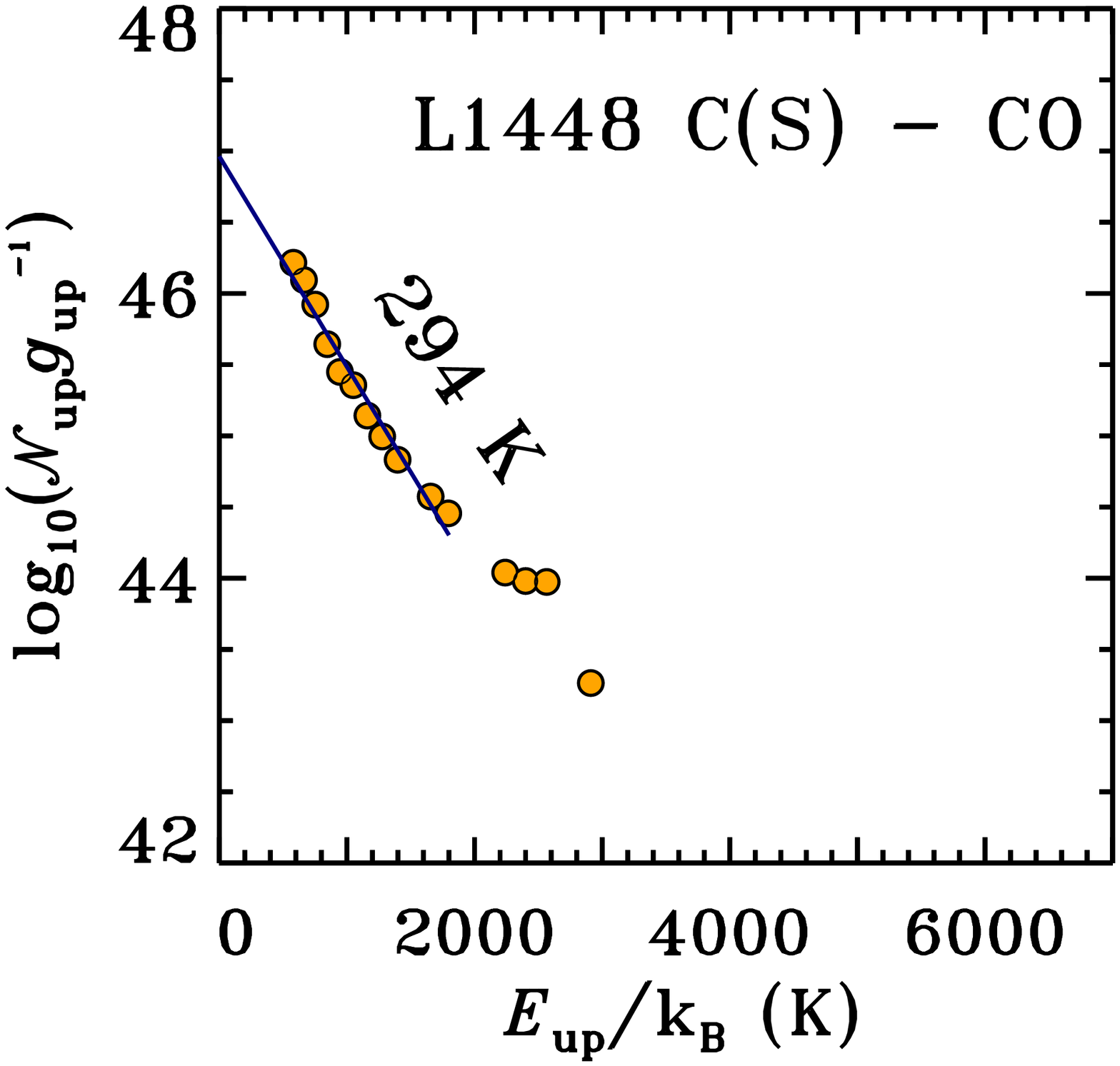} 
        \includegraphics[angle=90,height=4.8cm]{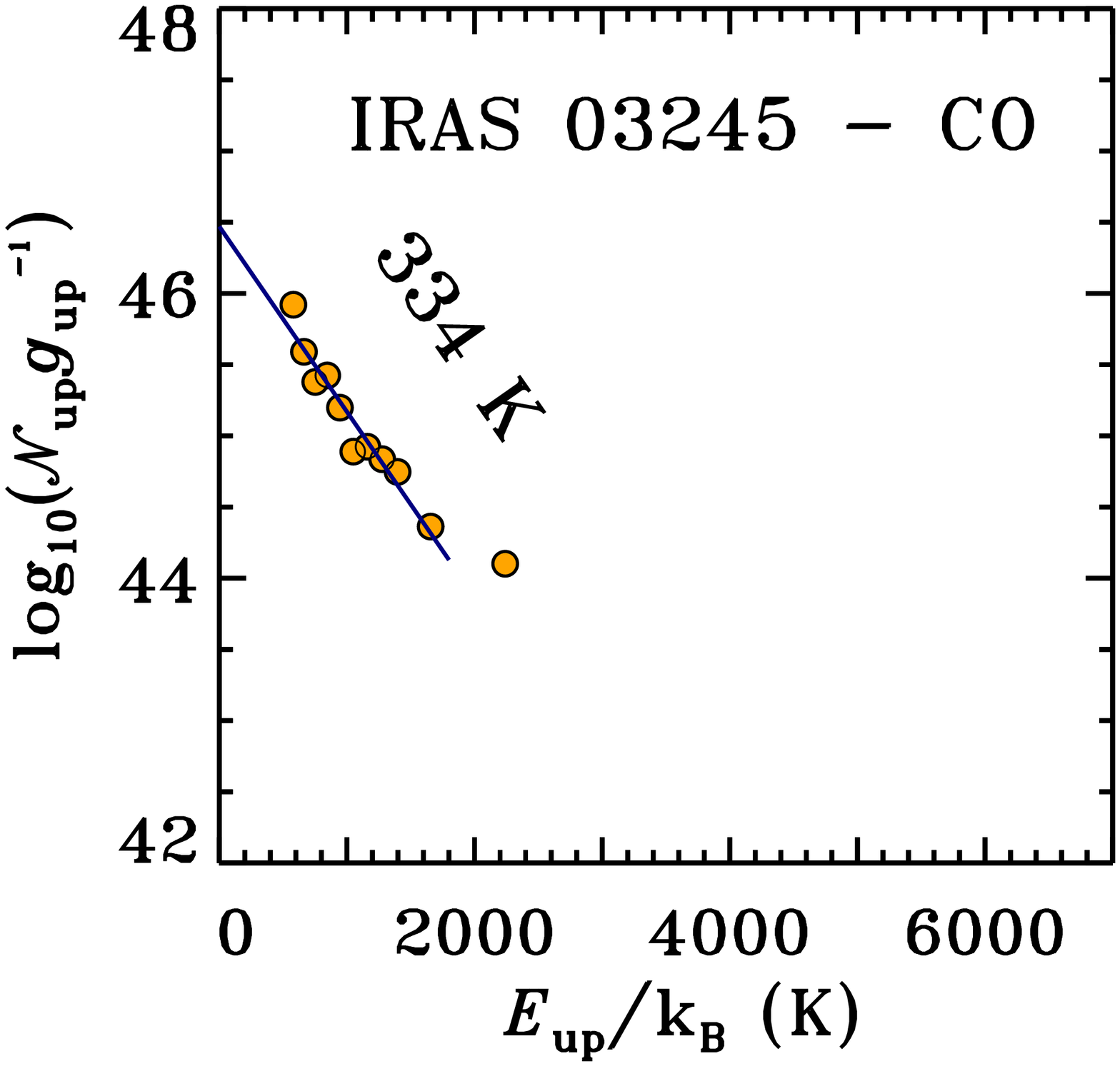} 
       \includegraphics[angle=90,height=4.8cm]{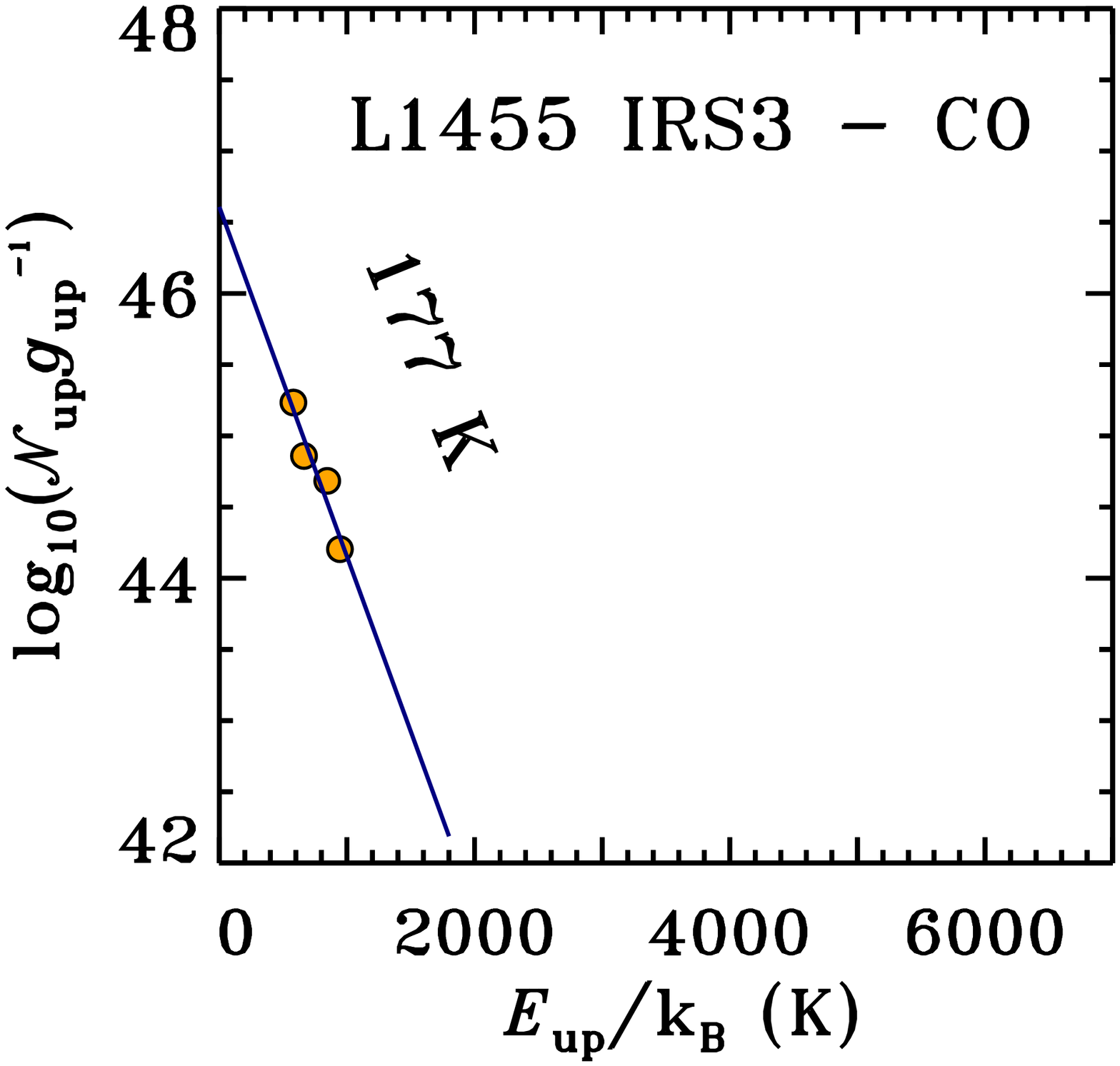} 
        \includegraphics[angle=90,height=4.8cm]{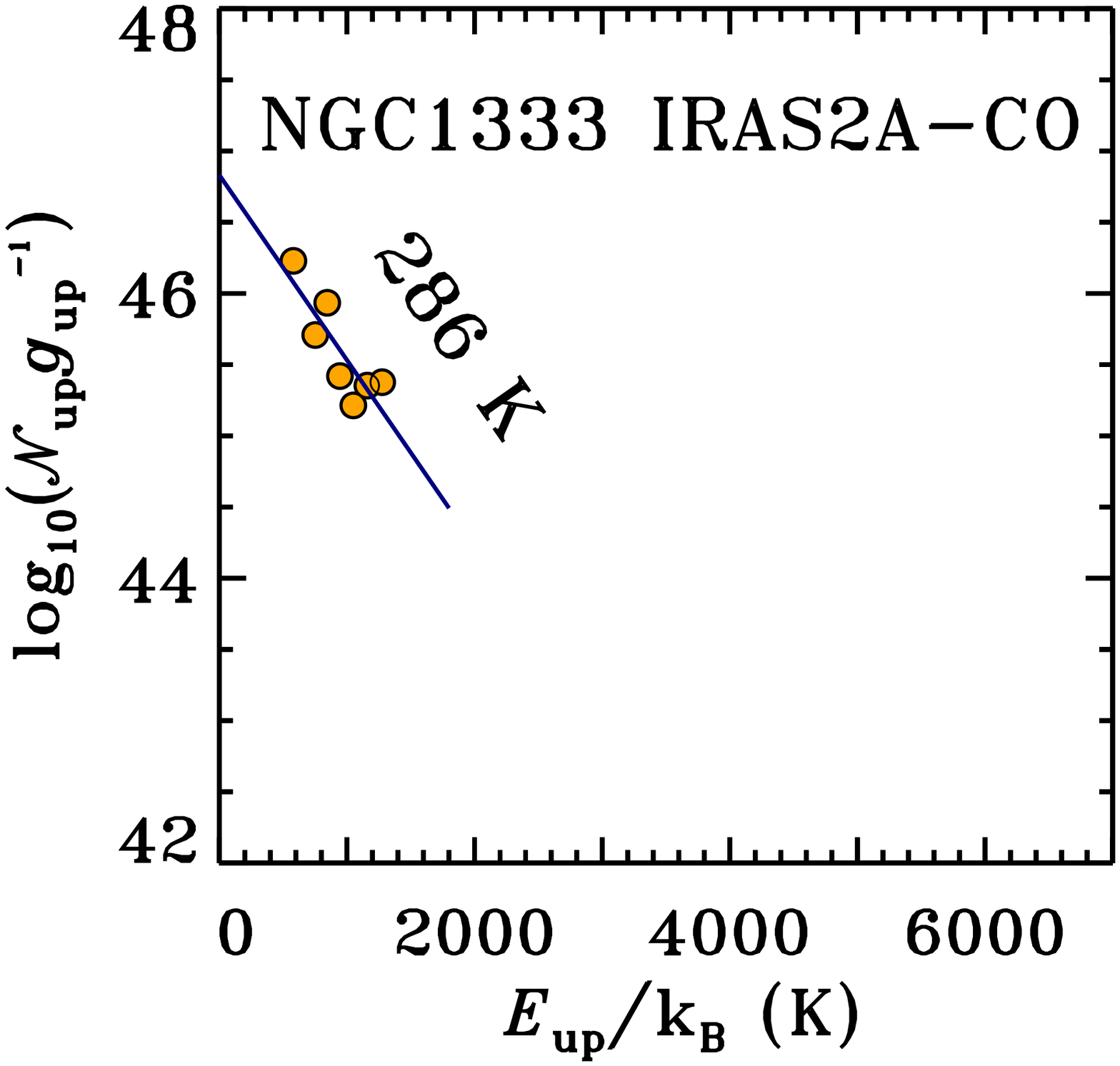} 
  \end{center}
  \end{minipage}
  \hfill
  \begin{minipage}[t]{.3\textwidth}
      \begin{center}
   	   \includegraphics[angle=90,height=4.8cm]{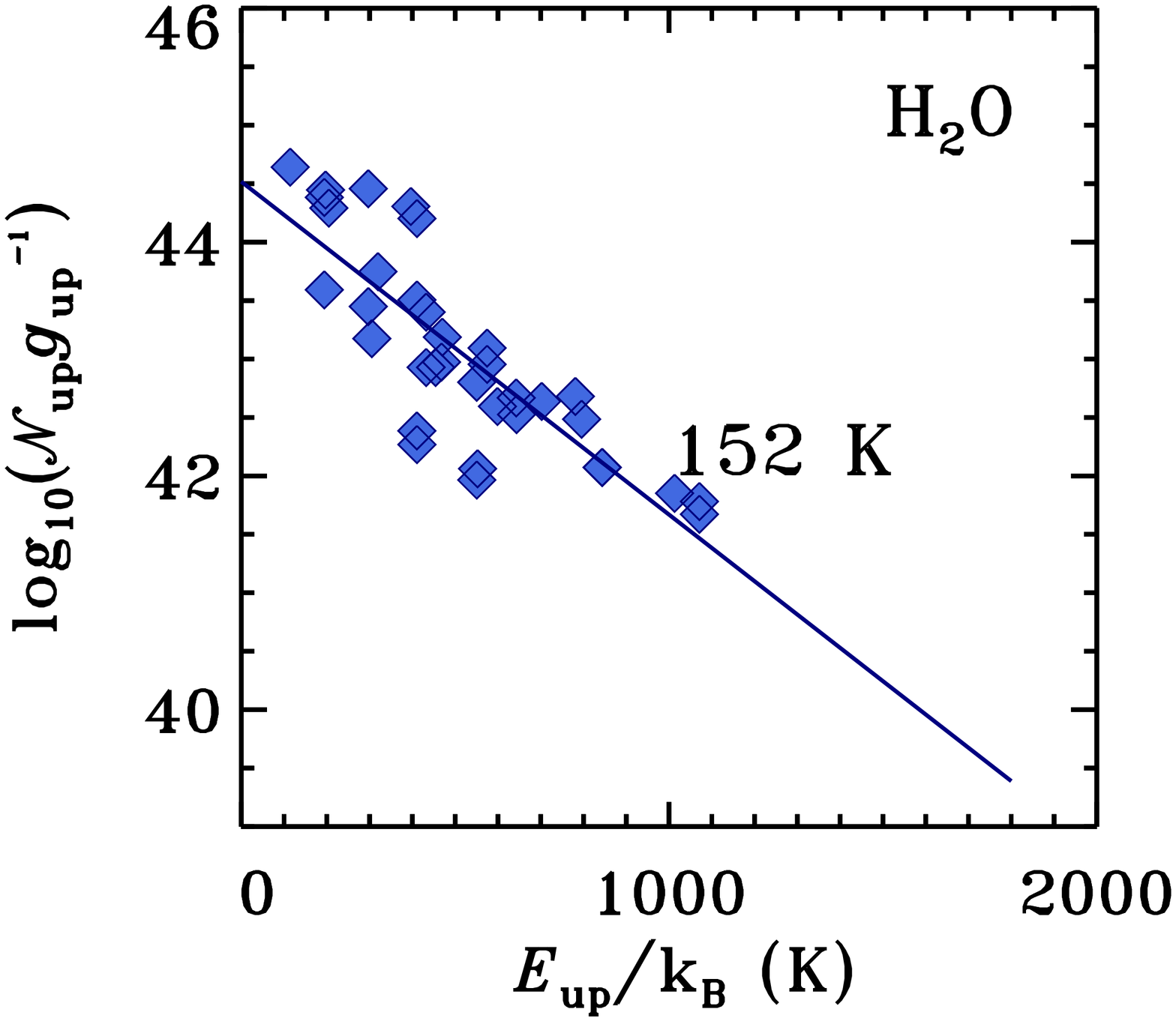} 
       \includegraphics[angle=90,height=4.8cm]{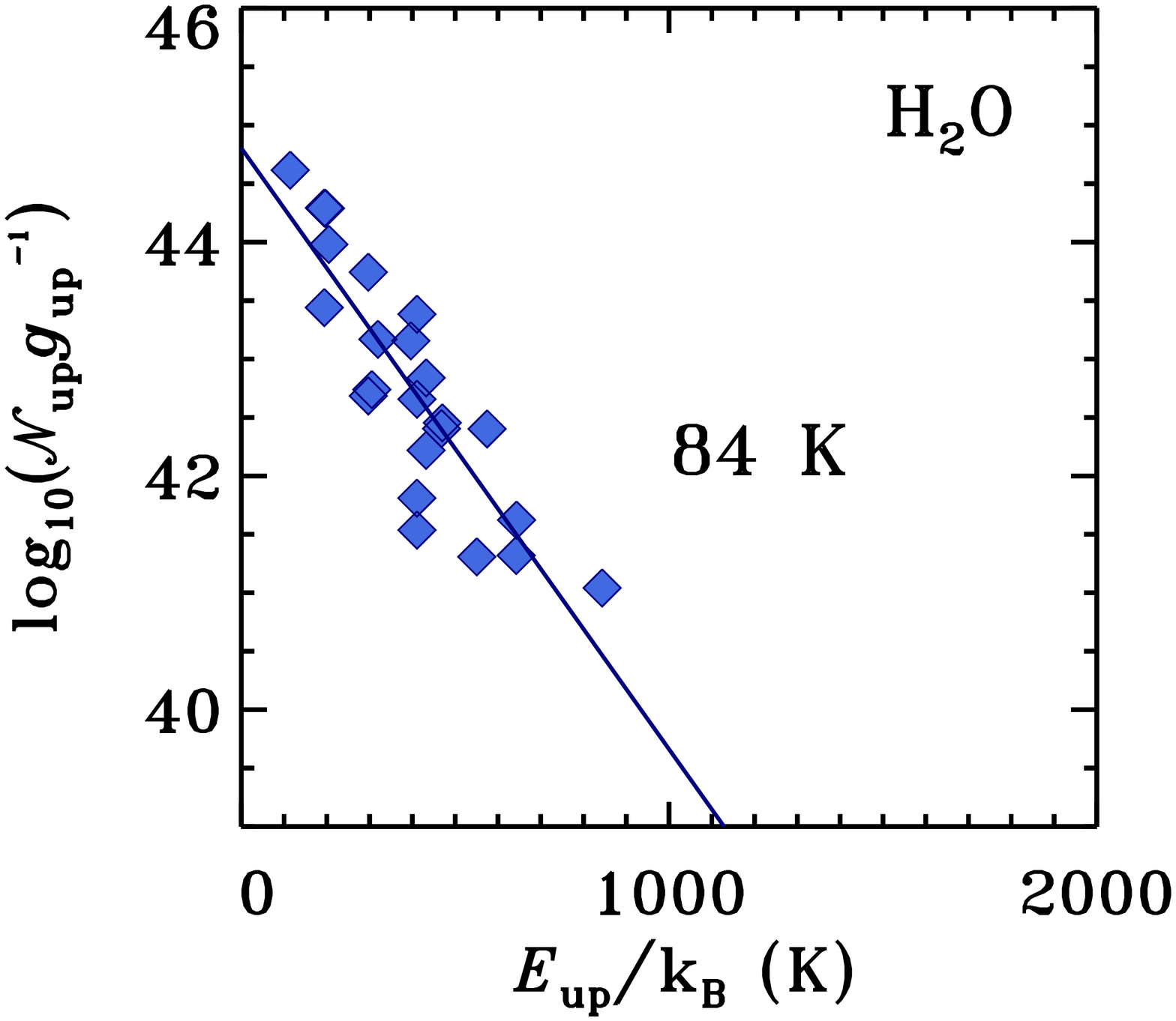} 
       \includegraphics[angle=90,height=4.8cm]{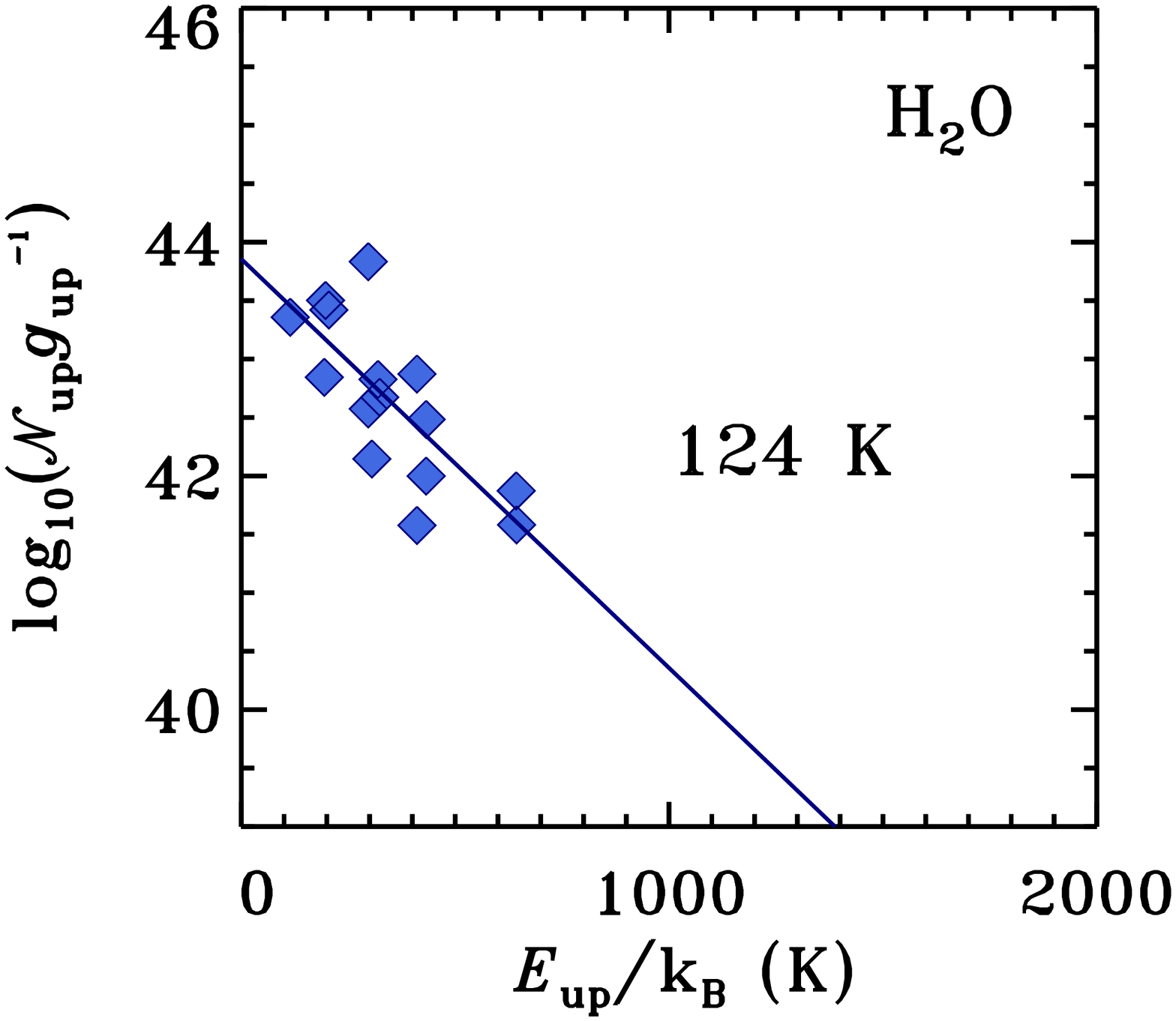} 
       \includegraphics[angle=90,height=4.8cm]{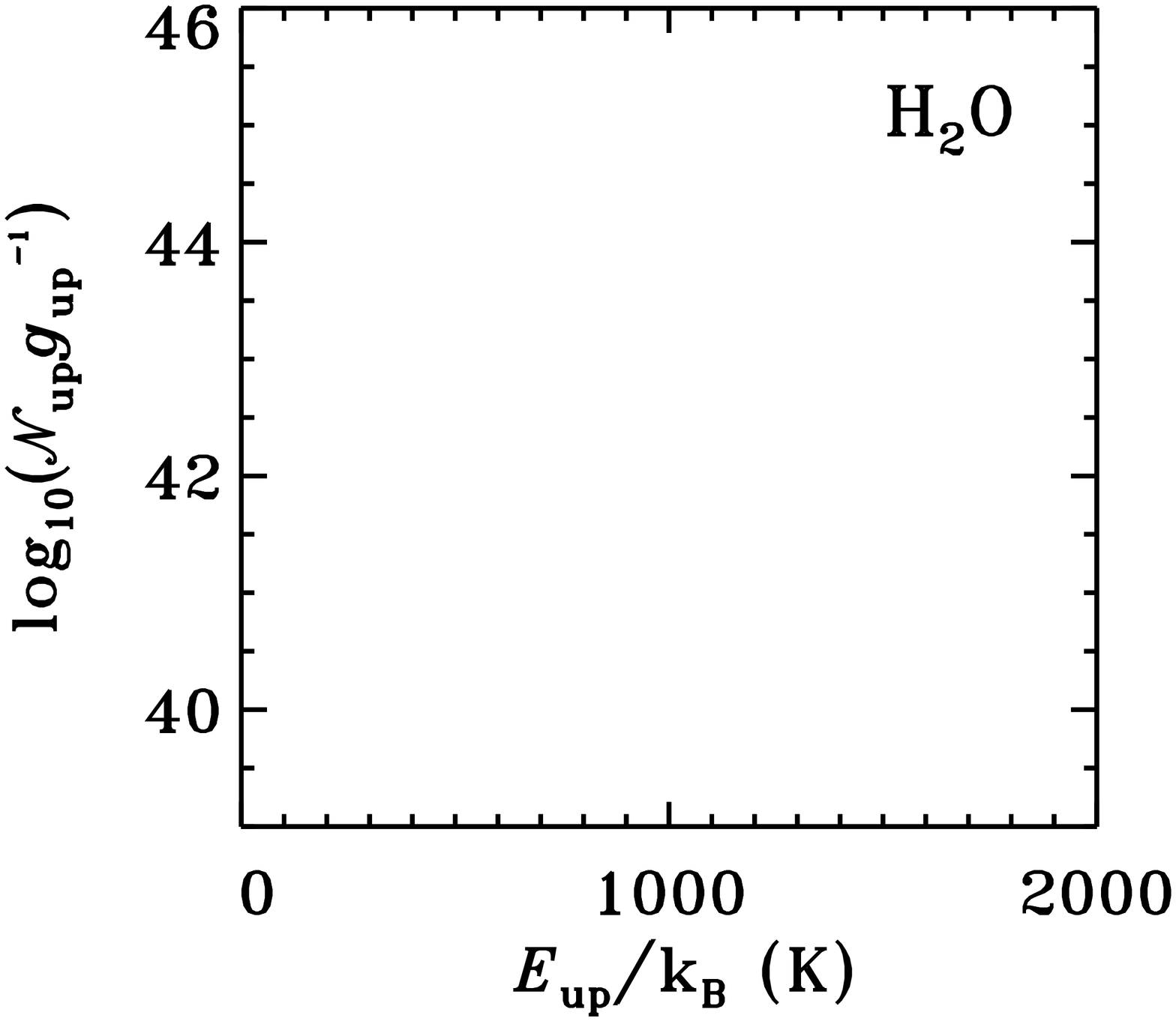} 
       \includegraphics[angle=90,height=4.8cm]{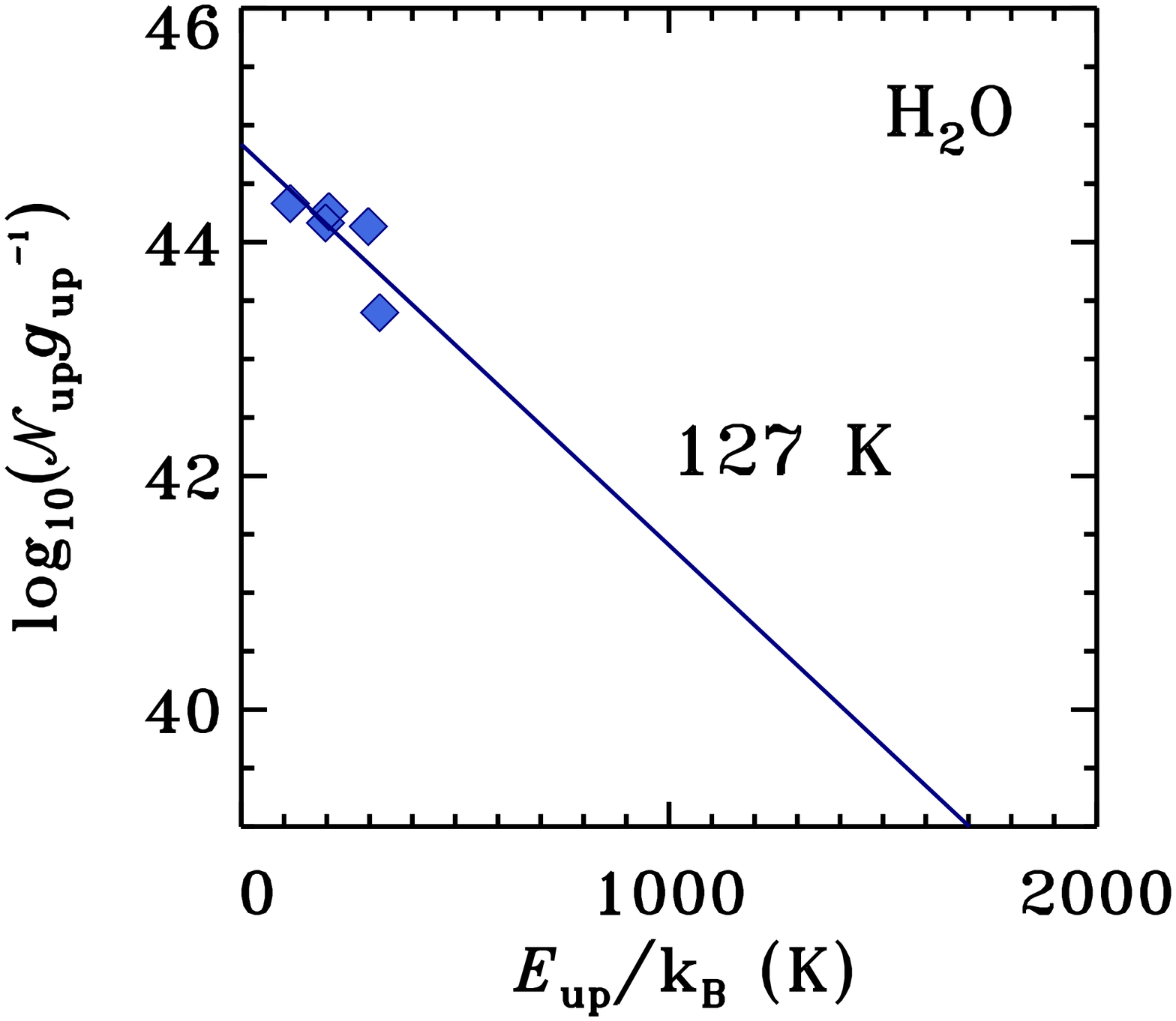} 
            
      \end{center}
  \end{minipage}
    \hfill
   \begin{minipage}[t]{.3\textwidth}
      \begin{center}
    	\includegraphics[angle=90,height=4.8cm]{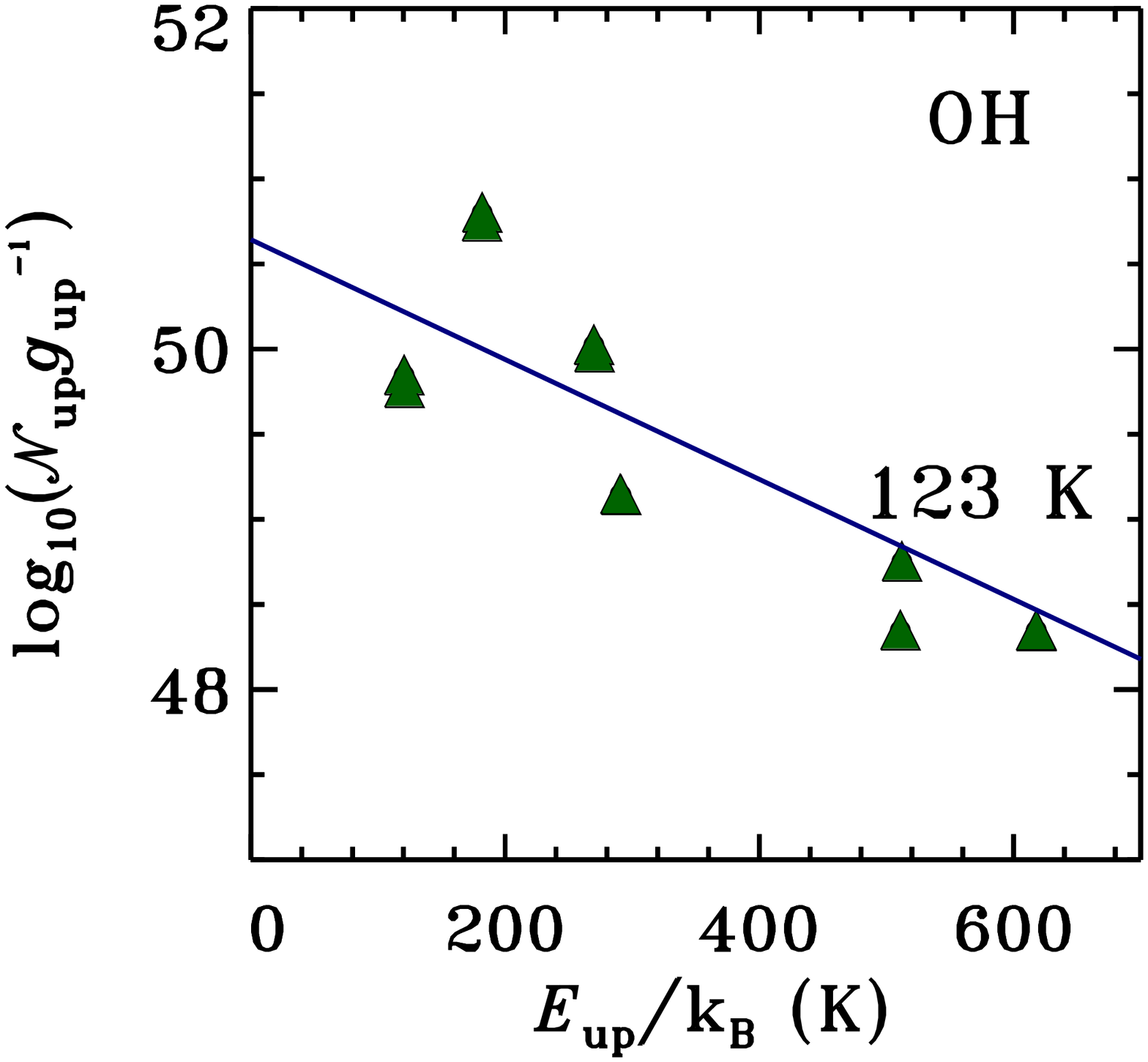} 
        \includegraphics[angle=90,height=4.8cm]{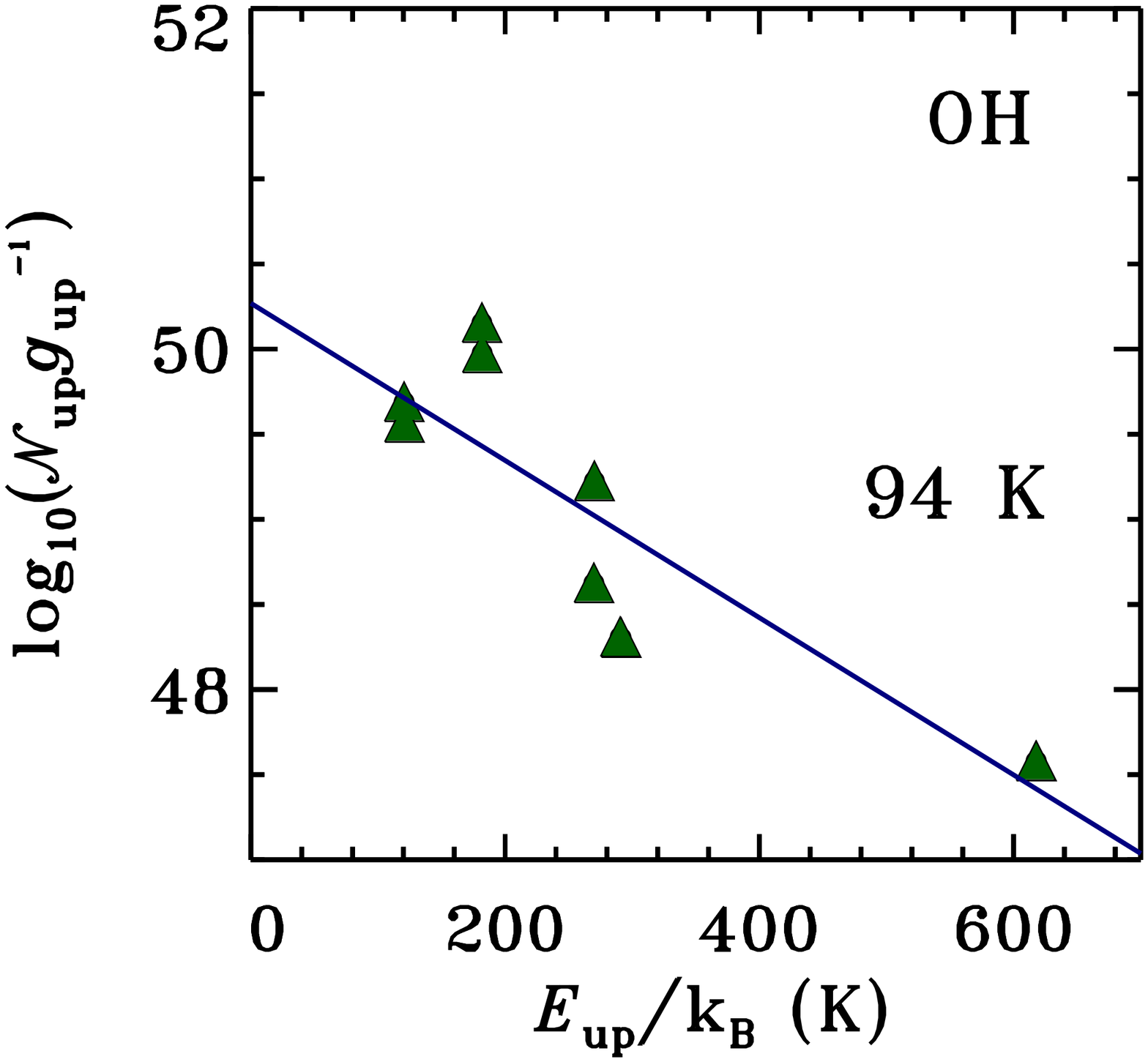} 
        \includegraphics[angle=90,height=4.8cm]{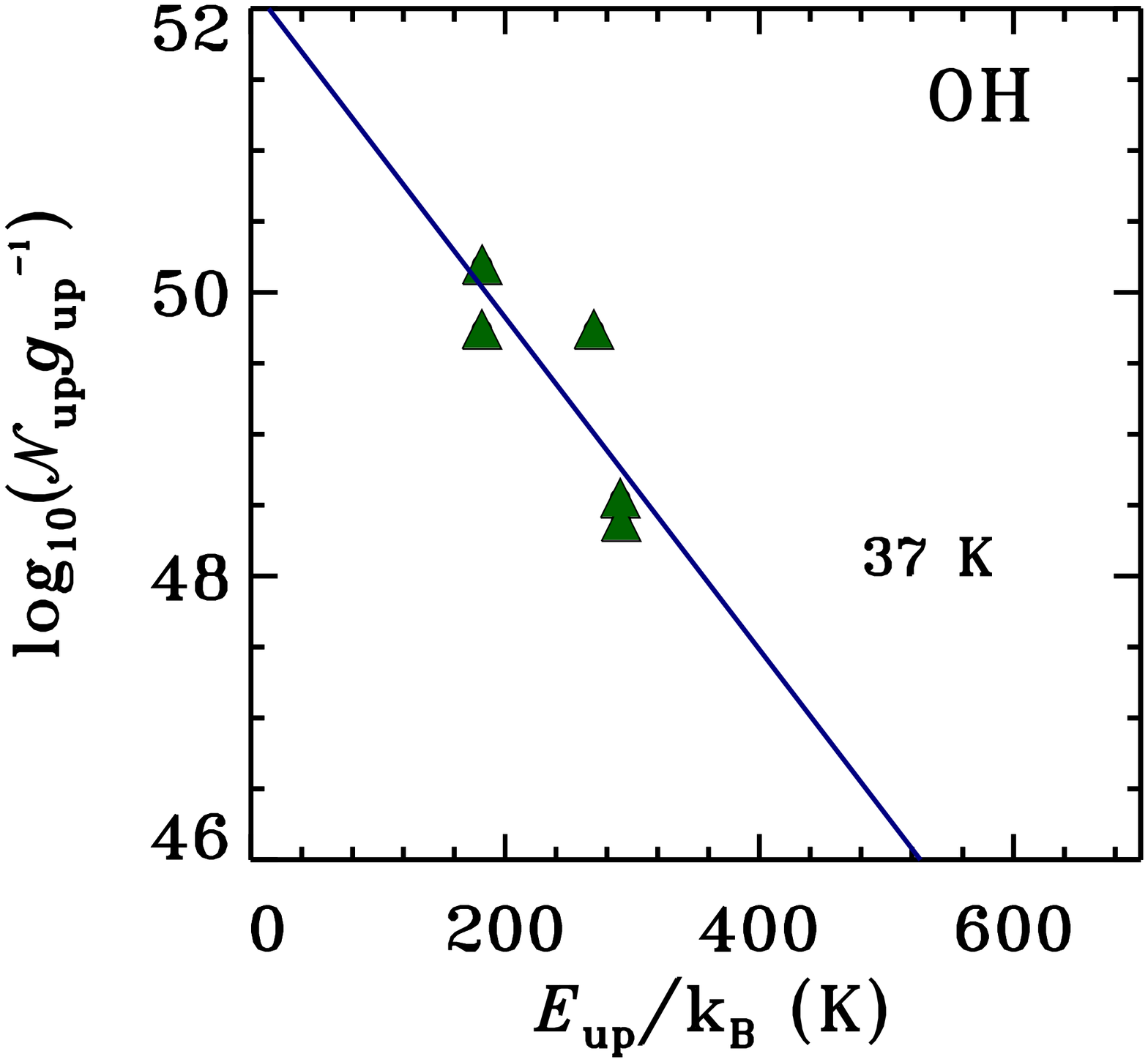} 
        \includegraphics[angle=90,height=4.8cm]{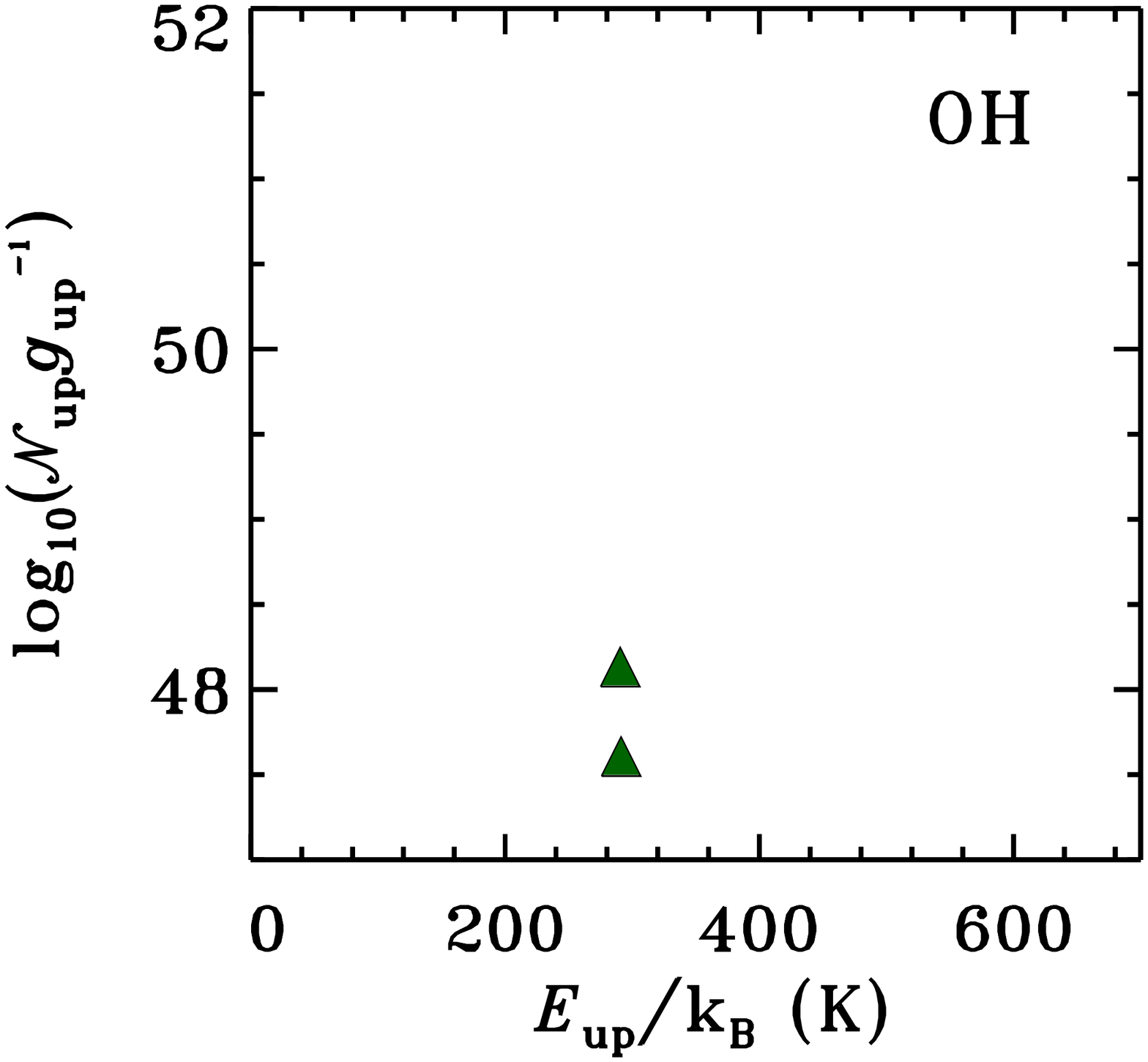} 
        \includegraphics[angle=90,height=4.8cm]{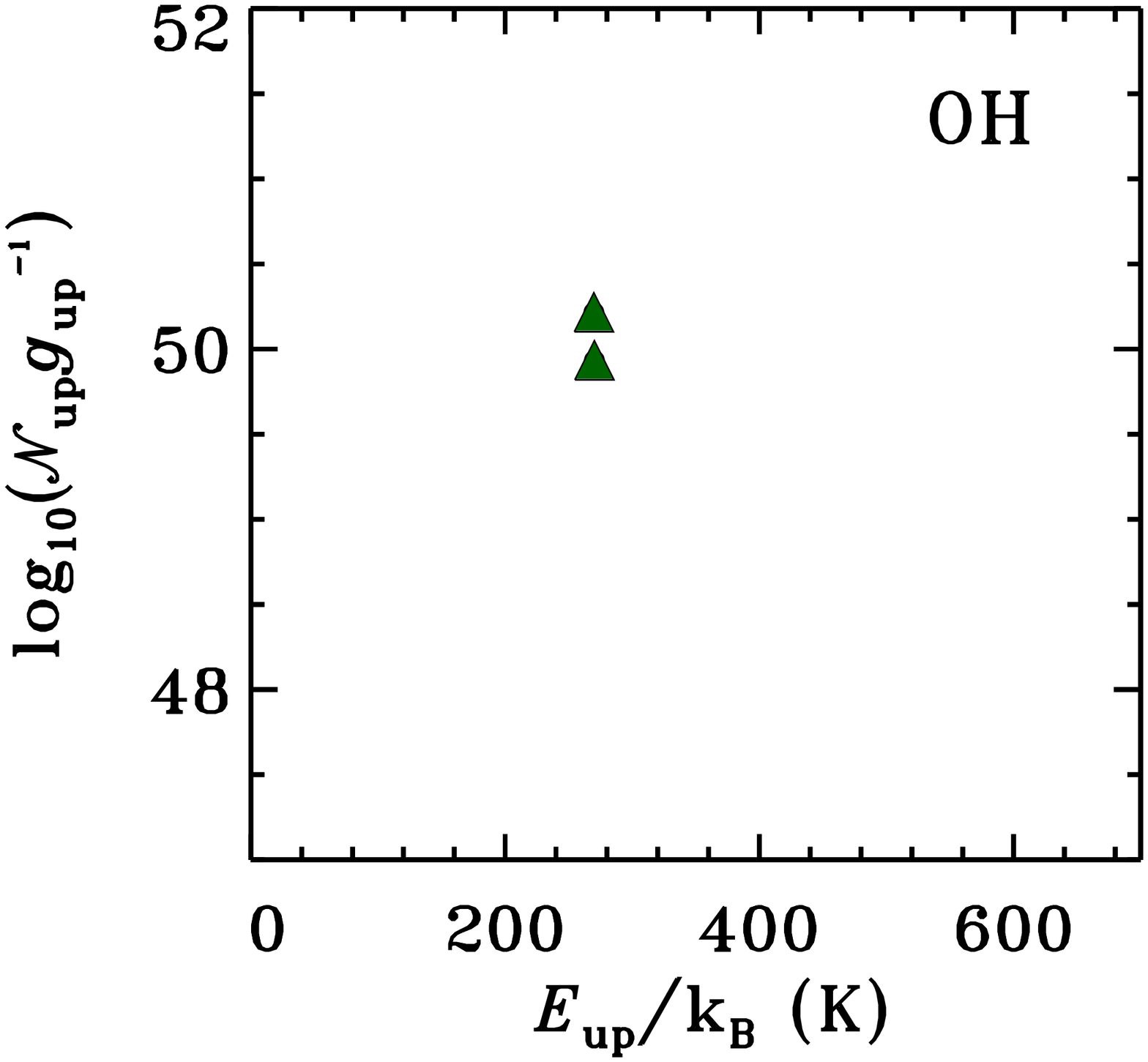} 
                            
      \end{center}
  \end{minipage}
      \hfill
        \caption{\label{dig1} Similar to Figure \ref{molexc}, but for L1448 C(N) and L1448 C(S) 
        - based on measurements from \citet{Lee13} - and for L1455 IRS1 (IRAS03245), L1455 IRS3 (see also Green et al. 2013), and 
        NGC1333 I2A (see also Karska et al. 2013).}
\end{figure*}
\renewcommand{\thefigure}{\thesection.\arabic{figure} (Cont.)}
\addtocounter{figure}{-1}   
\begin{figure*}[!tb]
  \begin{minipage}[t]{.3\textwidth}
  \begin{center}
       \includegraphics[angle=90,height=4.8cm]{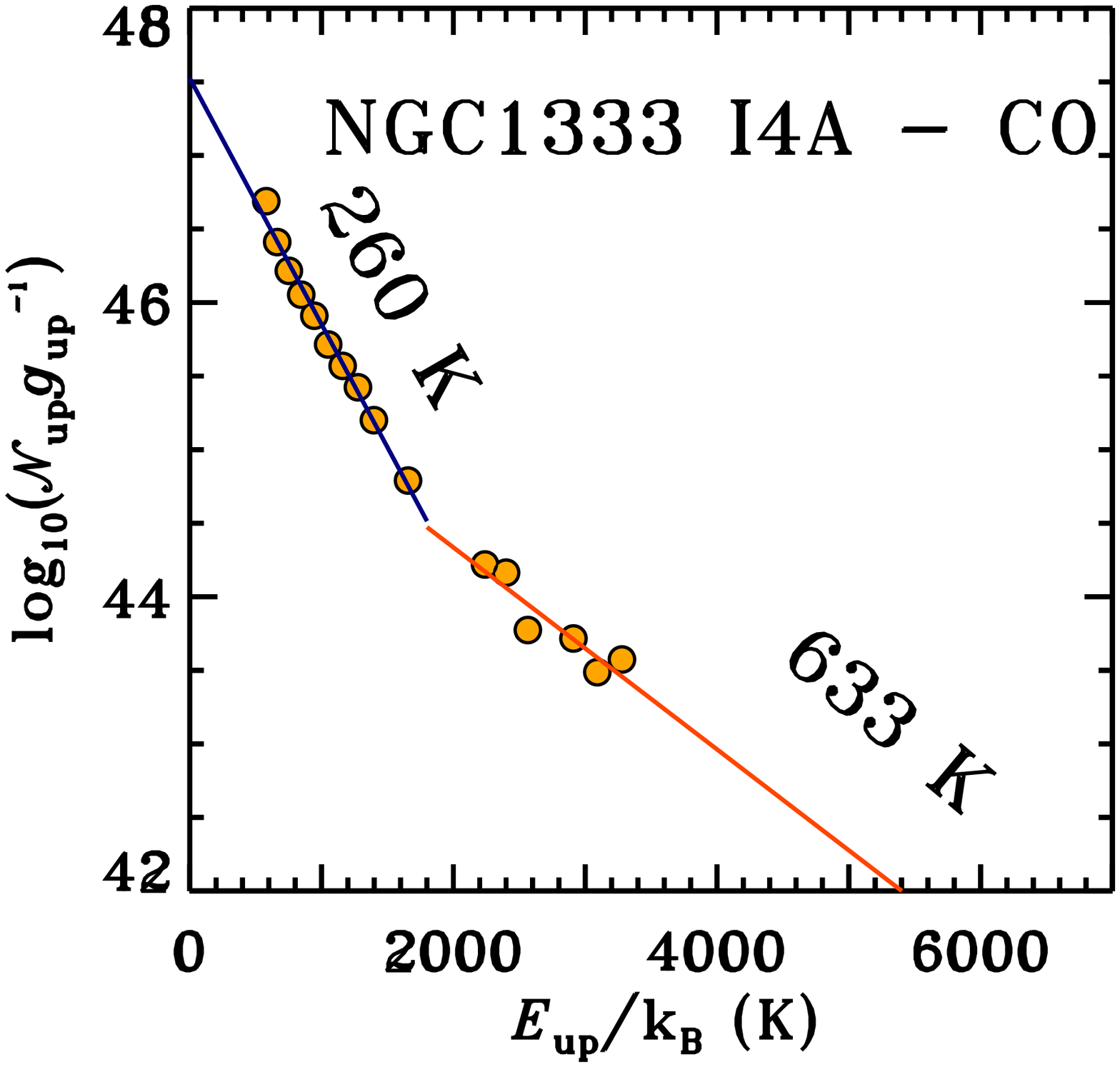} 
       \includegraphics[angle=90,height=4.8cm]{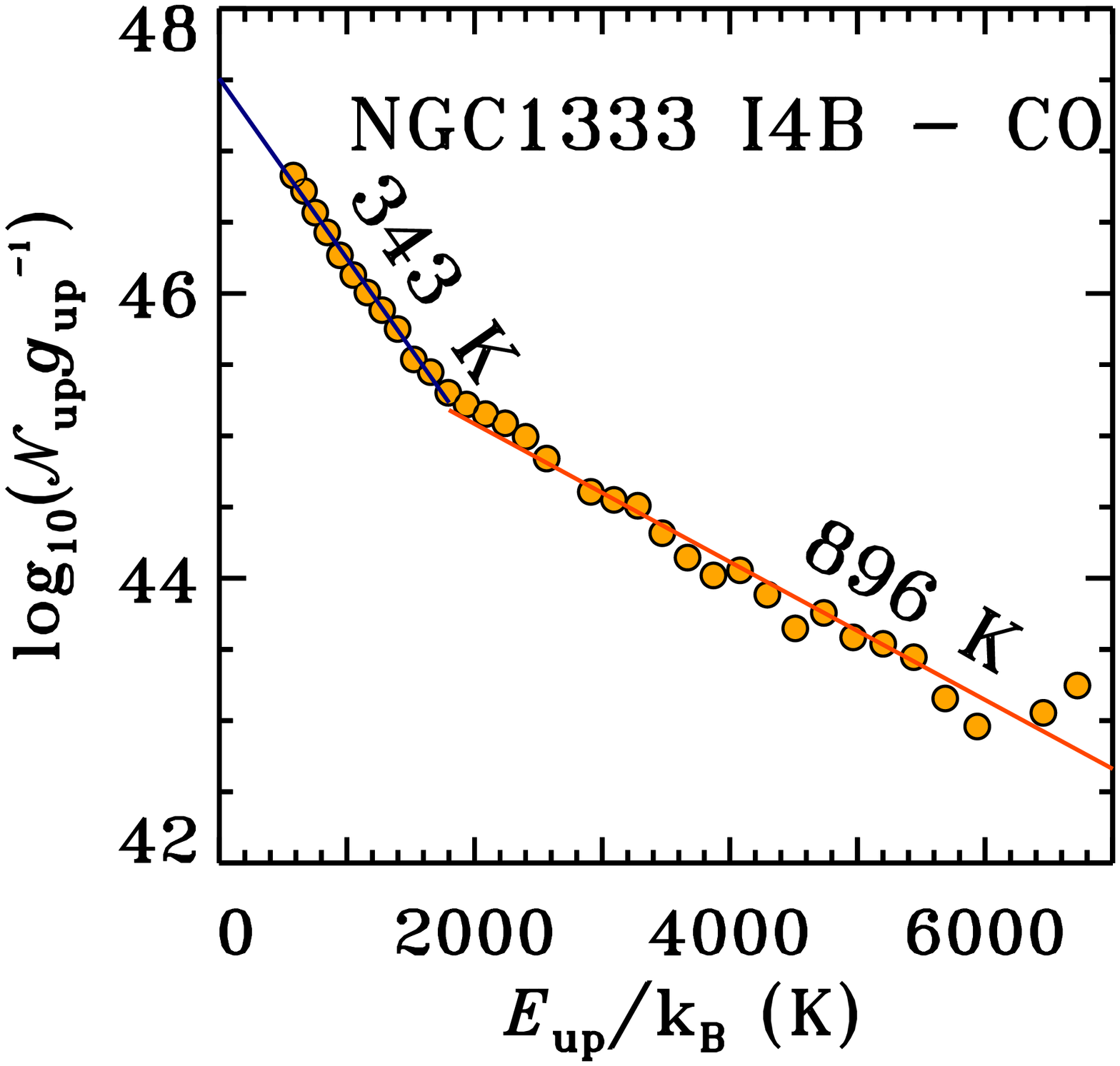} 
        \includegraphics[angle=90,height=4.8cm]{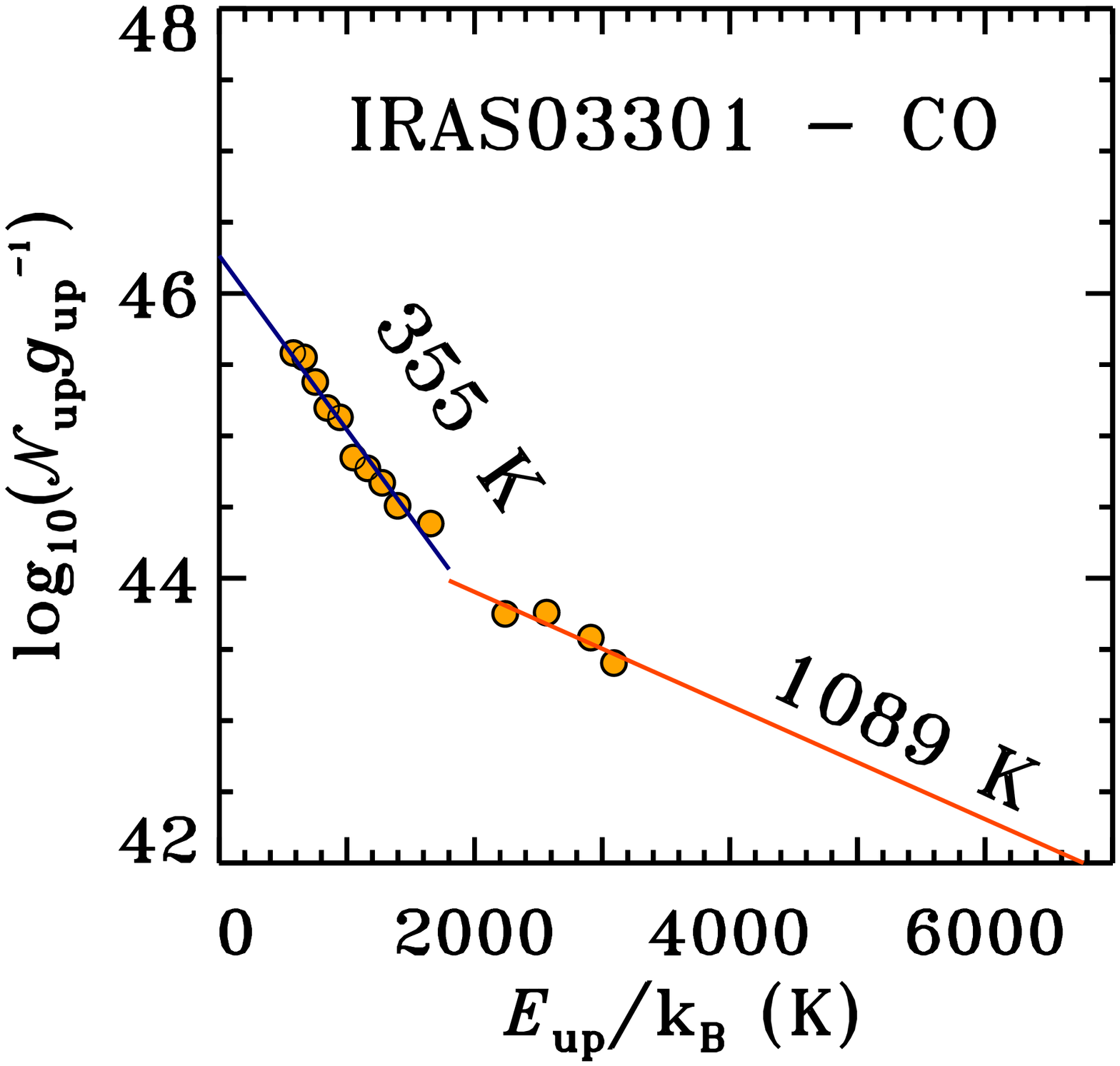} 
       \includegraphics[angle=90,height=4.8cm]{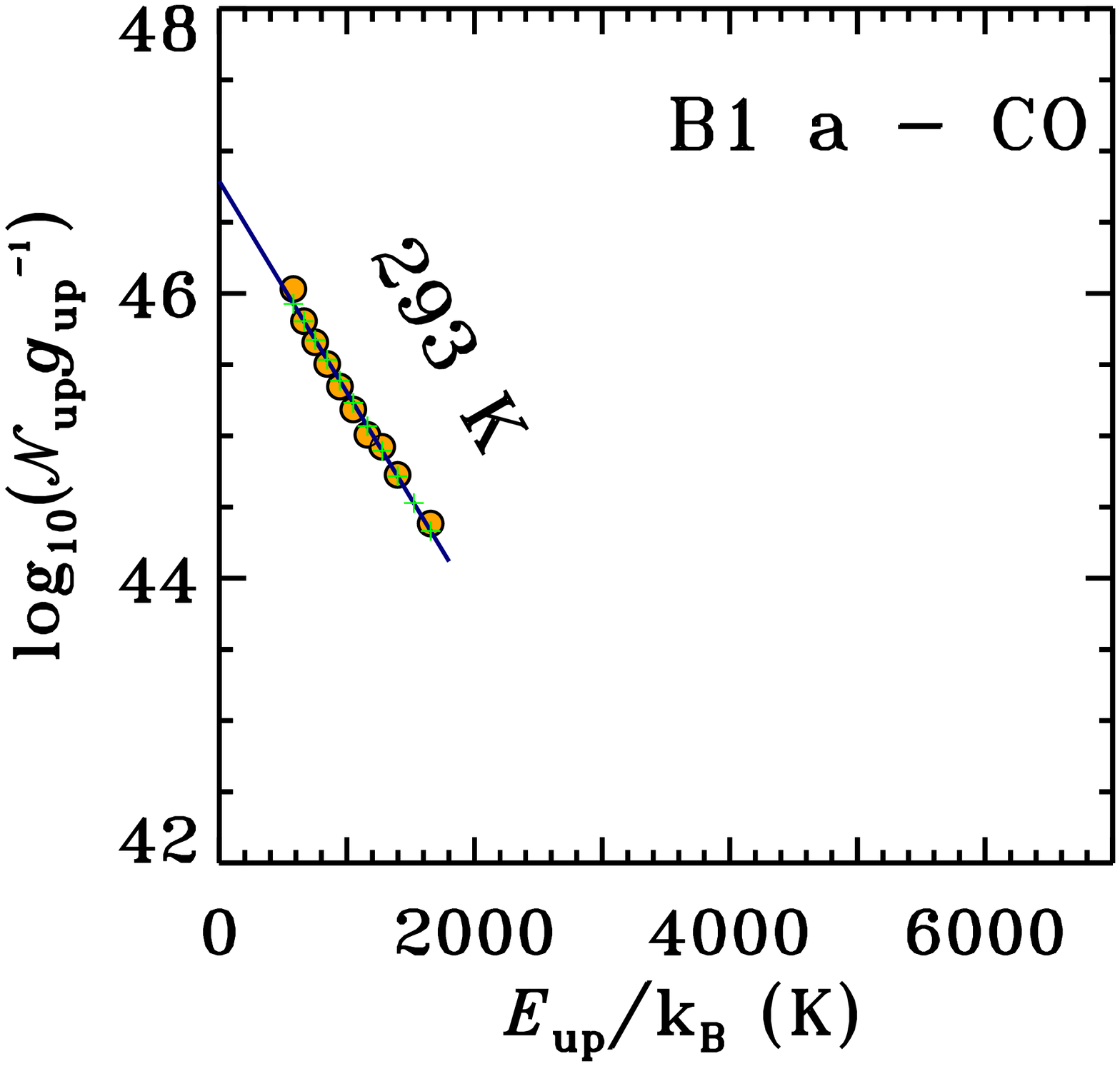} 
        \includegraphics[angle=90,height=4.8cm]{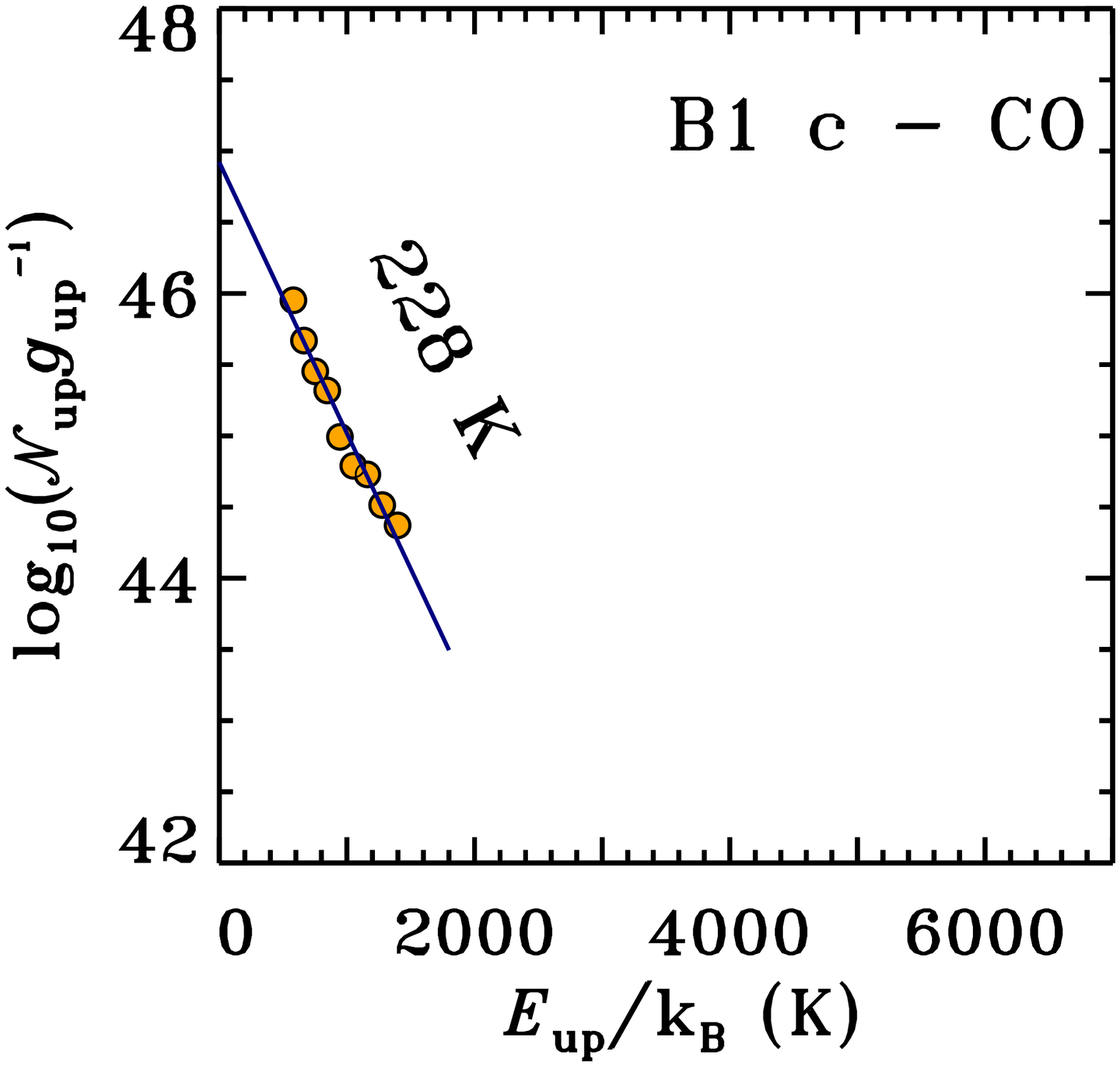} 
  \end{center}
  \end{minipage}
  \hfill
  \begin{minipage}[t]{.3\textwidth}
      \begin{center}
   	   \includegraphics[angle=90,height=4.8cm]{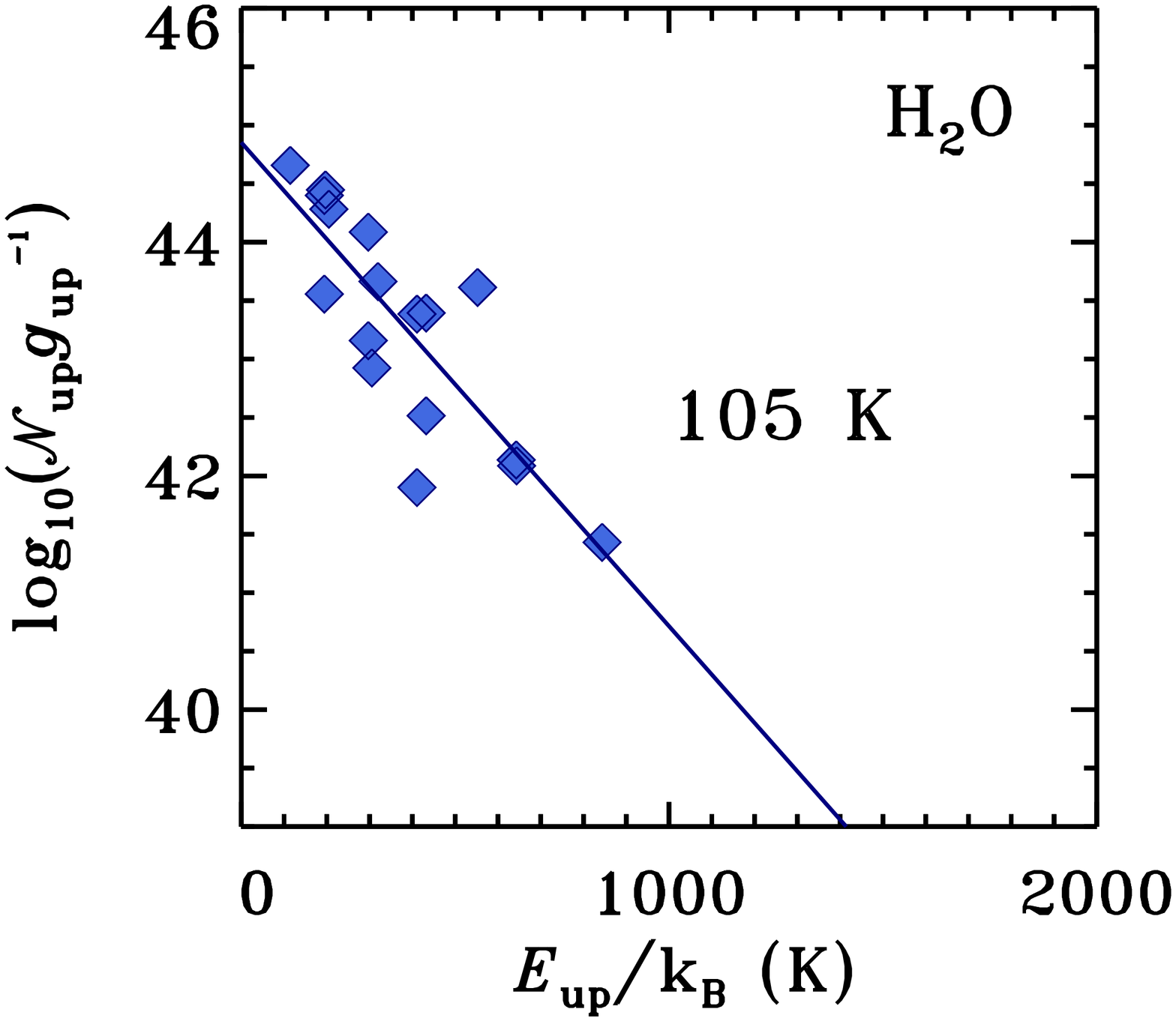} 
       \includegraphics[angle=90,height=4.8cm]{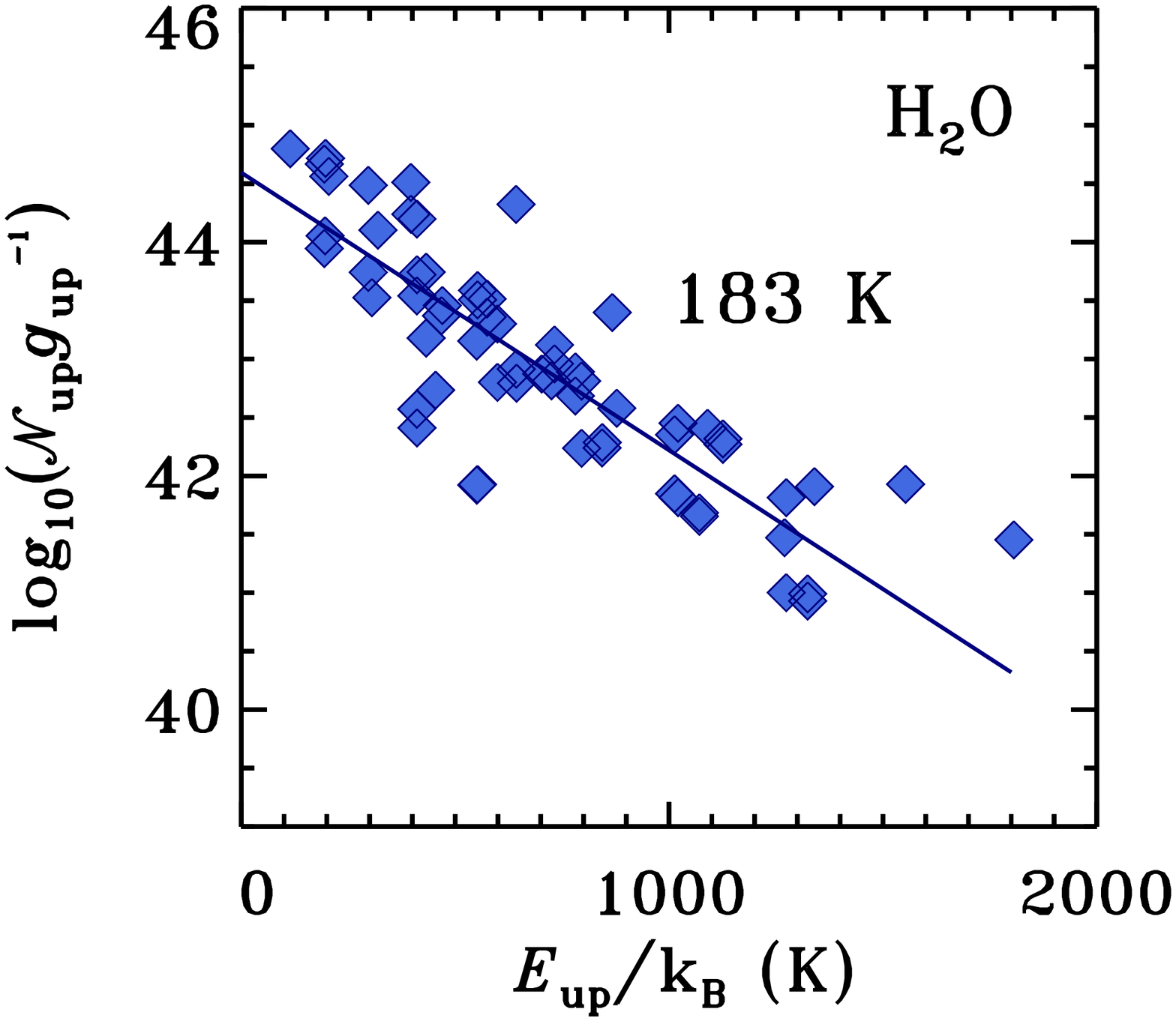} 
       \includegraphics[angle=90,height=4.8cm]{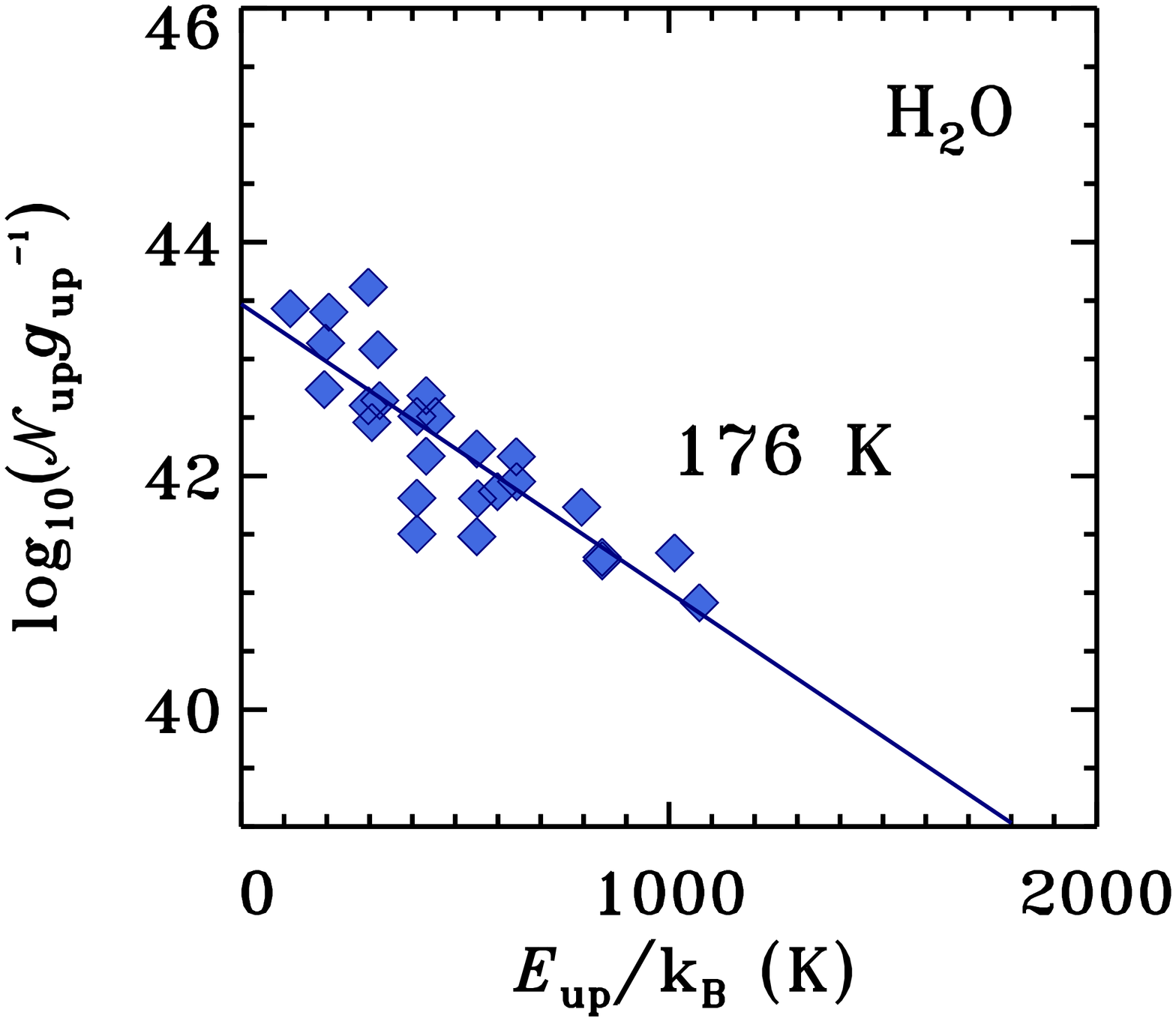} 
       \includegraphics[angle=90,height=4.8cm]{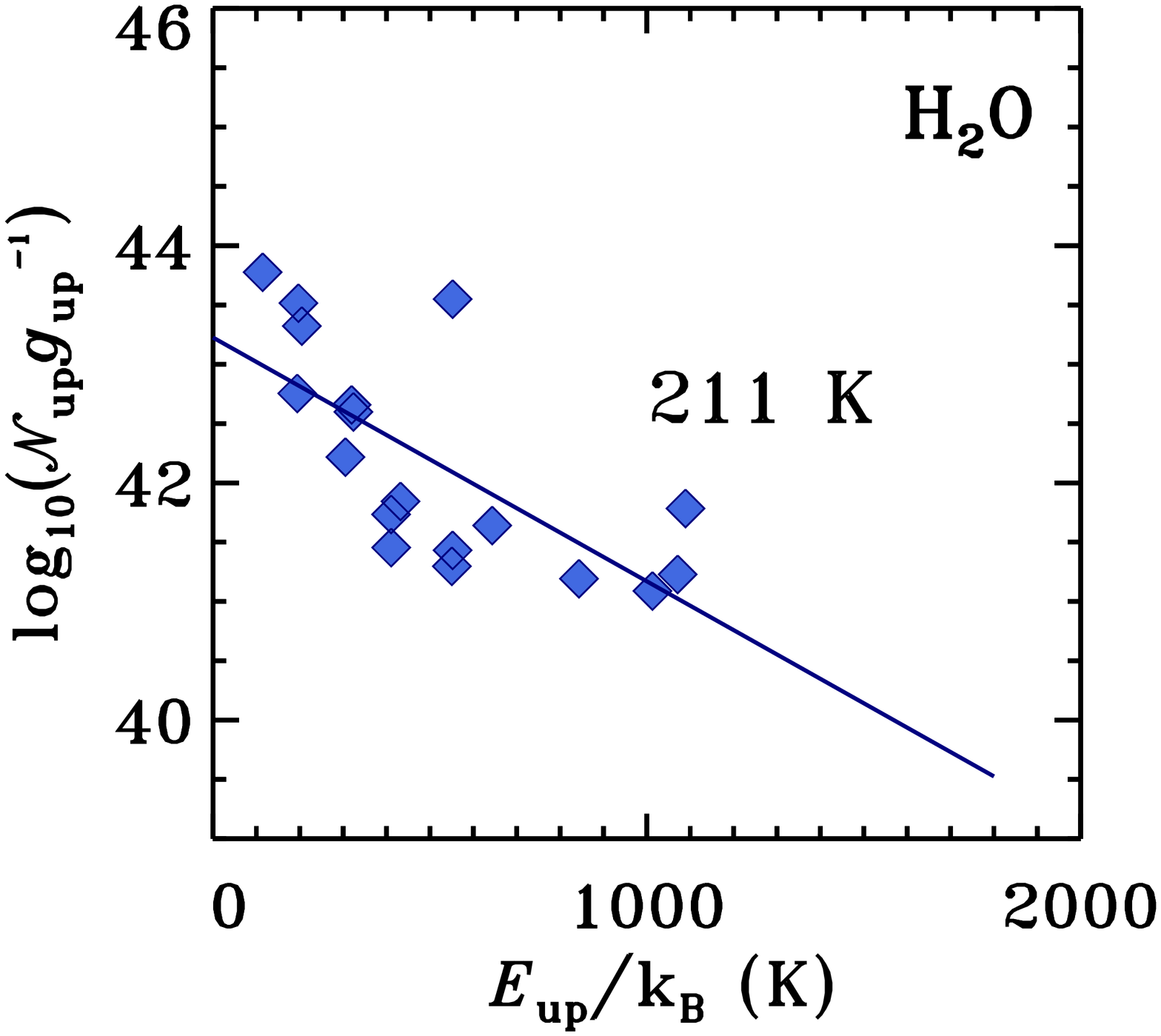} 
       \includegraphics[angle=90,height=4.8cm]{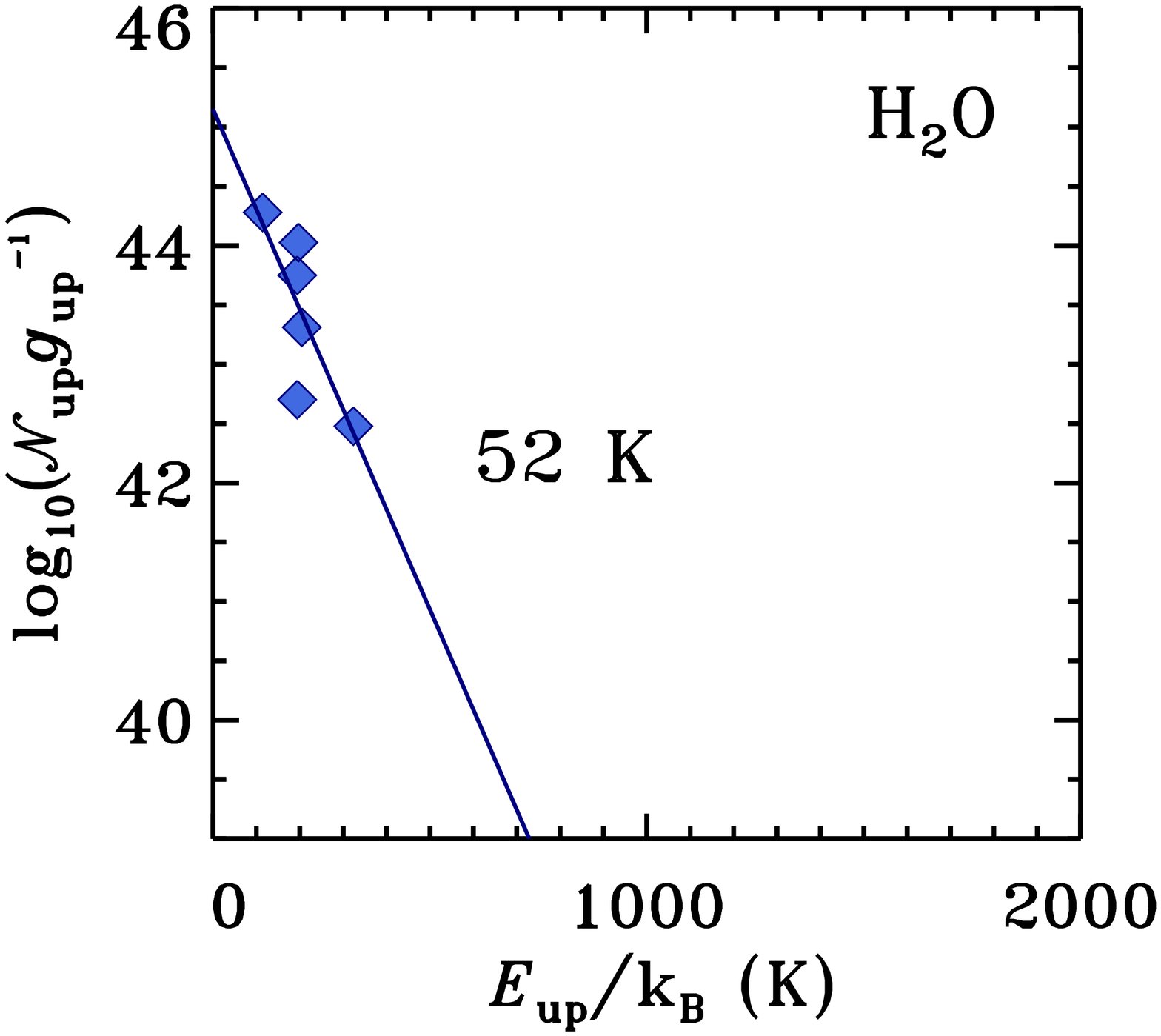} 
            
      \end{center}
  \end{minipage}
    \hfill
   \begin{minipage}[t]{.3\textwidth}
      \begin{center}
    	\includegraphics[angle=90,height=4.8cm]{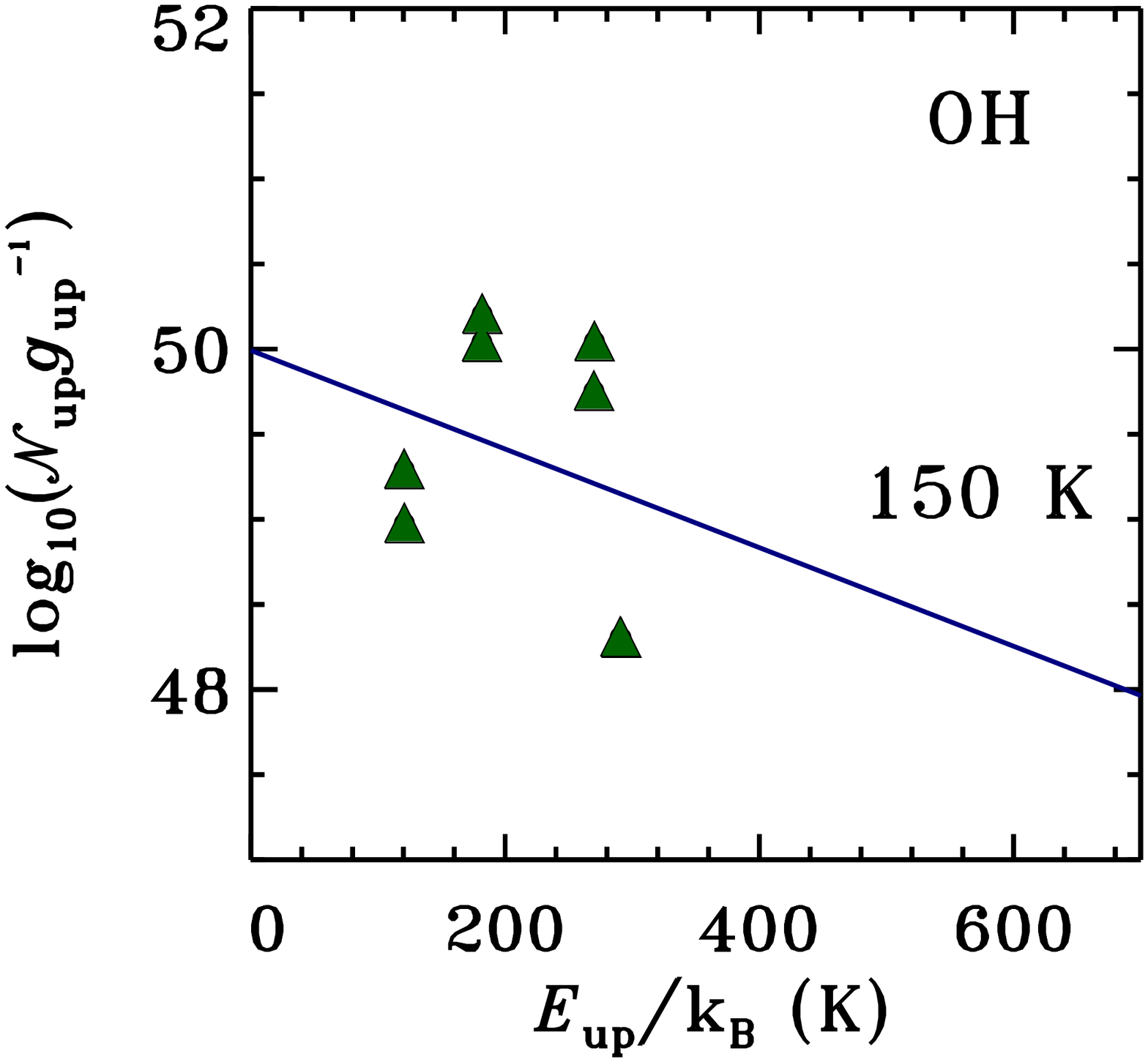} 
        \includegraphics[angle=90,height=4.8cm]{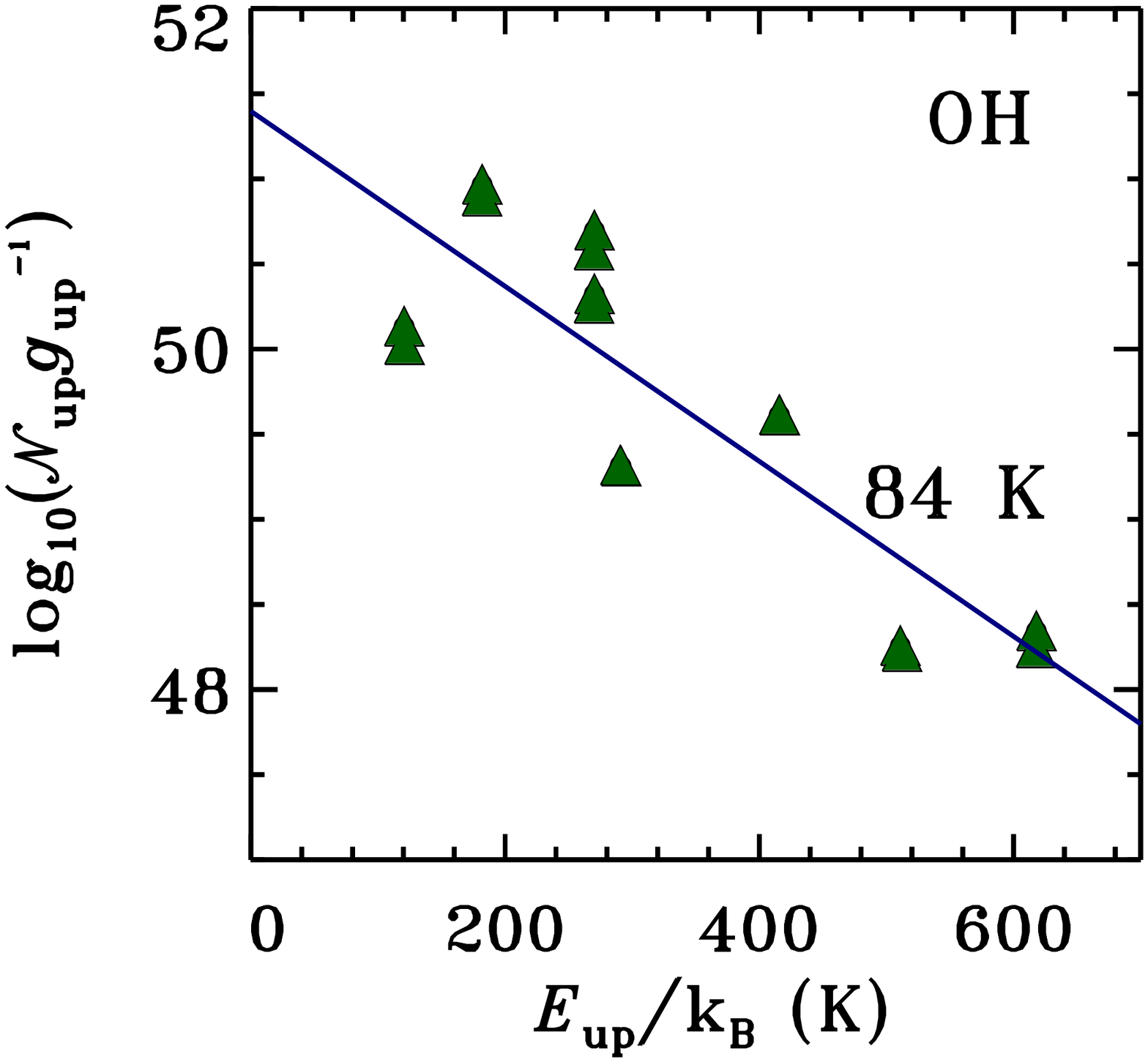} 
        \includegraphics[angle=90,height=4.8cm]{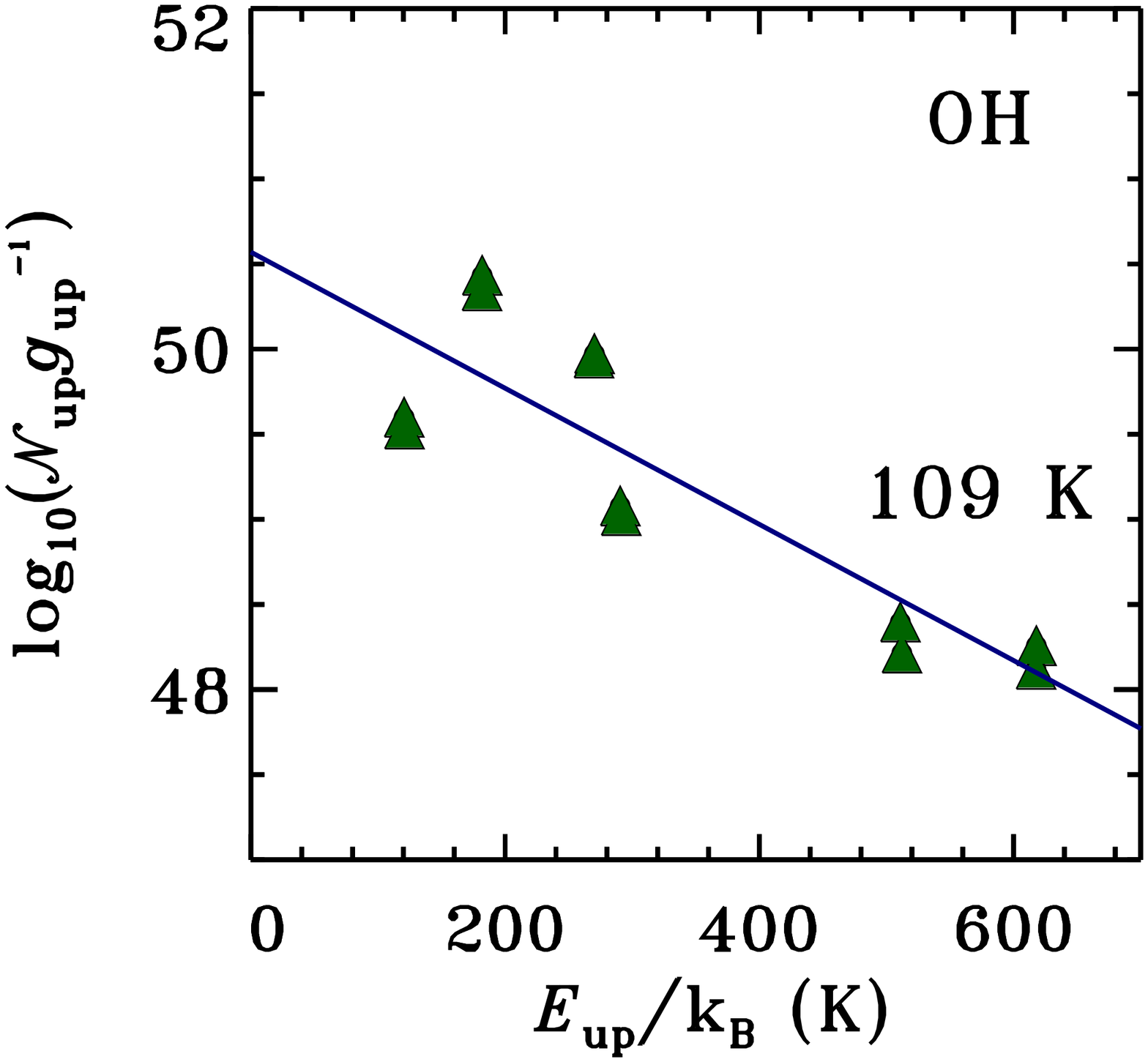} 
        \includegraphics[angle=90,height=4.8cm]{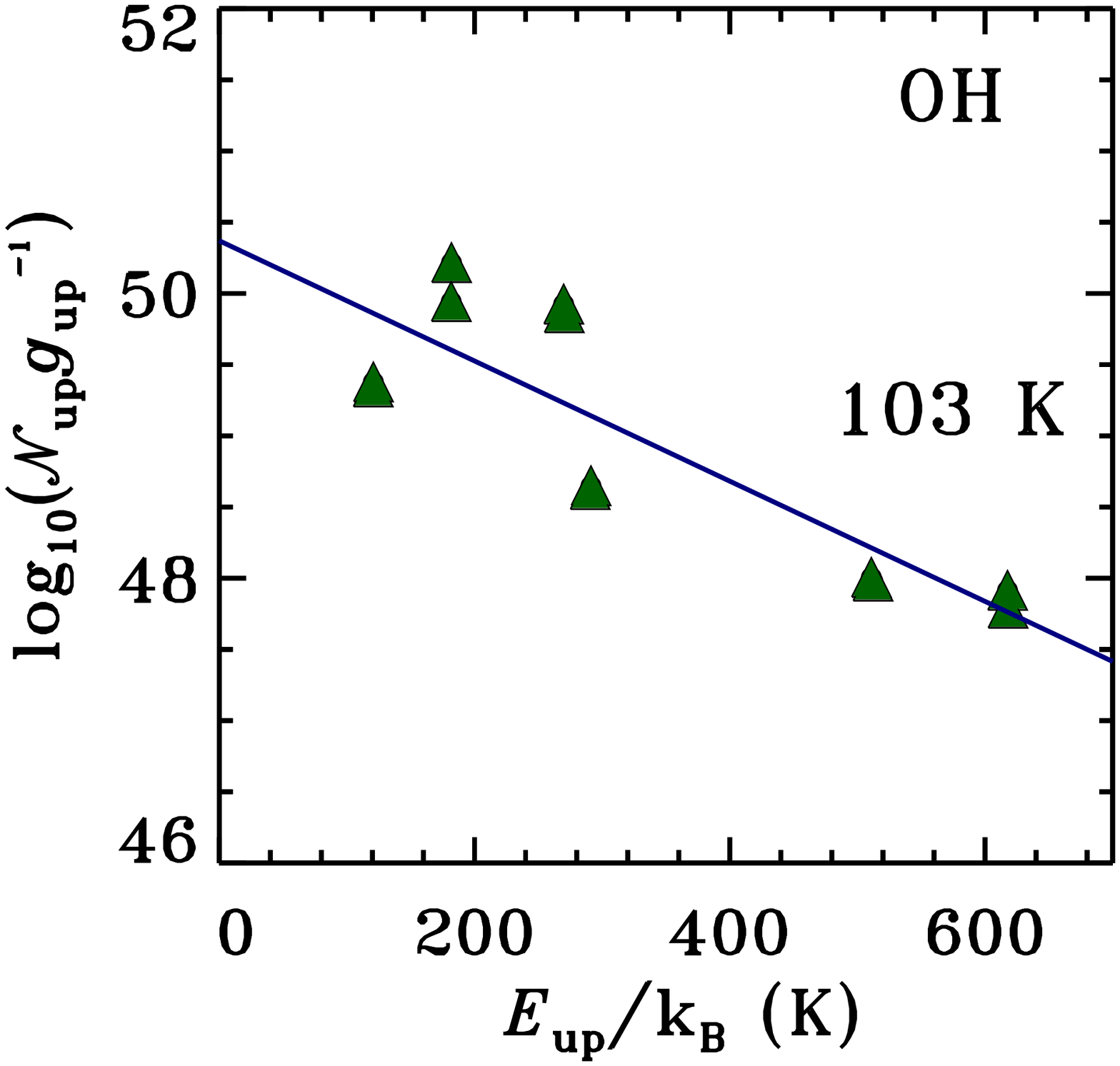} 
        \includegraphics[angle=90,height=4.8cm]{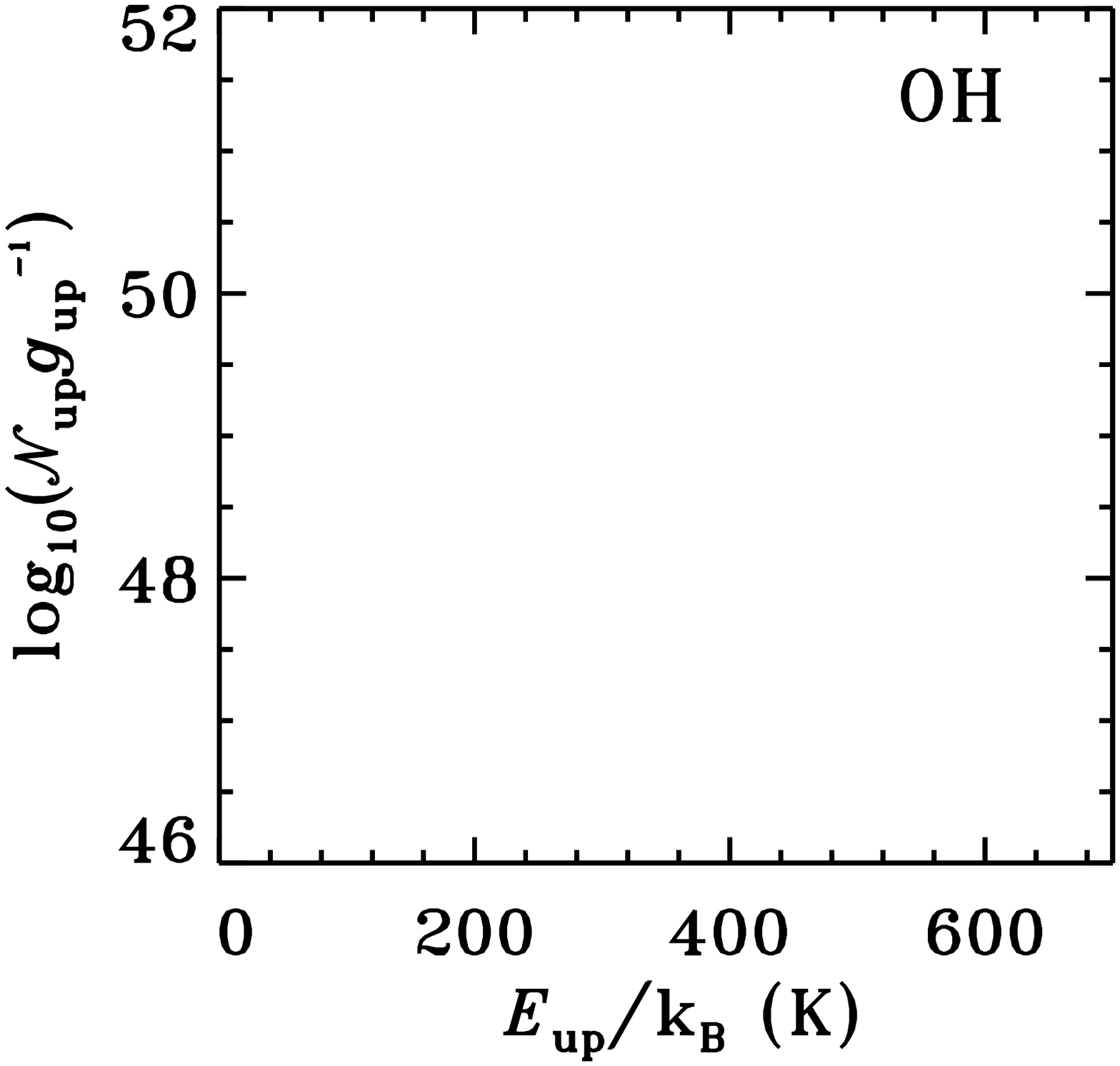} 
                            
      \end{center}
  \end{minipage}
      \hfill
        \caption{\label{dig1} Similar to Figure \ref{molexc}, but for 
        NGC1333 IRAS4A, NGC1333 IRAS4B (based on \citealt{He12}), IRAS 03301, 
        B1-a, and B1-c.}
\end{figure*}
\renewcommand{\thefigure}{\thesection.\arabic{figure} (Cont.)}
\addtocounter{figure}{-1}   
\begin{figure*}[!tb]
  \begin{minipage}[t]{.3\textwidth}
  \begin{center}
       \includegraphics[angle=90,height=4.8cm]{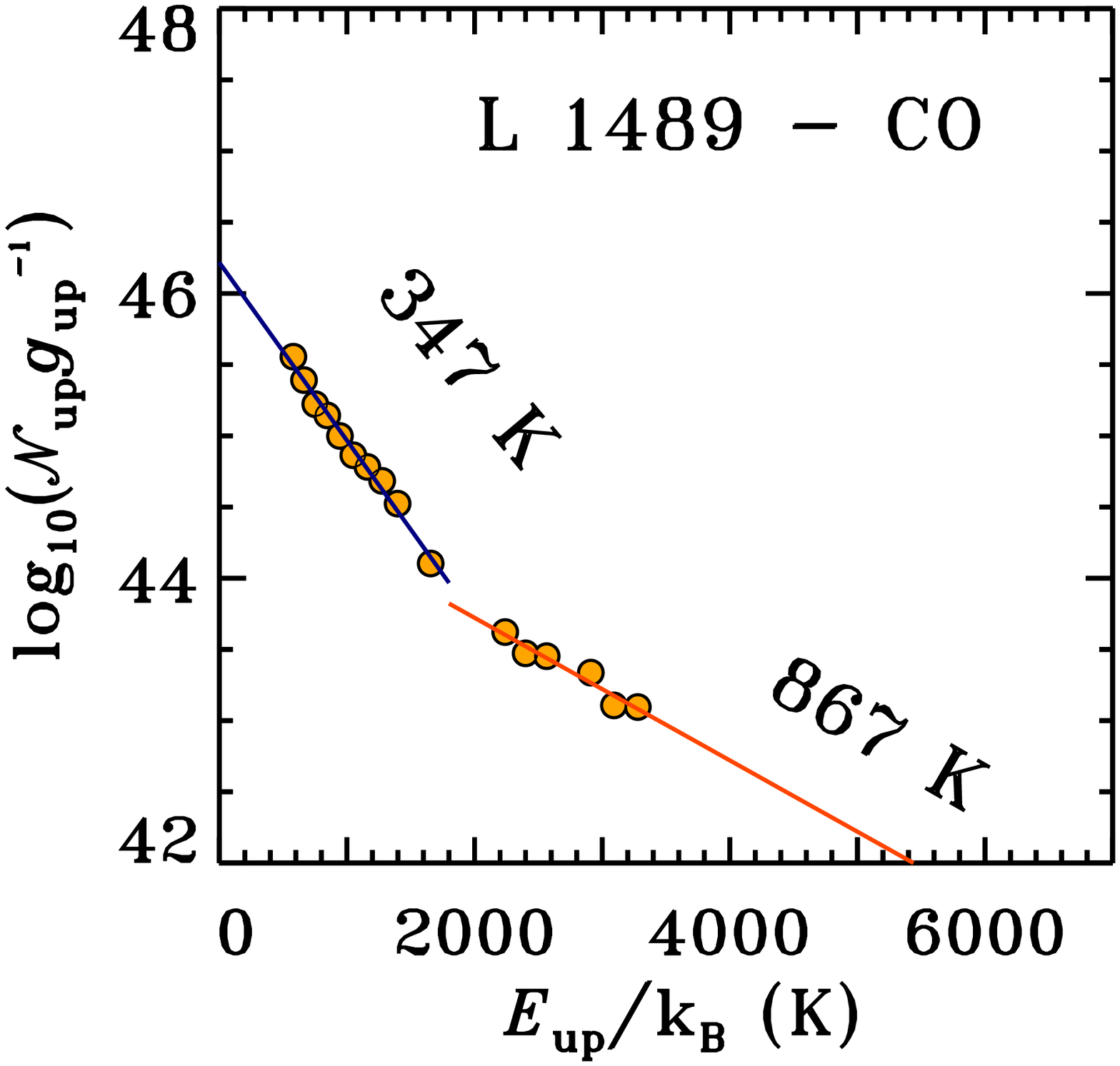} 
       \includegraphics[angle=90,height=4.8cm]{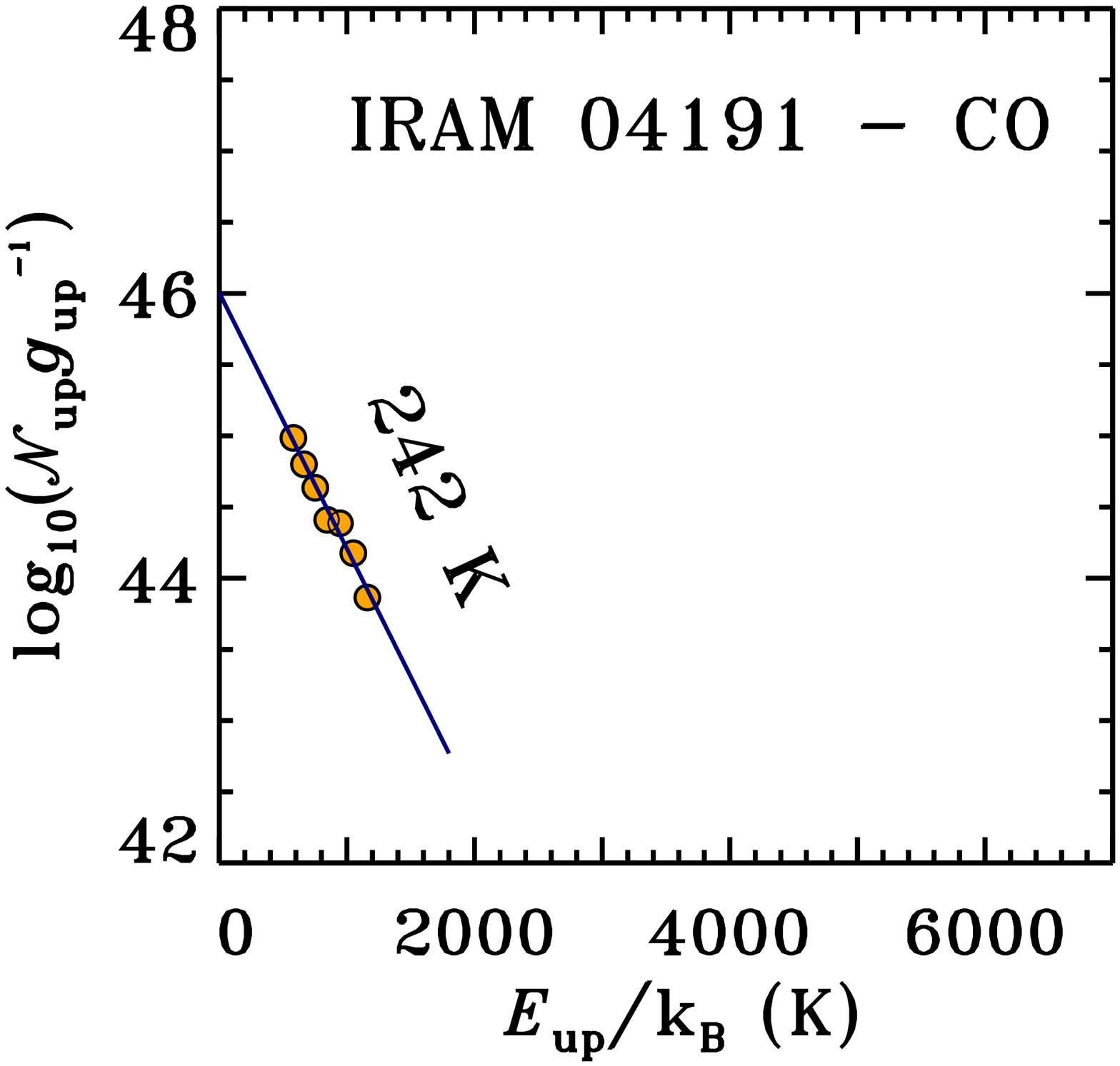} 
        \includegraphics[angle=90,height=4.8cm]{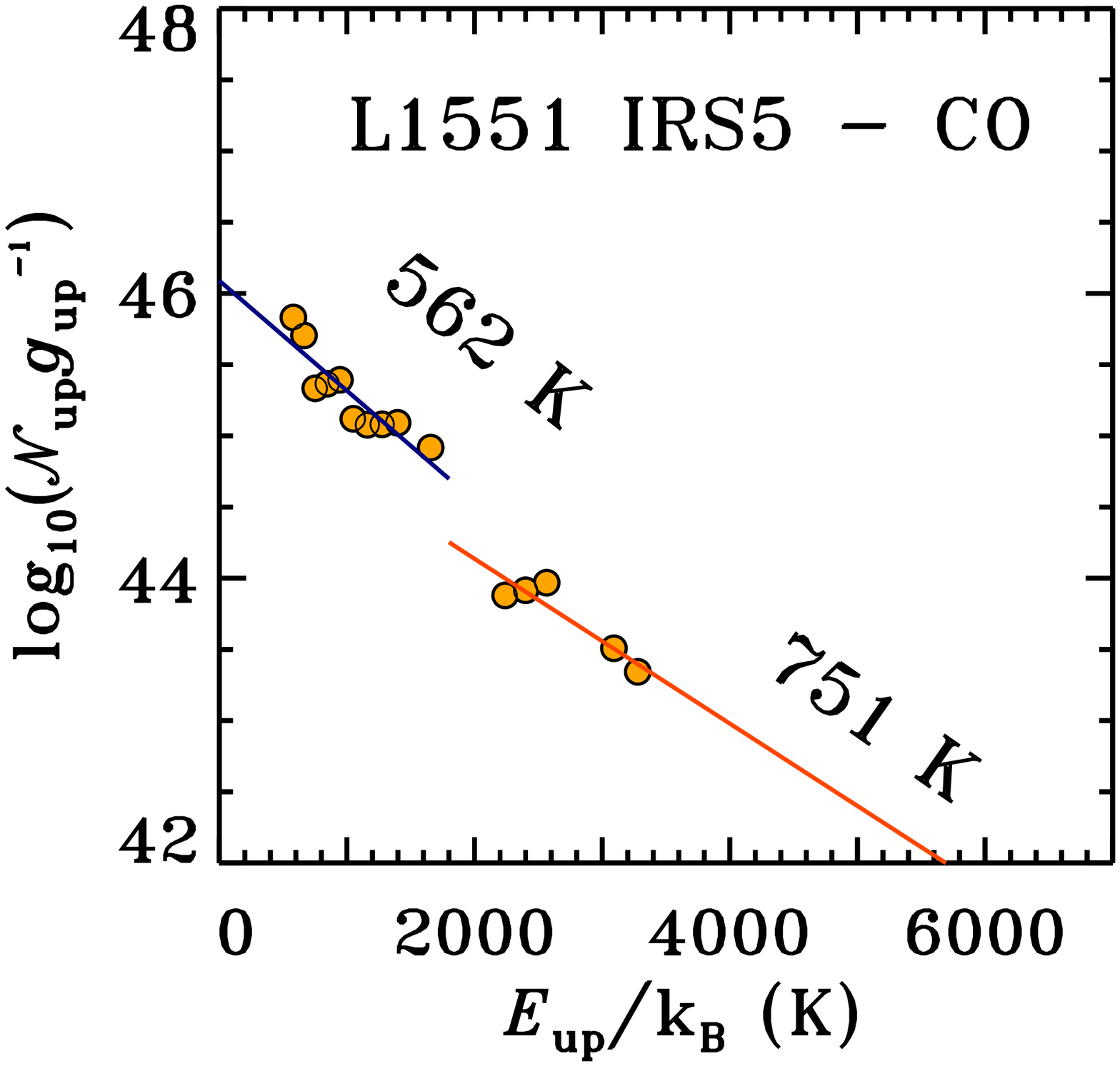} 
       \includegraphics[angle=90,height=4.8cm]{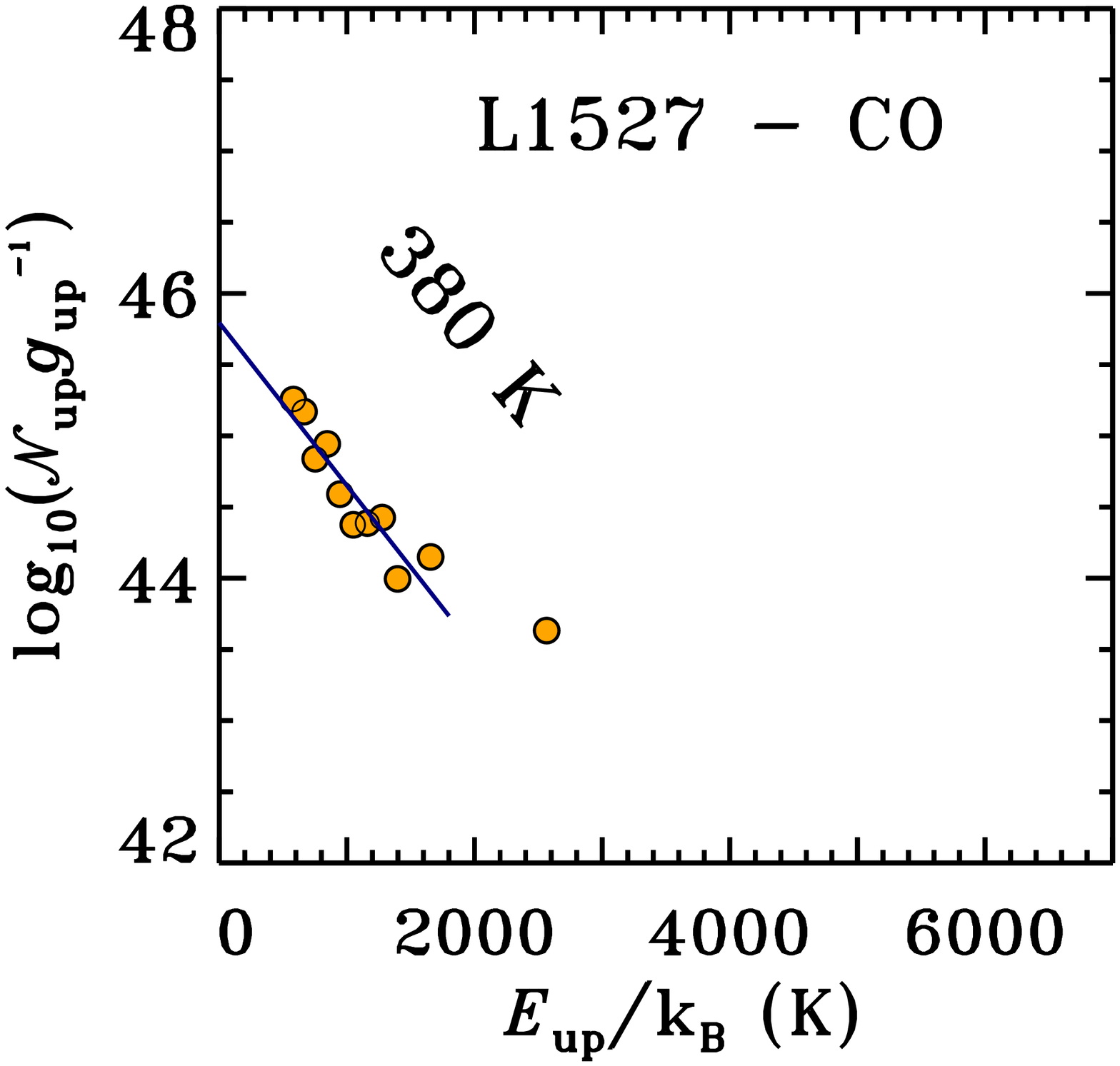} 
        \includegraphics[angle=90,height=4.8cm]{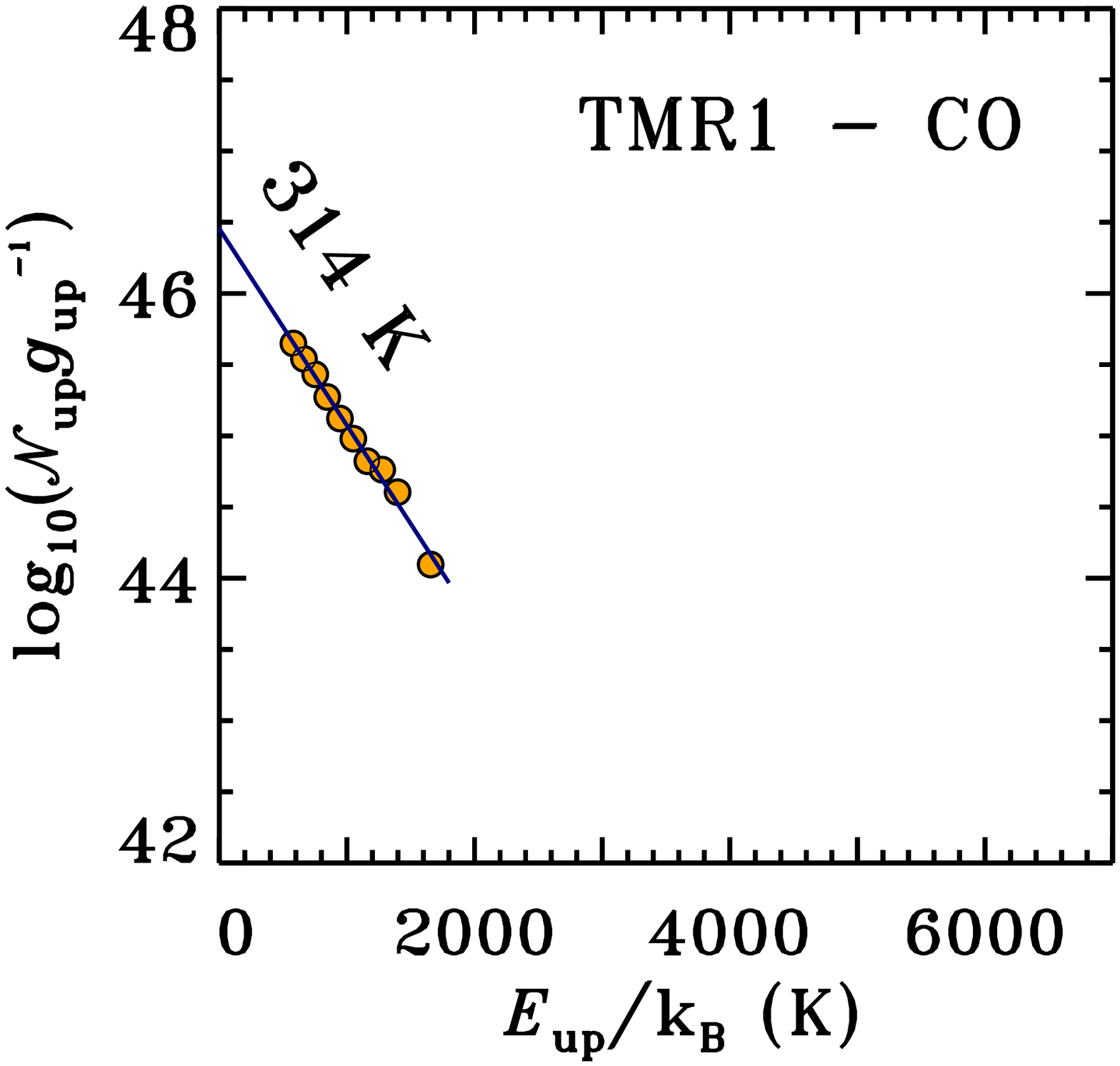} 
  \end{center}
  \end{minipage}
  \hfill
  \begin{minipage}[t]{.3\textwidth}
      \begin{center}
   	   \includegraphics[angle=90,height=4.8cm]{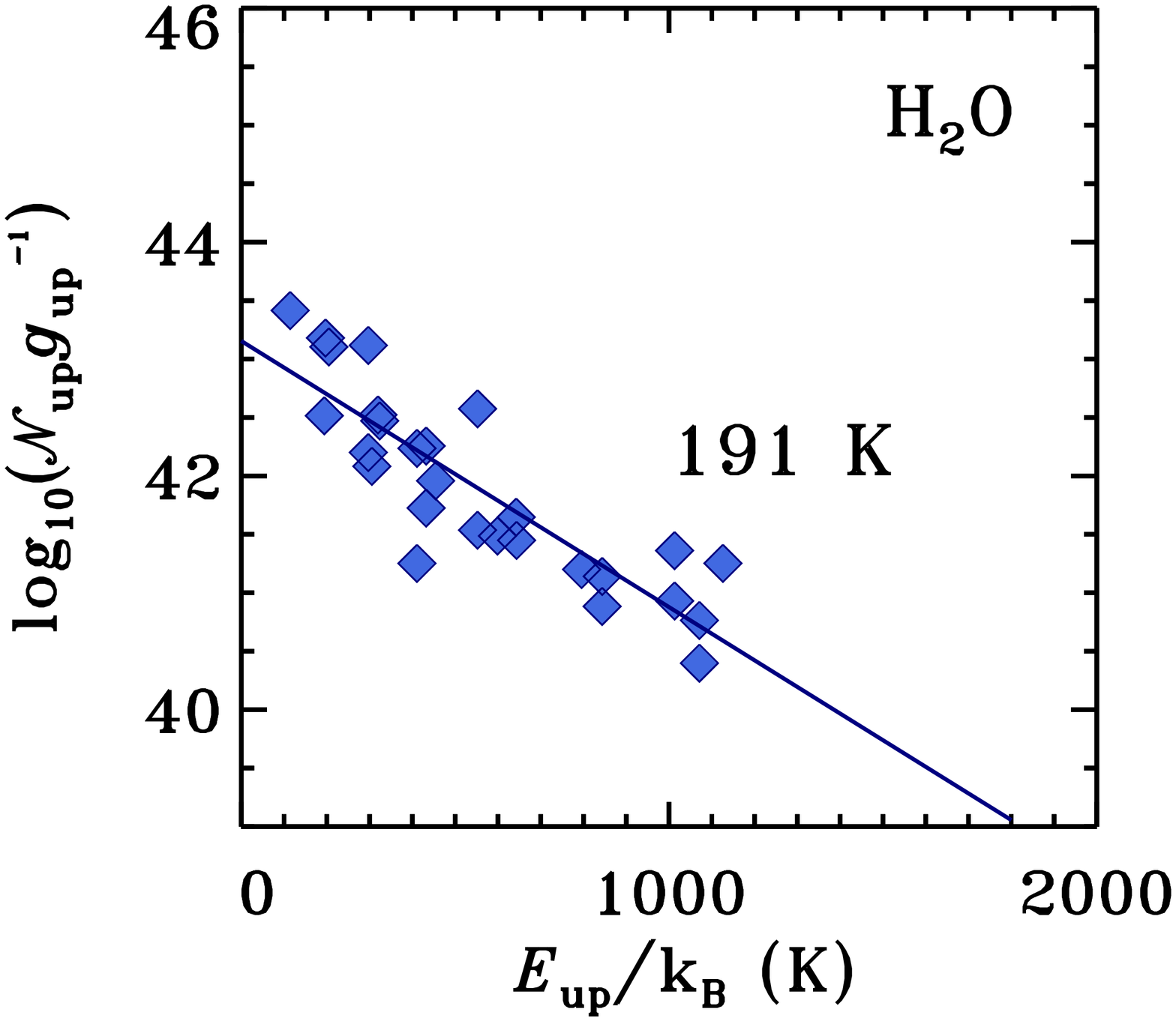} 
       \includegraphics[angle=90,height=4.8cm]{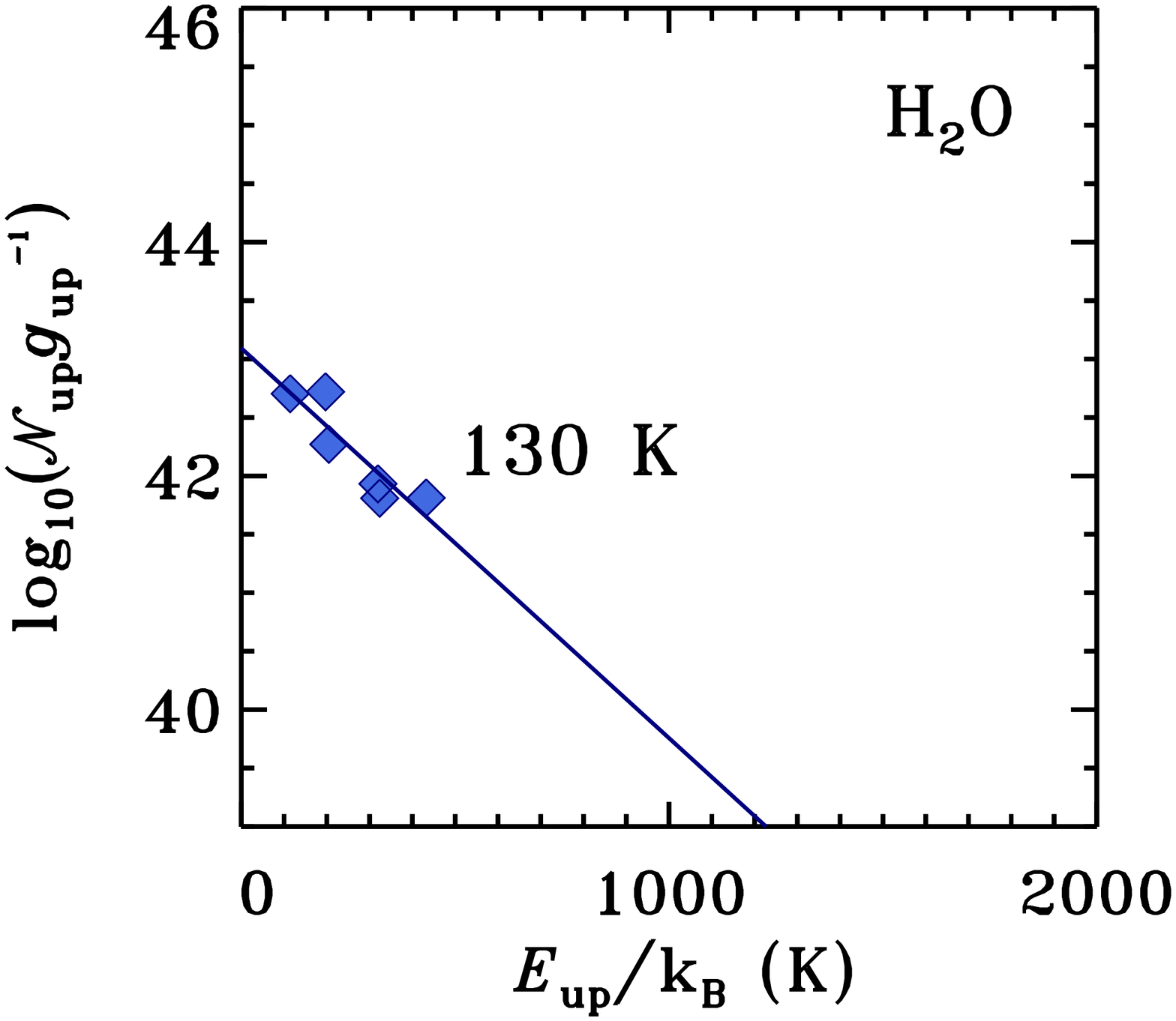} 
       \includegraphics[angle=90,height=4.8cm]{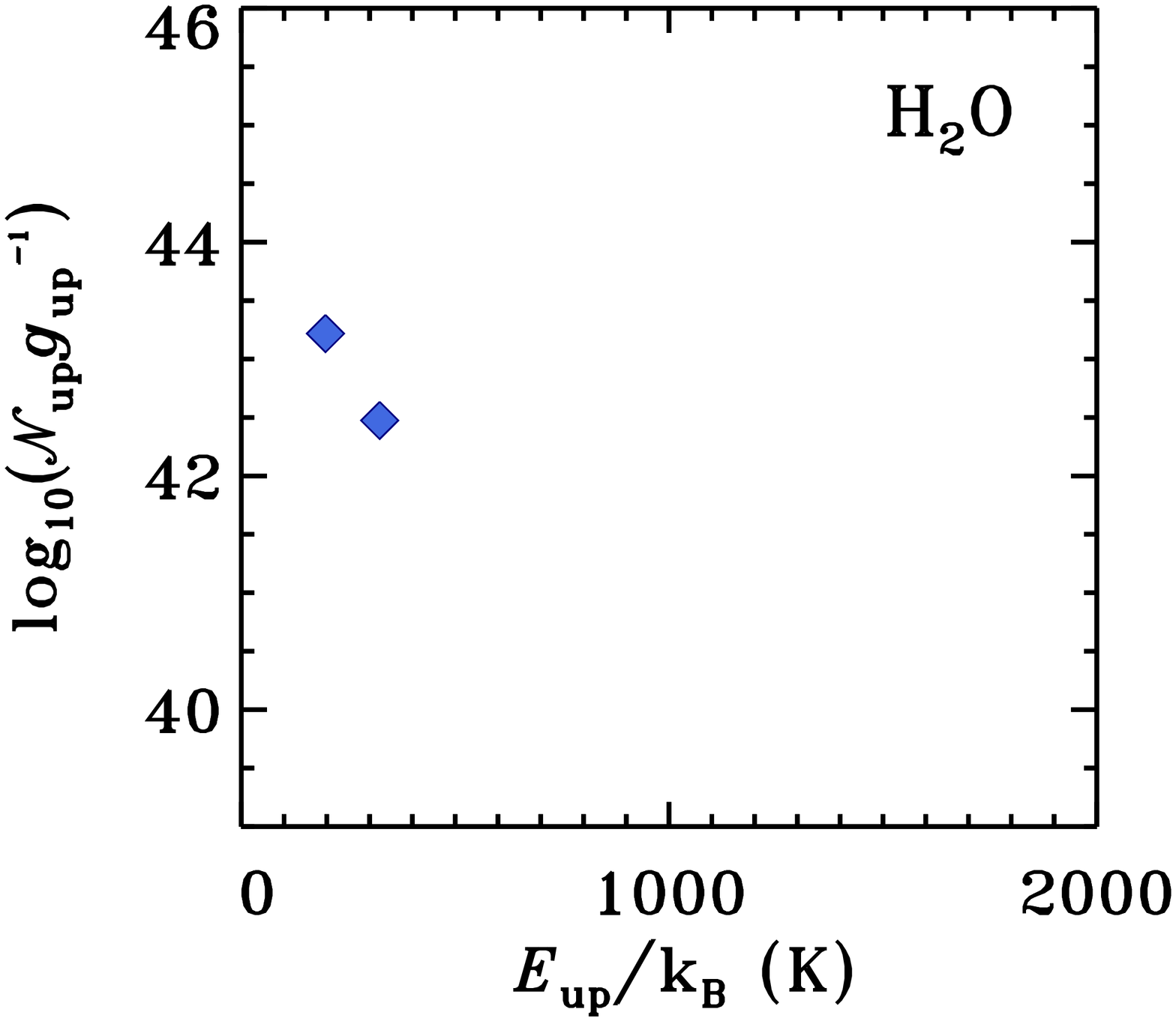} 
       \includegraphics[angle=90,height=4.8cm]{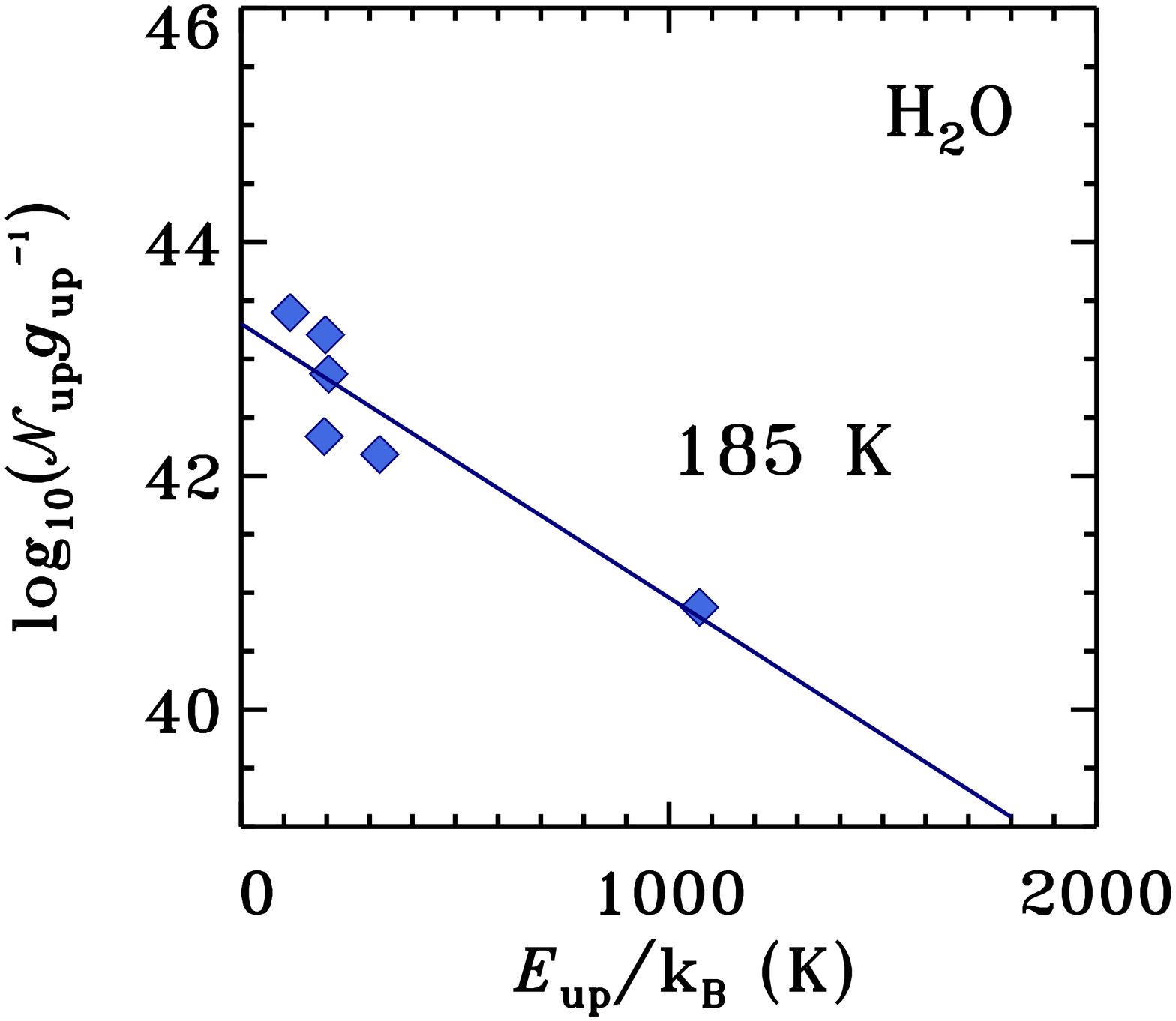} 
       \includegraphics[angle=90,height=4.8cm]{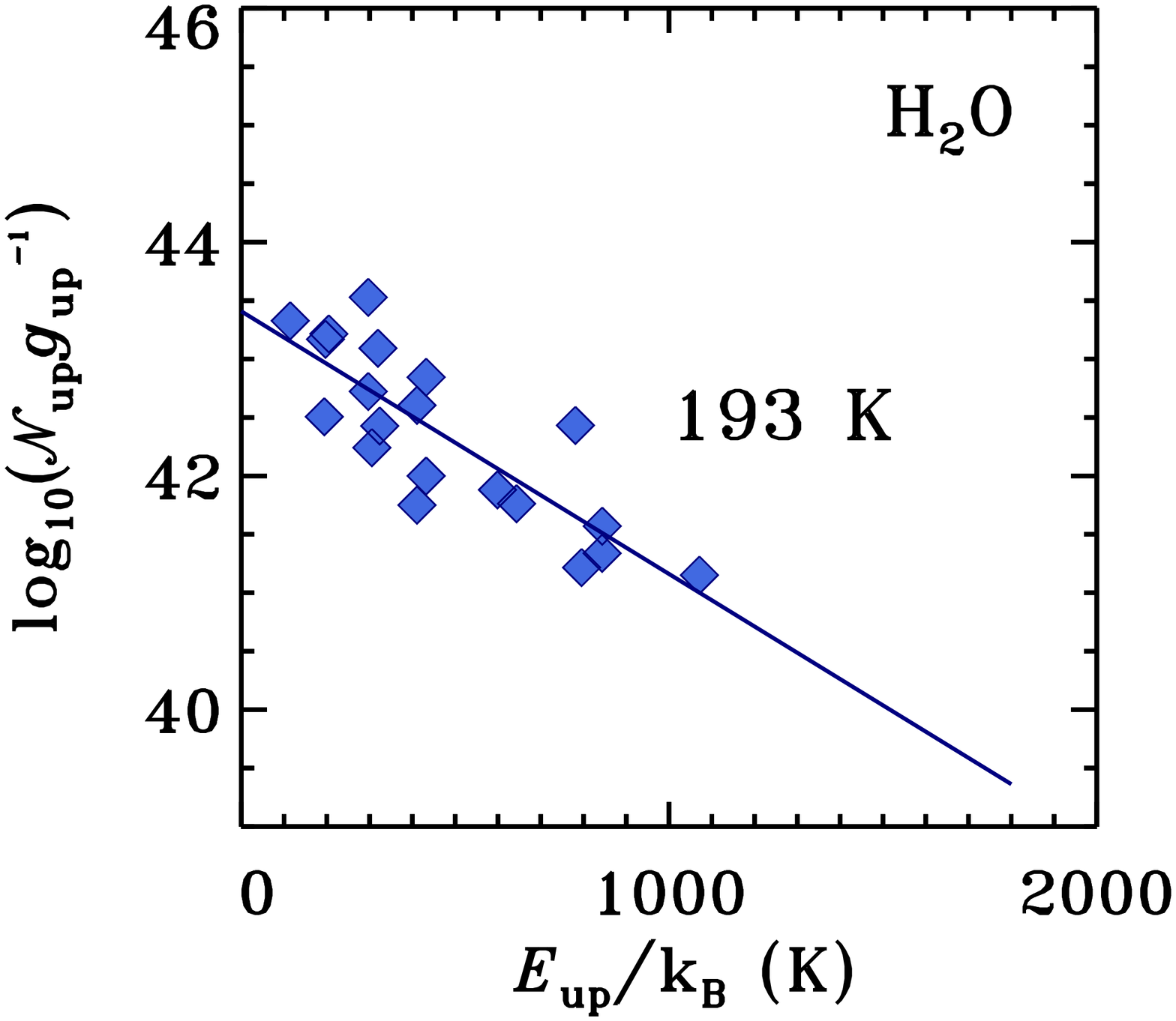} 
            
      \end{center}
  \end{minipage}
    \hfill
   \begin{minipage}[t]{.3\textwidth}
      \begin{center}
    	\includegraphics[angle=90,height=4.8cm]{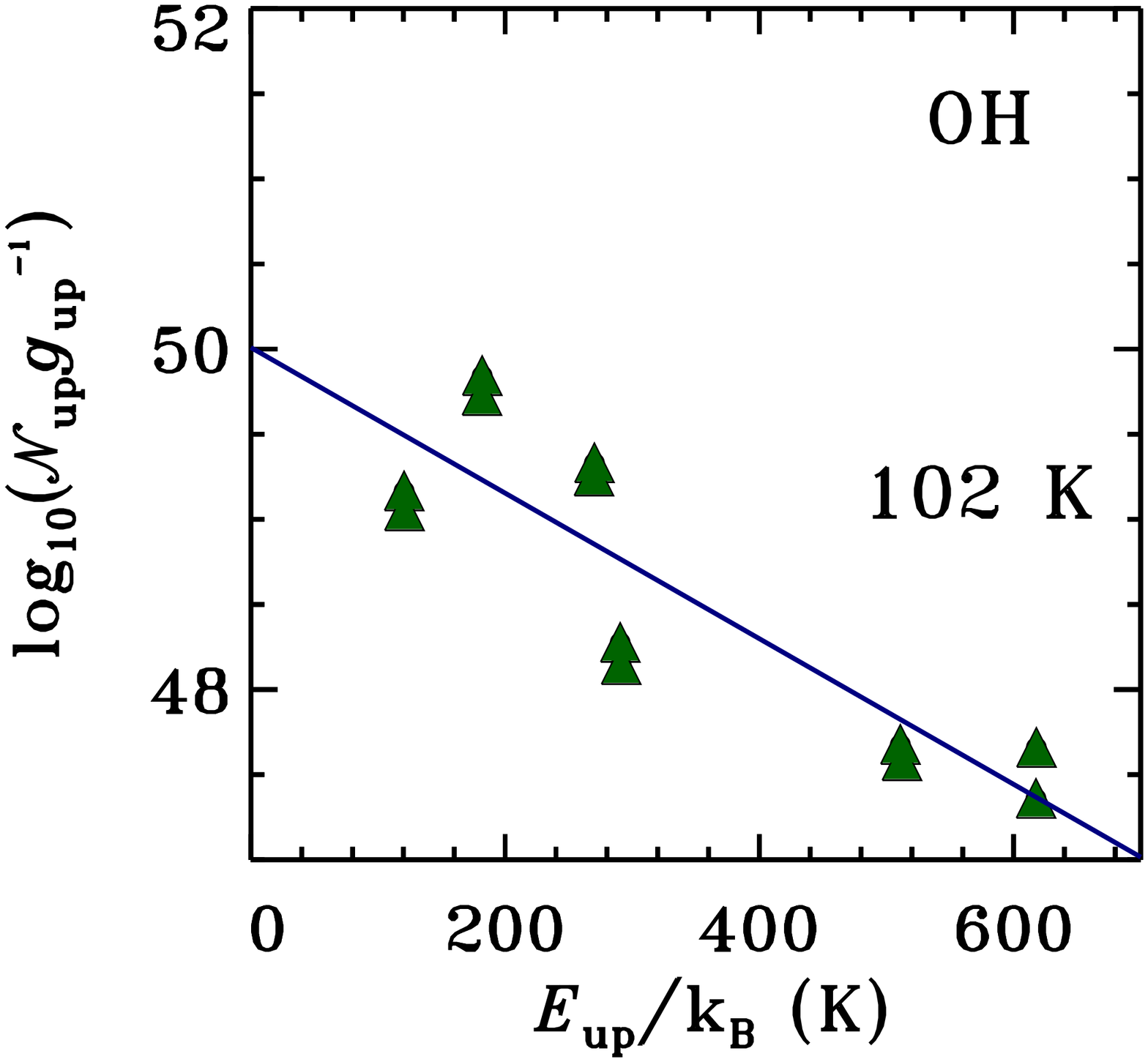} 
        \includegraphics[angle=90,height=4.8cm]{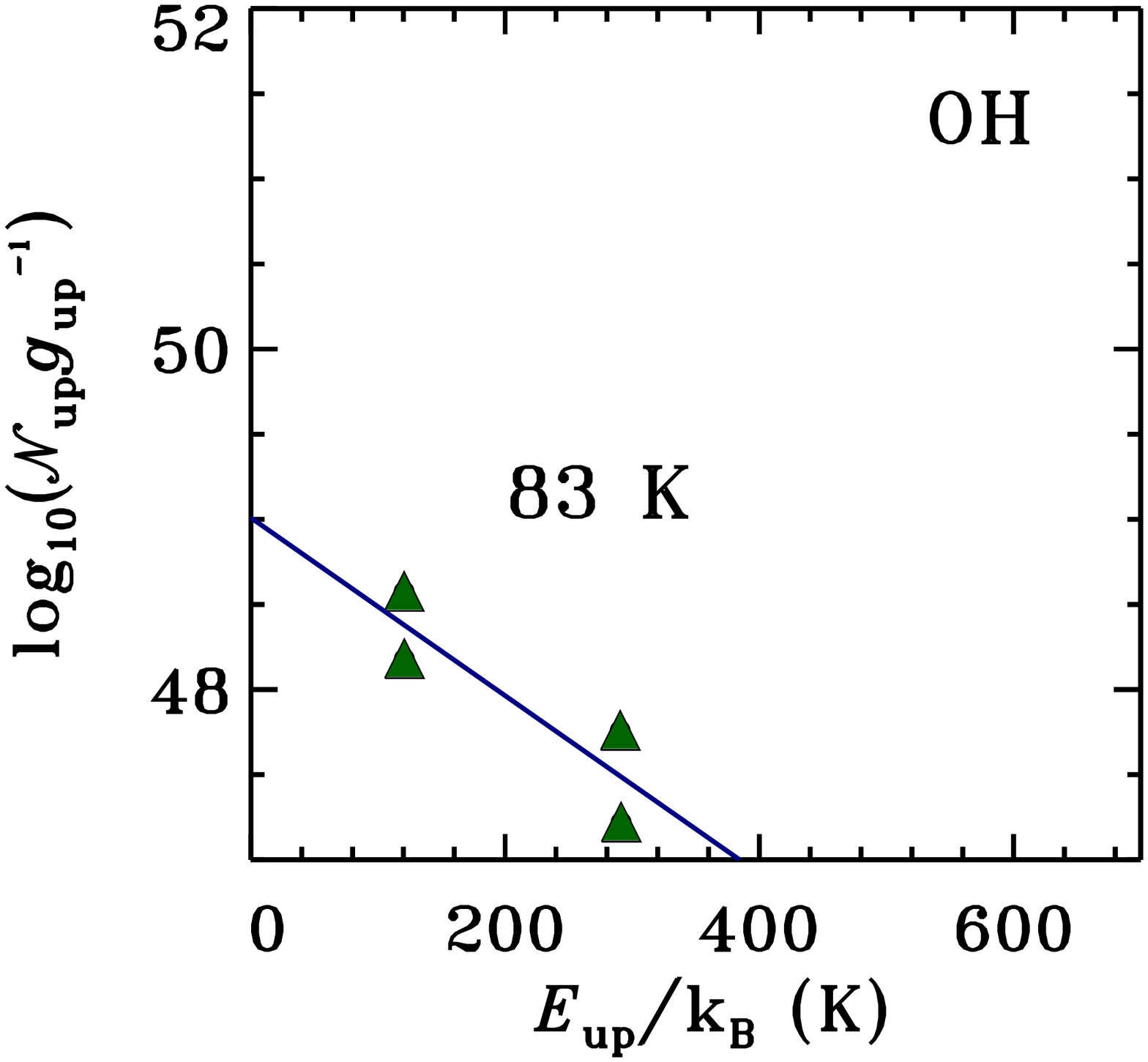} 
        \includegraphics[angle=90,height=4.8cm]{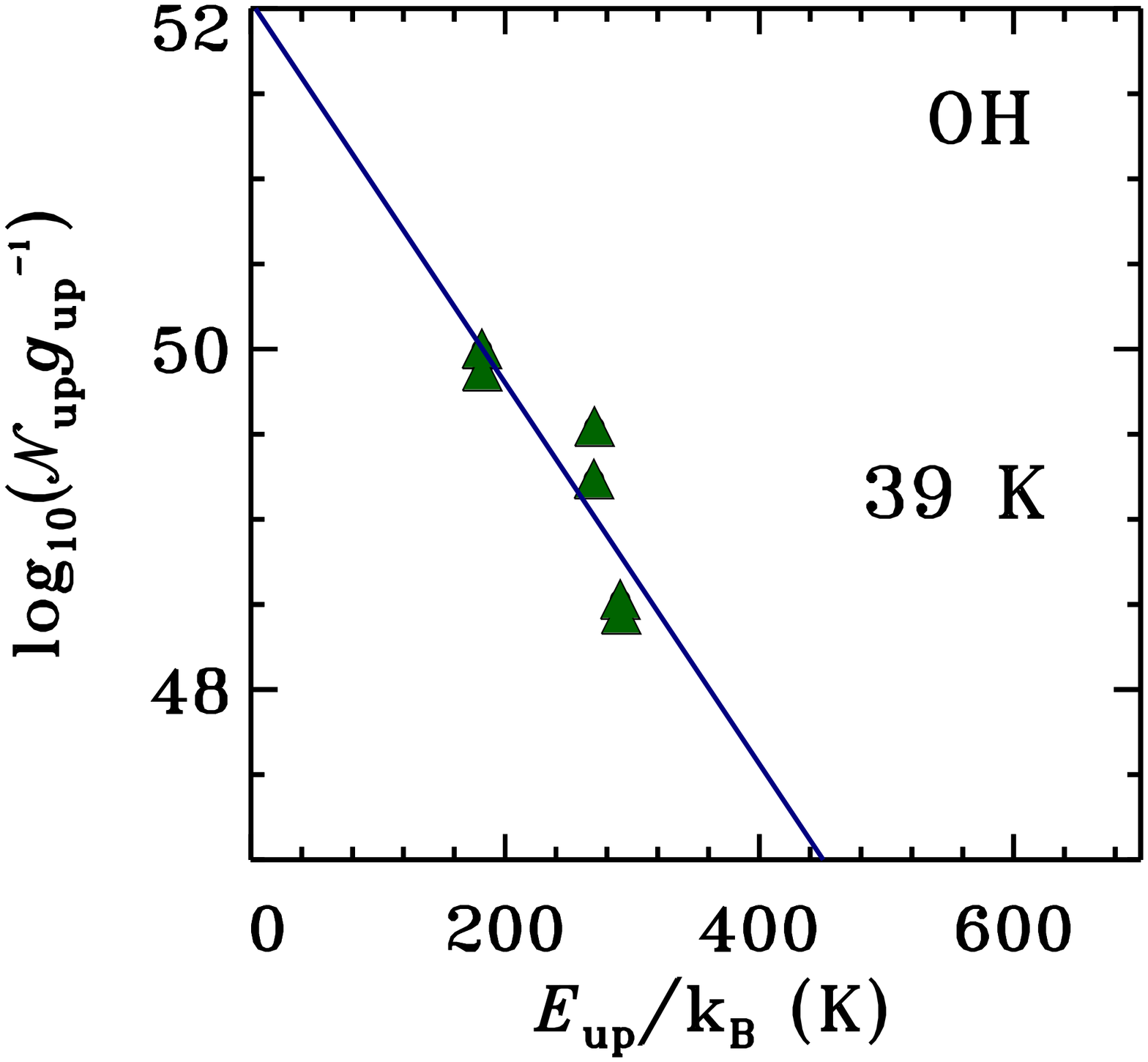} 
        \includegraphics[angle=90,height=4.8cm]{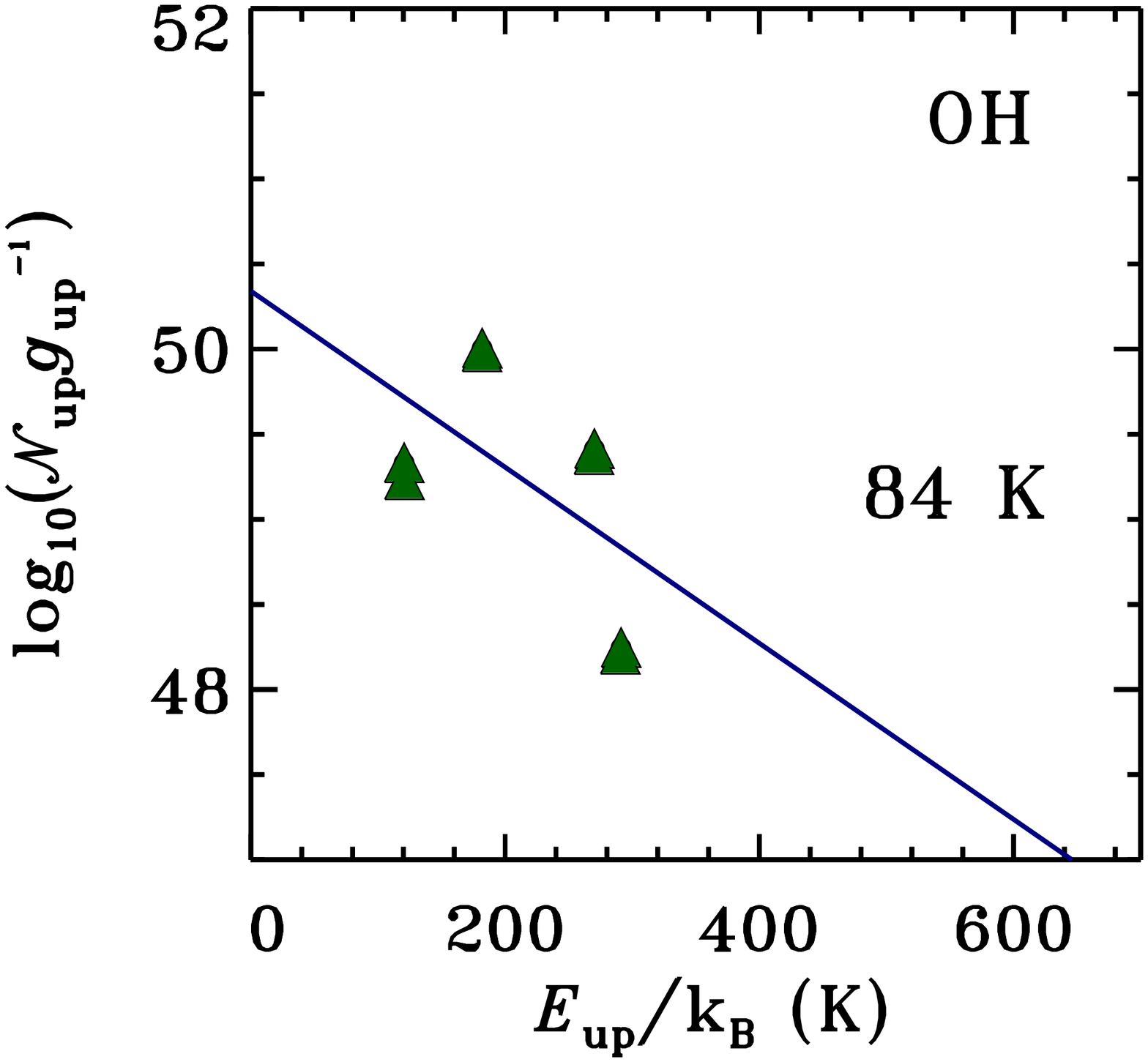} 
        \includegraphics[angle=90,height=4.8cm]{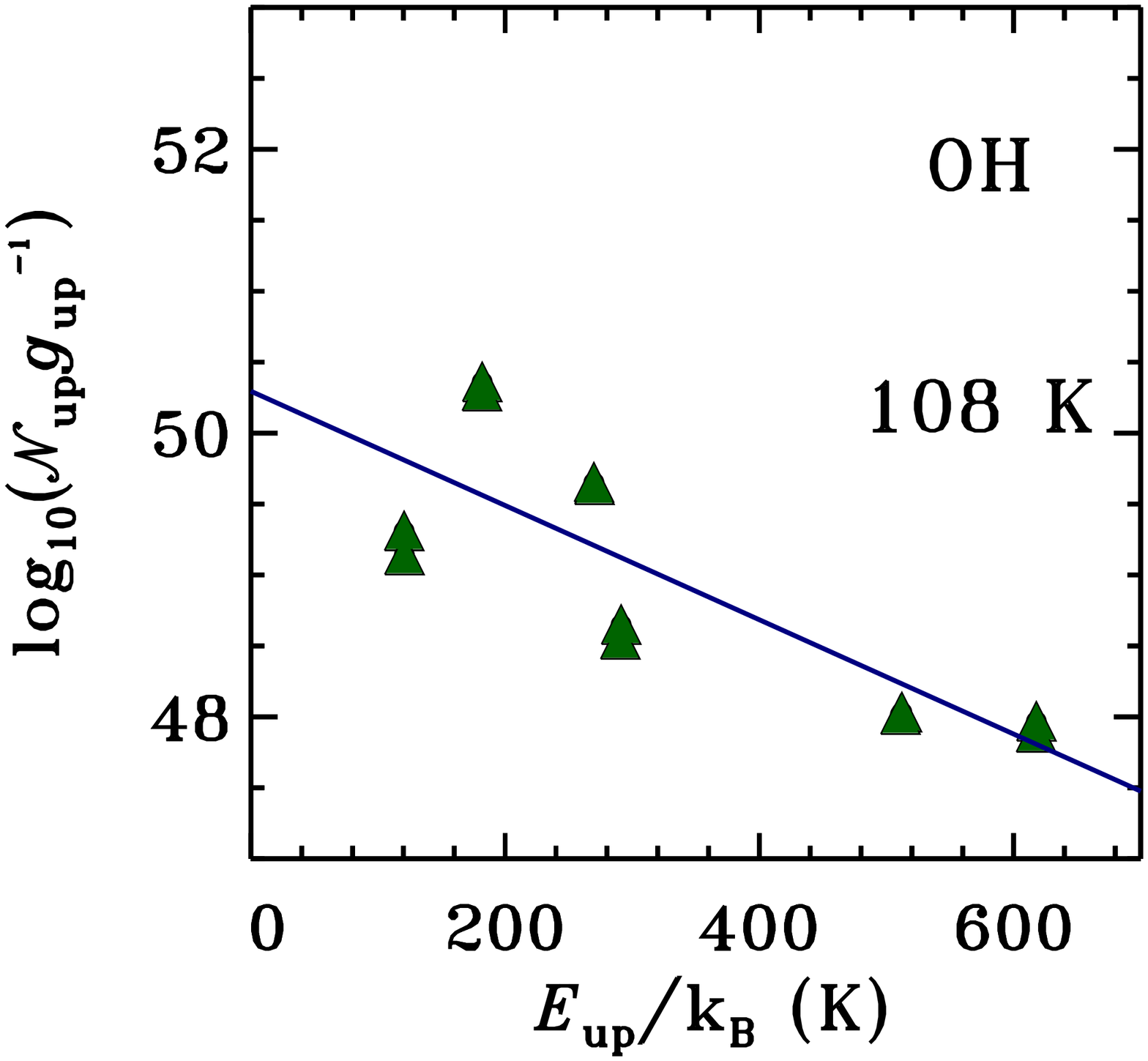} 
                            
      \end{center}
  \end{minipage}
      \hfill
        \caption{\label{dig2} Similar to Figure \ref{molexc}, but for 
        L1489, IRAM04191, L1551 IRS5, L1527, and TMR1.}
\end{figure*}
\renewcommand{\thefigure}{\thesection.\arabic{figure} (Cont.)}
\addtocounter{figure}{-1}   
\begin{figure*}[!tb]
  \begin{minipage}[t]{.3\textwidth}
  \begin{center}
       \includegraphics[angle=90,height=4.8cm]{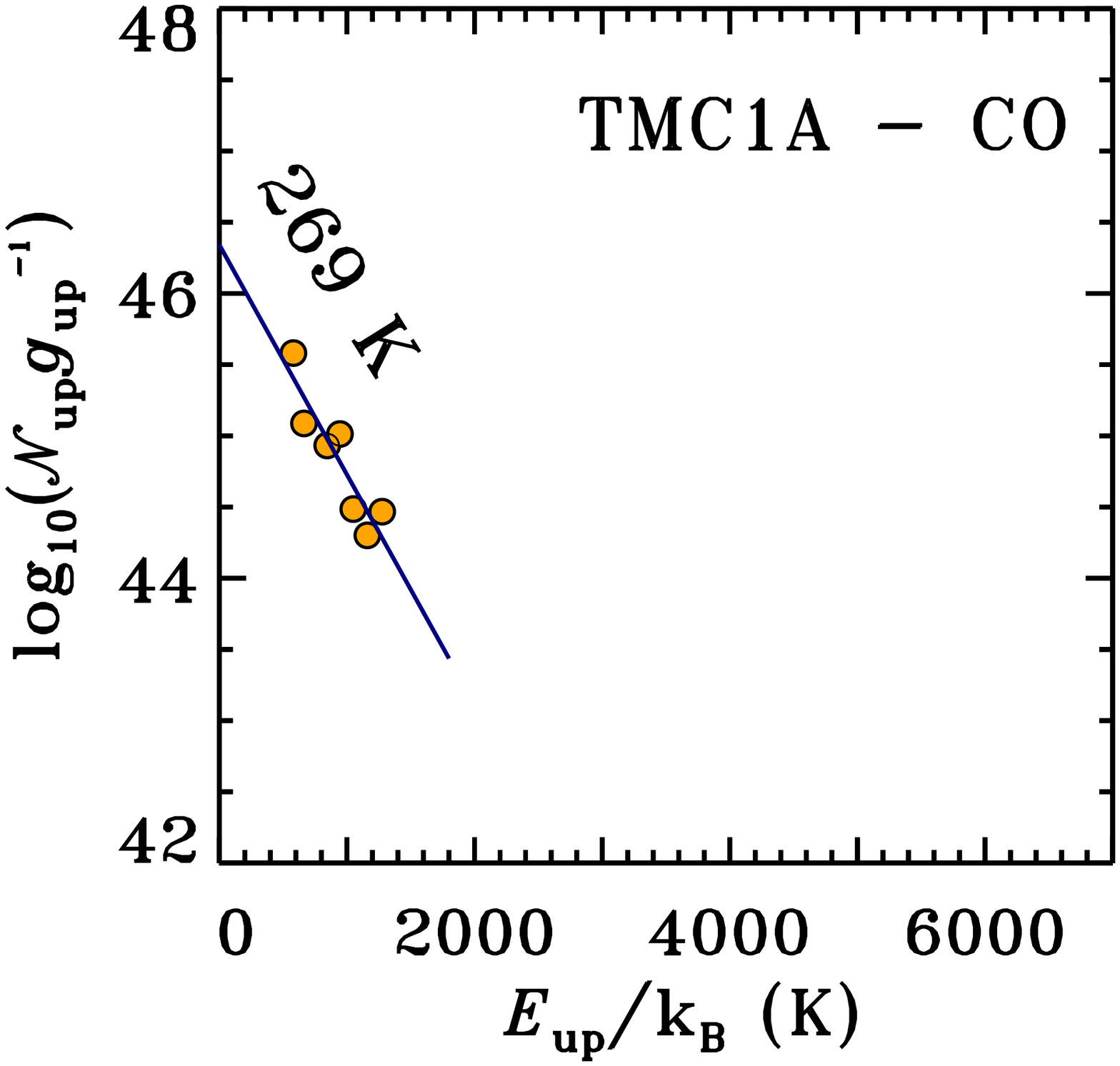} 
       \includegraphics[angle=90,height=4.8cm]{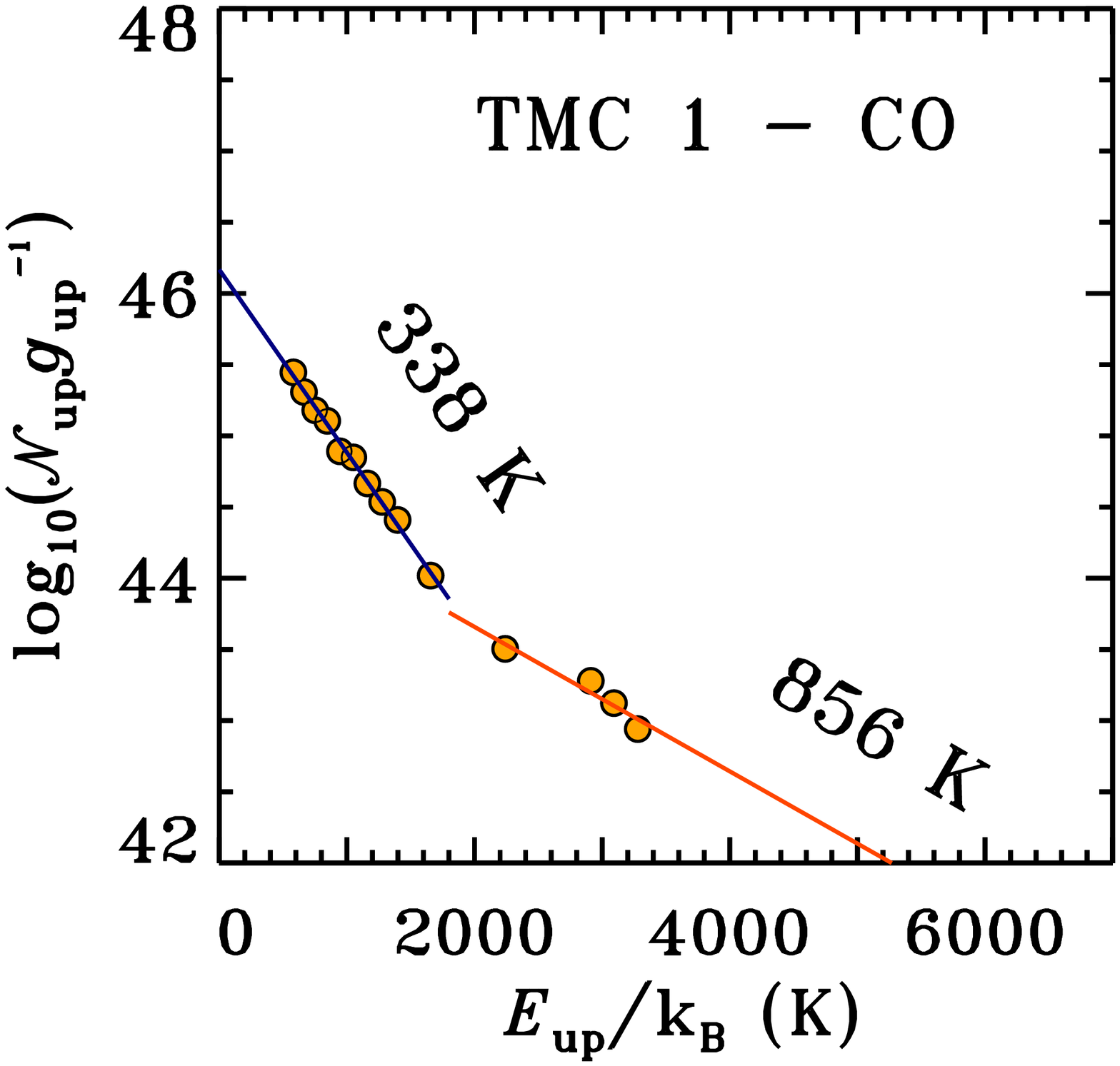} 
        \includegraphics[angle=90,height=4.8cm]{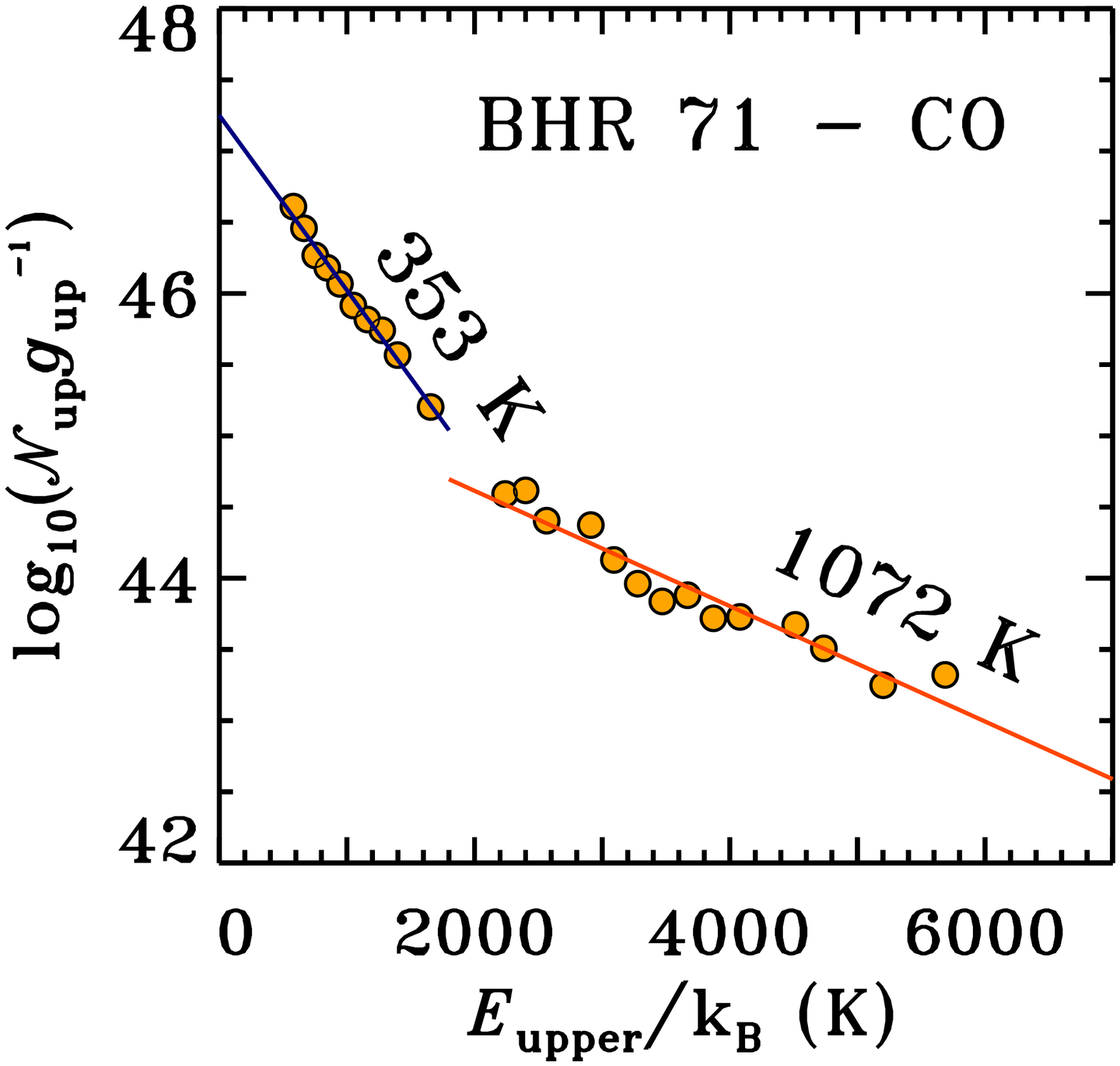} 
       \includegraphics[angle=90,height=4.8cm]{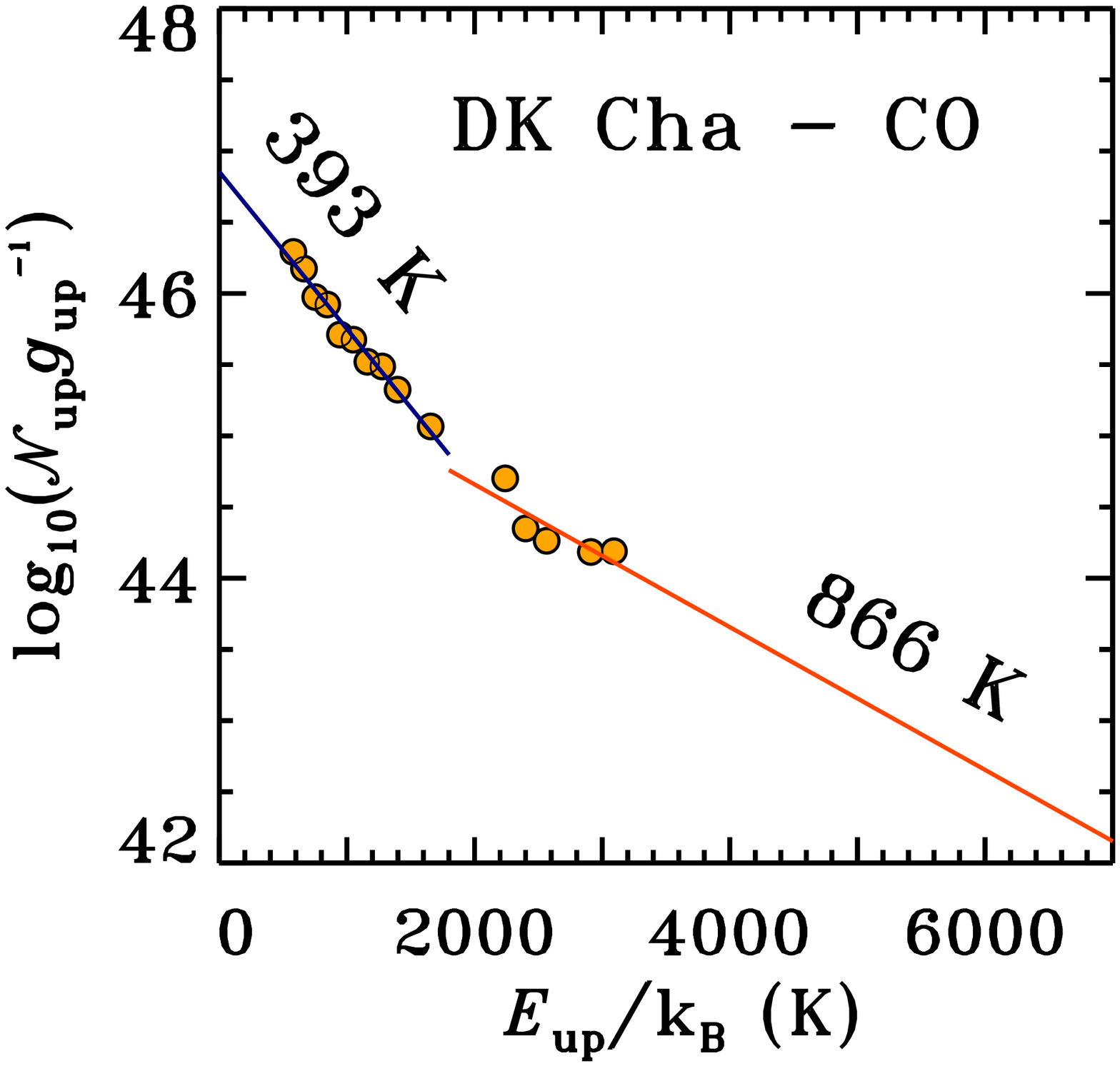} 
        \includegraphics[angle=90,height=4.8cm]{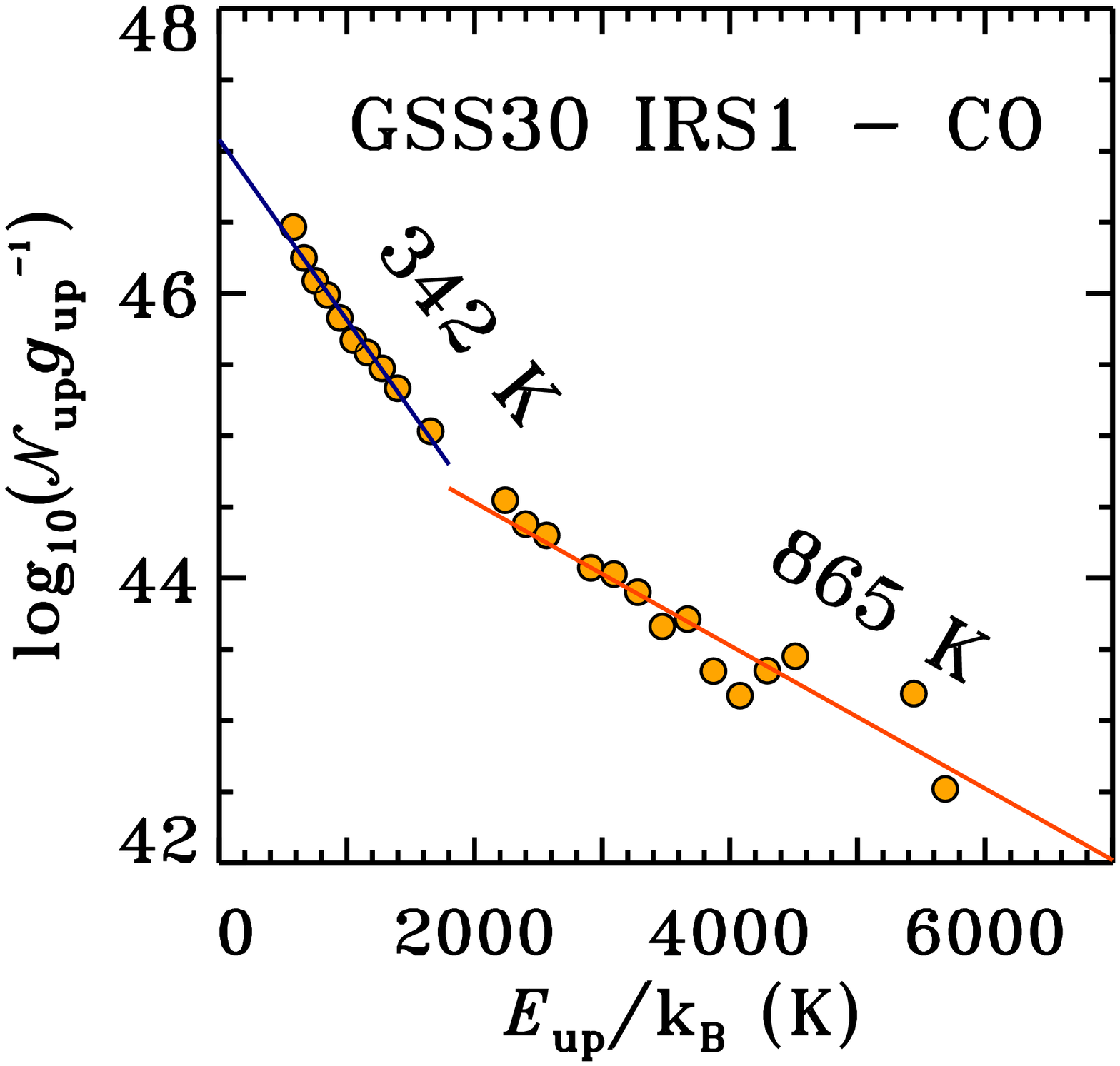} 
  \end{center}
  \end{minipage}
  \hfill
  \begin{minipage}[t]{.3\textwidth}
      \begin{center}
   	   \includegraphics[angle=90,height=4.8cm]{wdiag_l1455irs3.eps} 
       \includegraphics[angle=90,height=4.8cm]{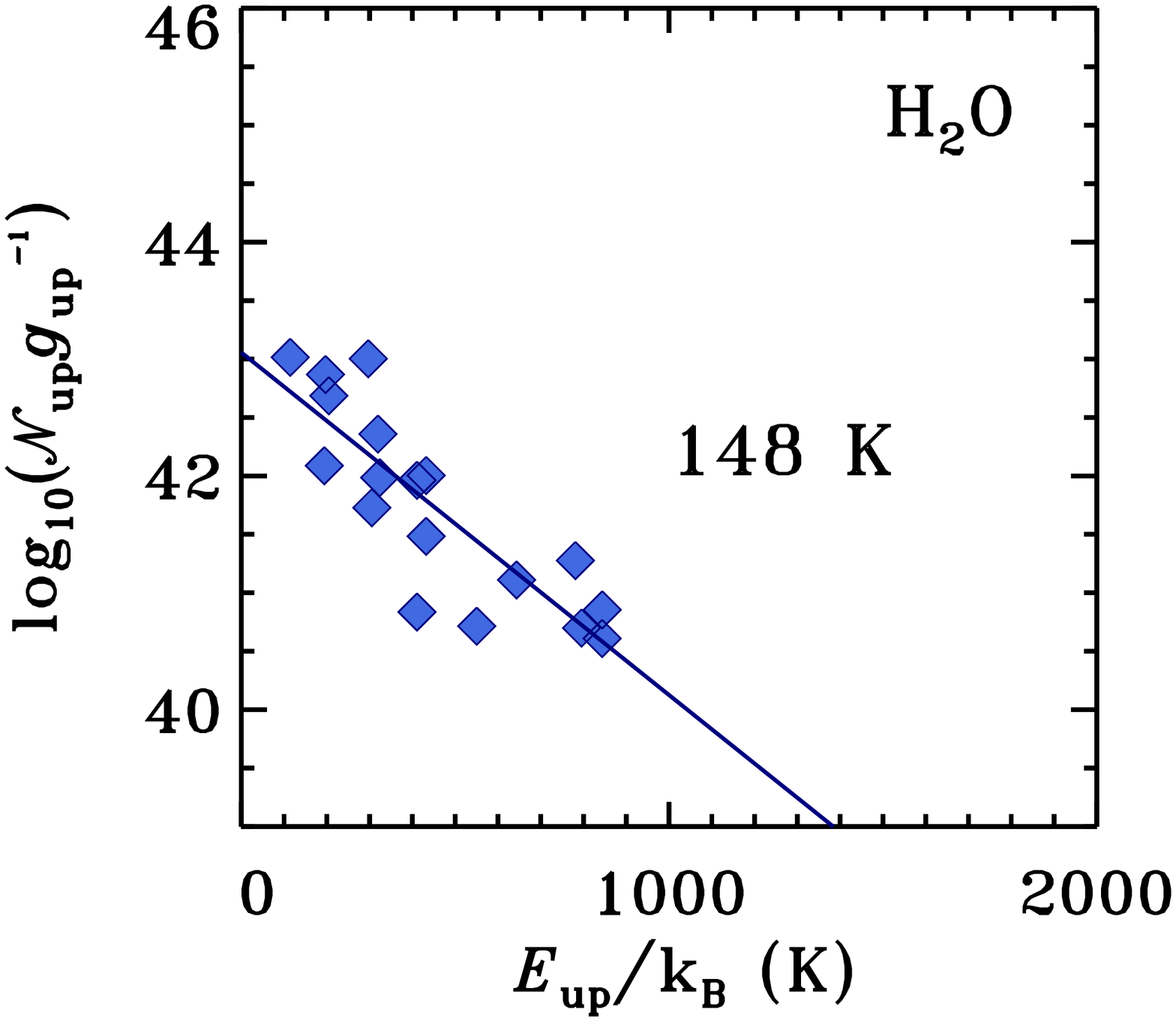} 
       \includegraphics[angle=90,height=4.8cm]{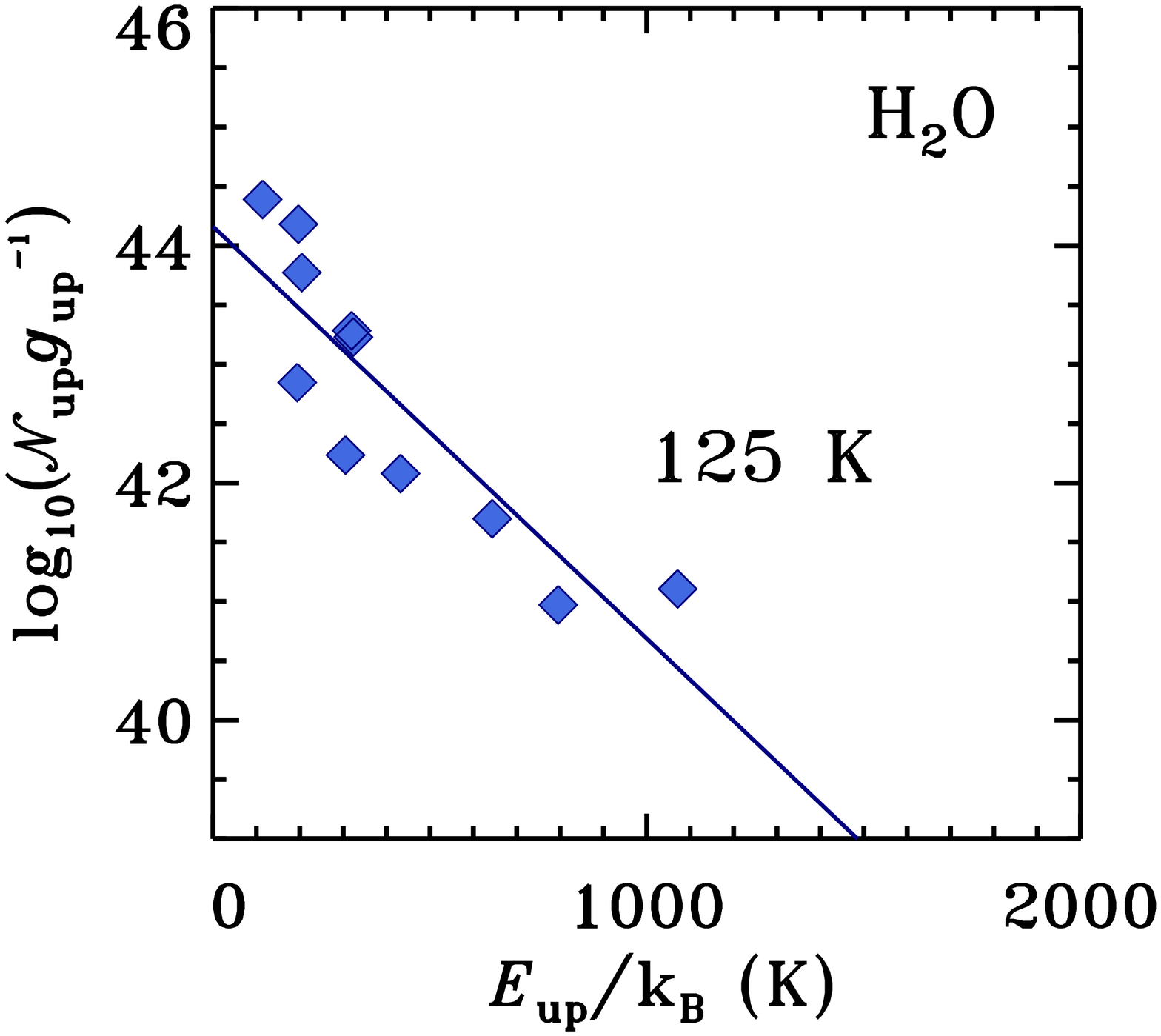} 
       \includegraphics[angle=90,height=4.8cm]{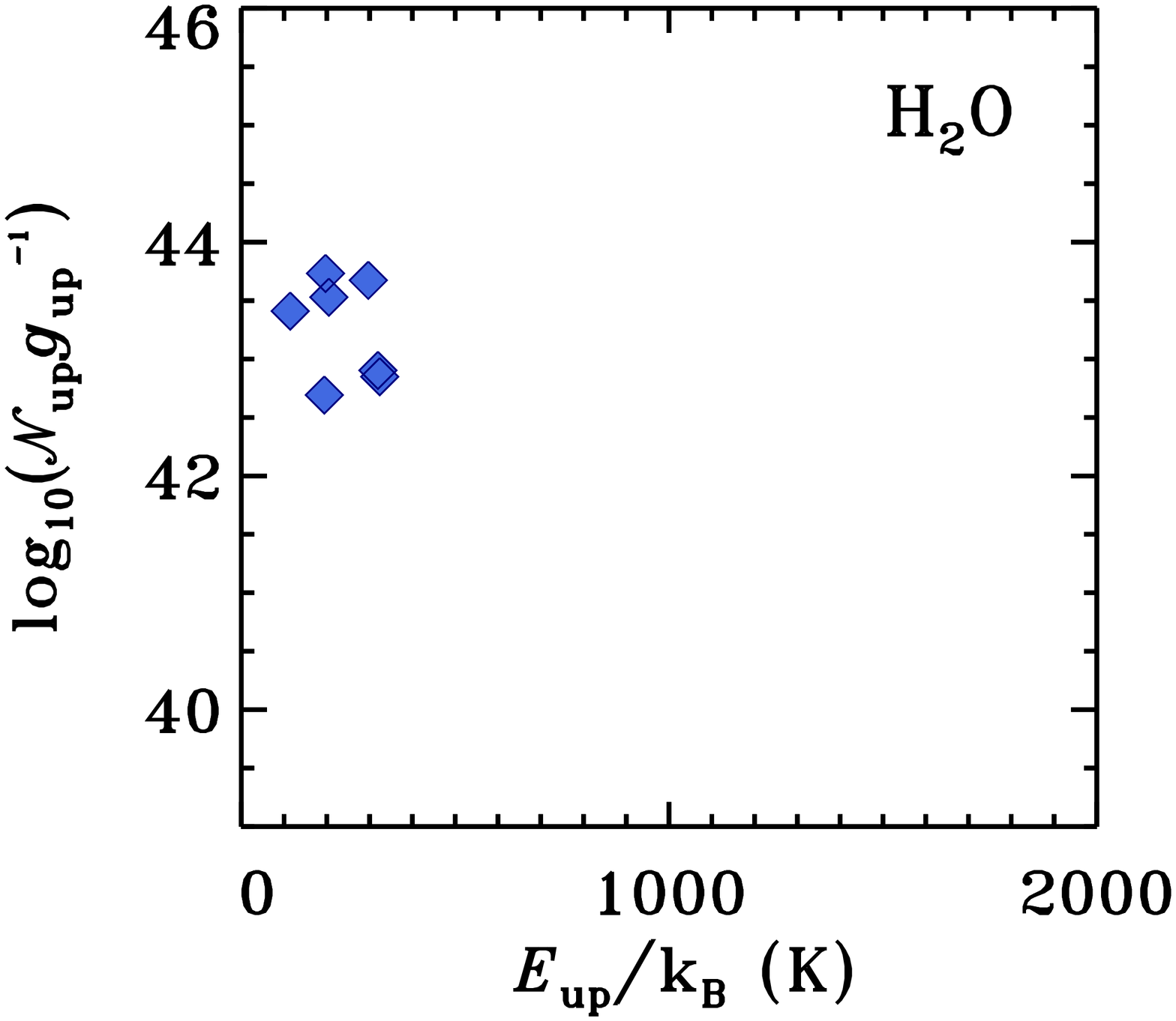} 
       \includegraphics[angle=90,height=4.8cm]{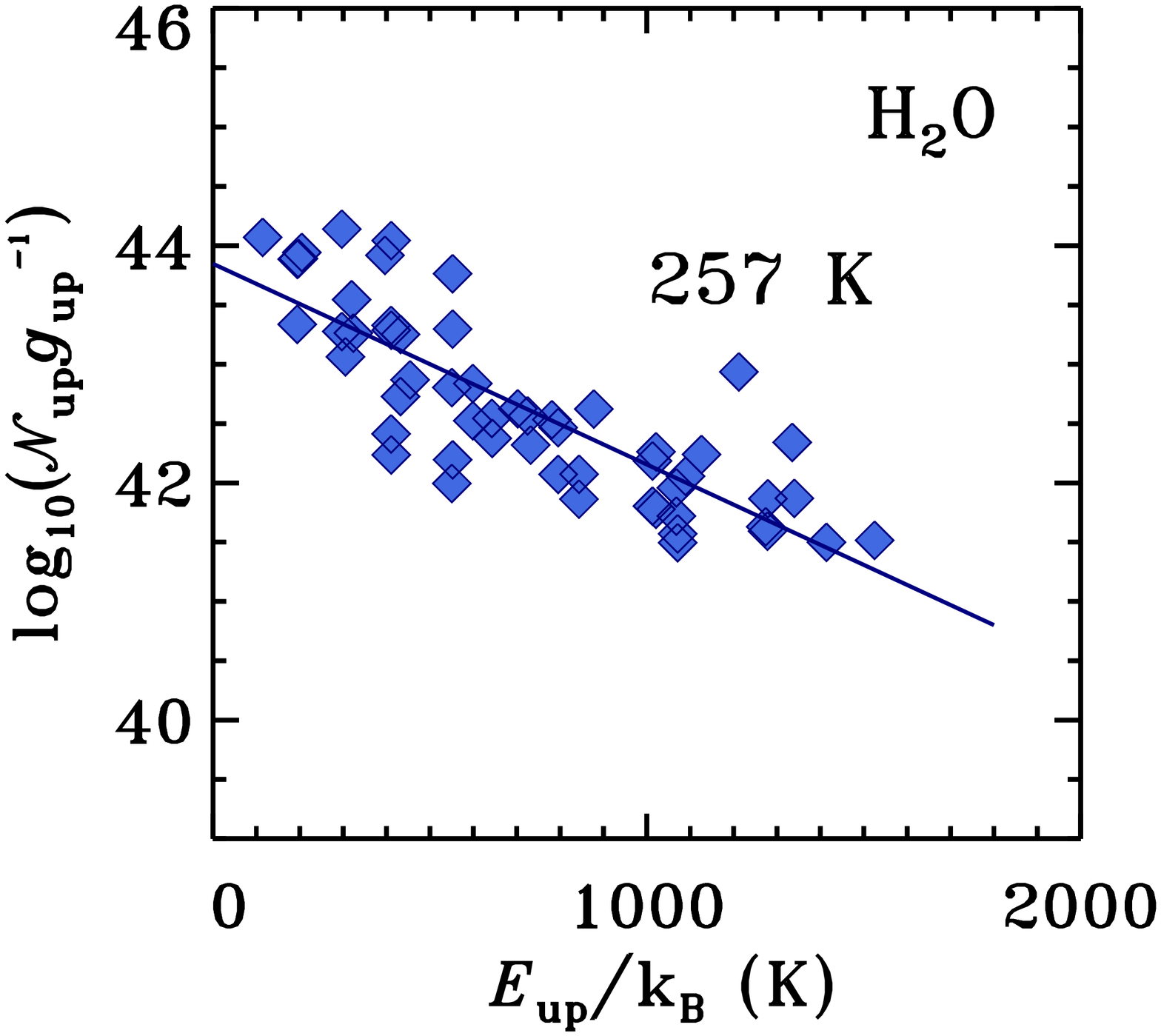} 
            
      \end{center}
  \end{minipage}
    \hfill
   \begin{minipage}[t]{.3\textwidth}
      \begin{center}
    	\includegraphics[angle=90,height=4.8cm]{ohdiag_b1c.eps} 
        \includegraphics[angle=90,height=4.8cm]{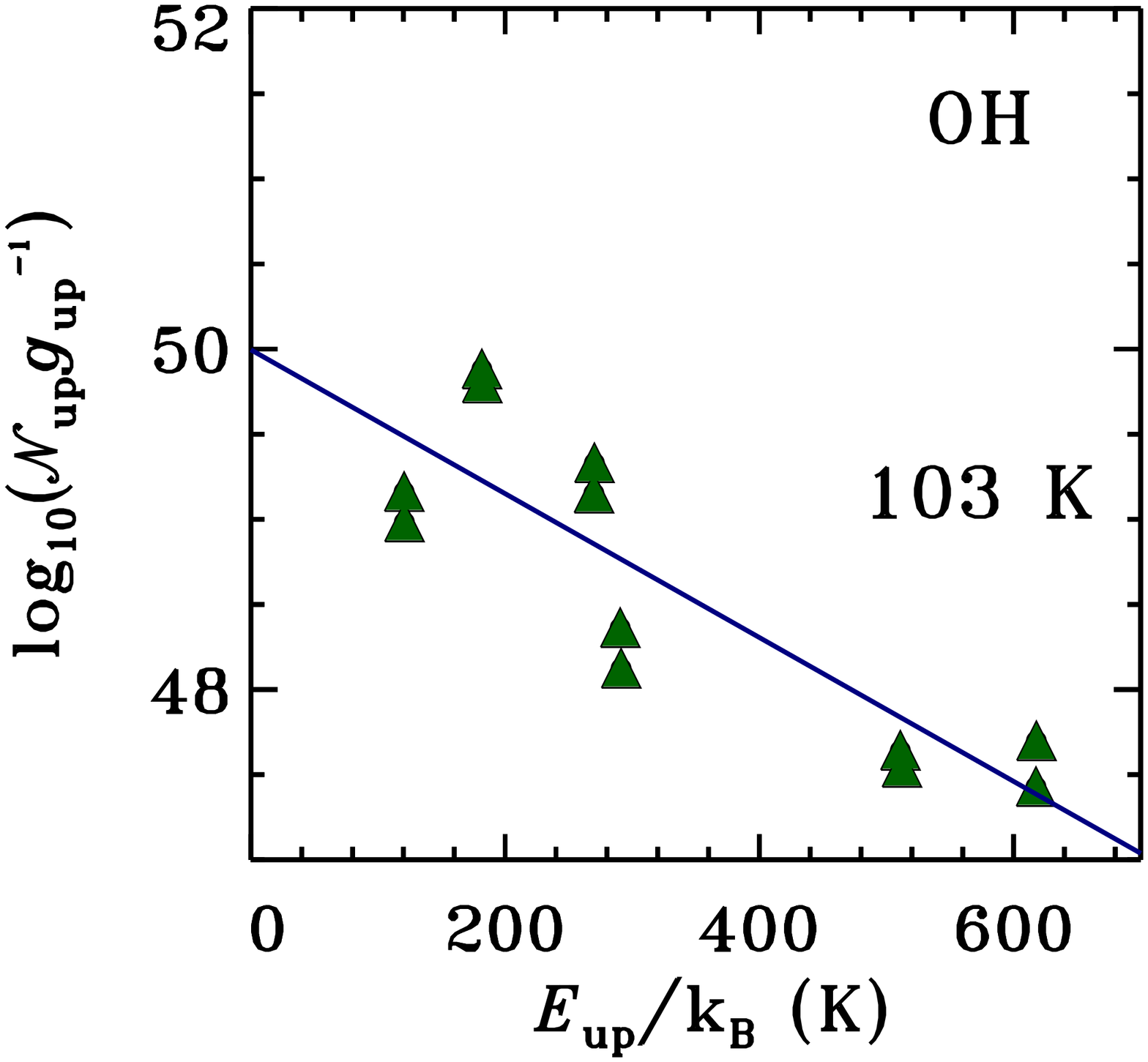} 
        \includegraphics[angle=90,height=4.8cm]{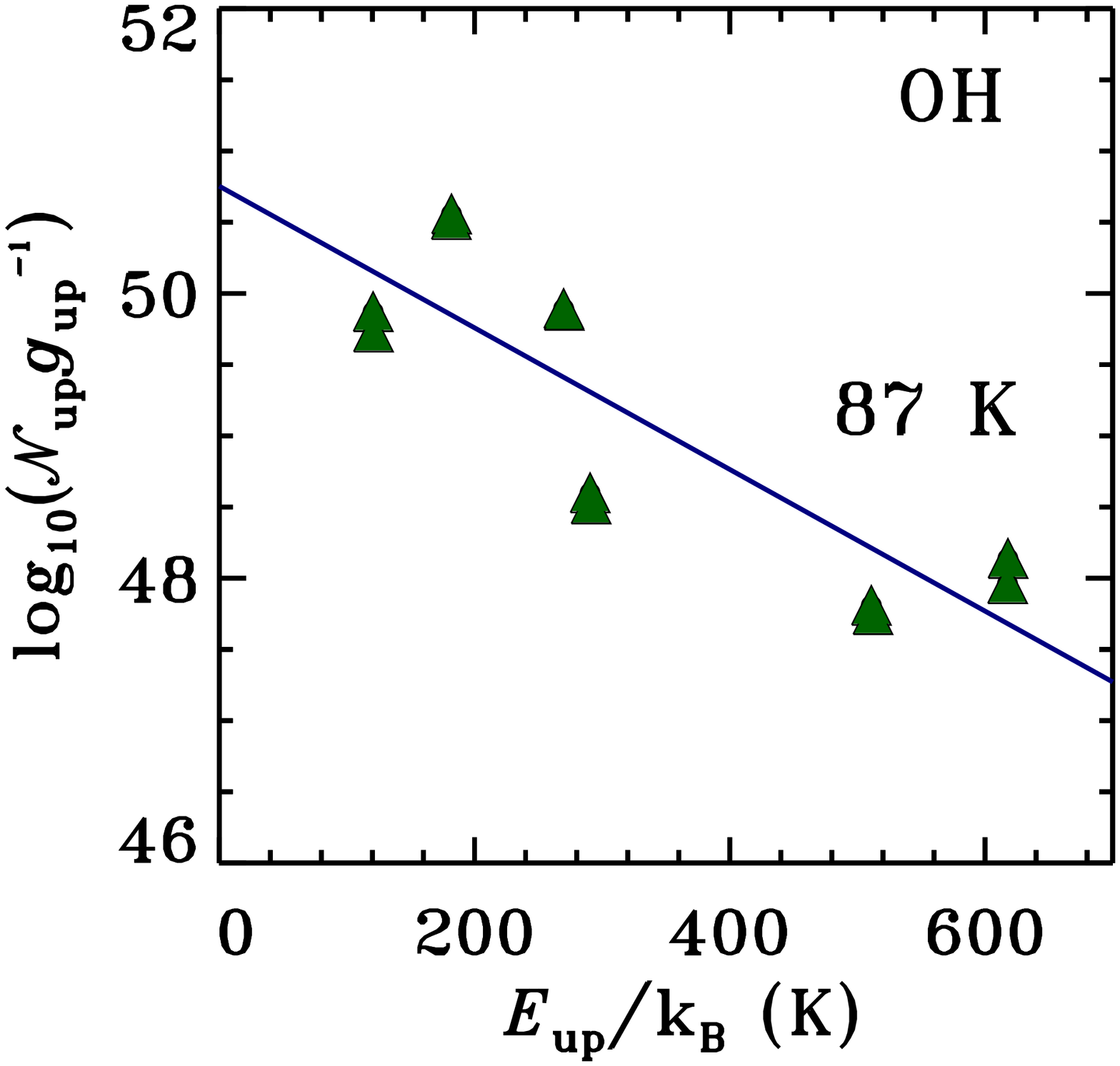} 
        \includegraphics[angle=90,height=4.8cm]{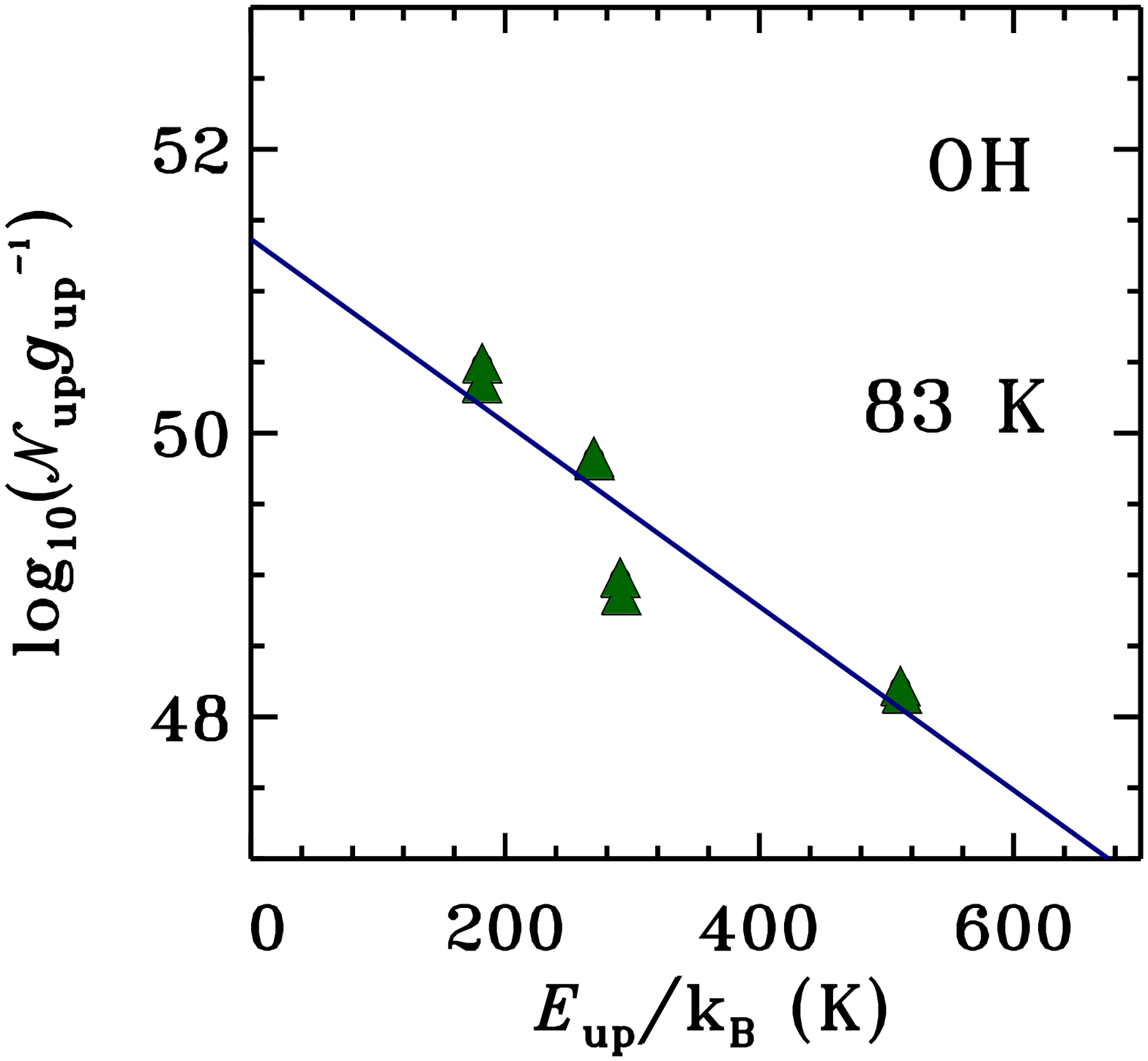} 
        \includegraphics[angle=90,height=4.8cm]{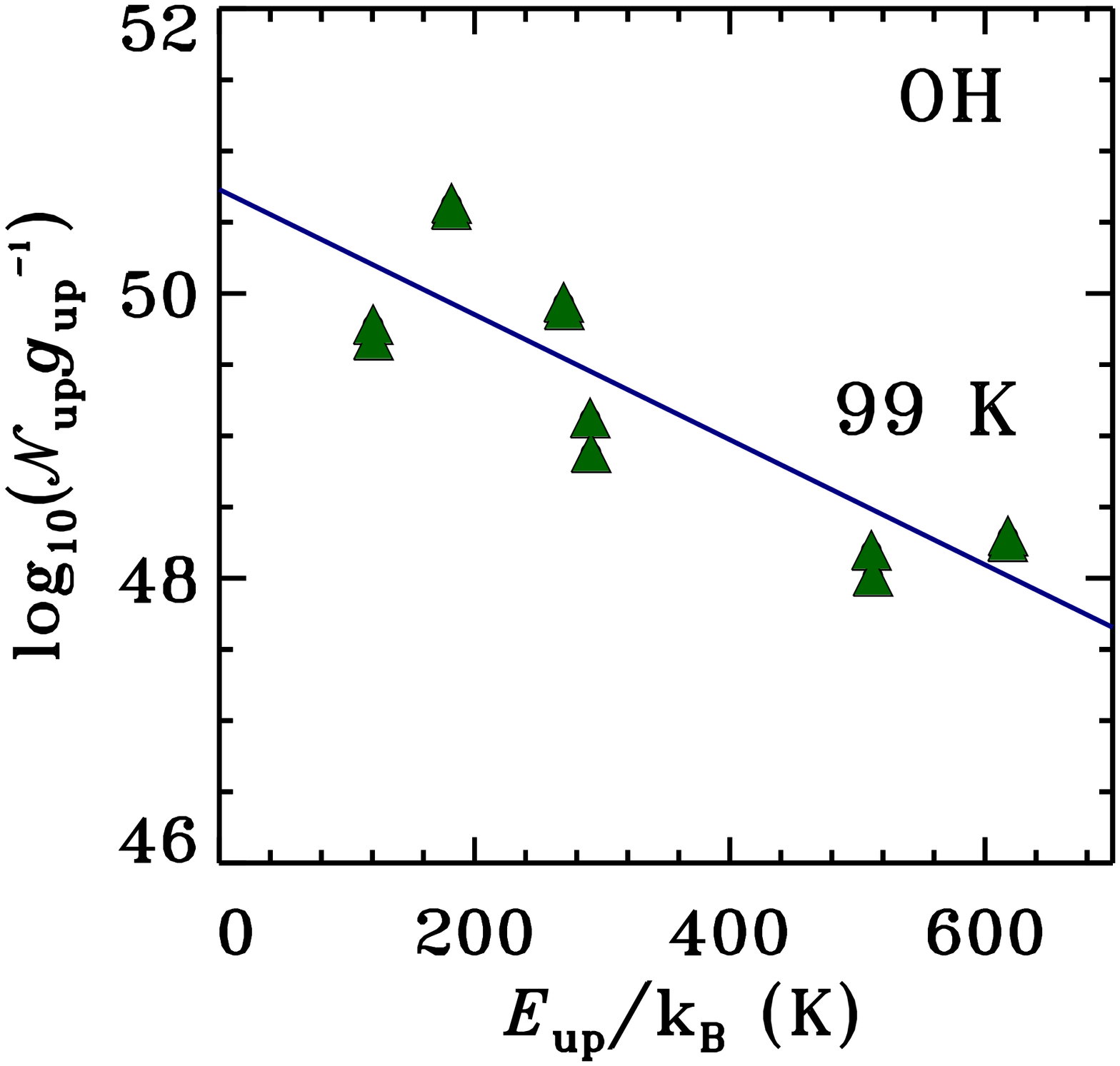} 
                            
      \end{center}
  \end{minipage}
      \hfill
        \caption{\label{dig1} Similar to Figure \ref{molexc}, but for TMC1 A, 
    TMC1, BHR 71, DK Cha, and GSS30 IRS1.}
\end{figure*}
\renewcommand{\thefigure}{\thesection.\arabic{figure} (Cont.)}
\addtocounter{figure}{-1}   
\begin{figure*}[!tb]
  \begin{minipage}[t]{.3\textwidth}
  \begin{center}
       \includegraphics[angle=90,height=4.8cm]{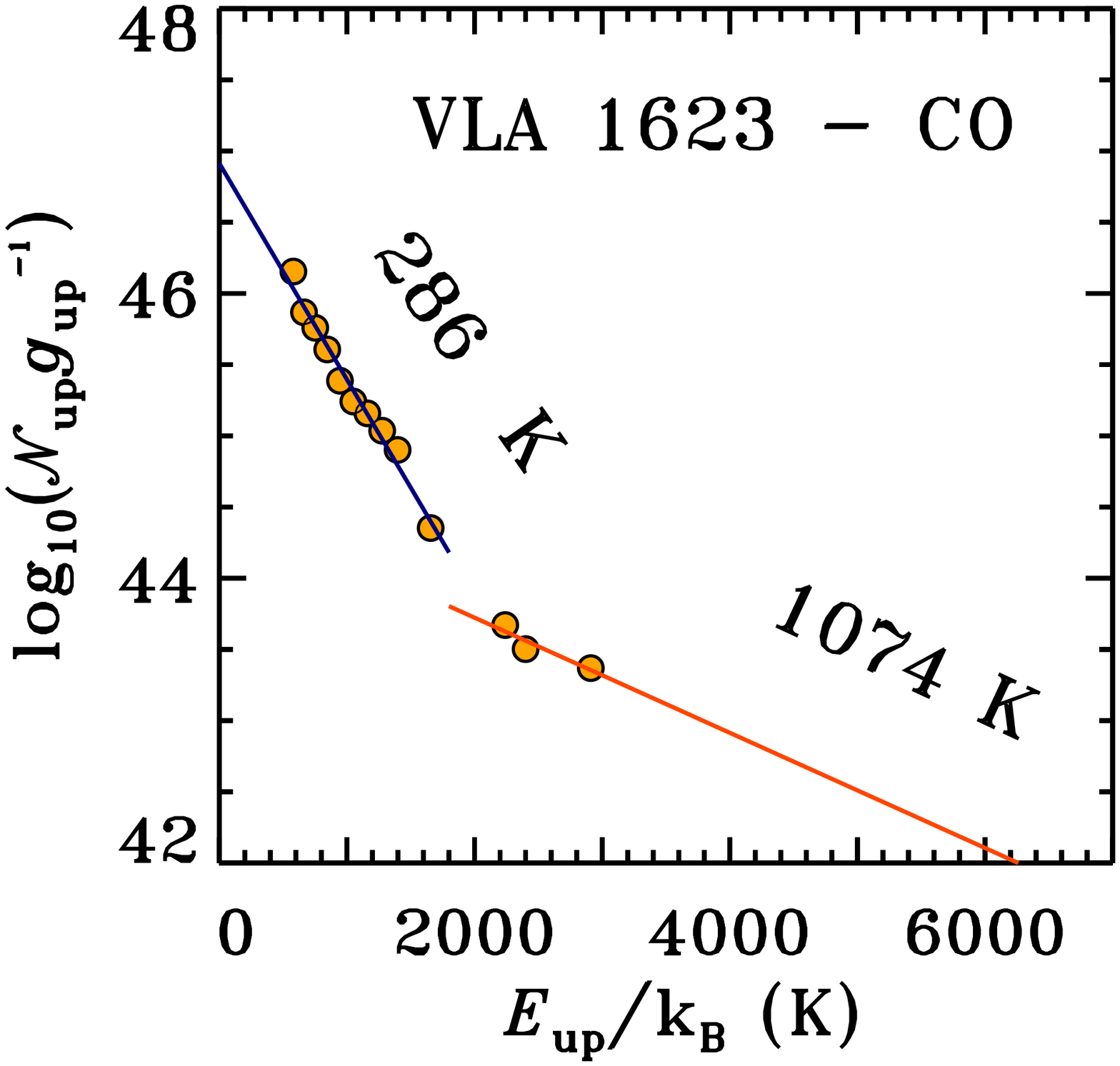} 
       \includegraphics[angle=90,height=4.8cm]{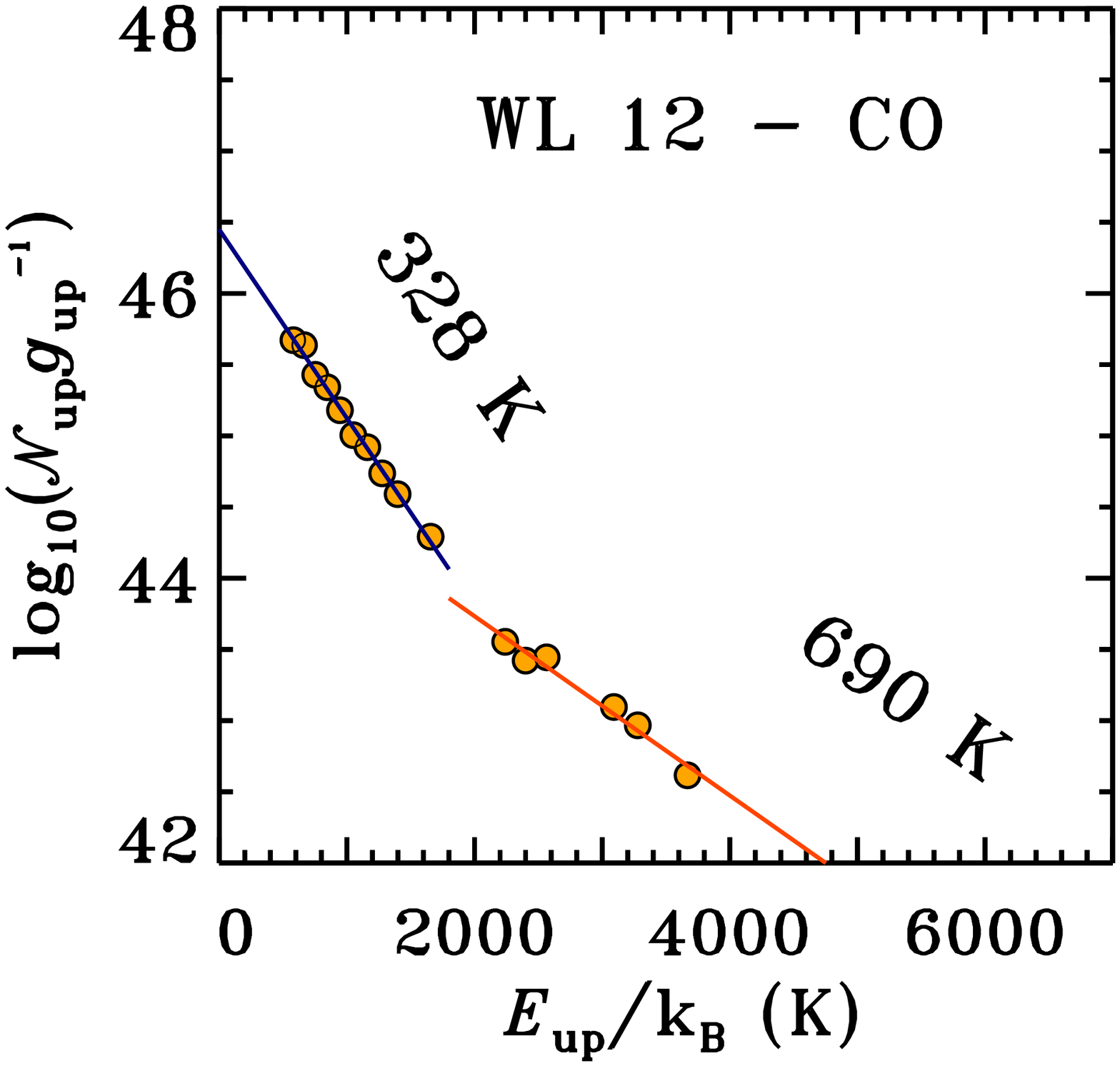} 
        \includegraphics[angle=90,height=4.8cm]{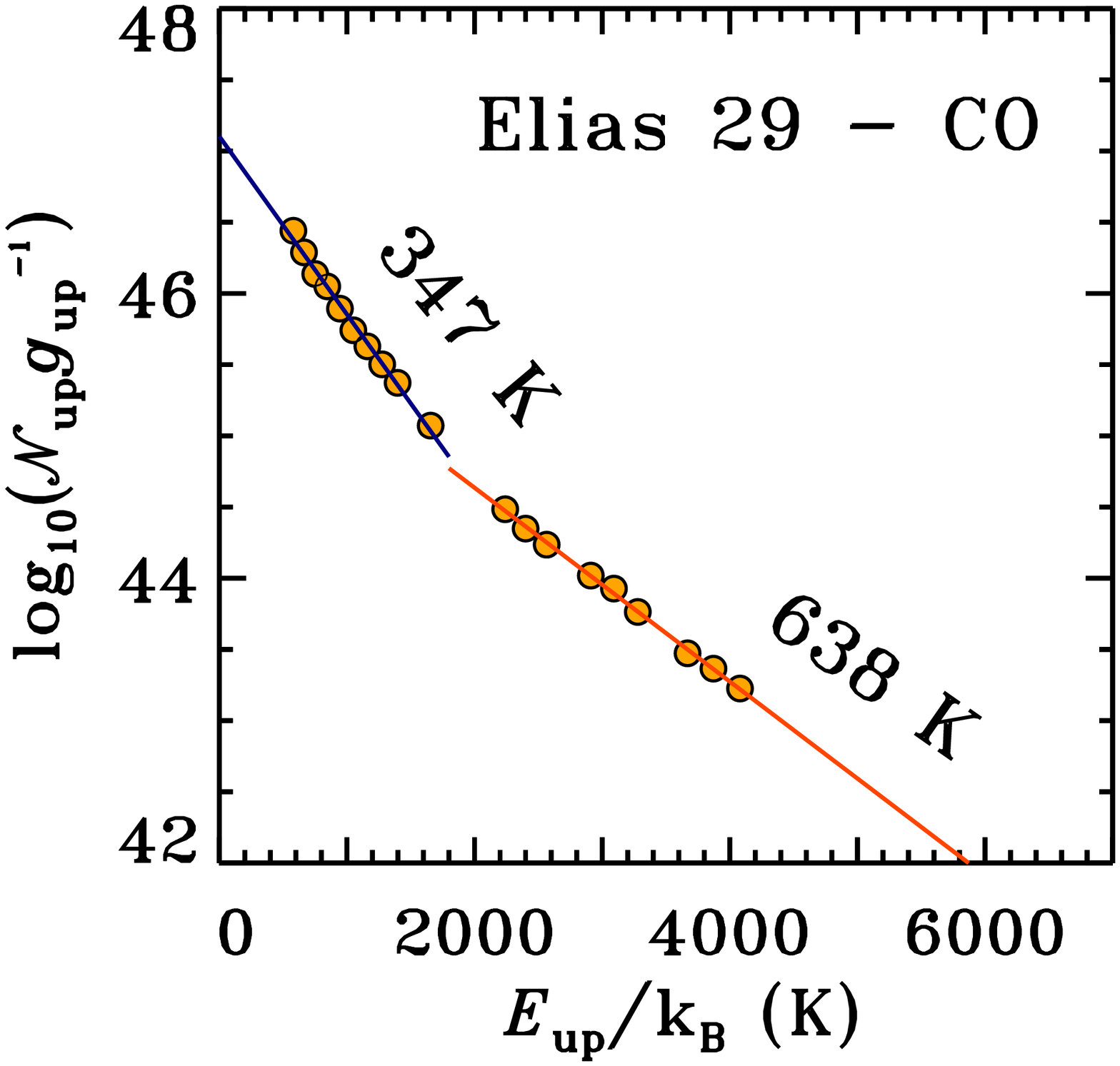} 
       \includegraphics[angle=90,height=4.8cm]{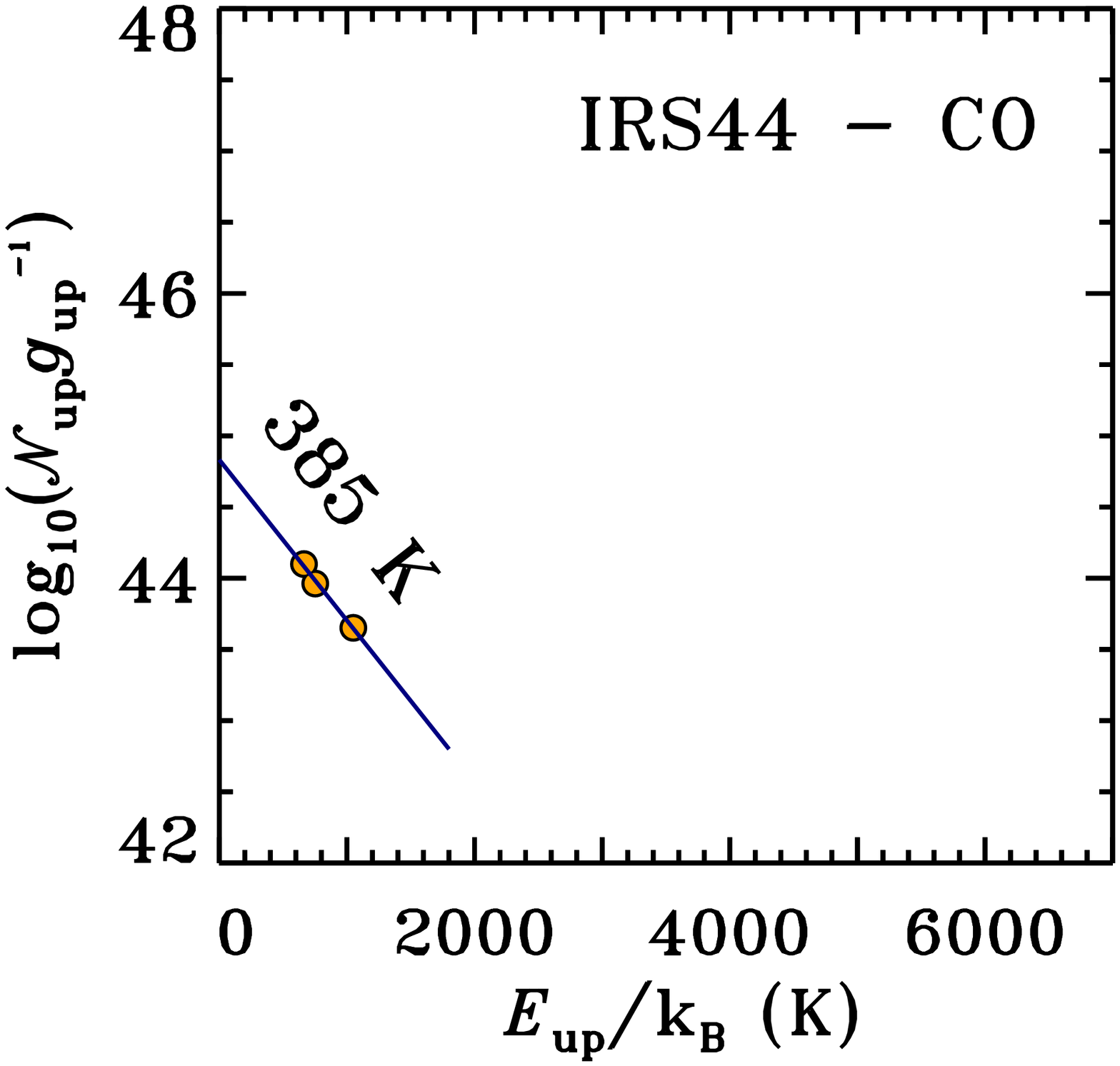} 
        \includegraphics[angle=90,height=4.8cm]{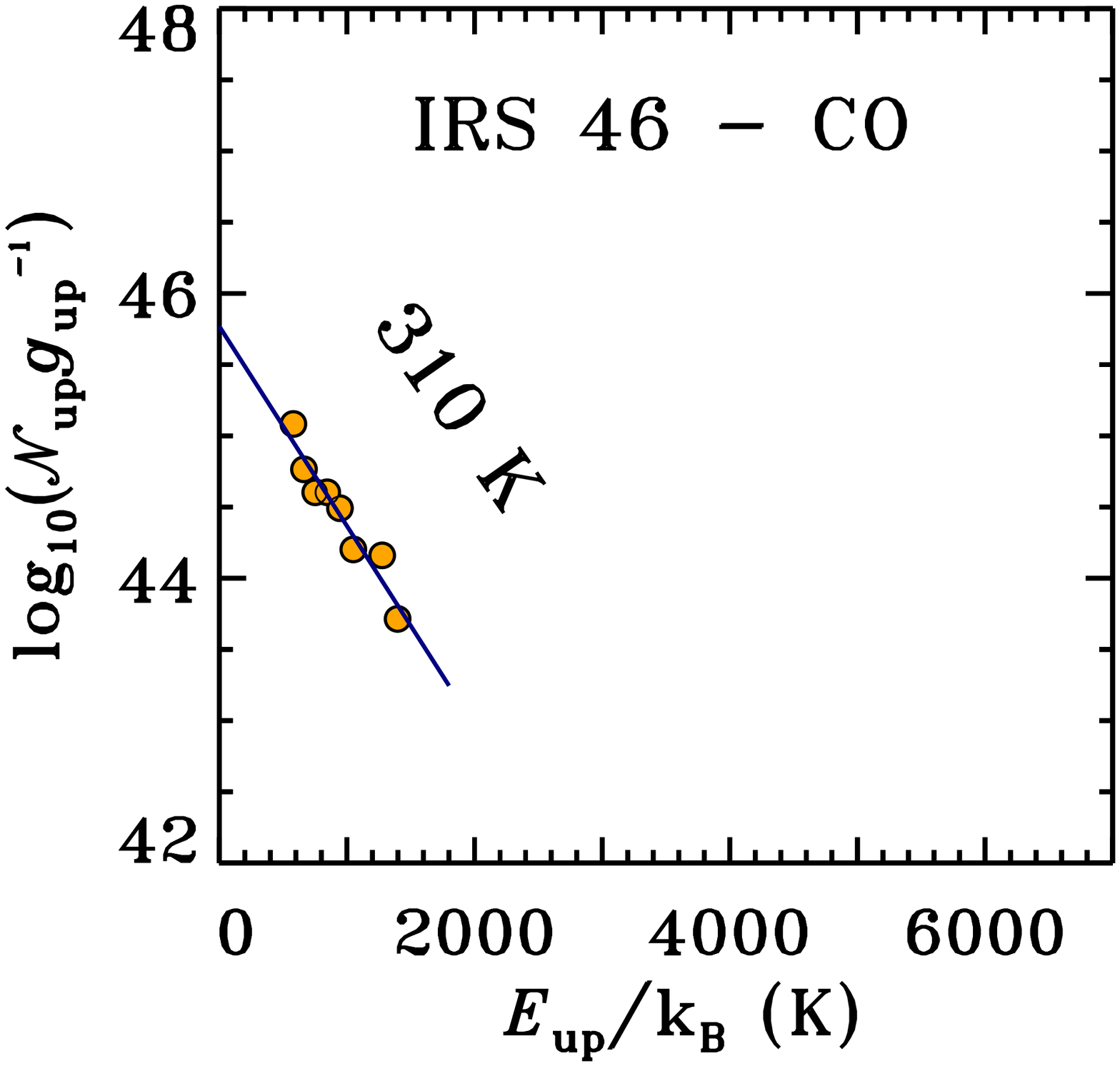} 
  \end{center}
  \end{minipage}
  \hfill
  \begin{minipage}[t]{.3\textwidth}
      \begin{center}
   	   \includegraphics[angle=90,height=4.8cm]{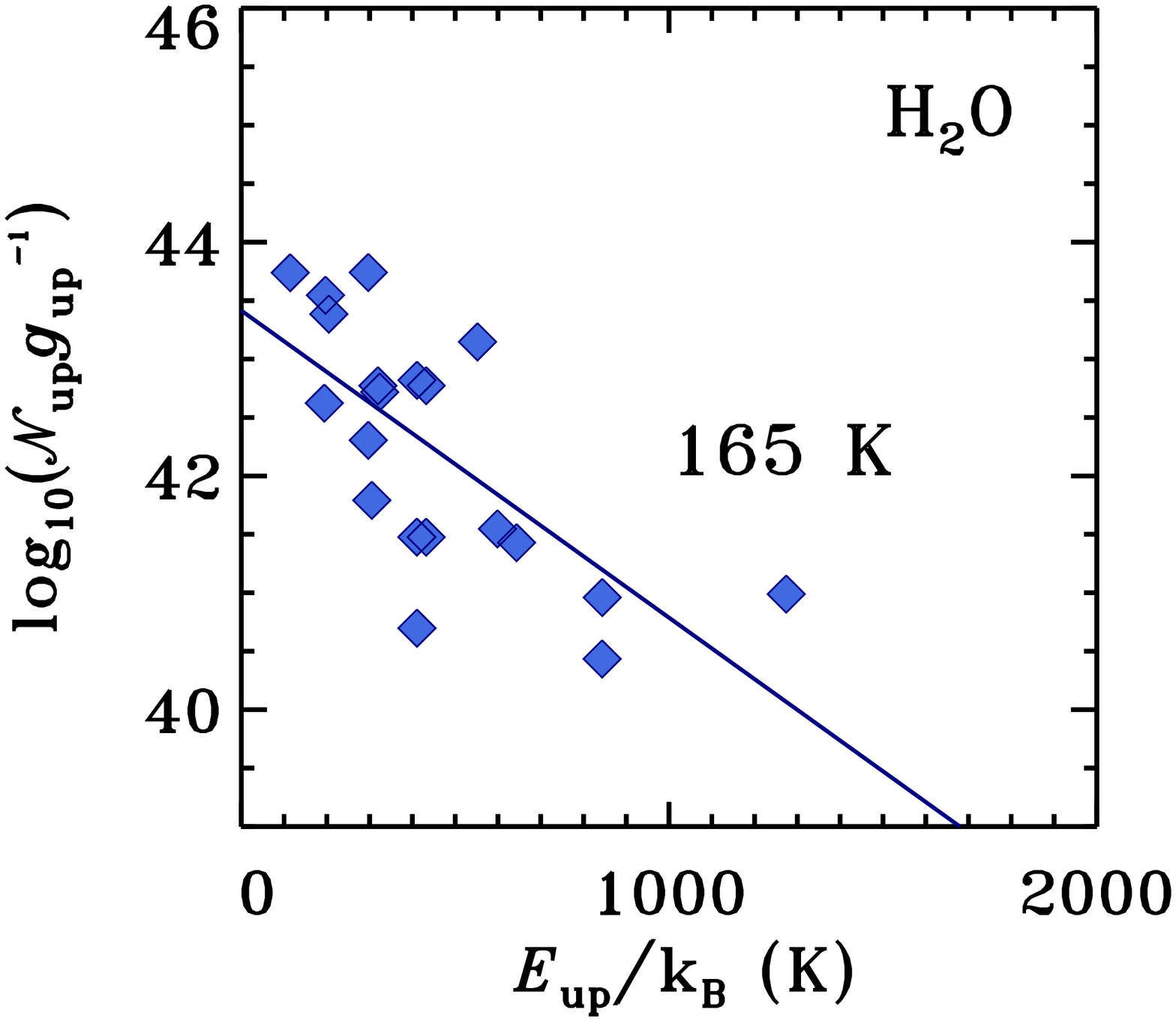} 
       \includegraphics[angle=90,height=4.8cm]{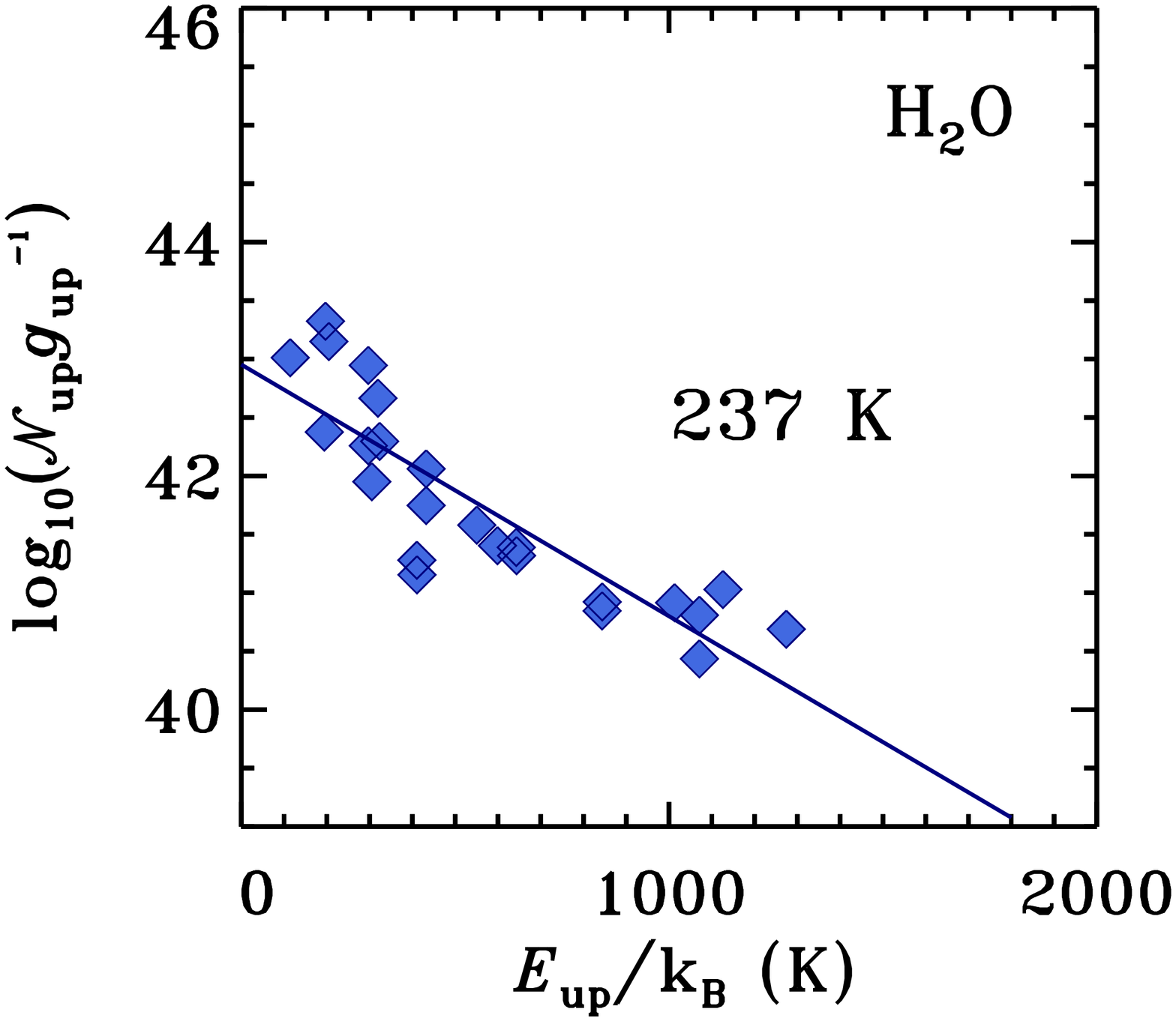} 
       \includegraphics[angle=90,height=4.8cm]{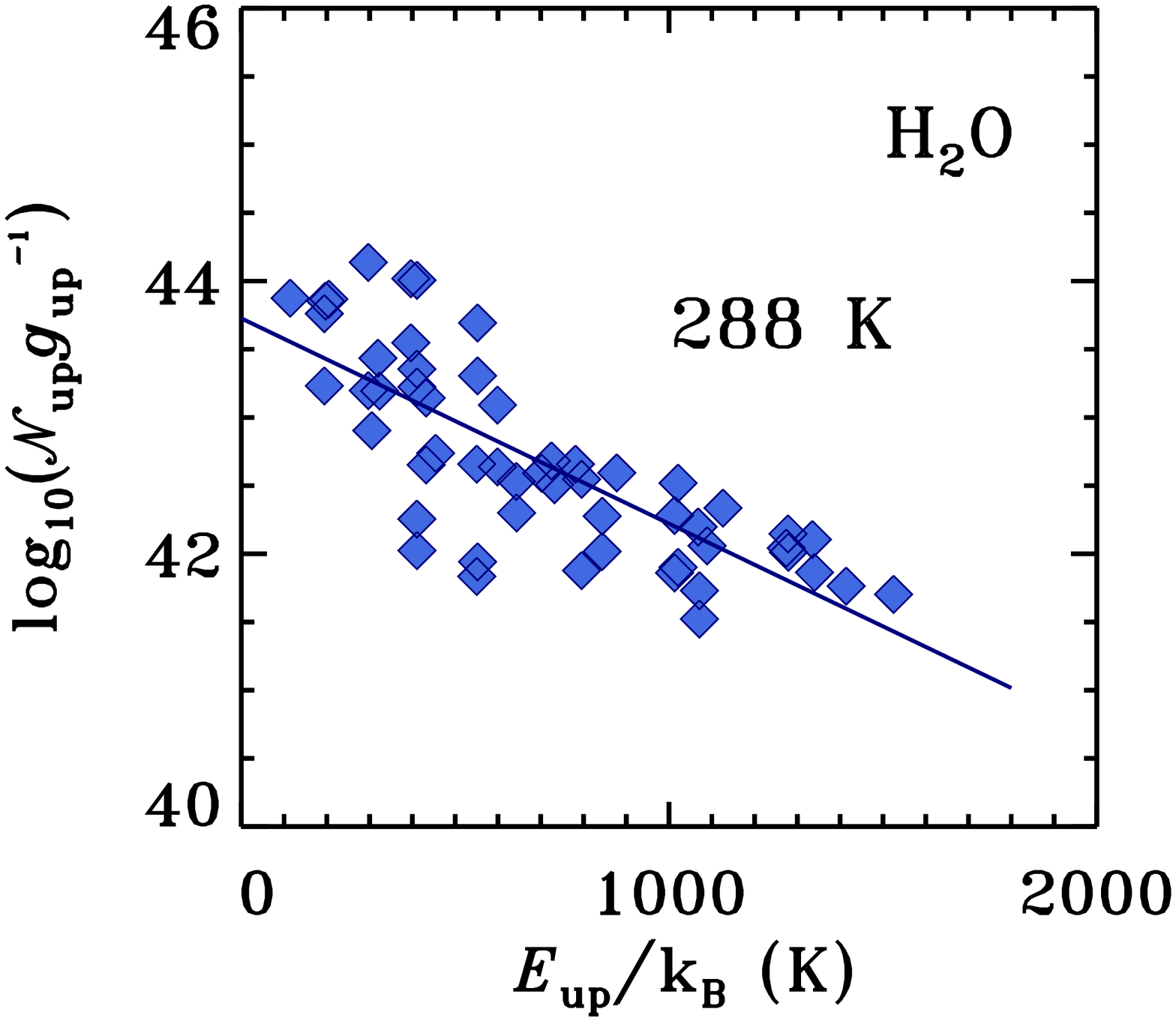} 
       \includegraphics[angle=90,height=4.8cm]{wdiag_iras2a.eps} 
       \includegraphics[angle=90,height=4.8cm]{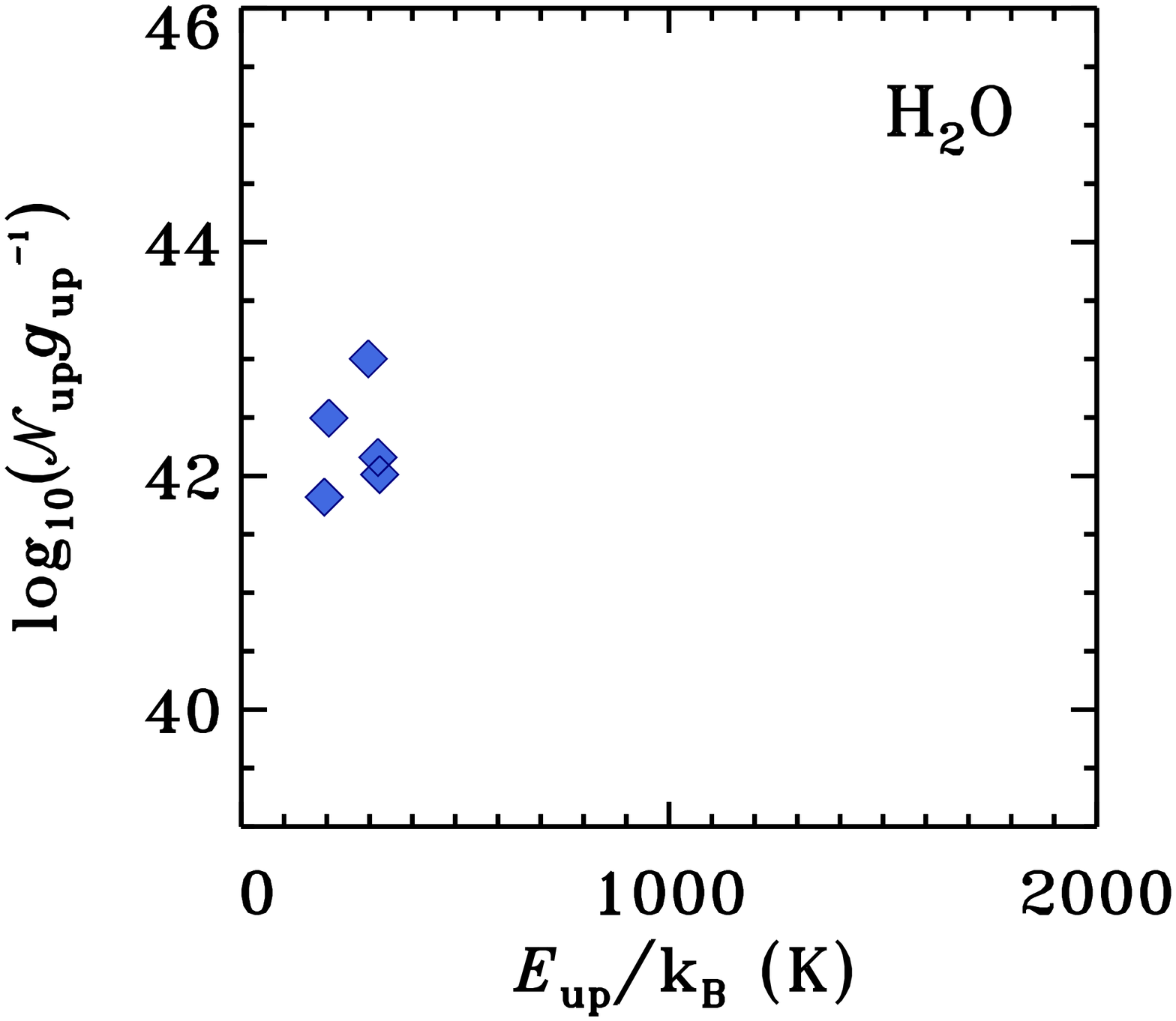} 
            
      \end{center}
  \end{minipage}
    \hfill
   \begin{minipage}[t]{.3\textwidth}
      \begin{center}
    	\includegraphics[angle=90,height=4.8cm]{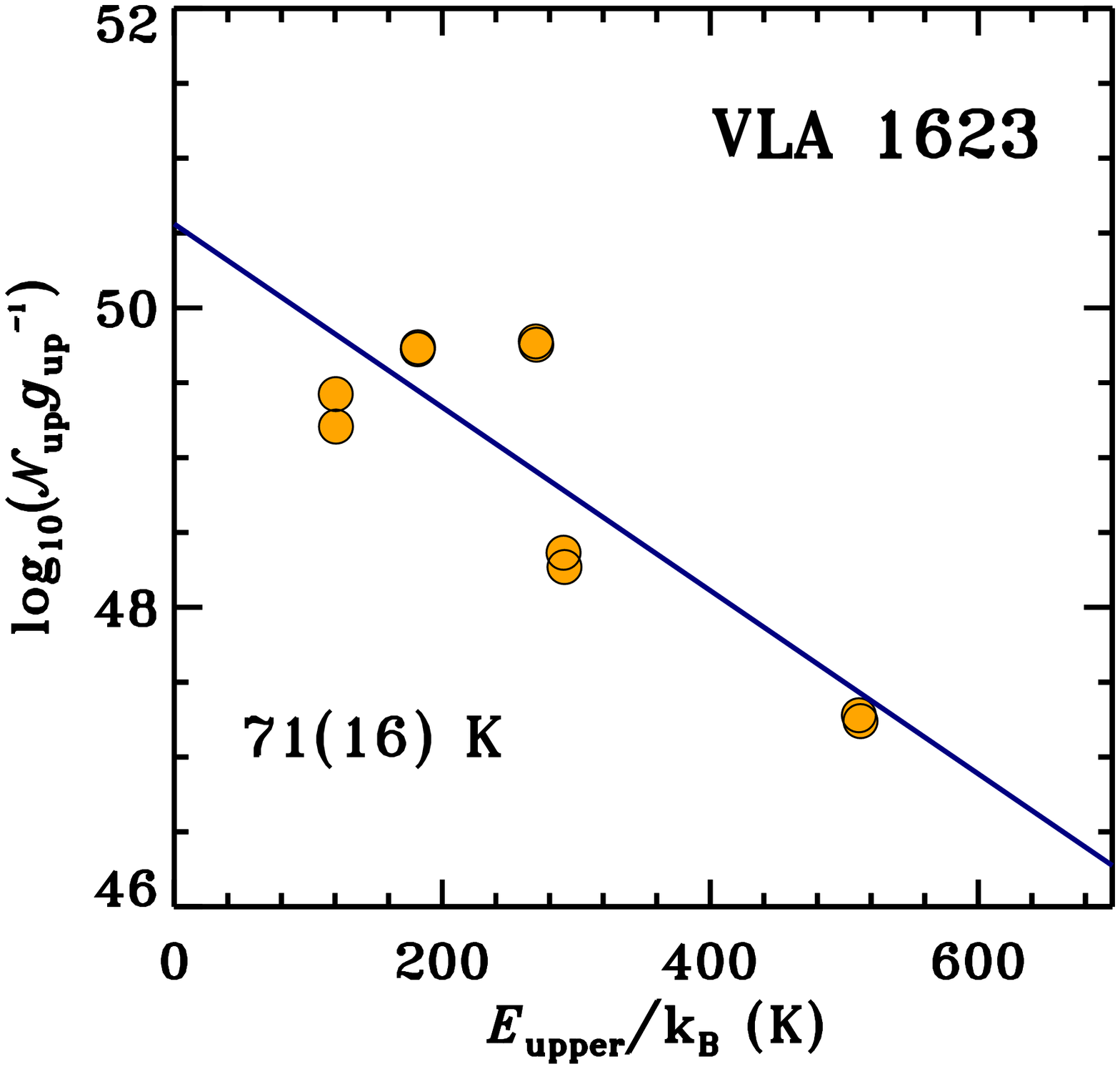} 
        \includegraphics[angle=90,height=4.8cm]{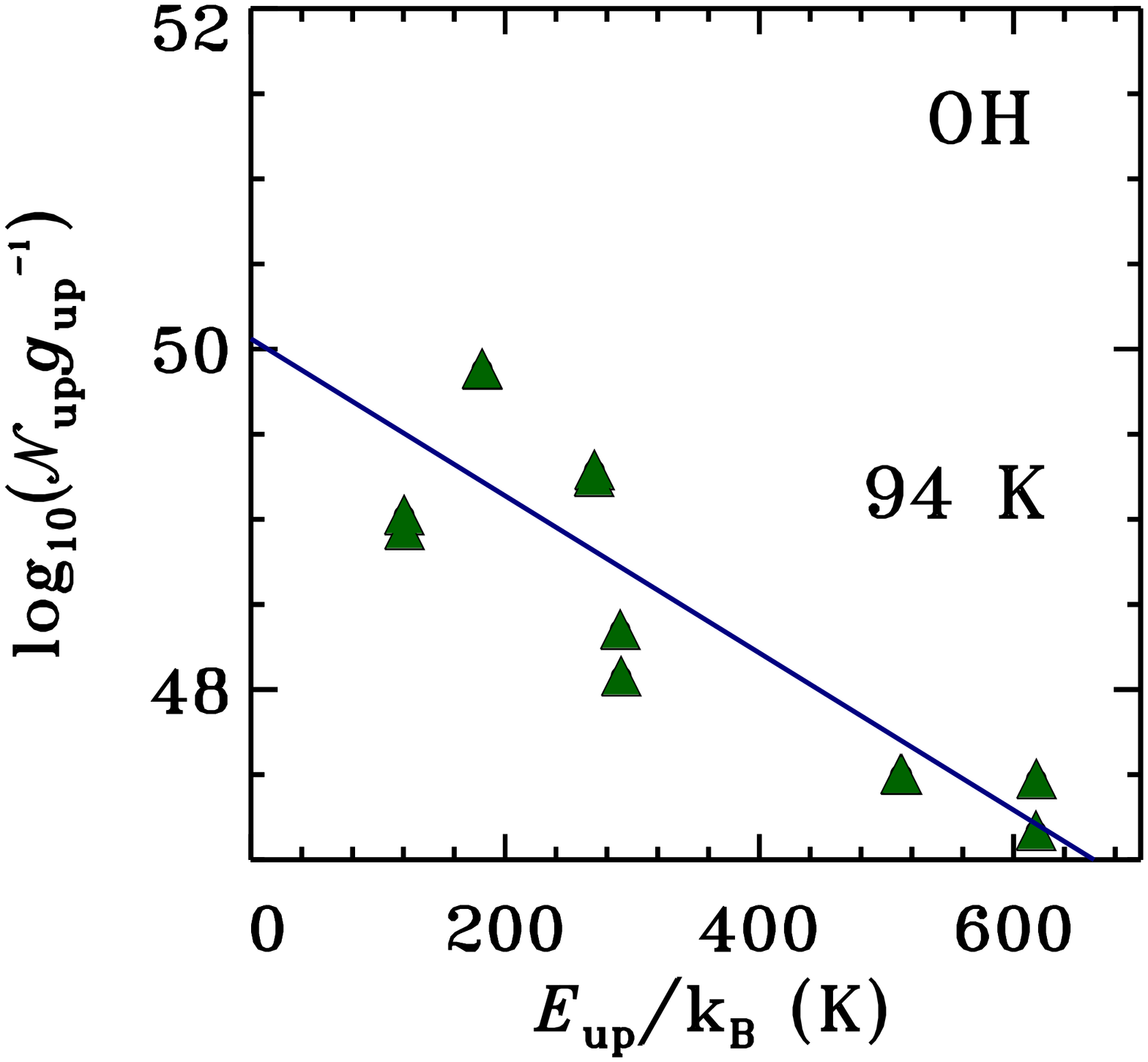} 
        \includegraphics[angle=90,height=4.8cm]{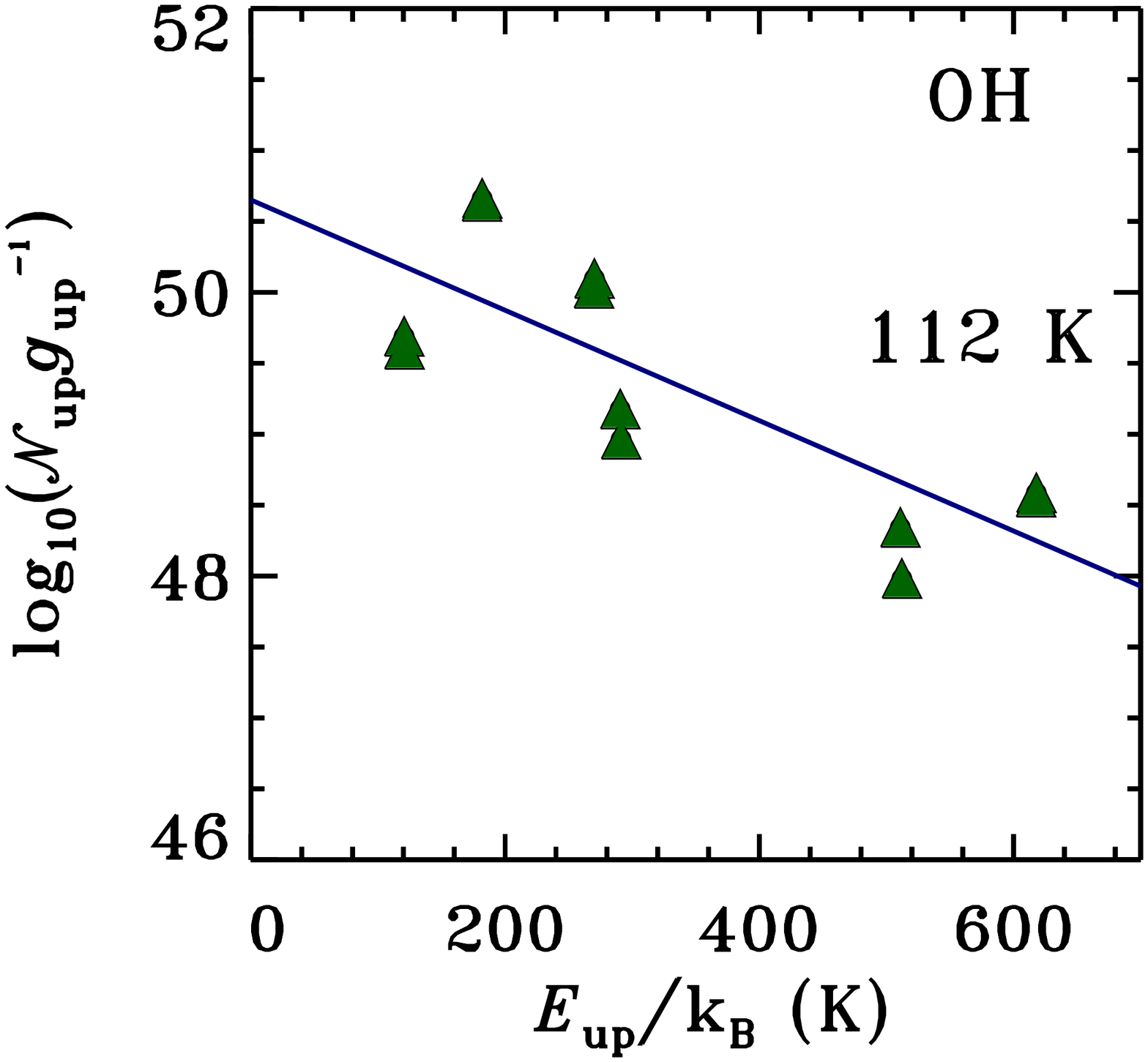} 
        \includegraphics[angle=90,height=4.8cm]{ohdiag_b1c.eps} 
        \includegraphics[angle=90,height=4.8cm]{ohdiag_b1c.eps} 
                            
      \end{center}
  \end{minipage}
      \hfill
        \caption{\label{dig2} Similar to Figure \ref{molexc}, but for 
        VLA 1623, WL 12, Elias 29, Oph IRS44, and Oph IRS 46.}
\end{figure*}
\renewcommand{\thefigure}{\thesection.\arabic{figure} (Cont.)}
\addtocounter{figure}{-1}   
\begin{figure*}[!tb]
  \begin{minipage}[t]{.3\textwidth}
  \begin{center}
       \includegraphics[angle=90,height=4.8cm]{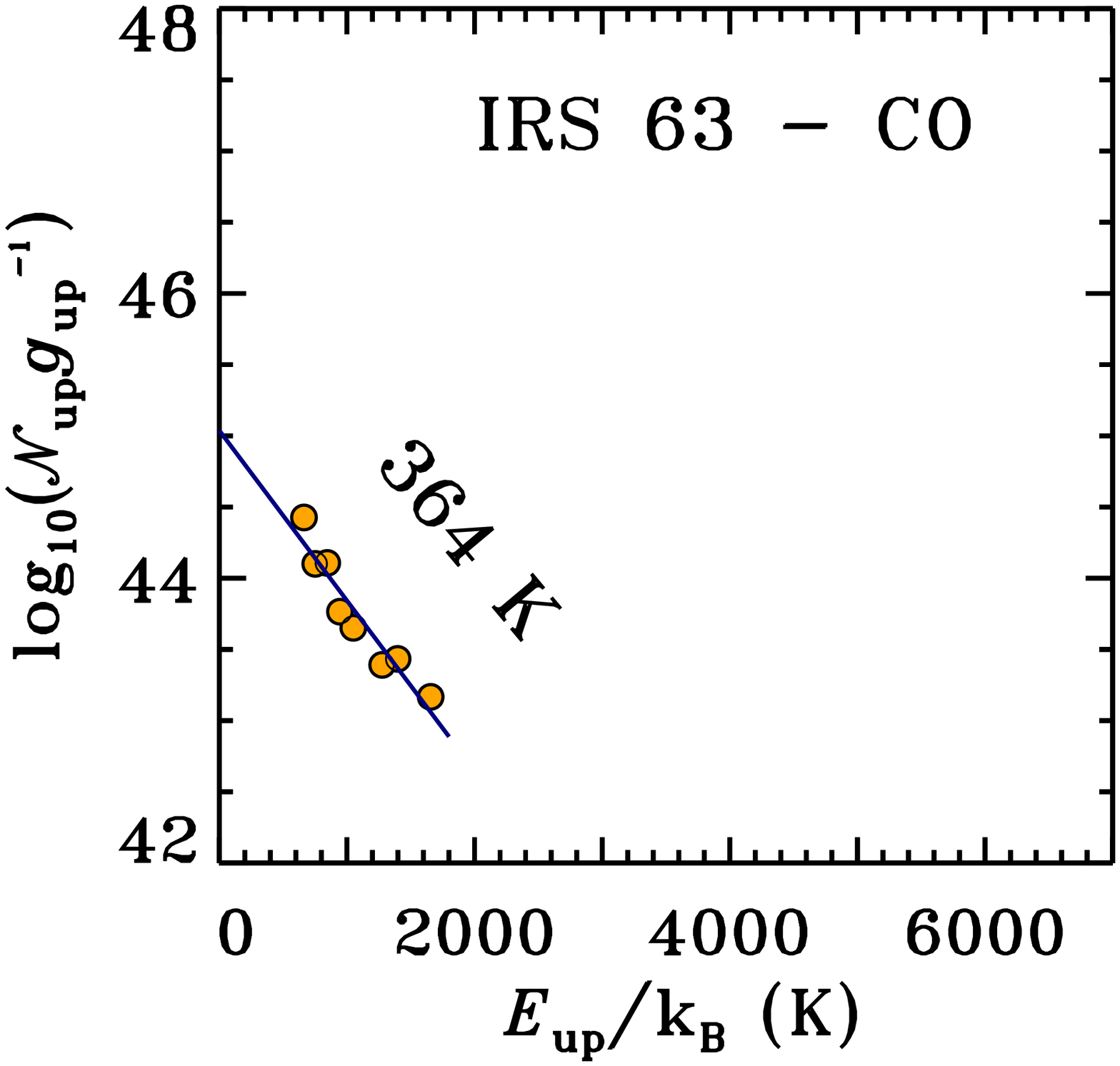} 
       \includegraphics[angle=90,height=4.8cm]{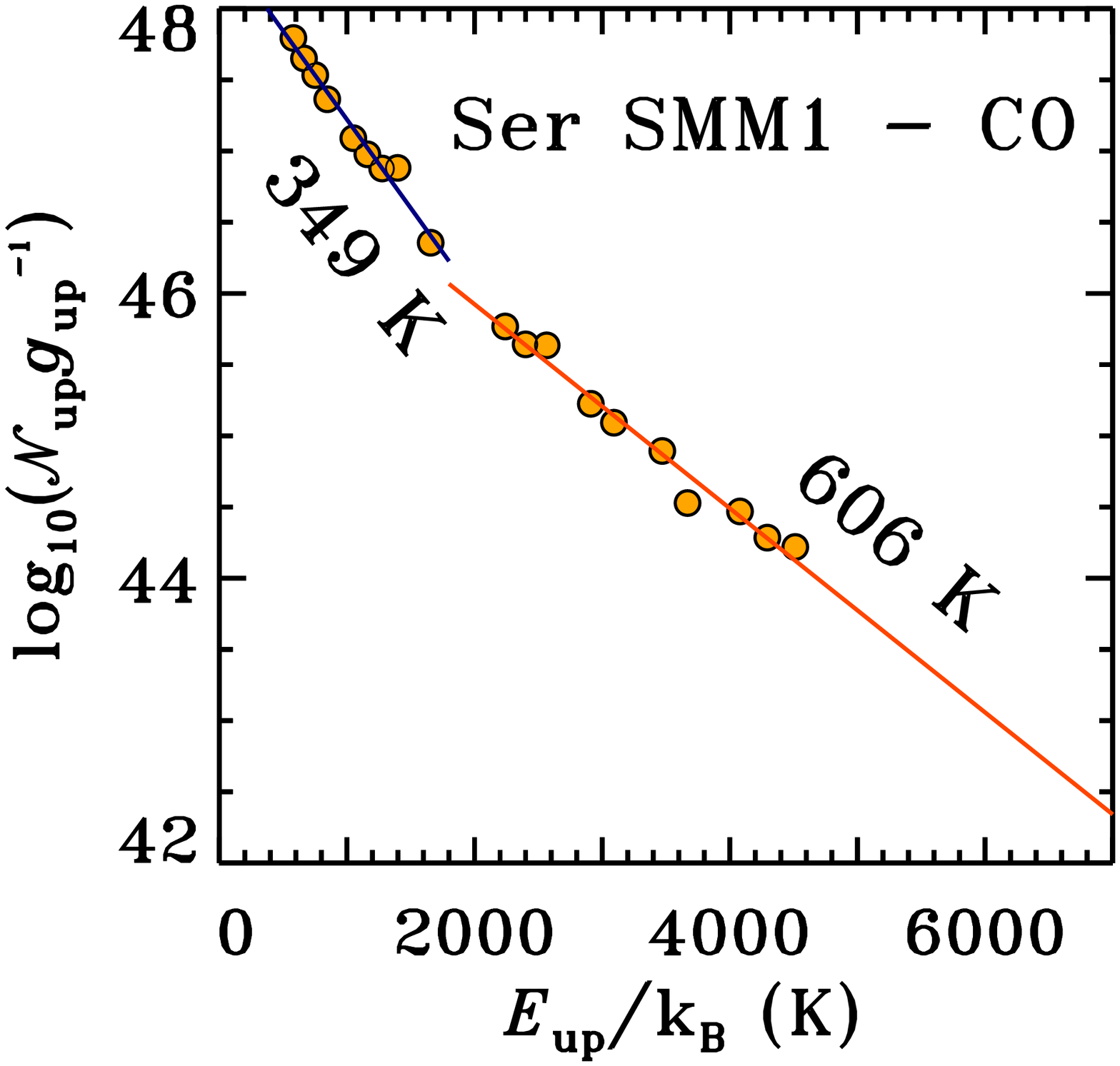} 
        \includegraphics[angle=90,height=4.8cm]{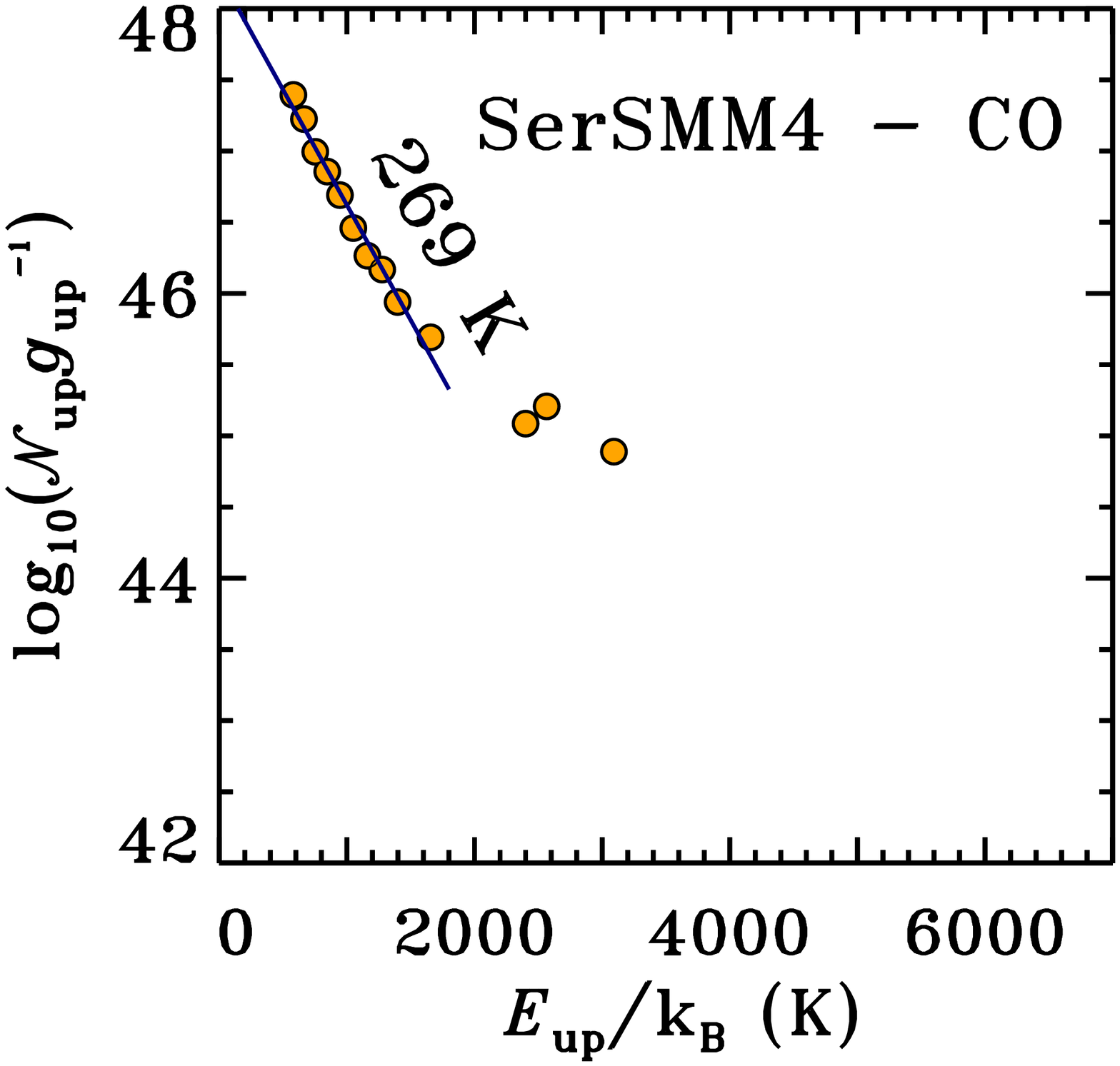} 
       \includegraphics[angle=90,height=4.8cm]{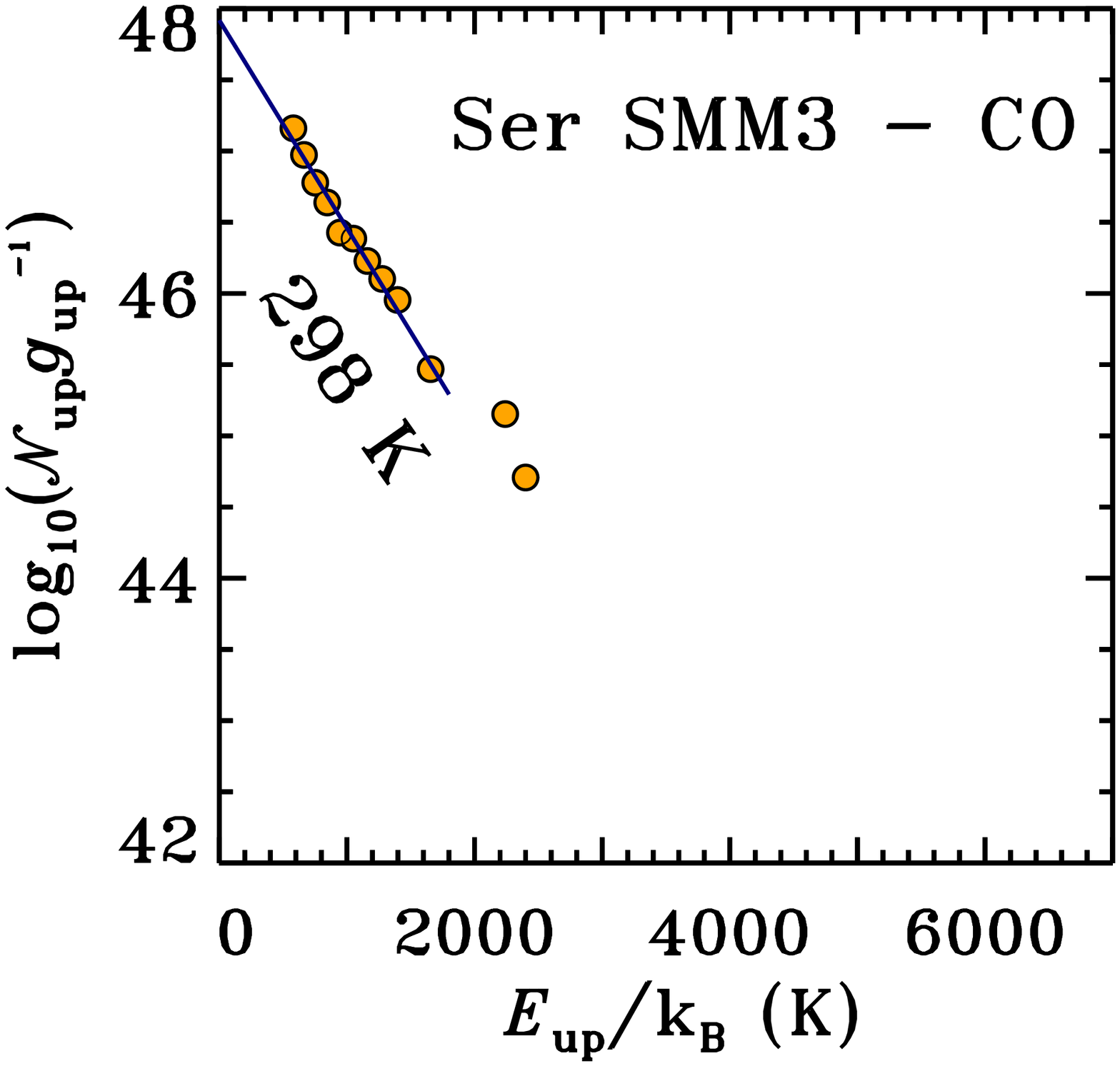} 
        \includegraphics[angle=90,height=4.8cm]{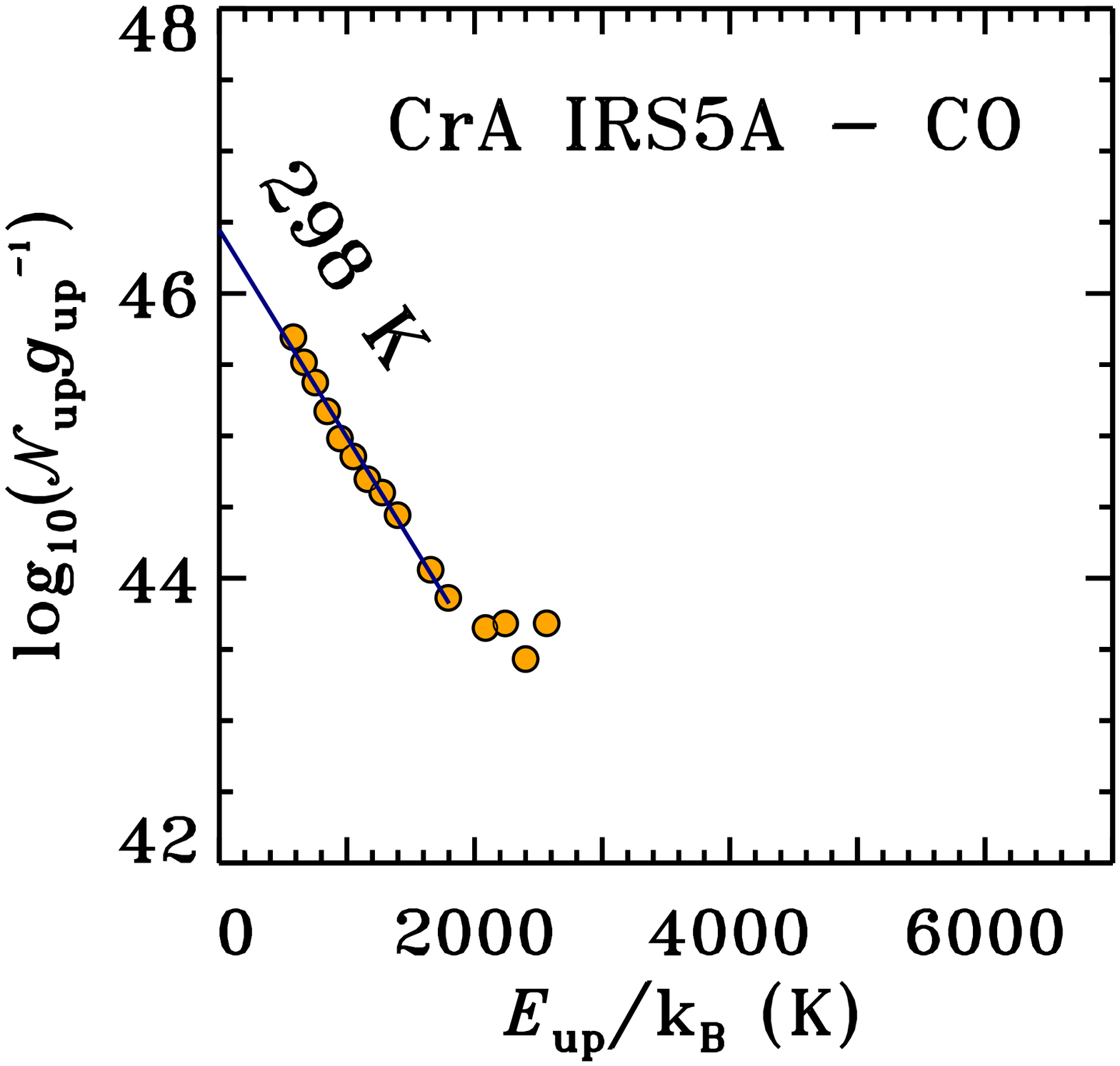} 
  \end{center}
  \end{minipage}
  \hfill
  \begin{minipage}[t]{.3\textwidth}
      \begin{center}
   	   \includegraphics[angle=90,height=4.8cm]{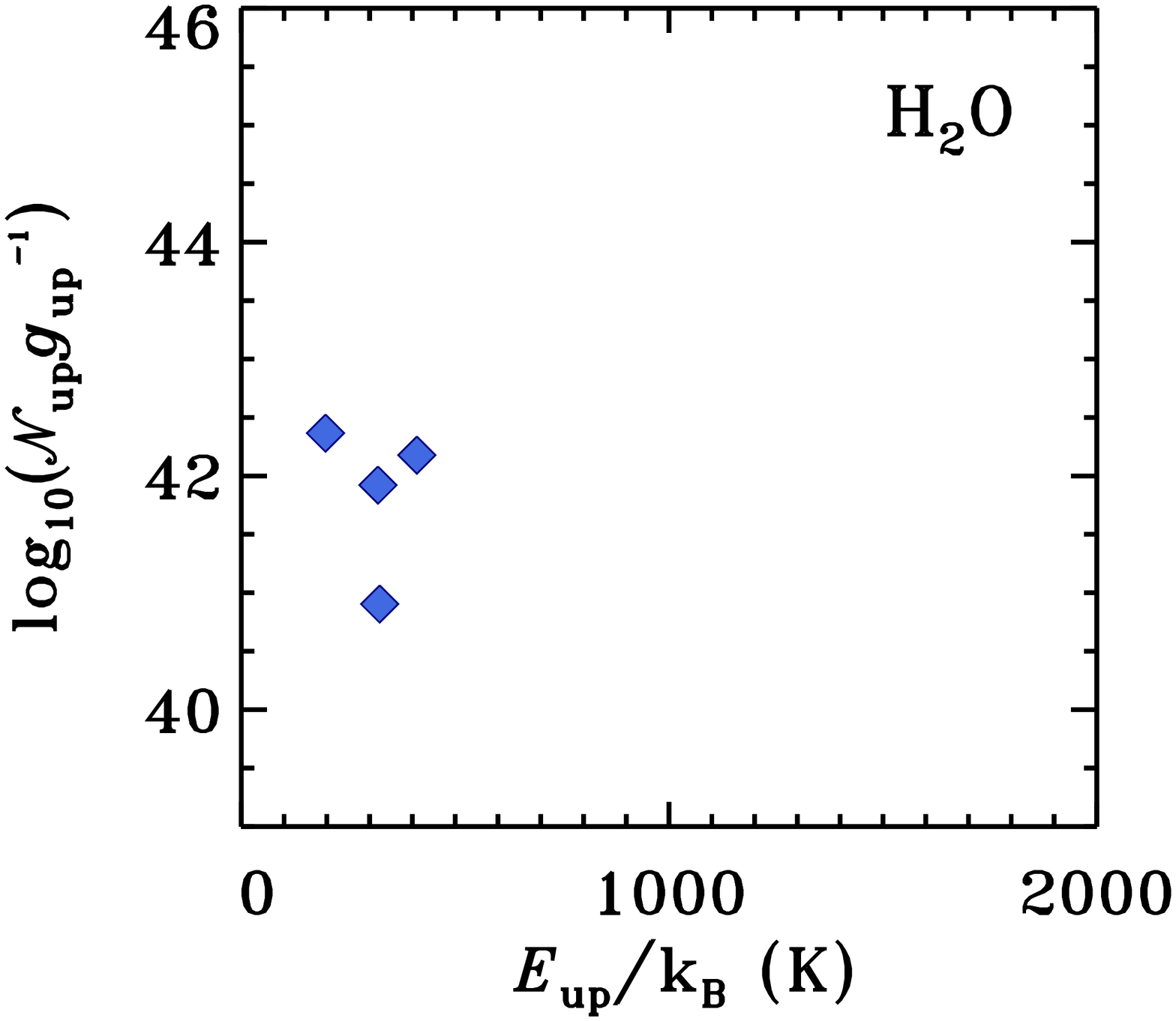} 
       \includegraphics[angle=90,height=4.8cm]{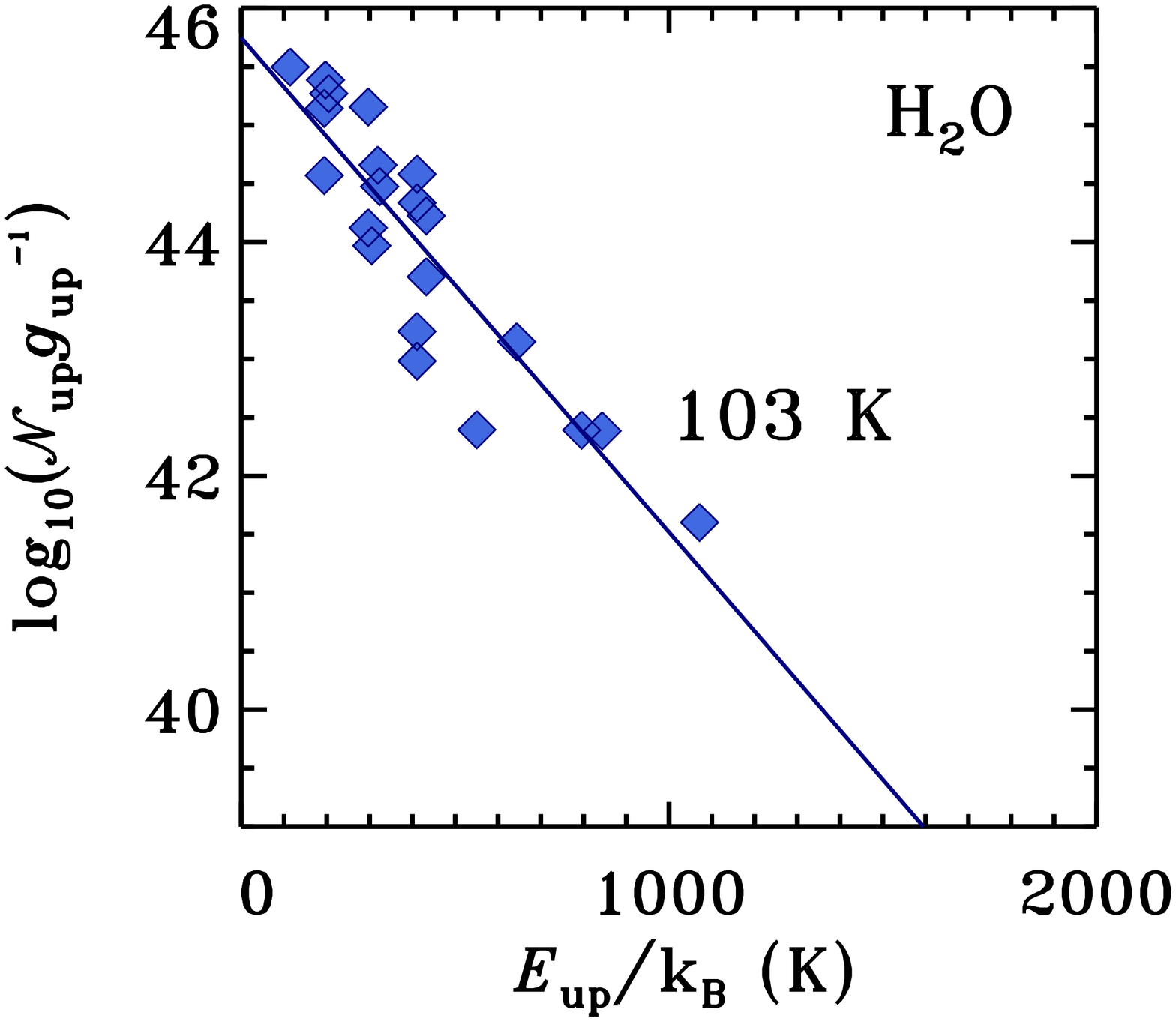} 
       \includegraphics[angle=90,height=4.8cm]{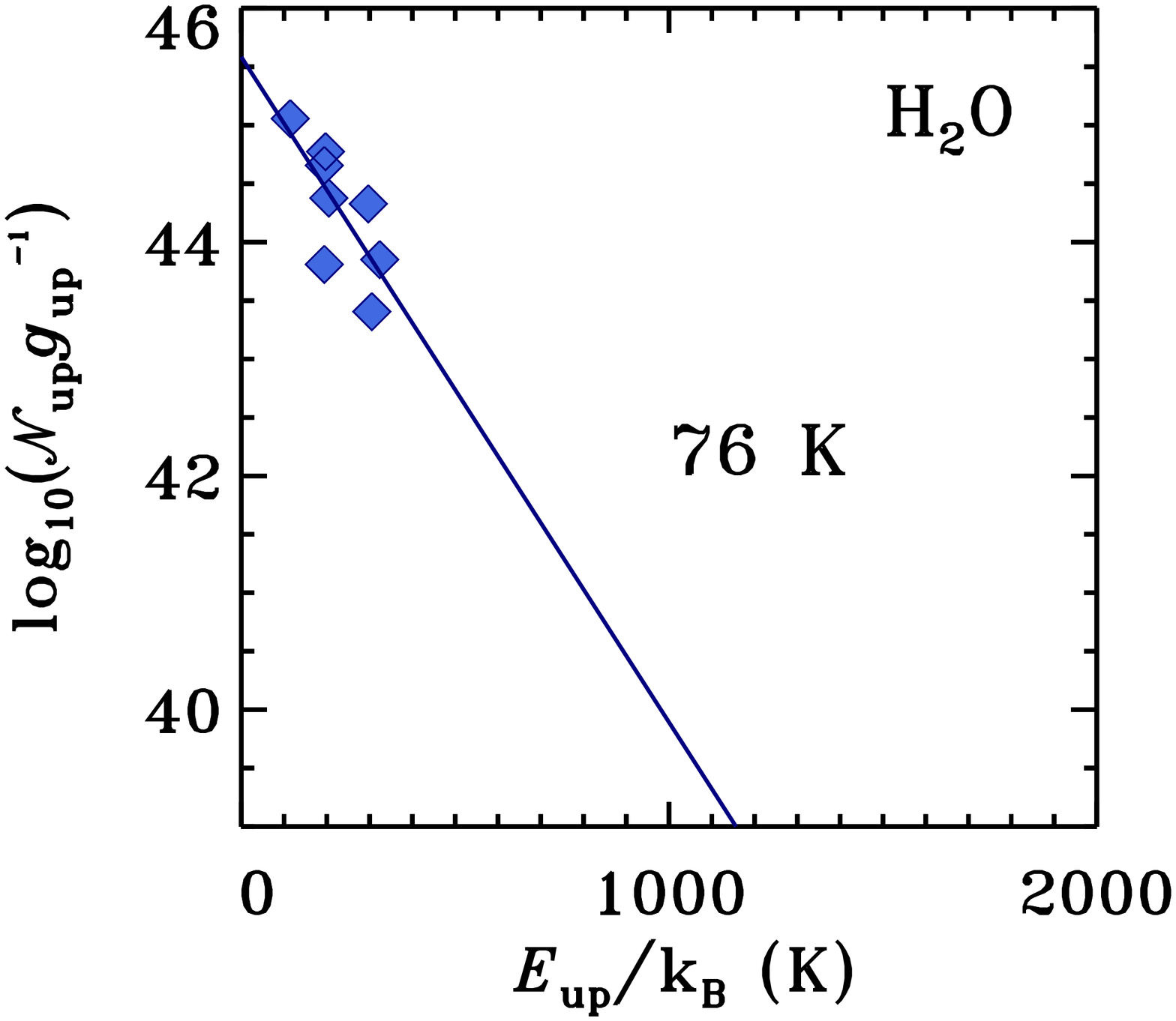} 
       \includegraphics[angle=90,height=4.8cm]{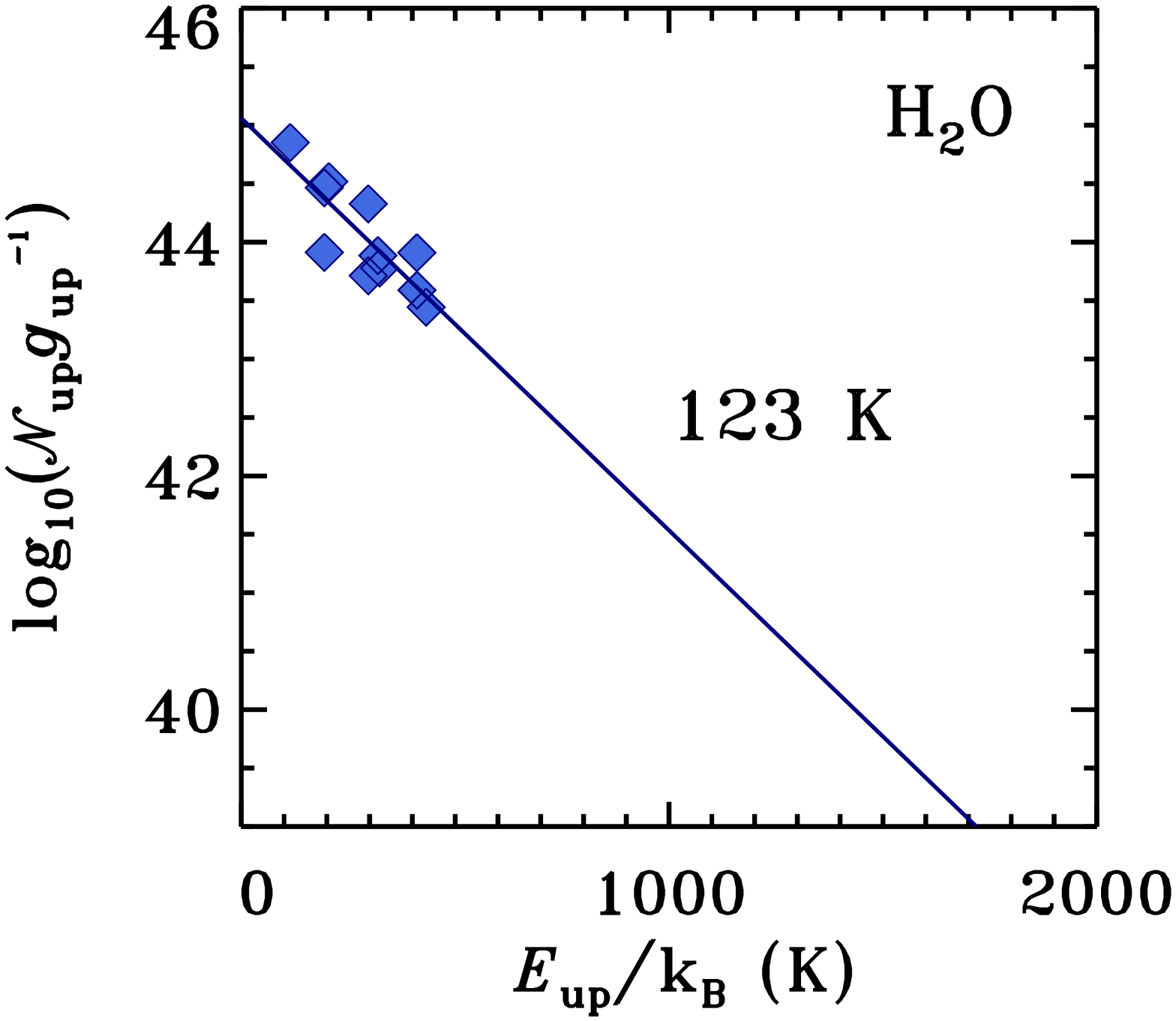} 
       \includegraphics[angle=90,height=4.8cm]{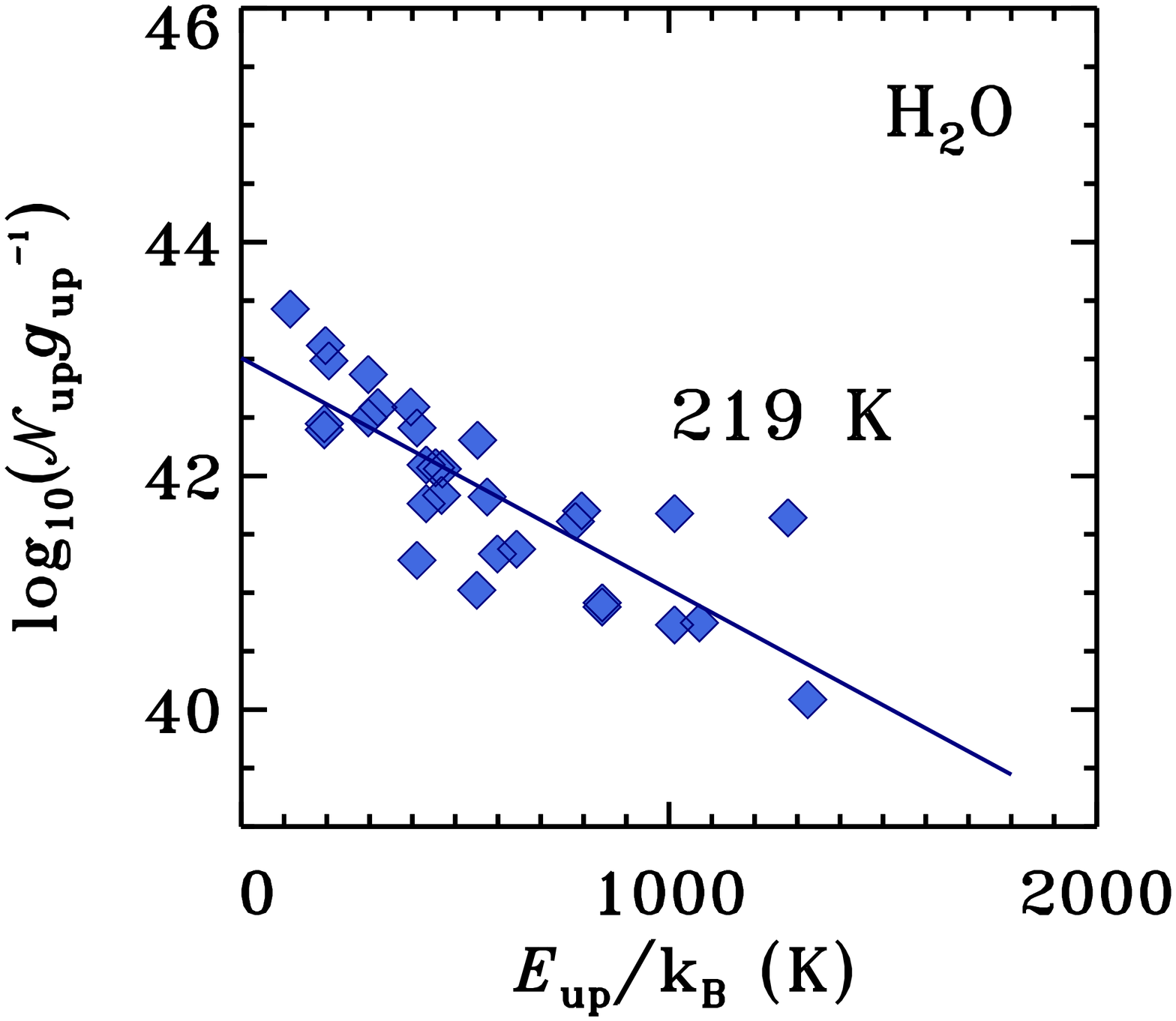} 
            
      \end{center}
  \end{minipage}
    \hfill
   \begin{minipage}[t]{.3\textwidth}
      \begin{center}
    	\includegraphics[angle=90,height=4.8cm]{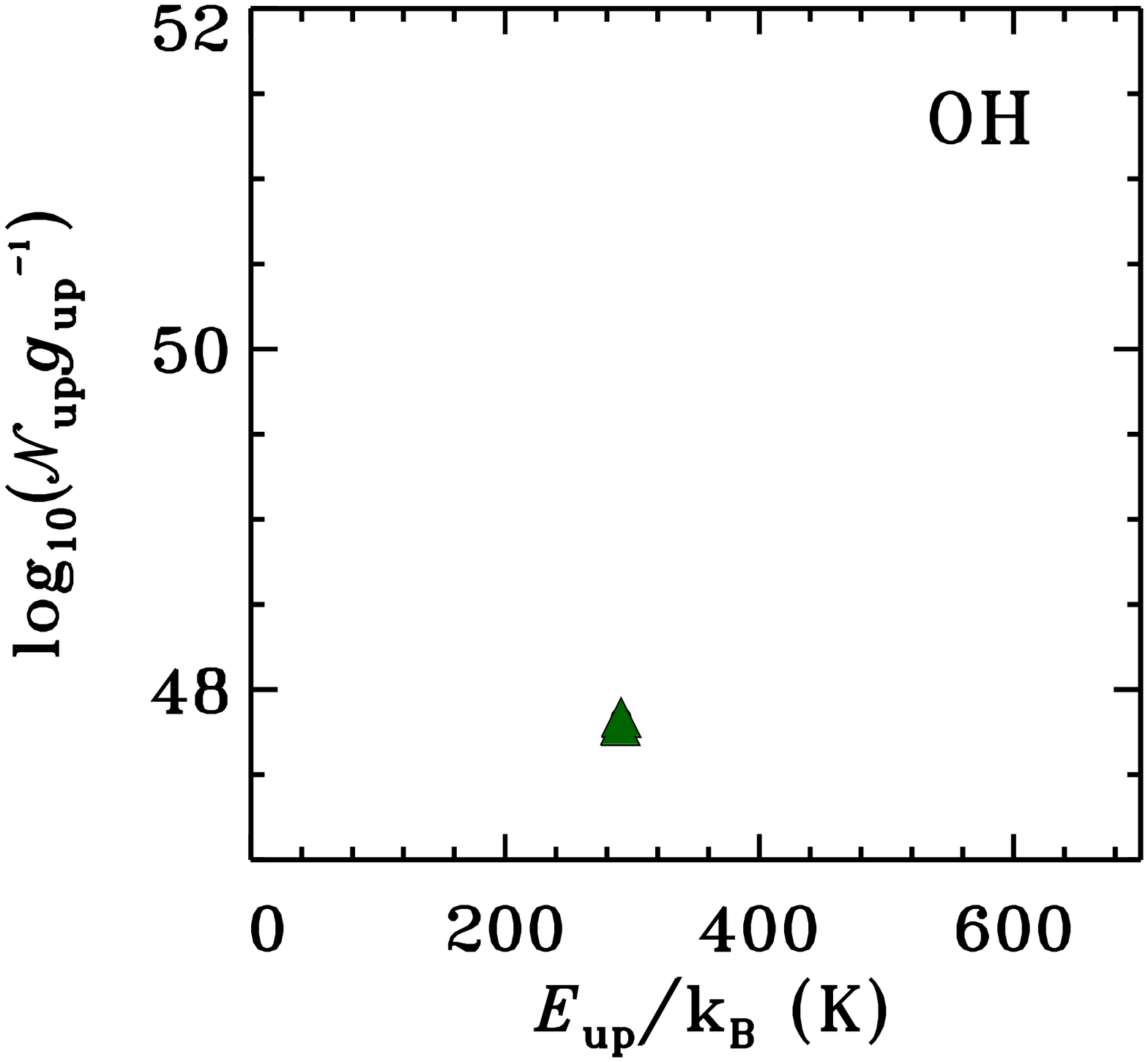} 
        \includegraphics[angle=90,height=4.8cm]{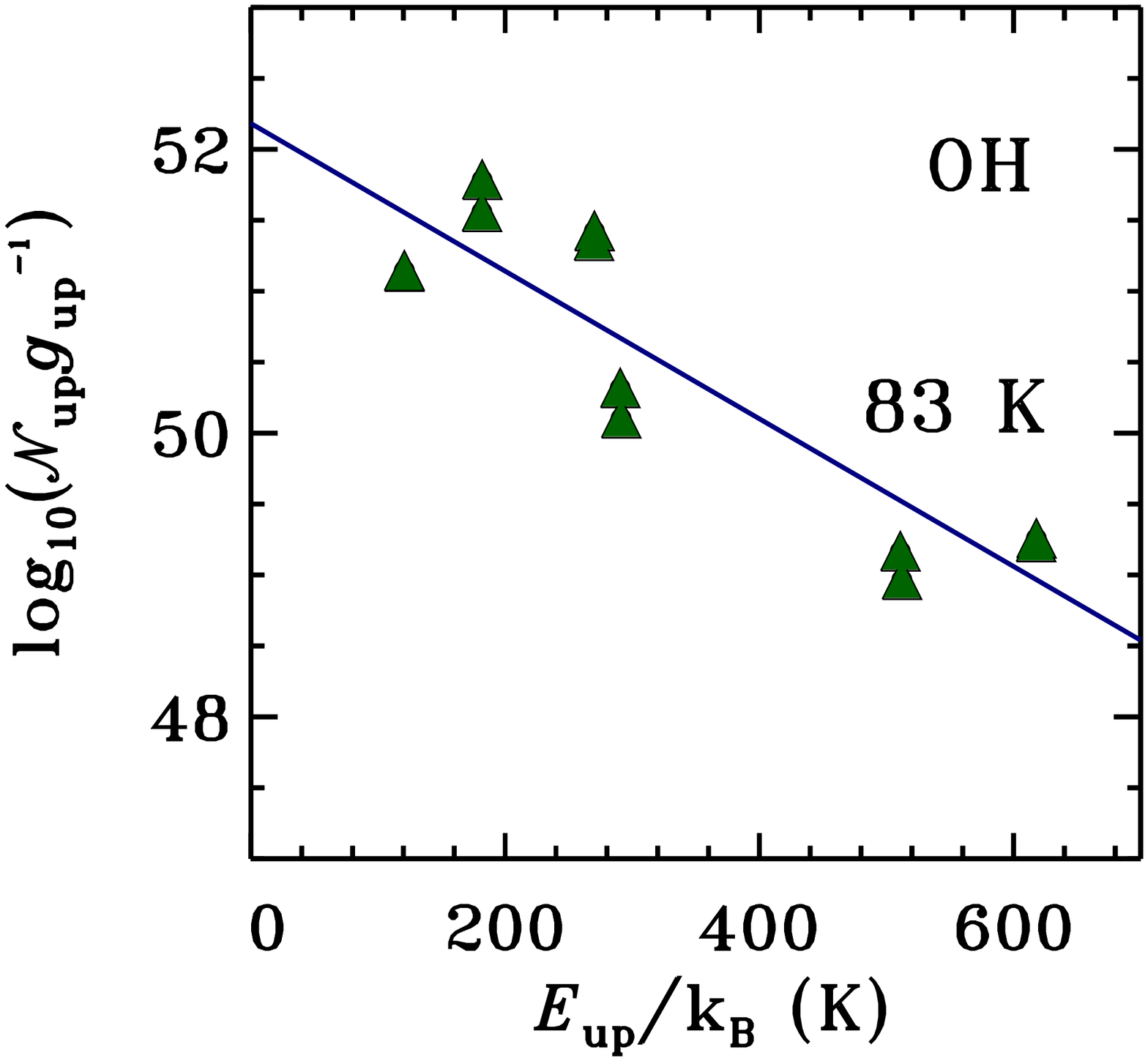} 
        \includegraphics[angle=90,height=4.8cm]{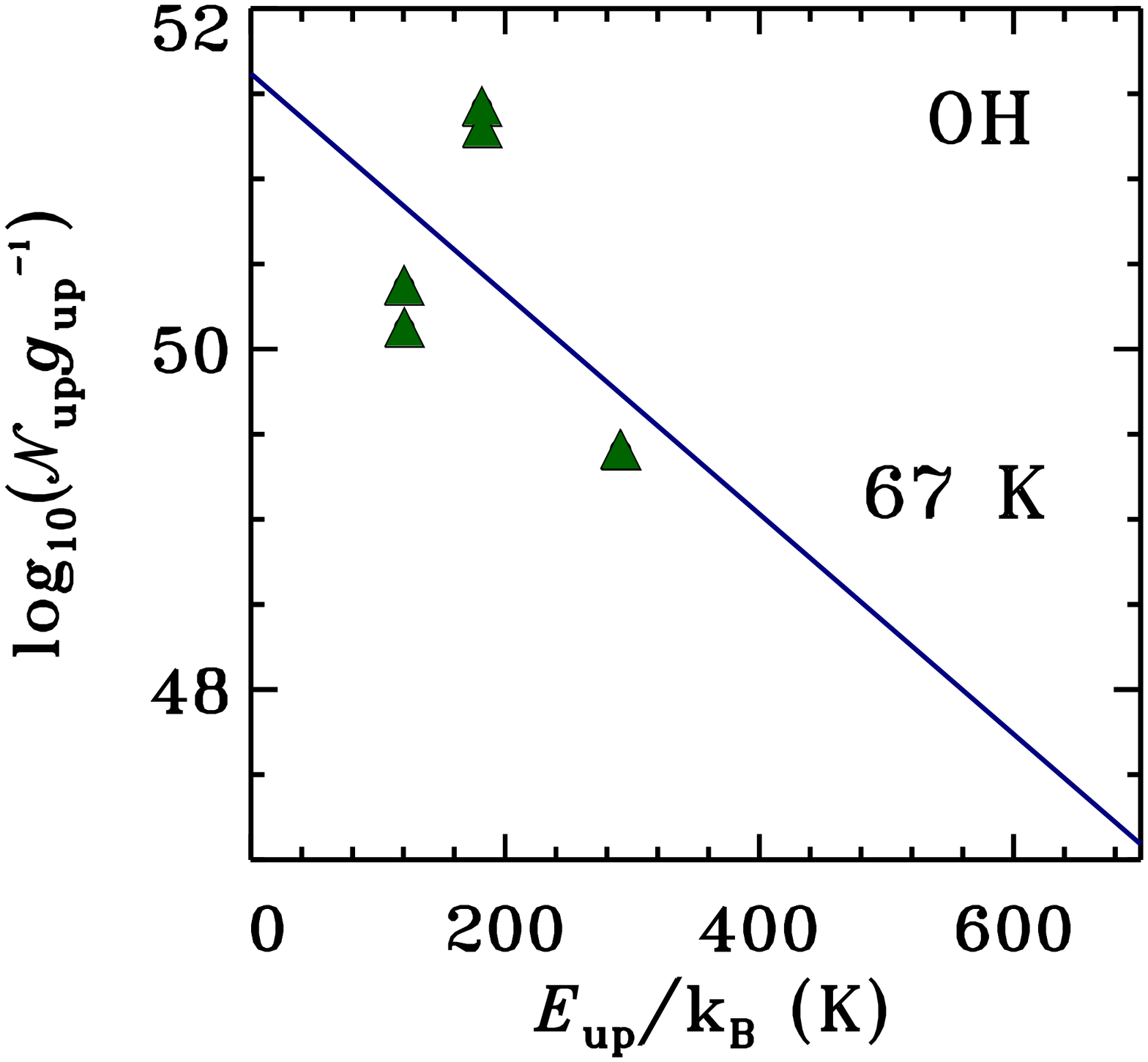} 
        \includegraphics[angle=90,height=4.8cm]{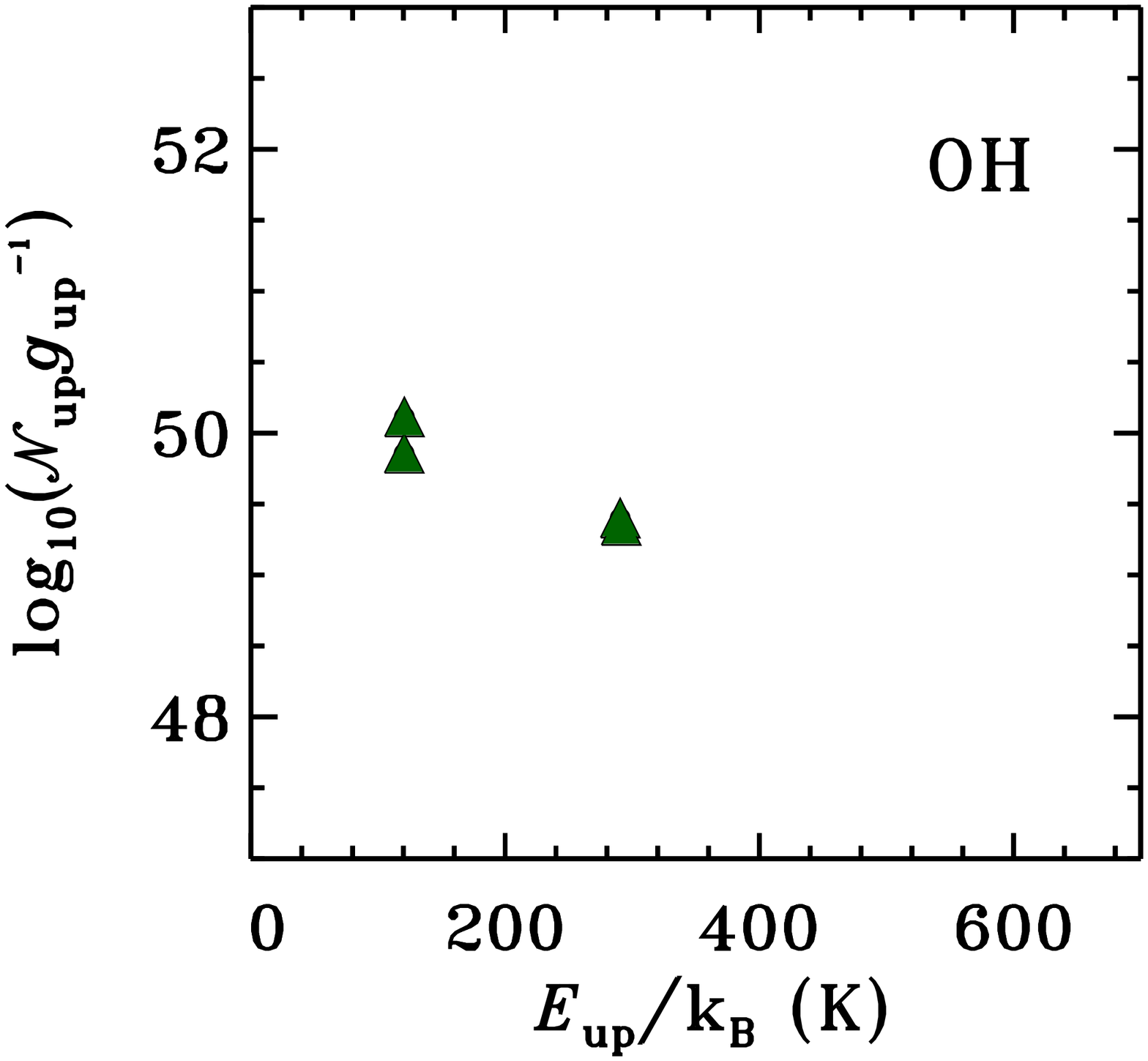} 
        \includegraphics[angle=90,height=4.8cm]{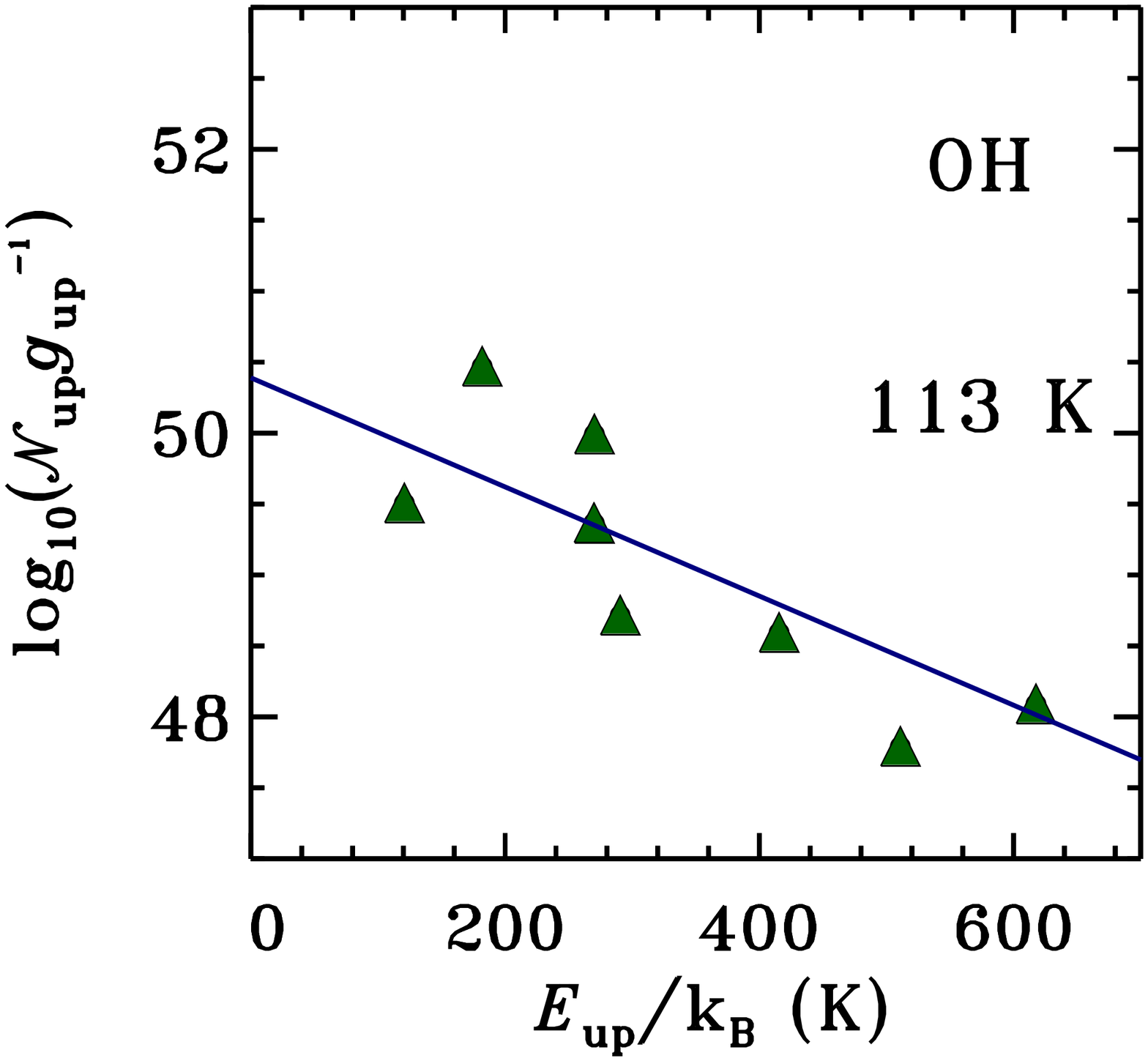} 
                    
      \end{center}
  \end{minipage}
      \hfill
        \caption{\label{dig2} Similar to Figure \ref{molexc}, but for 
        Oph IRS 63, Ser SMM1 (see also \citealt{Go12}), Ser SMM4 and 3 
        (see also \citealt{Di13}), and CrA IRS 5A (based on measurements 
        from \citealt{Li14}).}
\end{figure*}
\renewcommand{\thefigure}{\thesection.\arabic{figure} (Cont.)}
\addtocounter{figure}{-1}   
\begin{figure*}[!tb]
  \begin{minipage}[t]{.3\textwidth}
  \begin{center}
       \includegraphics[angle=90,height=4.8cm]{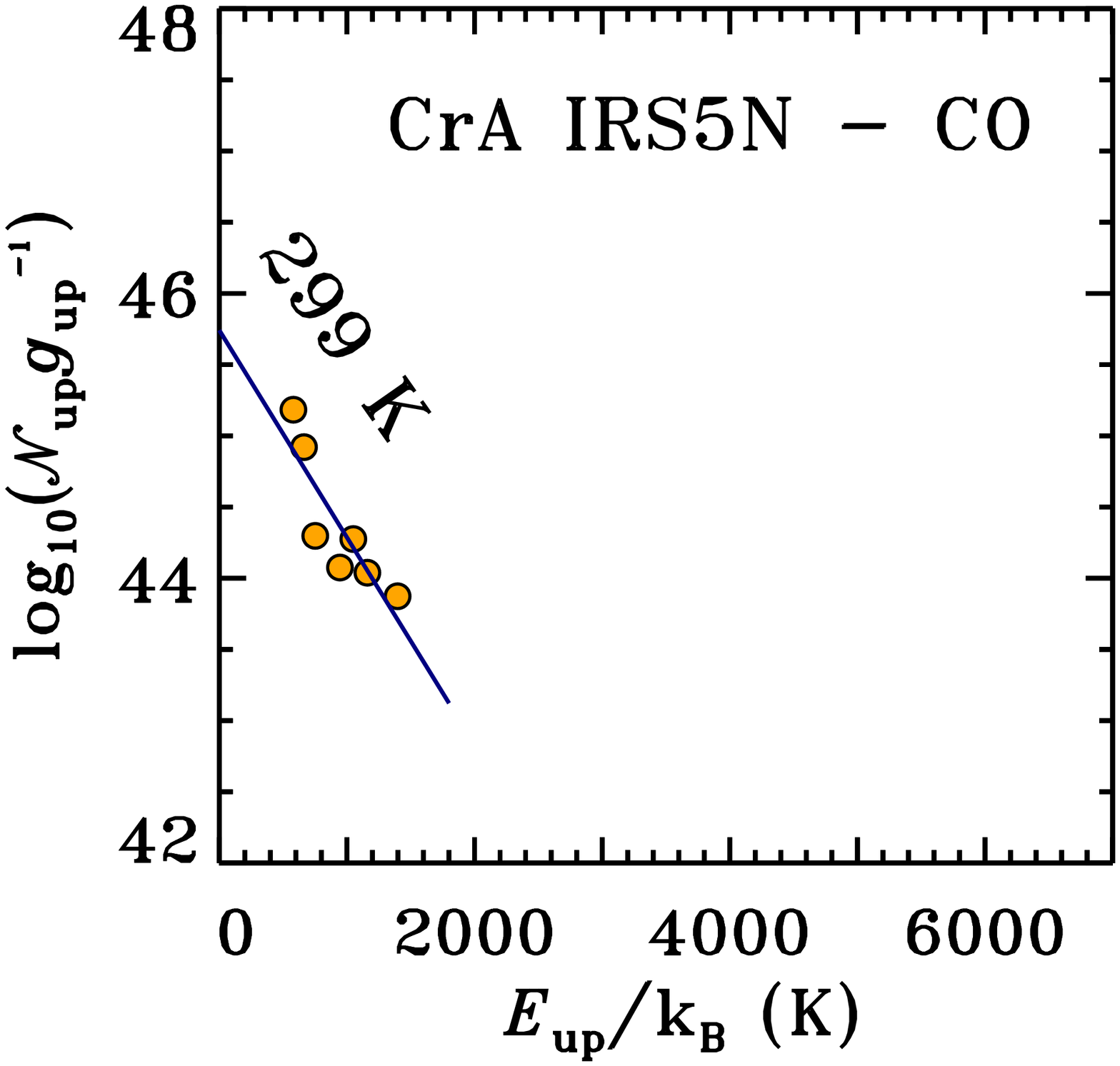} 
       \includegraphics[angle=90,height=4.8cm]{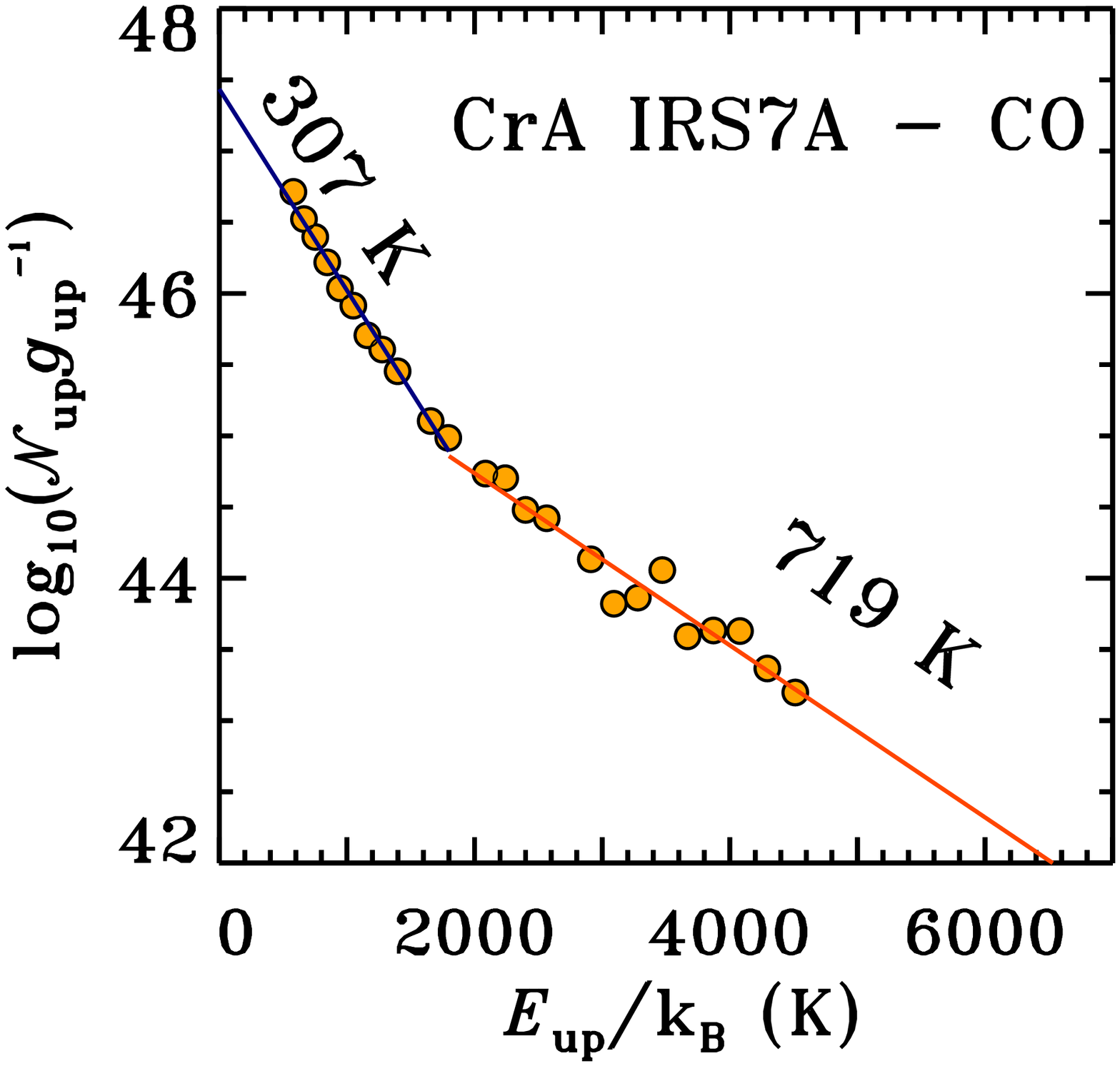} 
        \includegraphics[angle=90,height=4.8cm]{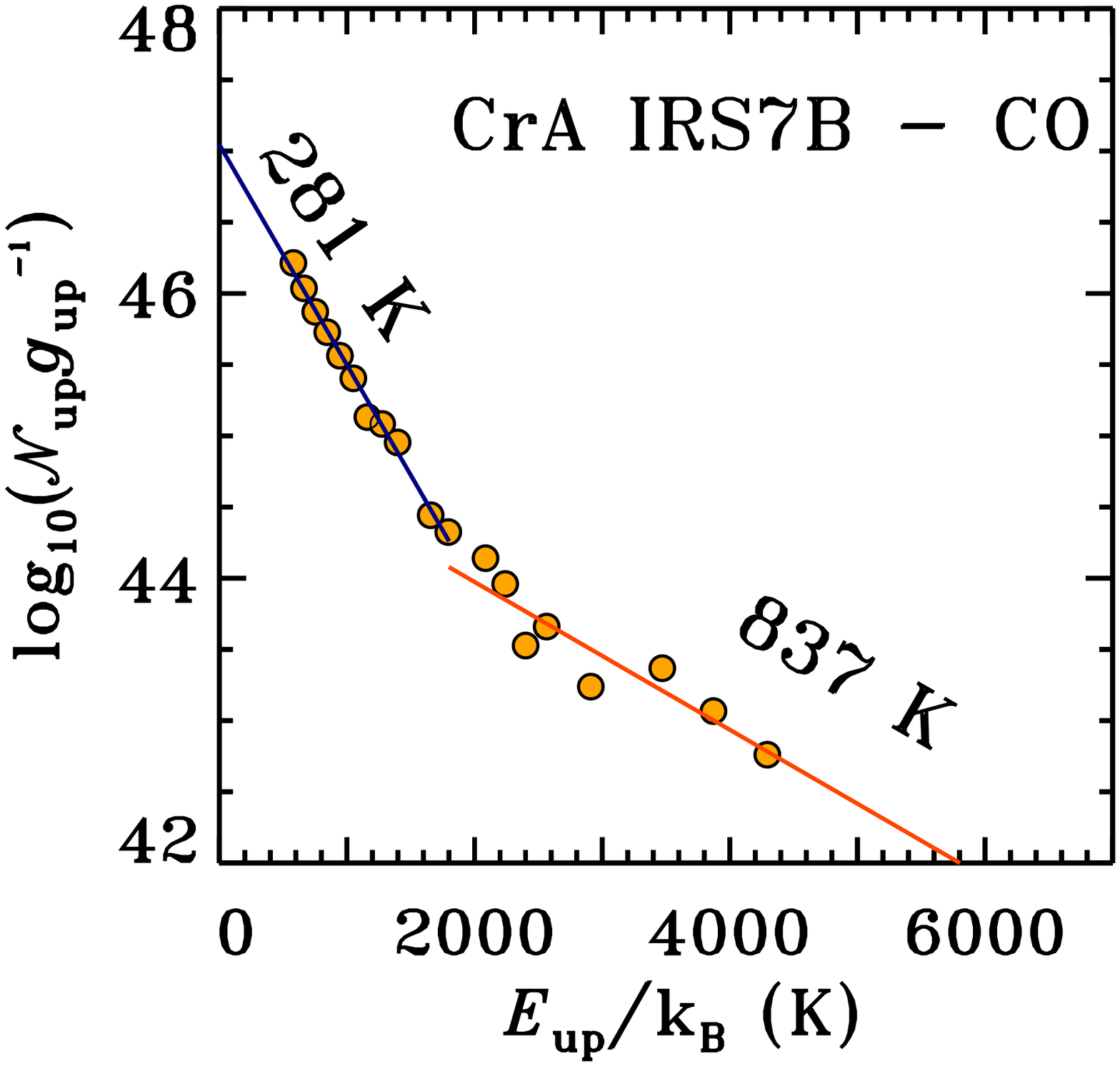} 
       \includegraphics[angle=90,height=4.8cm]{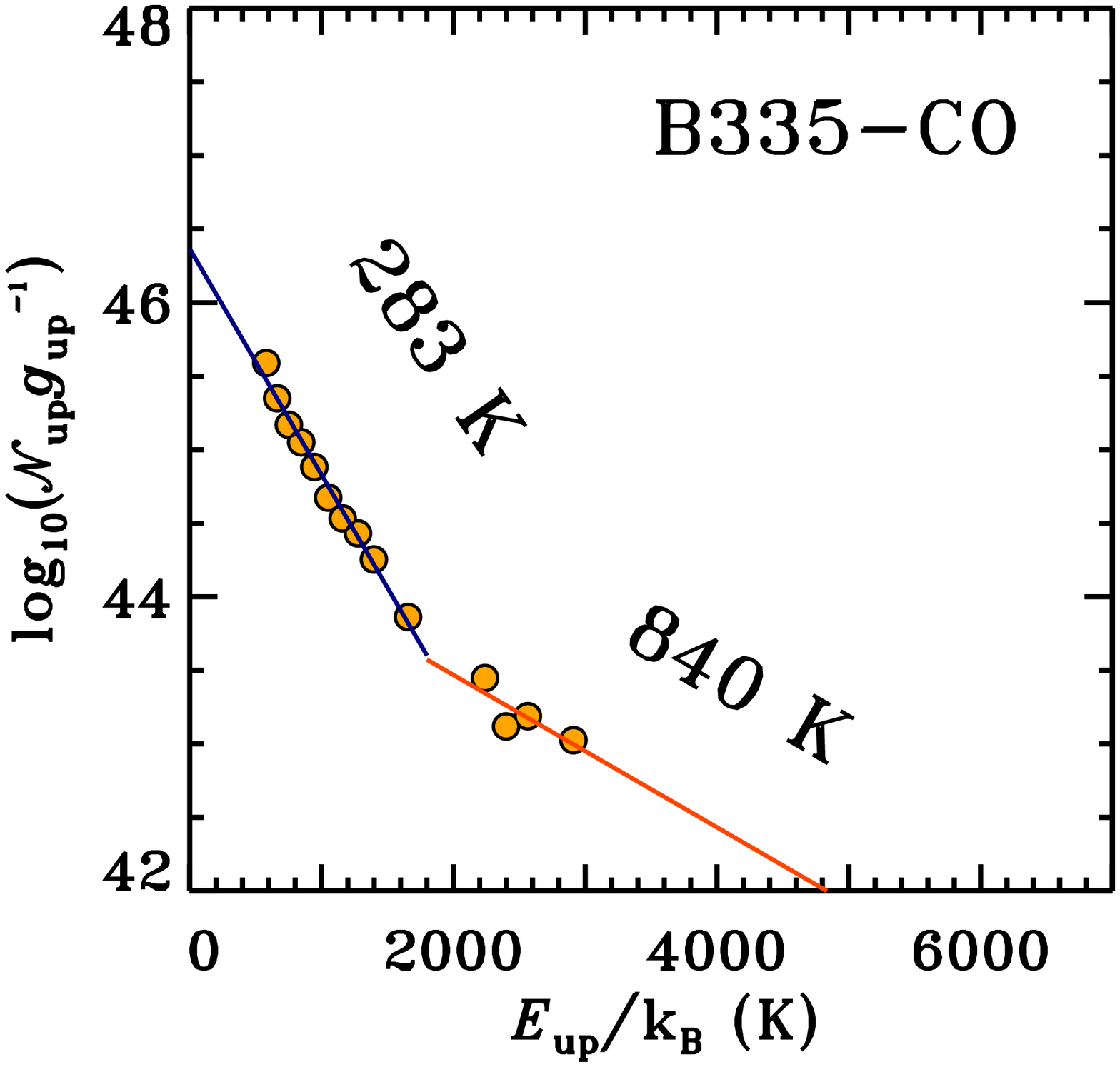} 
        \includegraphics[angle=90,height=4.8cm]{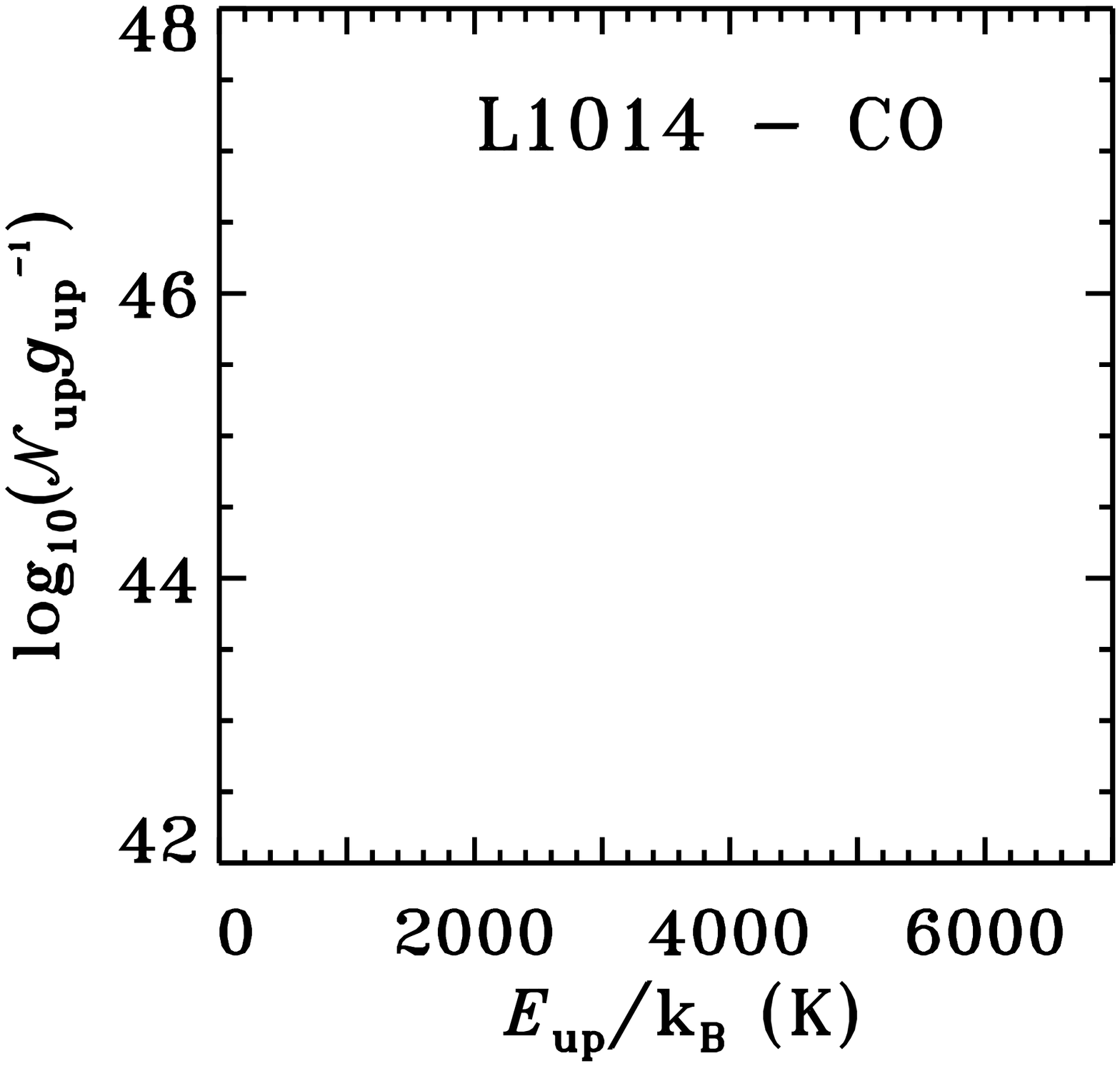} 
  \end{center}
  \end{minipage}
  \hfill
  \begin{minipage}[t]{.3\textwidth}
      \begin{center}
   	   \includegraphics[angle=90,height=4.8cm]{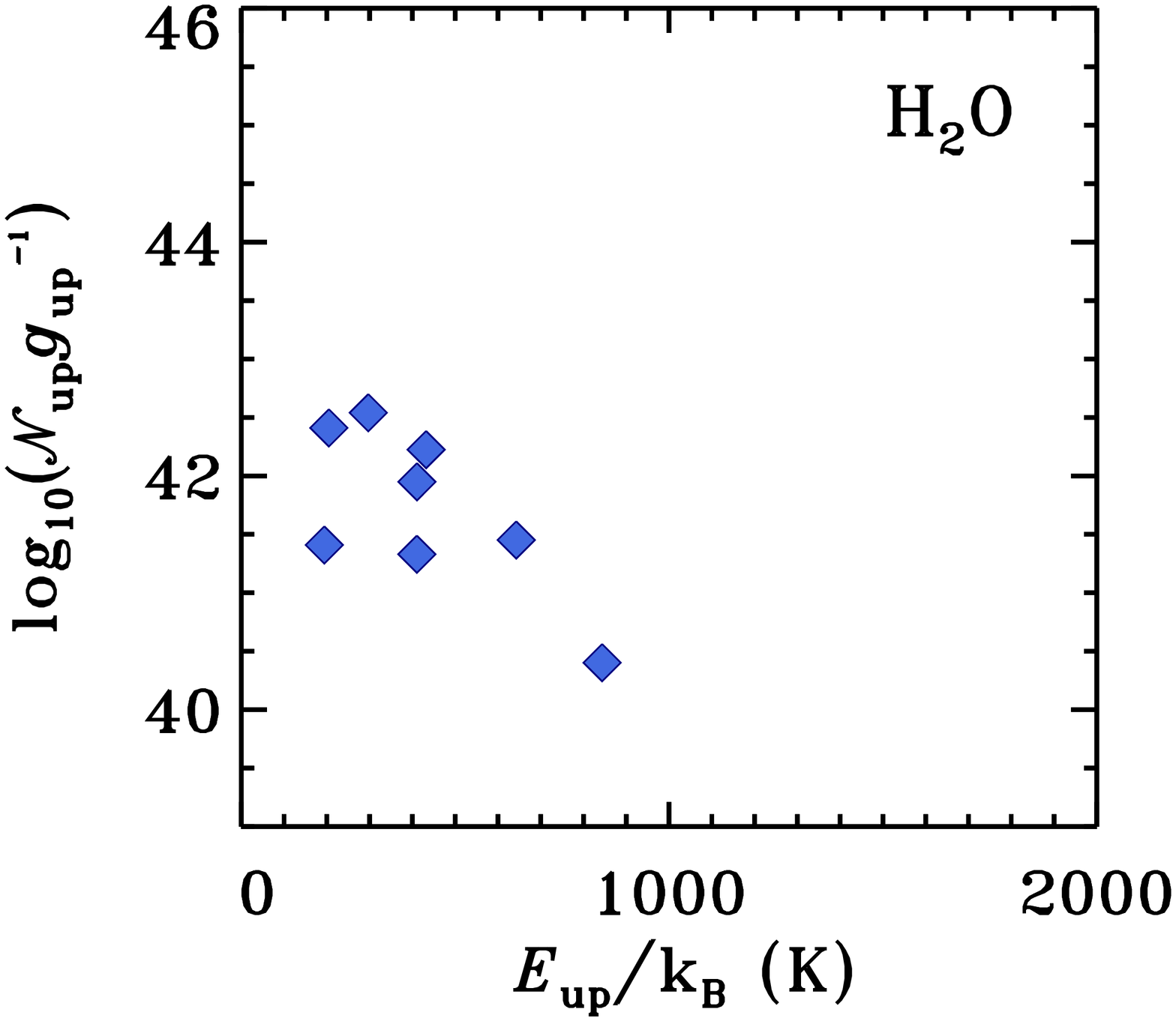} 
       \includegraphics[angle=90,height=4.8cm]{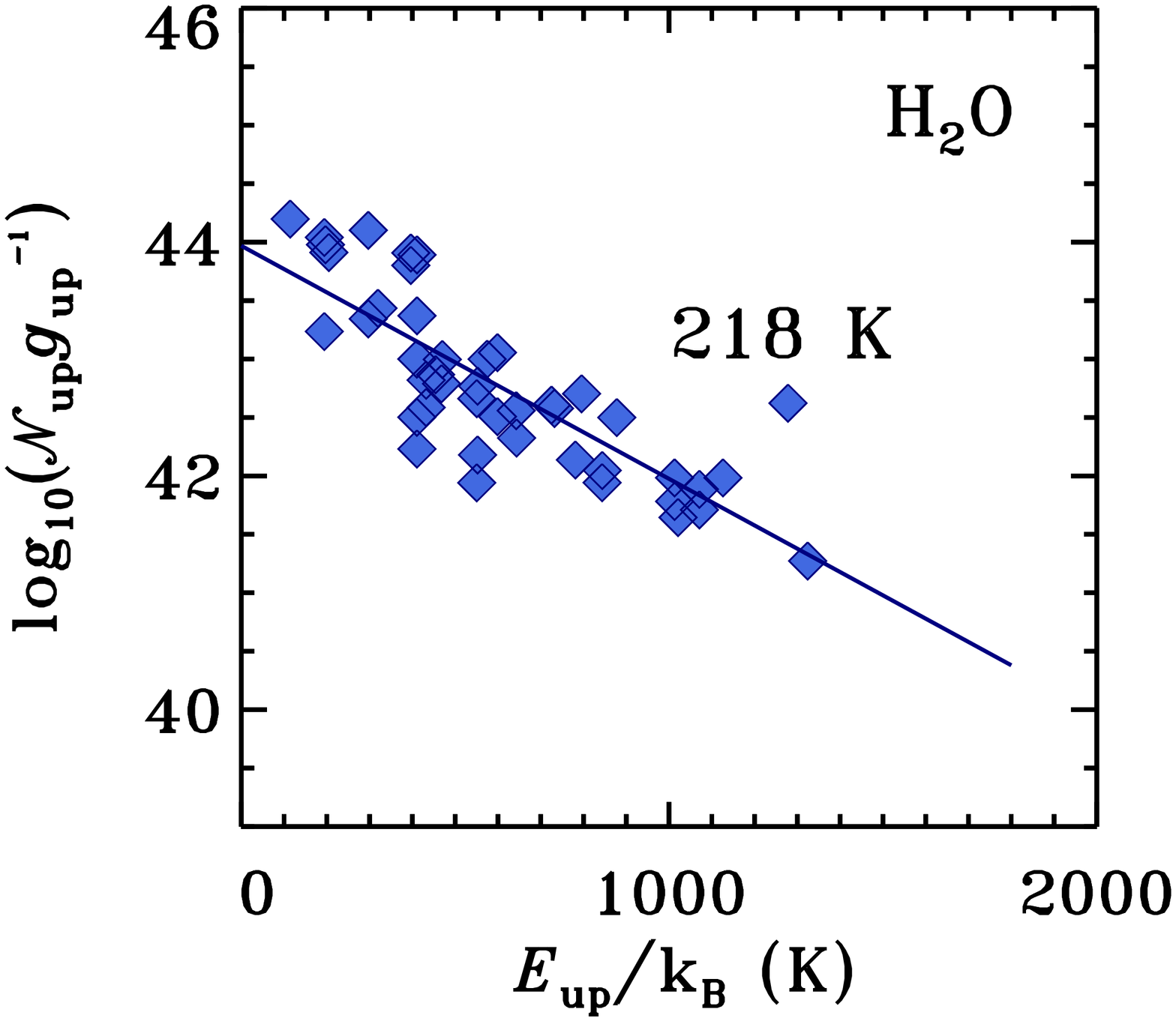} 
       \includegraphics[angle=90,height=4.8cm]{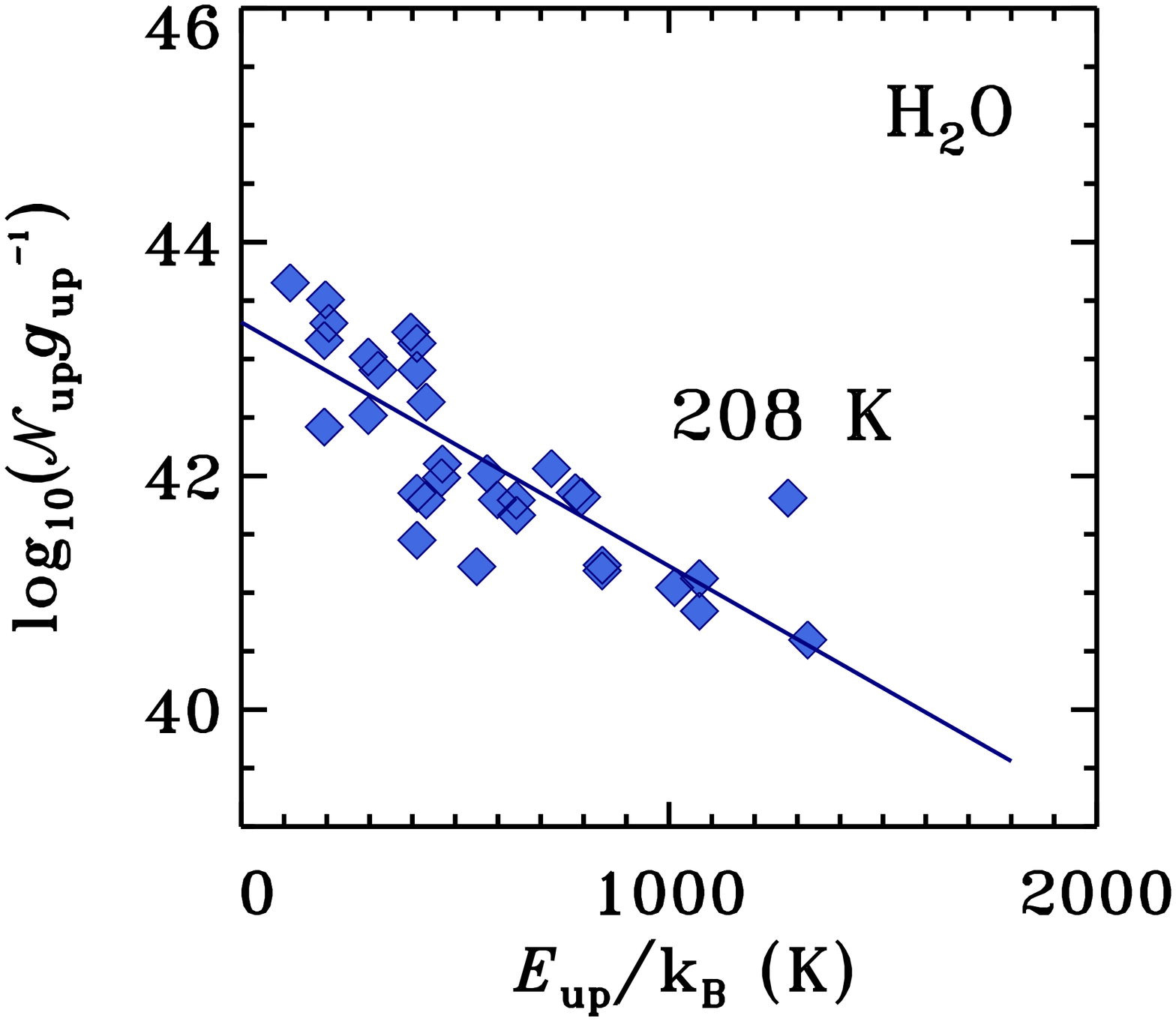} 
       \includegraphics[angle=90,height=4.8cm]{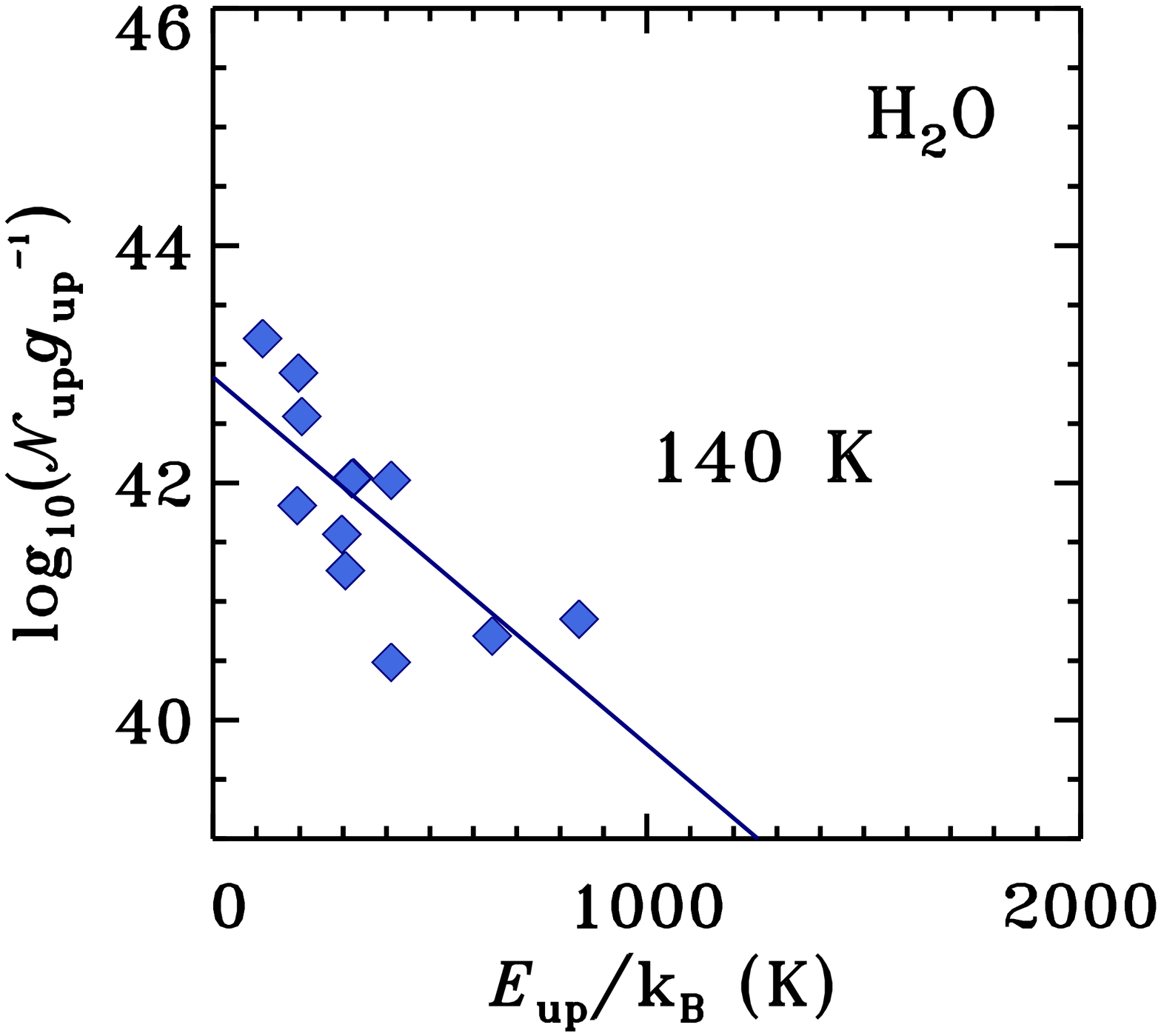} 
       \includegraphics[angle=90,height=4.8cm]{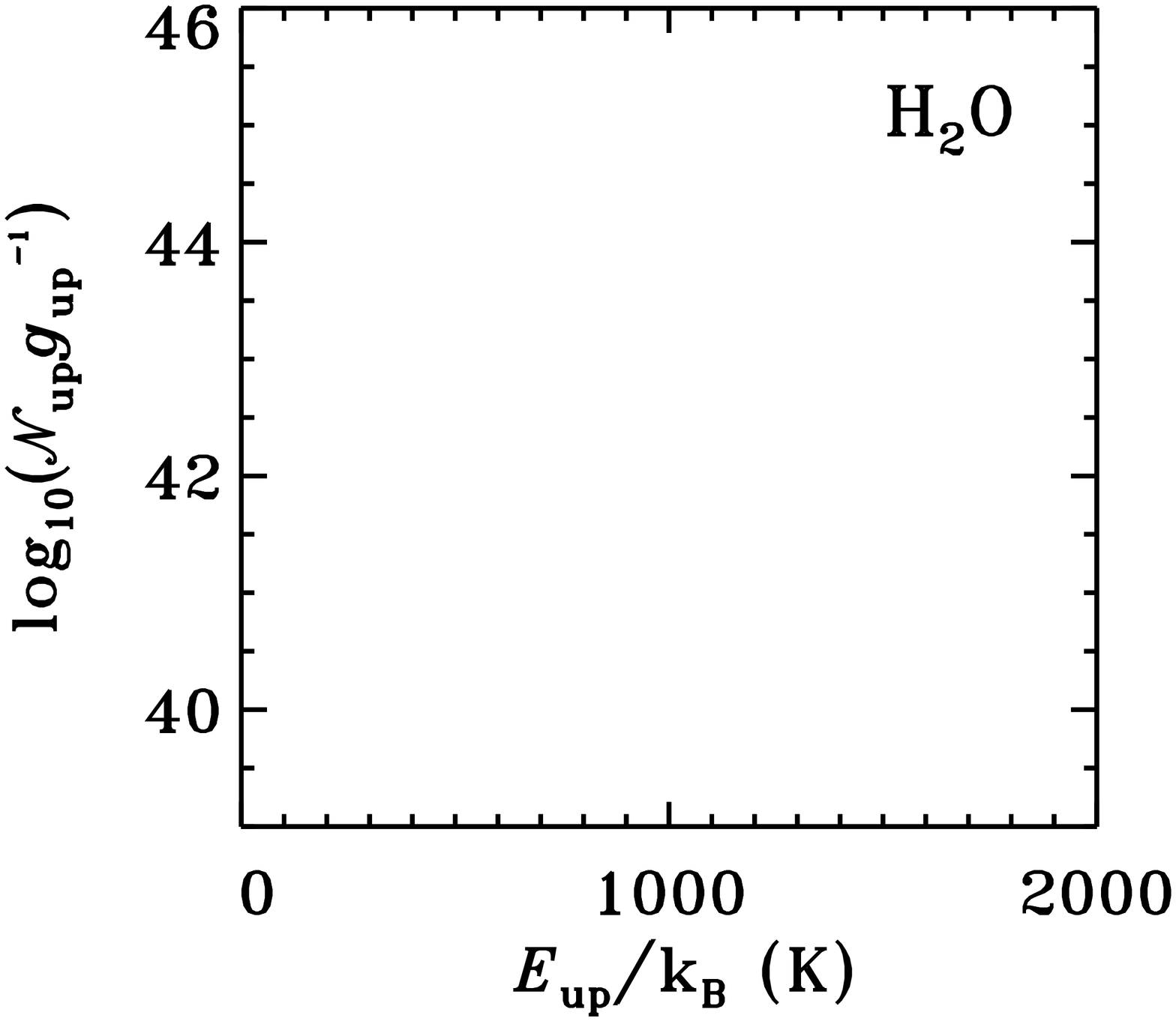} 
            
      \end{center}
  \end{minipage}
    \hfill
   \begin{minipage}[t]{.3\textwidth}
      \begin{center}
    	\includegraphics[angle=90,height=4.8cm]{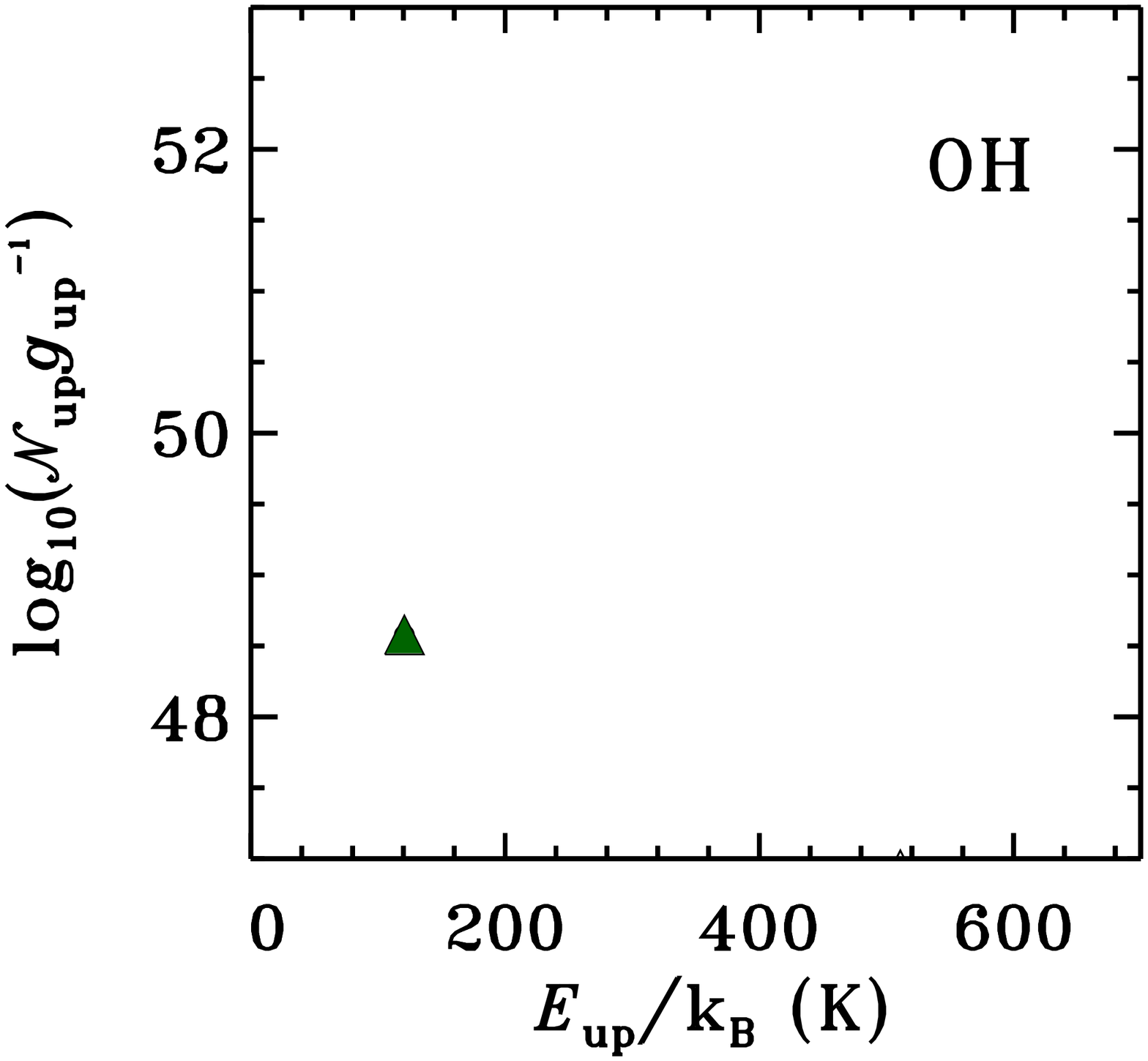} 
        \includegraphics[angle=90,height=4.8cm]{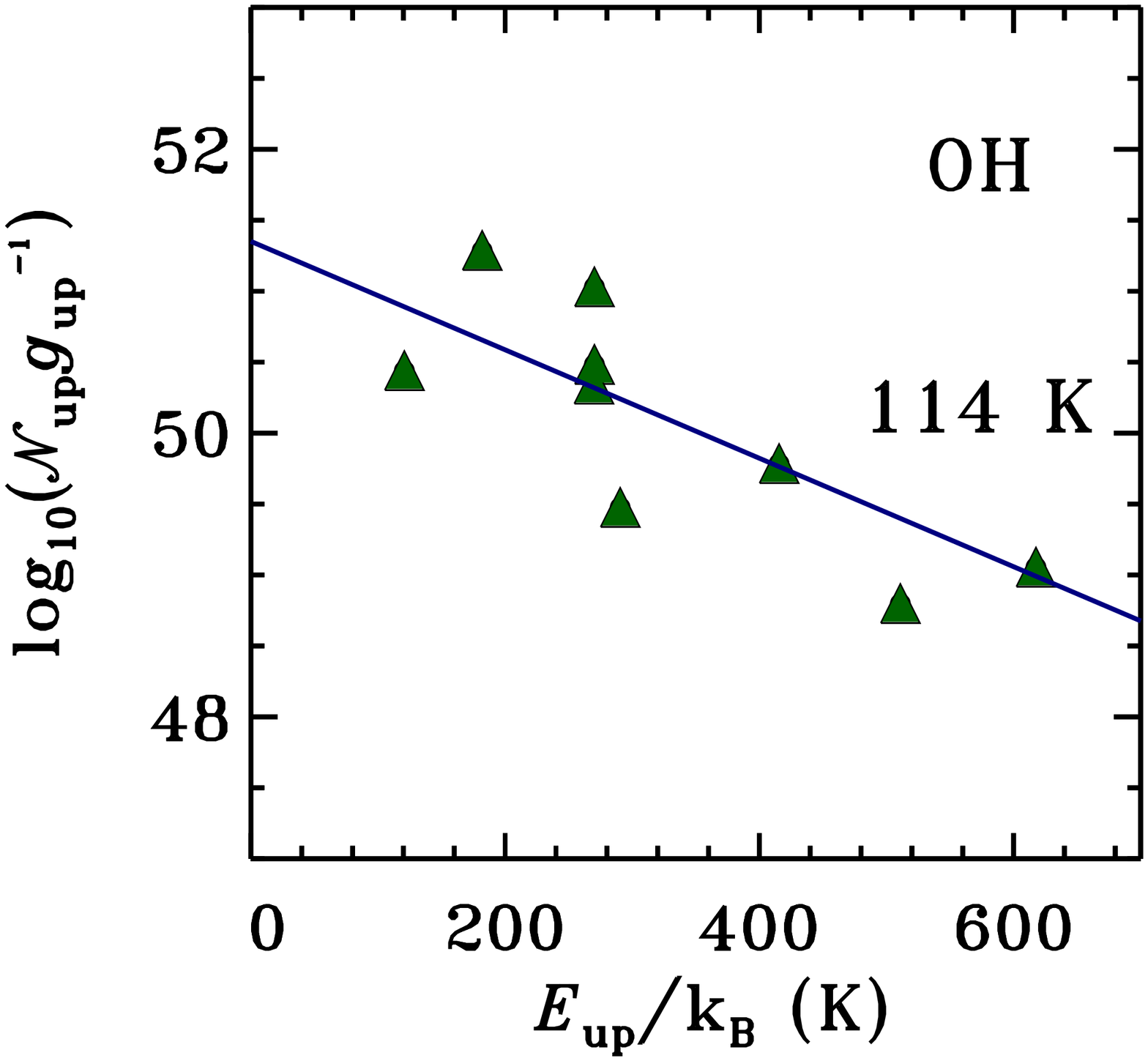} 
        \includegraphics[angle=90,height=4.8cm]{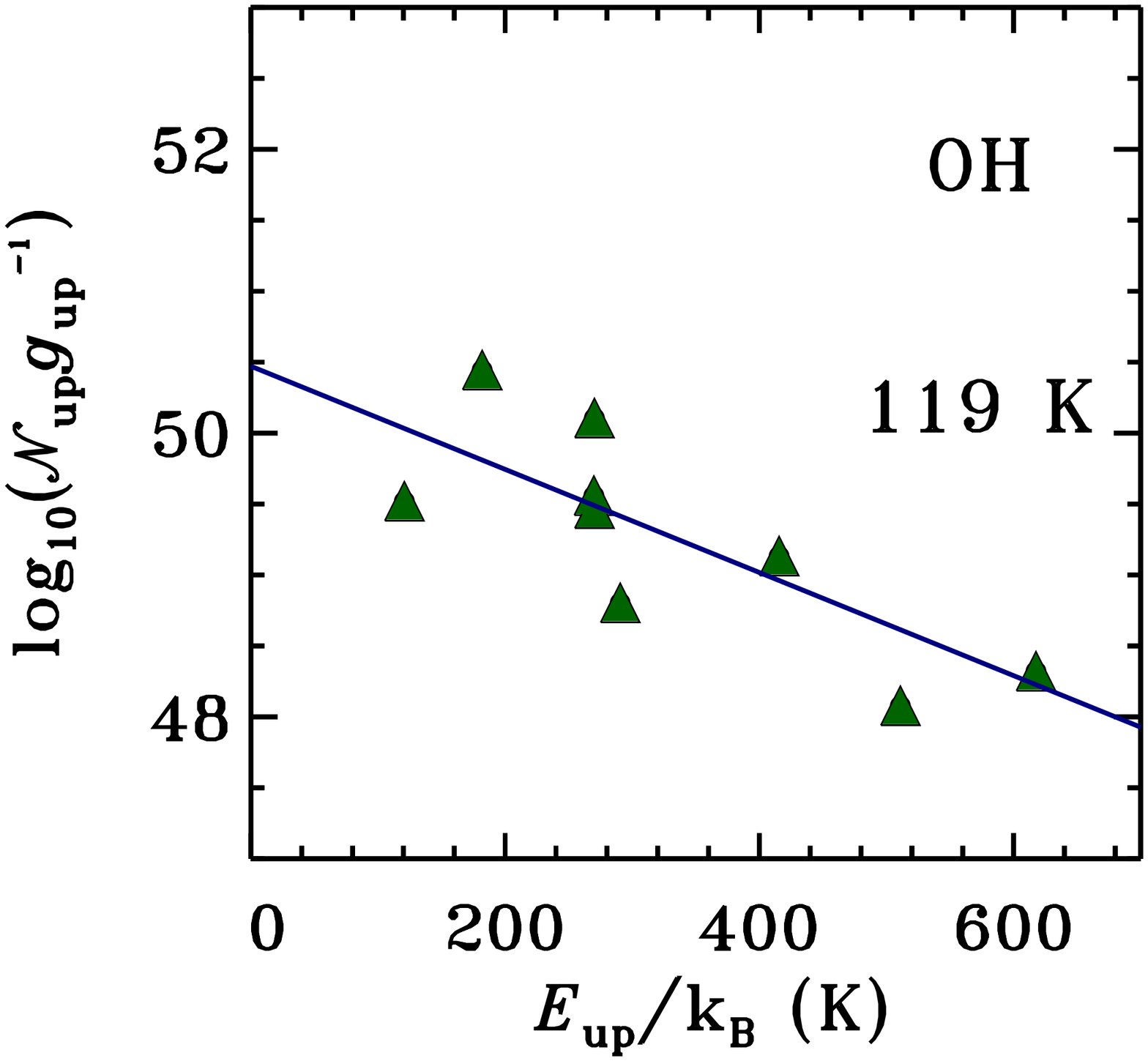} 
        \includegraphics[angle=90,height=4.8cm]{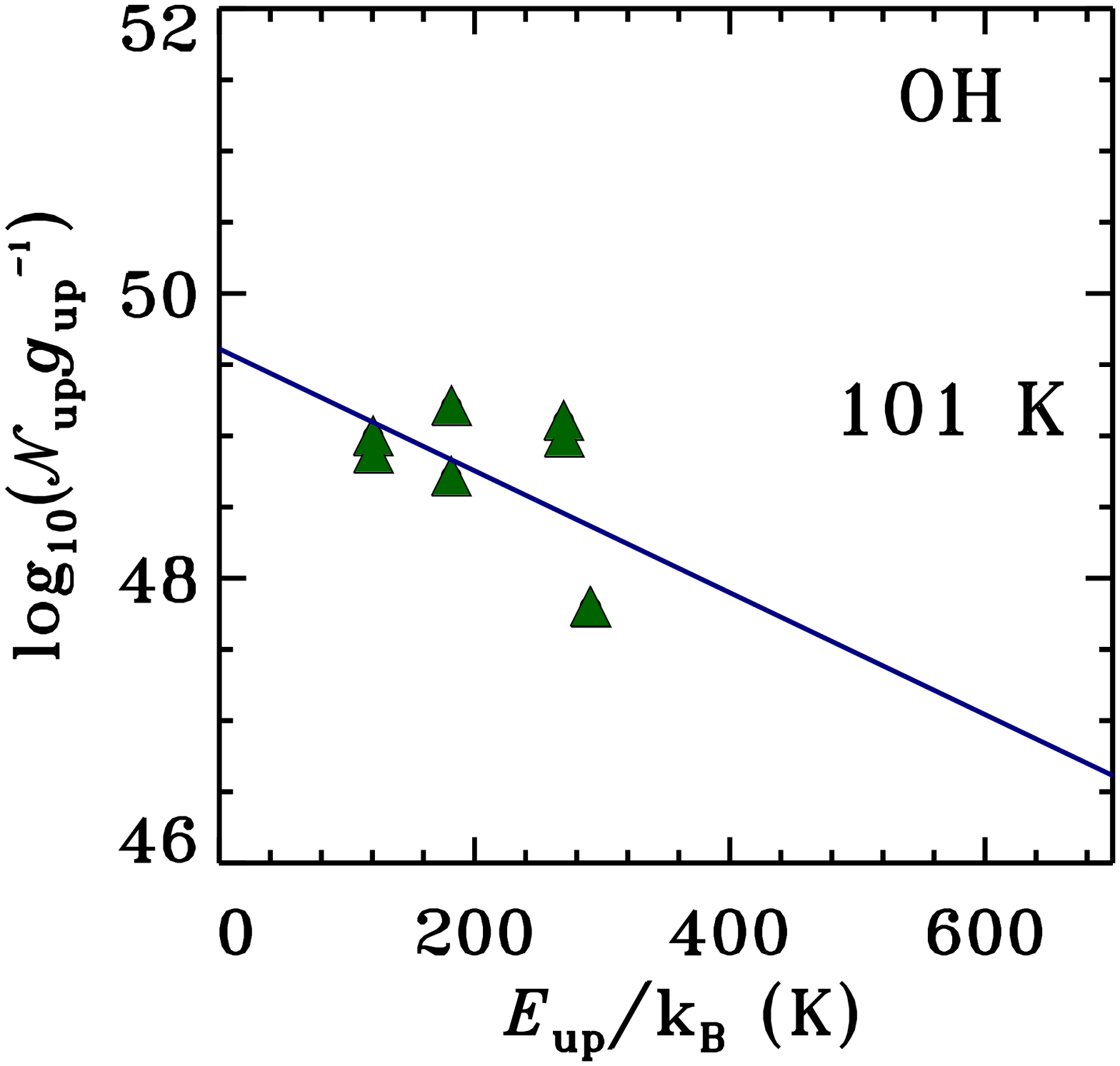} 
        \includegraphics[angle=90,height=4.8cm]{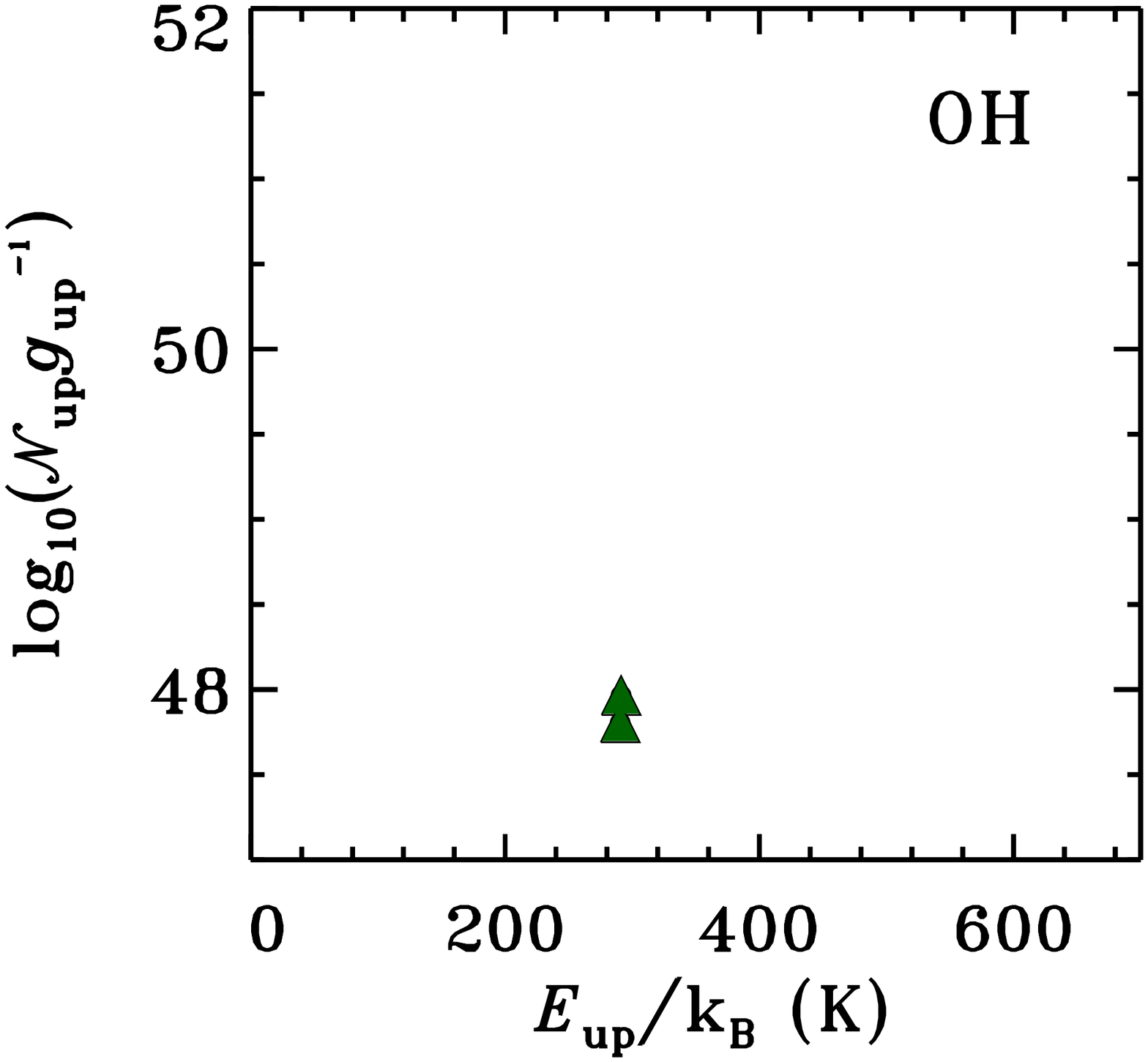} 
                    
      \end{center}
  \end{minipage}
      \hfill
        \caption{\label{dig2} Similar to Figure \ref{molexc}, but for 
        CrA IRS 5N, 7A, 7B (based on measurements 
        from \citealt{Li14}), B335, and L1014.}
\end{figure*}
\renewcommand{\thefigure}{\thesection.\arabic{figure} (Cont.)}
\addtocounter{figure}{-1}   
\clearpage
\begin{figure*}[!tb]
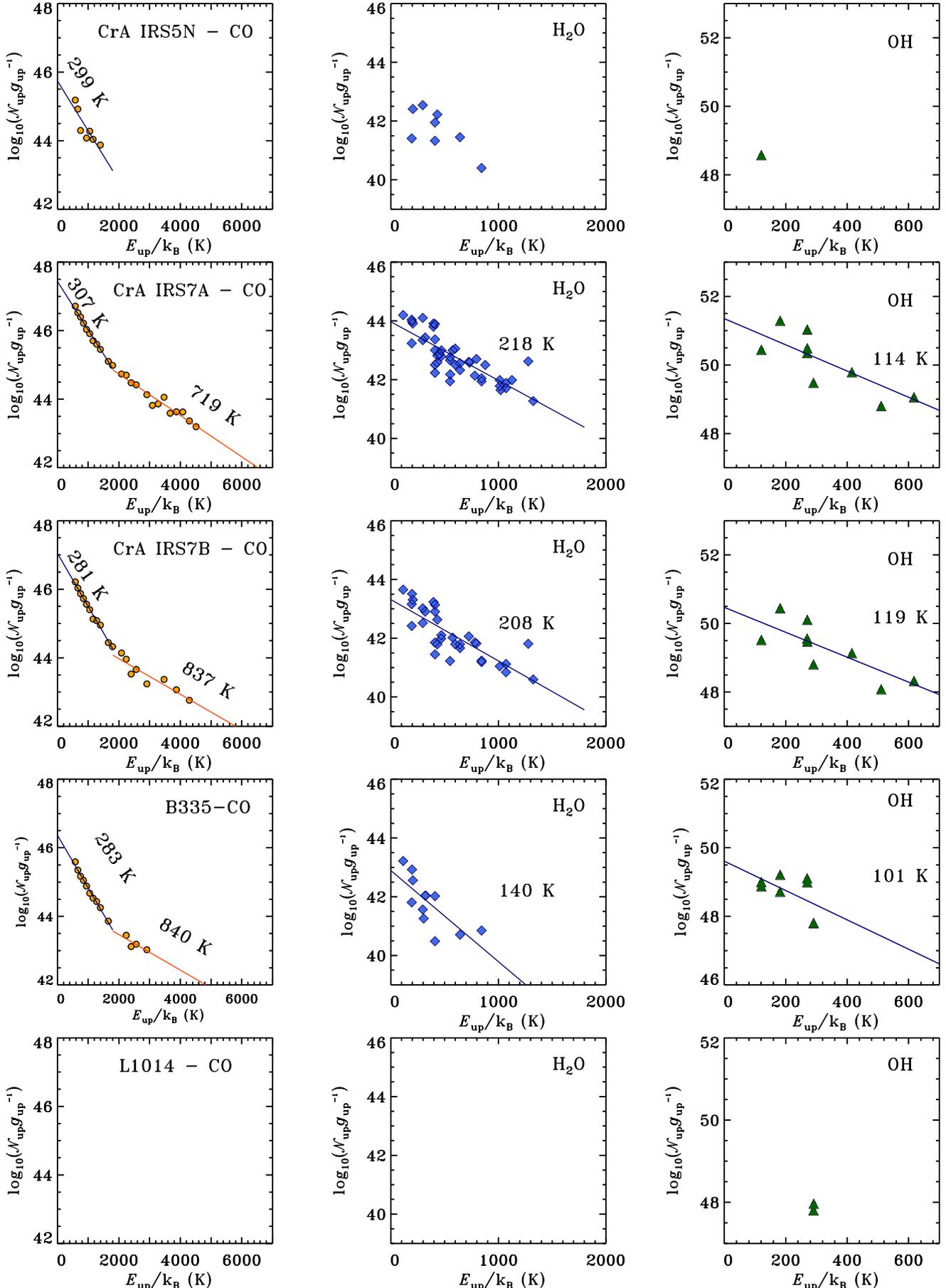

  \begin{minipage}[t]{.3\textwidth}
  \begin{center}
       \includegraphics[angle=90,height=4.8cm]{codiag_crairs5n.eps} 
       \includegraphics[angle=90,height=4.8cm]{codiag_crairs7a.eps} 
        \includegraphics[angle=90,height=4.8cm]{codiag_crairs7b.eps} 
       \includegraphics[angle=90,height=4.8cm]{codiag_b335.eps} 
        \includegraphics[angle=90,height=4.8cm]{codiag_l1014.eps} 
  \end{center}
  \end{minipage}
  \hfill
  \begin{minipage}[t]{.3\textwidth}
      \begin{center}
   	   \includegraphics[angle=90,height=4.8cm]{wdiag_crairs5n.eps} 
       \includegraphics[angle=90,height=4.8cm]{wdiag_crairs7a.eps} 
       \includegraphics[angle=90,height=4.8cm]{wdiag_crairs7b.eps} 
       \includegraphics[angle=90,height=4.8cm]{wdiag_b335.eps} 
       \includegraphics[angle=90,height=4.8cm]{wdiag_l1014.eps} 
            
      \end{center}
  \end{minipage}
    \hfill
   \begin{minipage}[t]{.3\textwidth}
      \begin{center}
    	\includegraphics[angle=90,height=4.8cm]{ohdiag_crairs5n.eps} 
        \includegraphics[angle=90,height=4.8cm]{ohdiag_crairs7a.eps} 
        \includegraphics[angle=90,height=4.8cm]{ohdiag_crairs7b.eps} 
        \includegraphics[angle=90,height=4.8cm]{ohdiag_b335.eps} 
        \includegraphics[angle=90,height=4.8cm]{ohdiag_l1014.eps} 
                    
      \end{center}
  \end{minipage}
      \hfill
        \caption{\label{dig2} Similar to Figure \ref{molexc}, but for 
        CrA IRS 5N, 7A, 7B (based on measurements 
        from \citealt{Li14}), B335, and L1014.}
\end{figure*}
\renewcommand{\thefigure}{\thesection.\arabic{figure} (Cont.)}
\addtocounter{figure}{-1}   
\begin{figure*}[!tb]
  \begin{minipage}[t]{.3\textwidth}
  \begin{center}
       \includegraphics[angle=90,height=4.8cm]{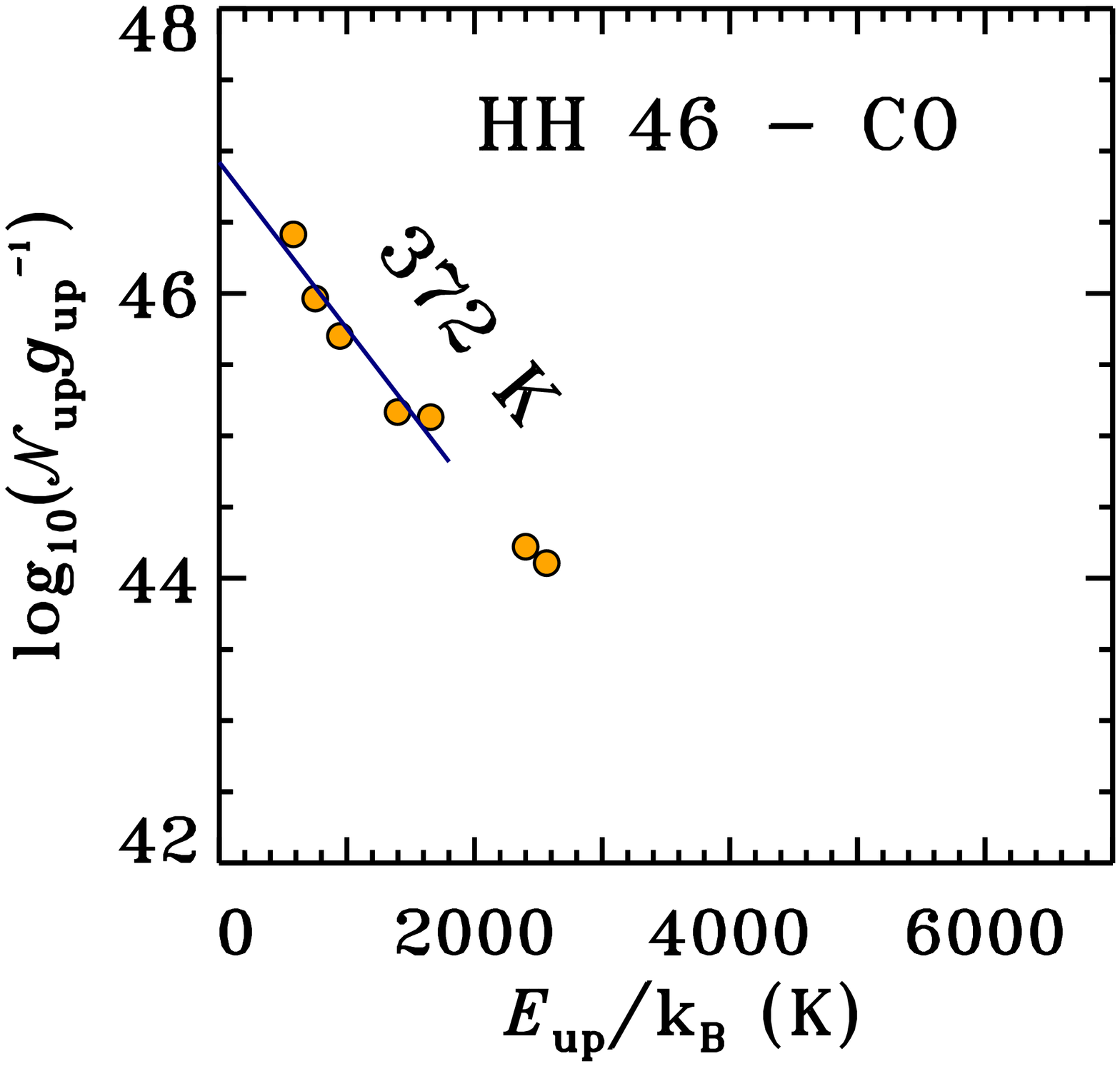} 
       \includegraphics[angle=90,height=4.8cm]{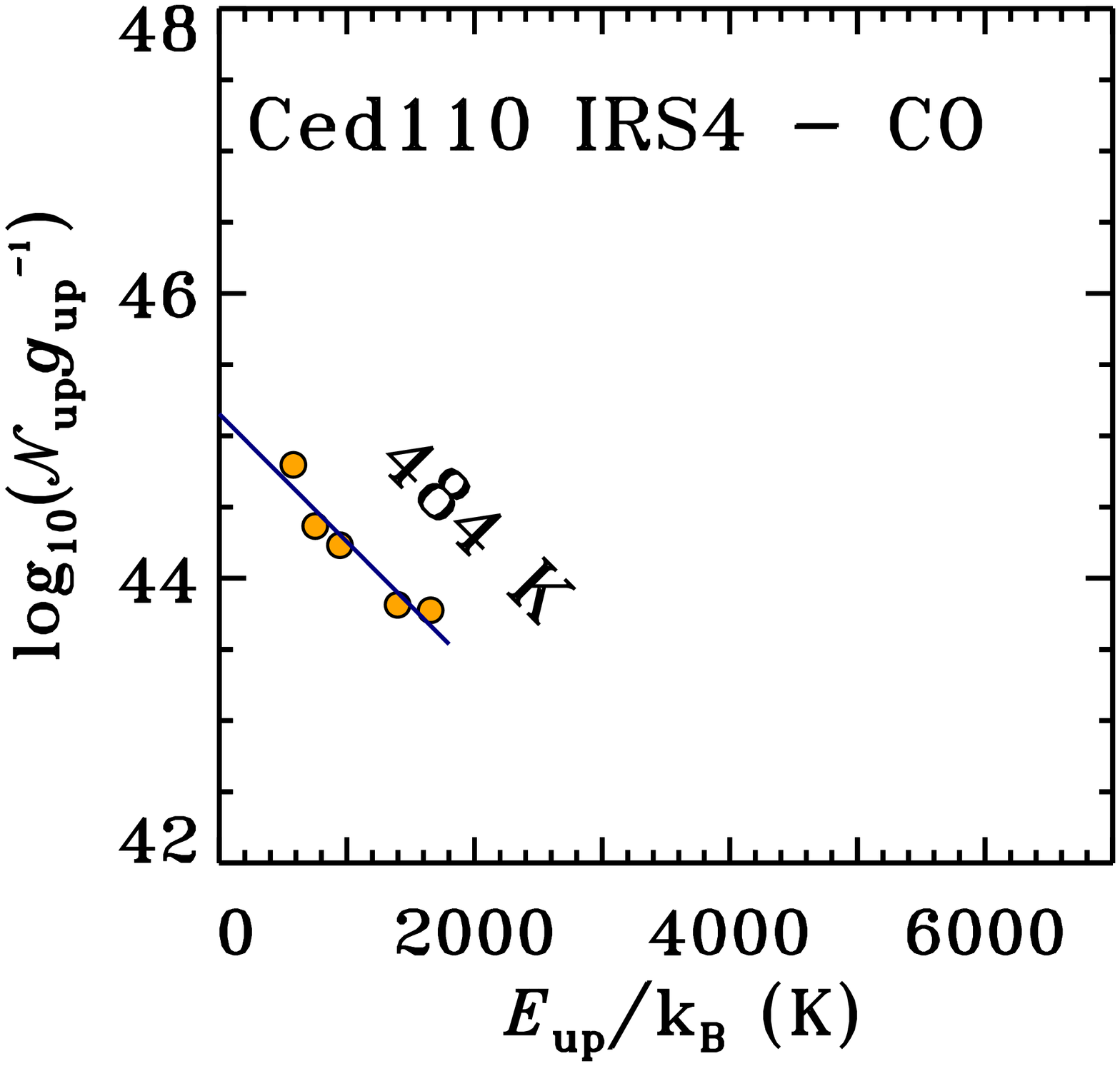} 
        \includegraphics[angle=90,height=4.8cm]{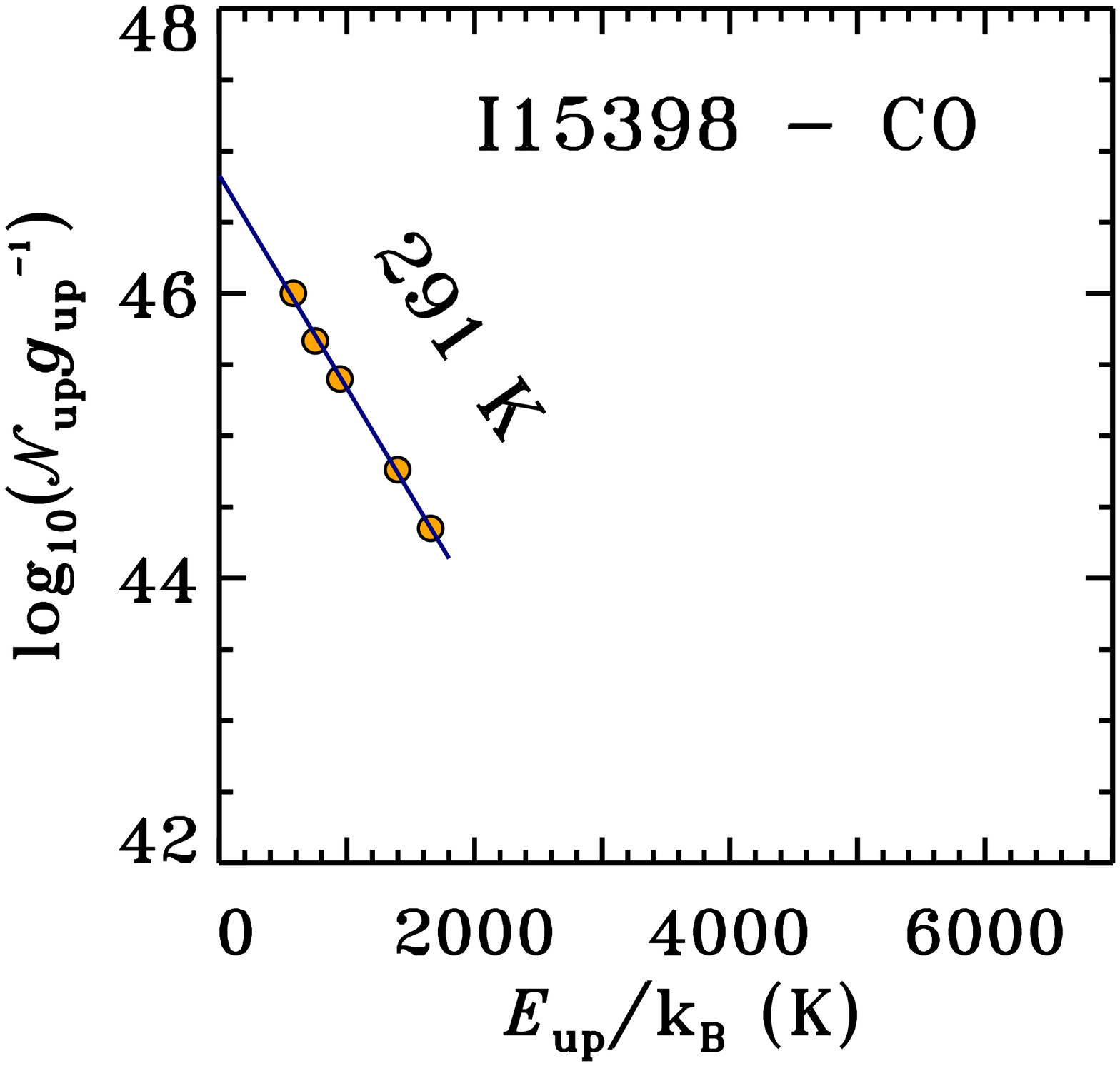} 
  \end{center}
  \end{minipage}
  \hfill
  \begin{minipage}[t]{.3\textwidth}
      \begin{center}
   	   \includegraphics[angle=90,height=4.8cm]{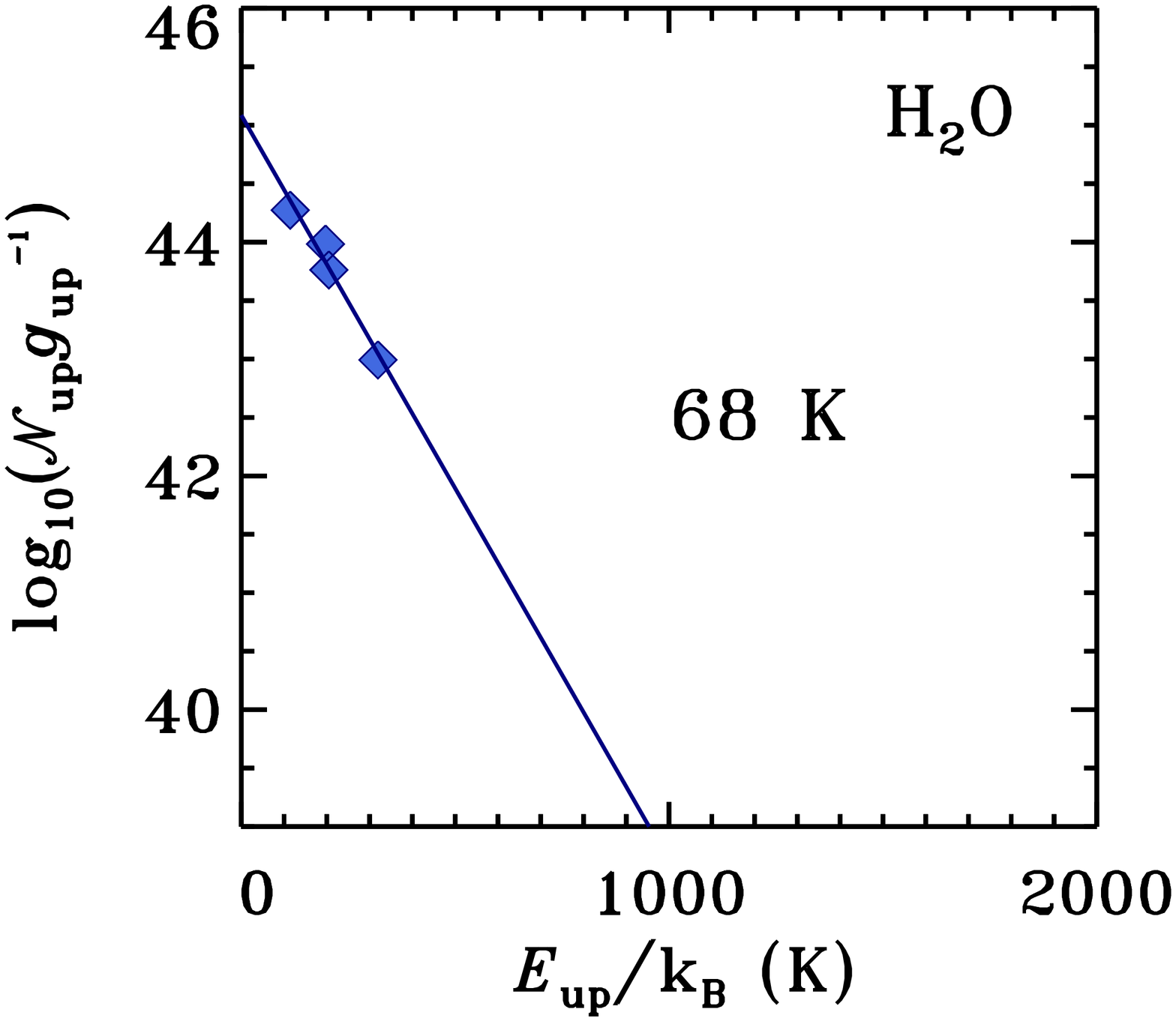} 
       \includegraphics[angle=90,height=4.8cm]{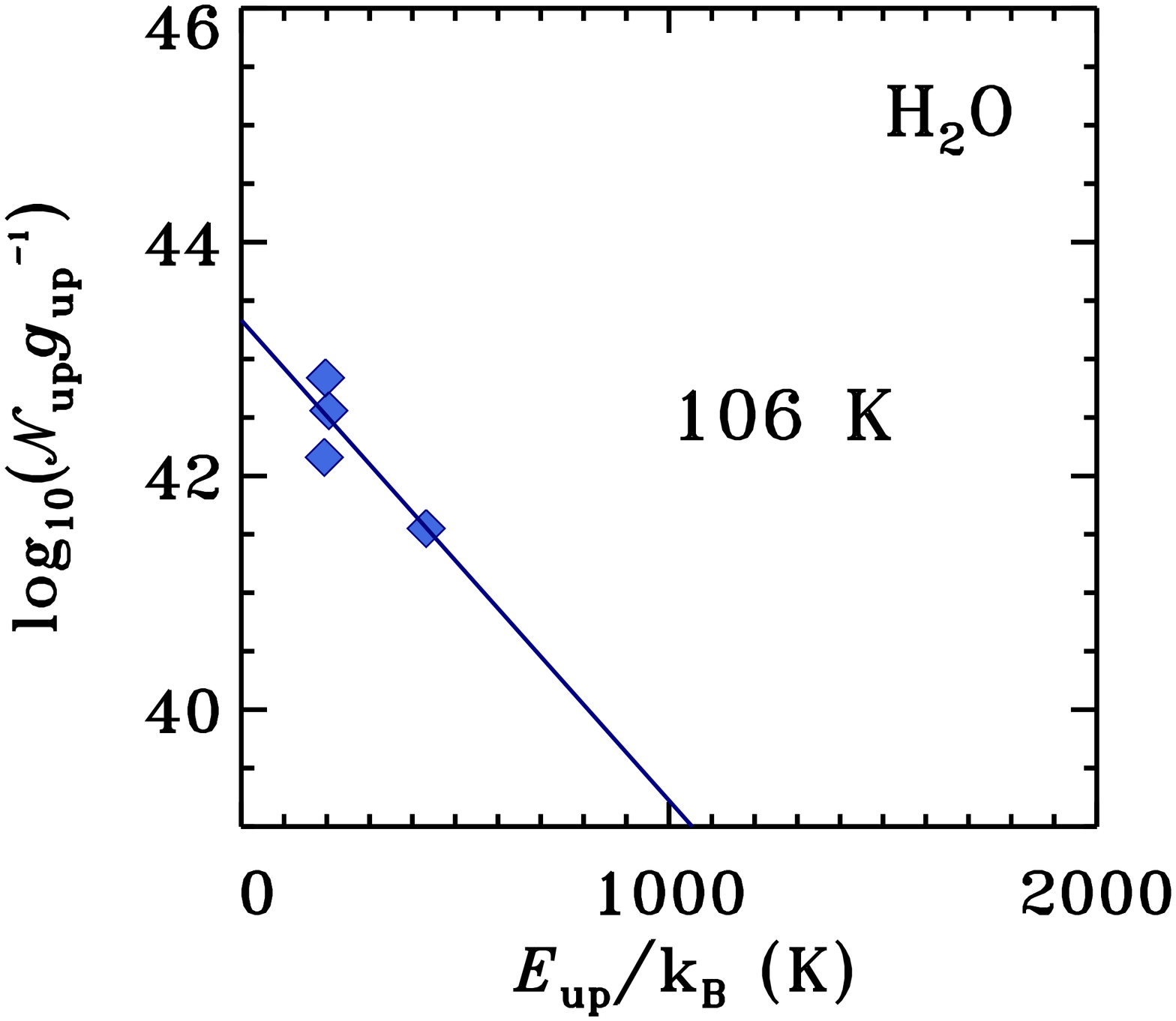} 
       \includegraphics[angle=90,height=4.8cm]{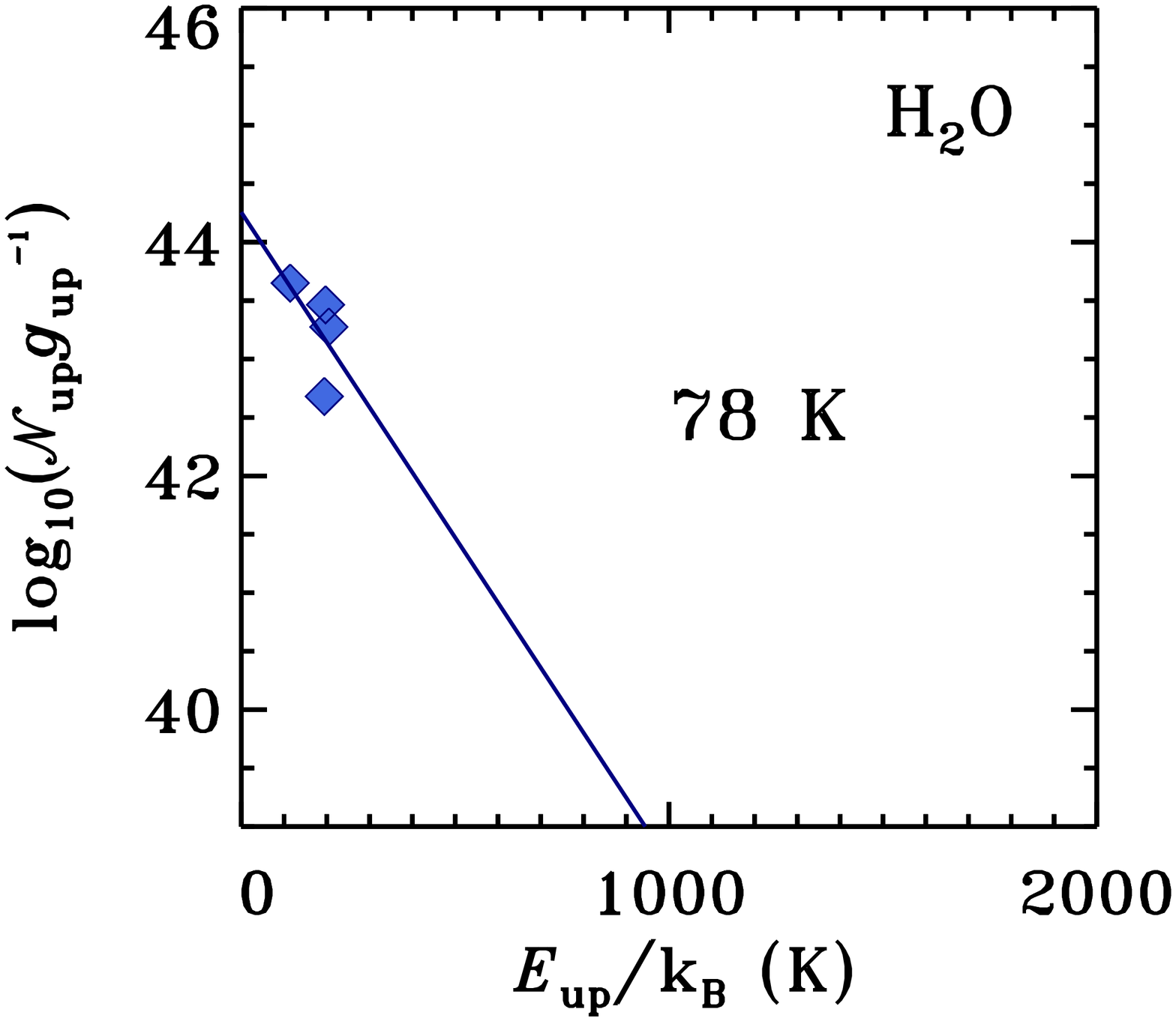}             
      \end{center}
  \end{minipage}
    \hfill
   \begin{minipage}[t]{.3\textwidth}
      \begin{center}
    	\includegraphics[angle=90,height=4.8cm]{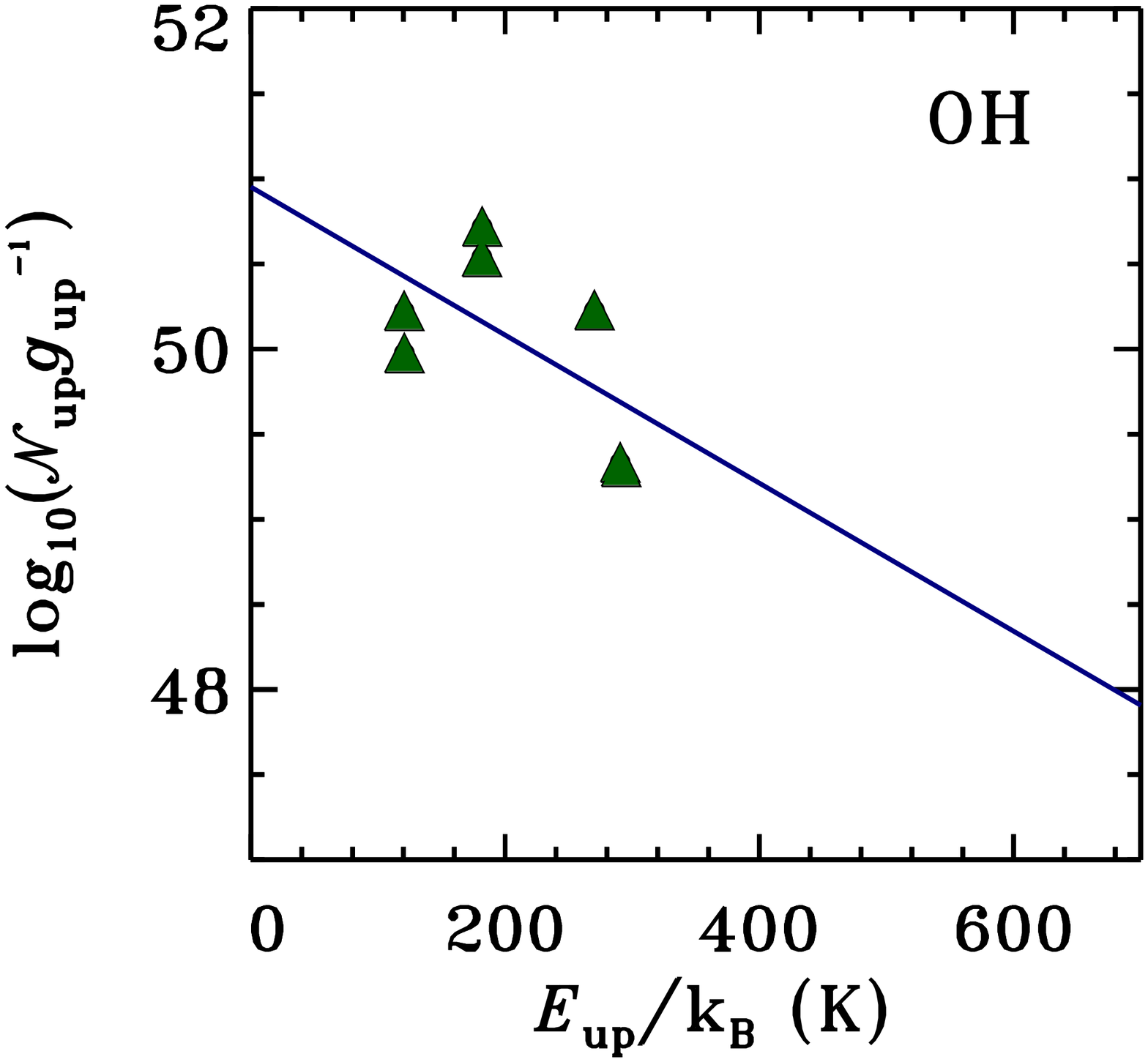} 
        \includegraphics[angle=90,height=4.8cm]{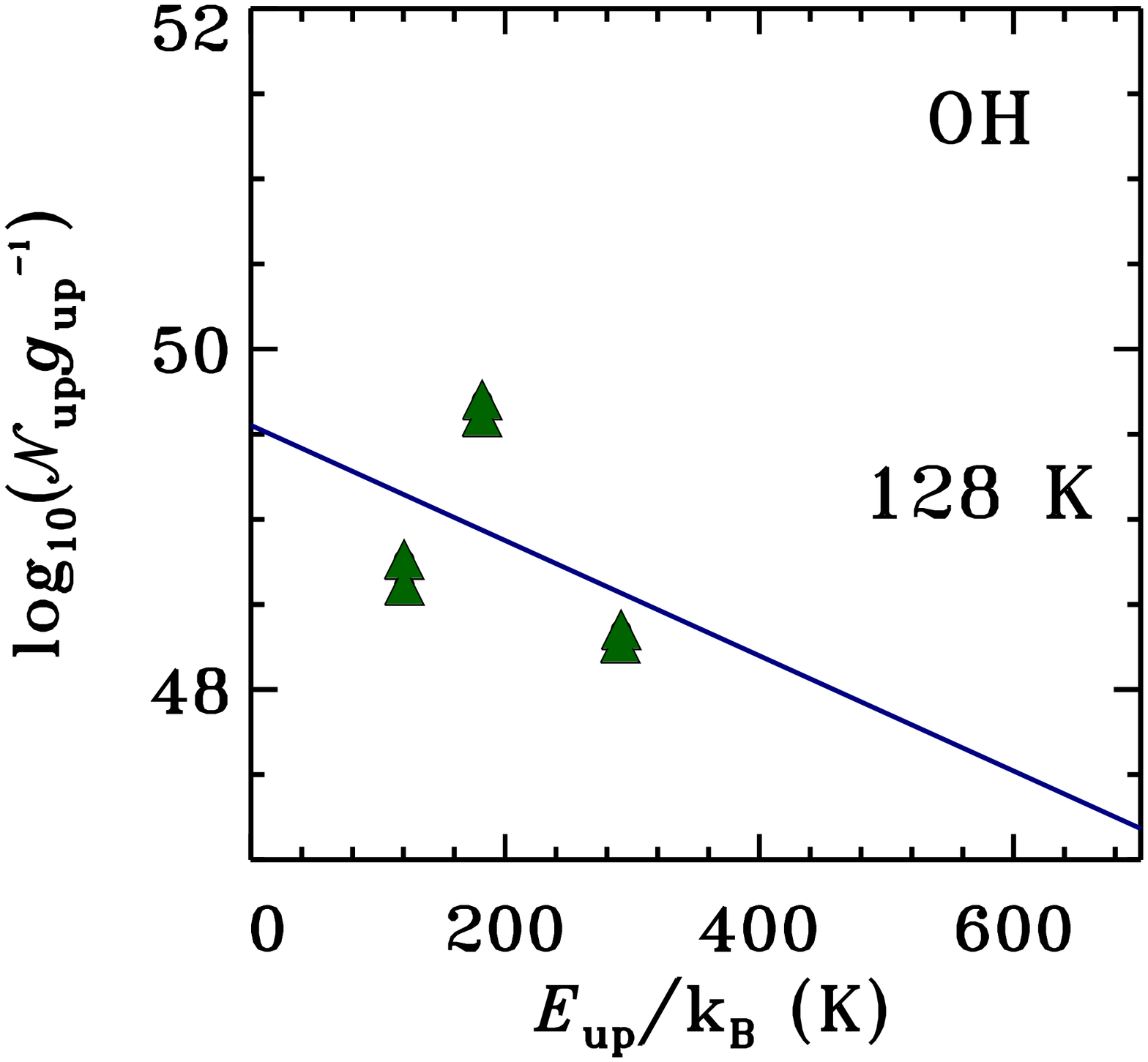} 
        \includegraphics[angle=90,height=4.8cm]{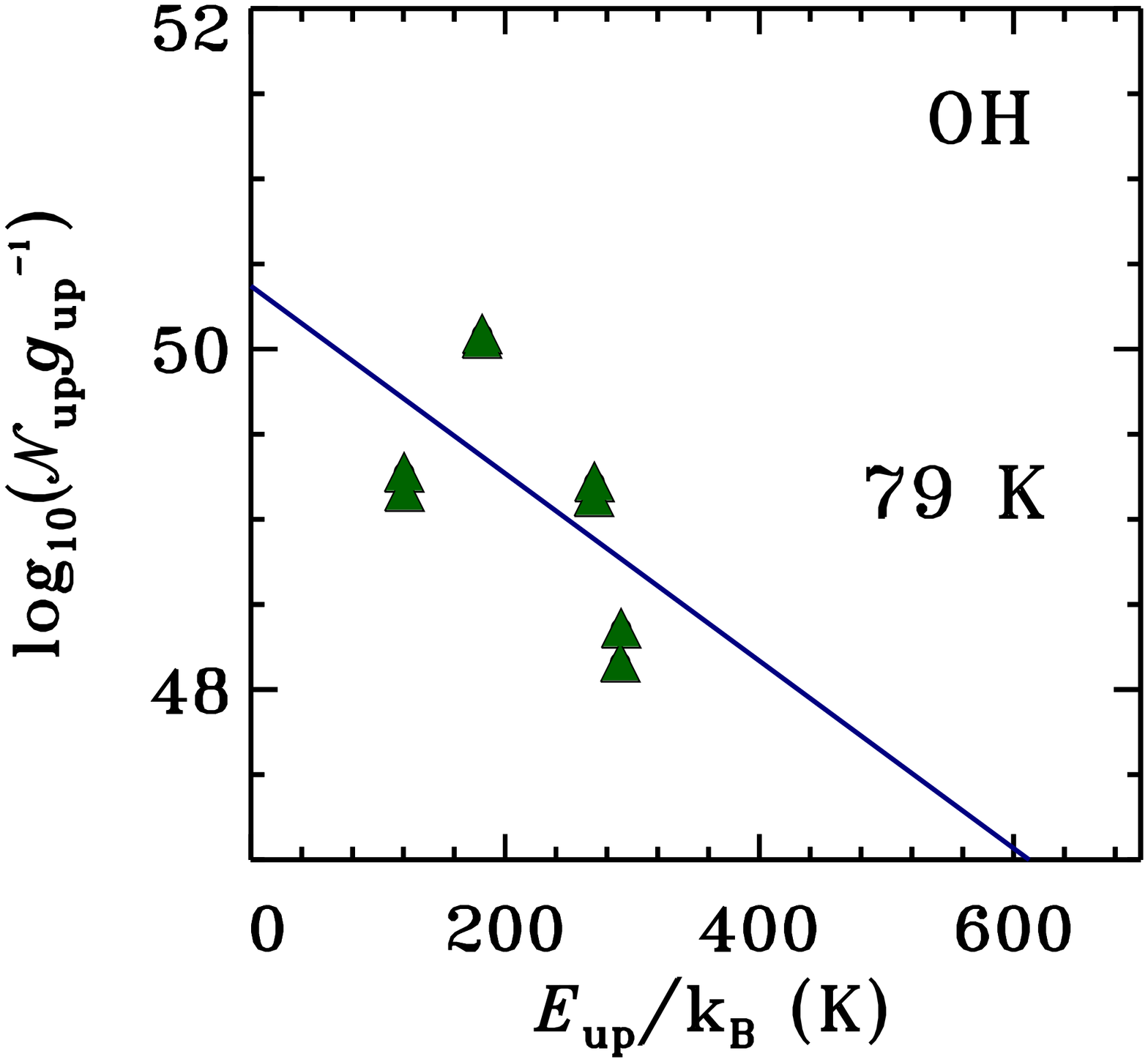} 
                    
      \end{center}
  \end{minipage}
      \hfill
        \caption{\label{wh1} Similar to Figure \ref{molexc}, but for 
        line scan observations of the WISH sources: HH46, Ced110 IRS4, 
        and L723 (see also \citealt{Ka13}).}
\end{figure*}
\clearpage
\renewcommand{\thefigure}{\thesection.\arabic{figure} (Cont.)}
\addtocounter{figure}{-1}   
\begin{figure*}[!tb]
  \begin{minipage}[t]{.3\textwidth}
  \begin{center}
       \includegraphics[angle=90,height=4.8cm]{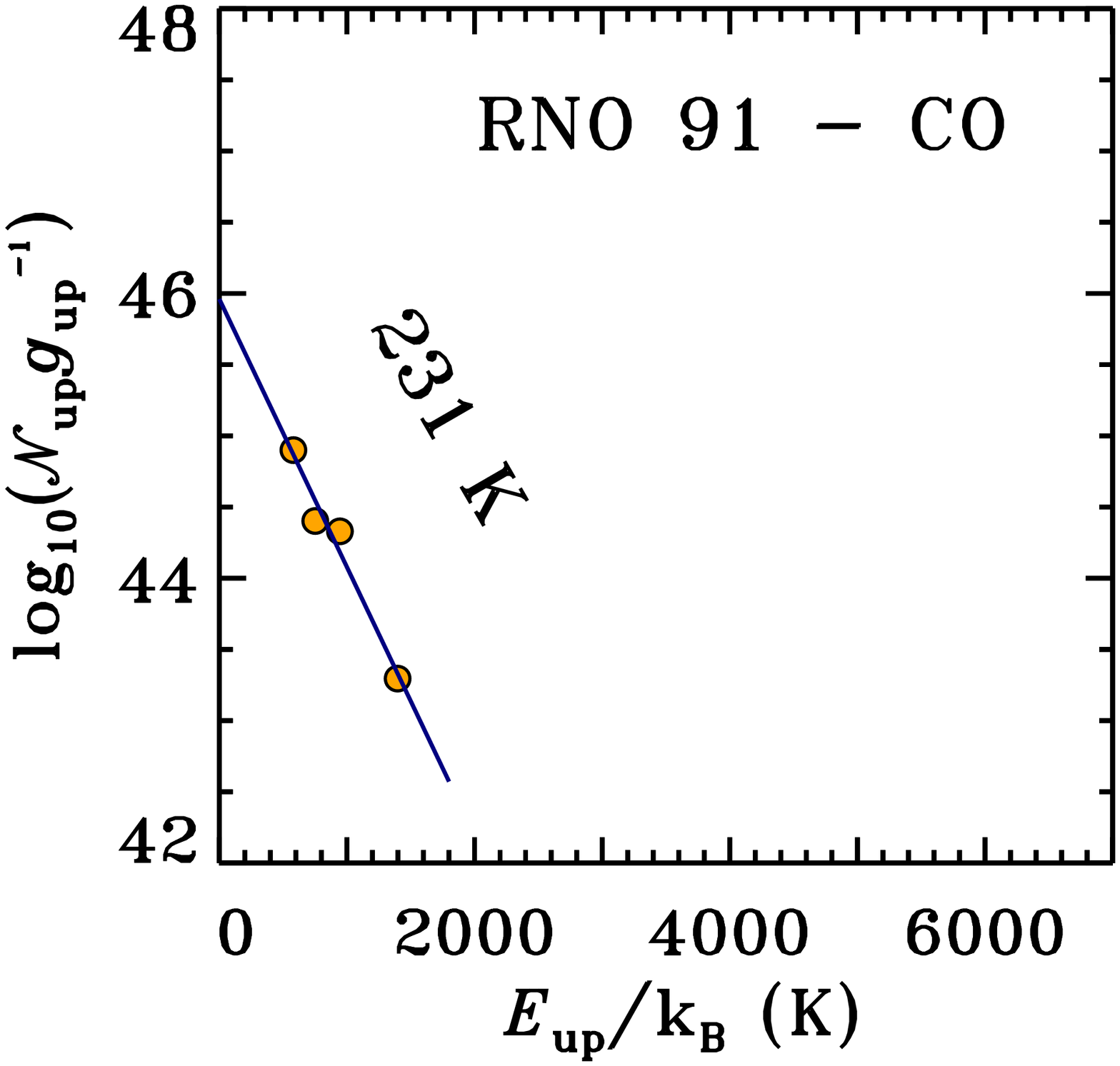} 
       \includegraphics[angle=90,height=4.8cm]{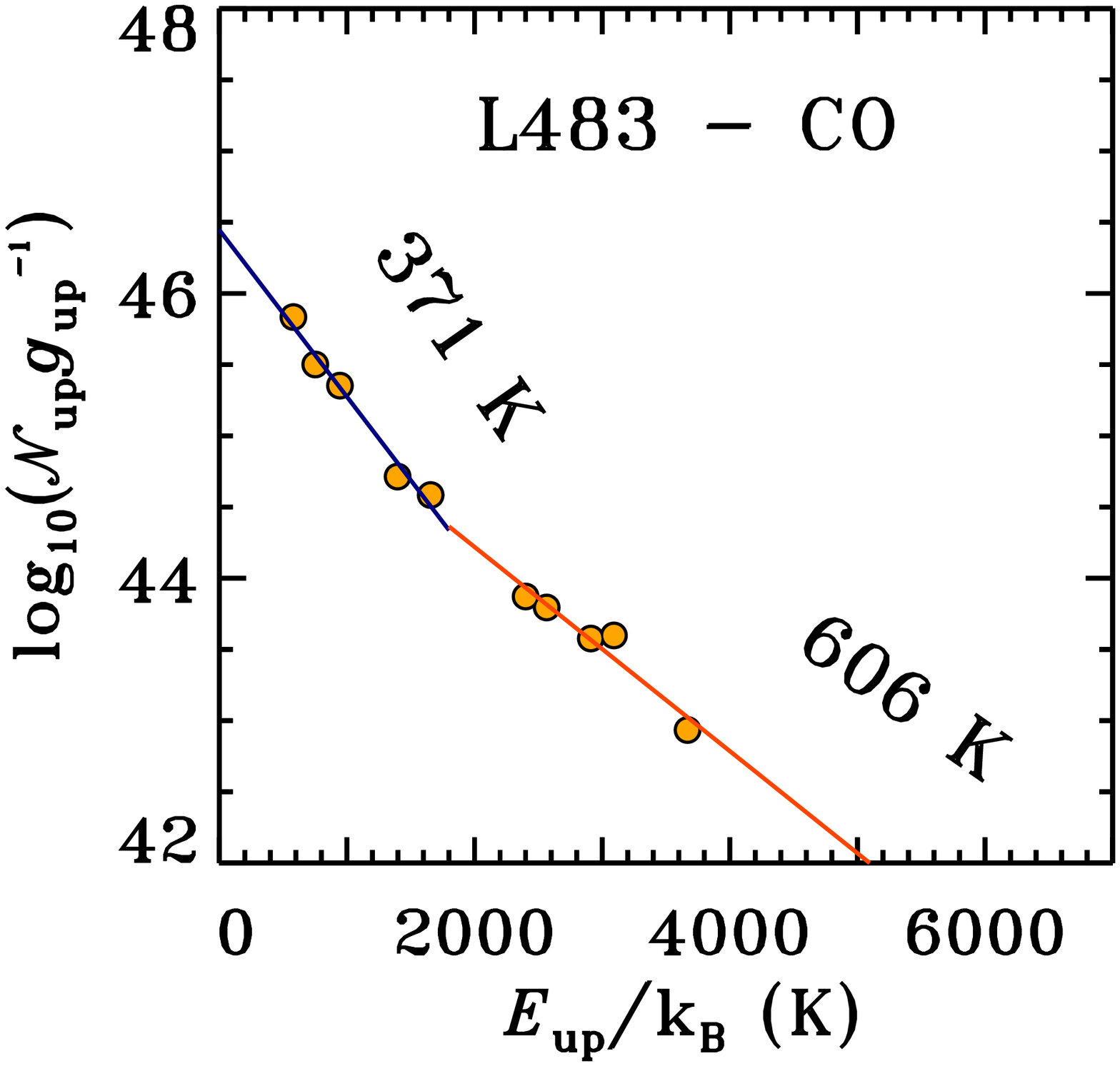} 
        \includegraphics[angle=90,height=4.8cm]{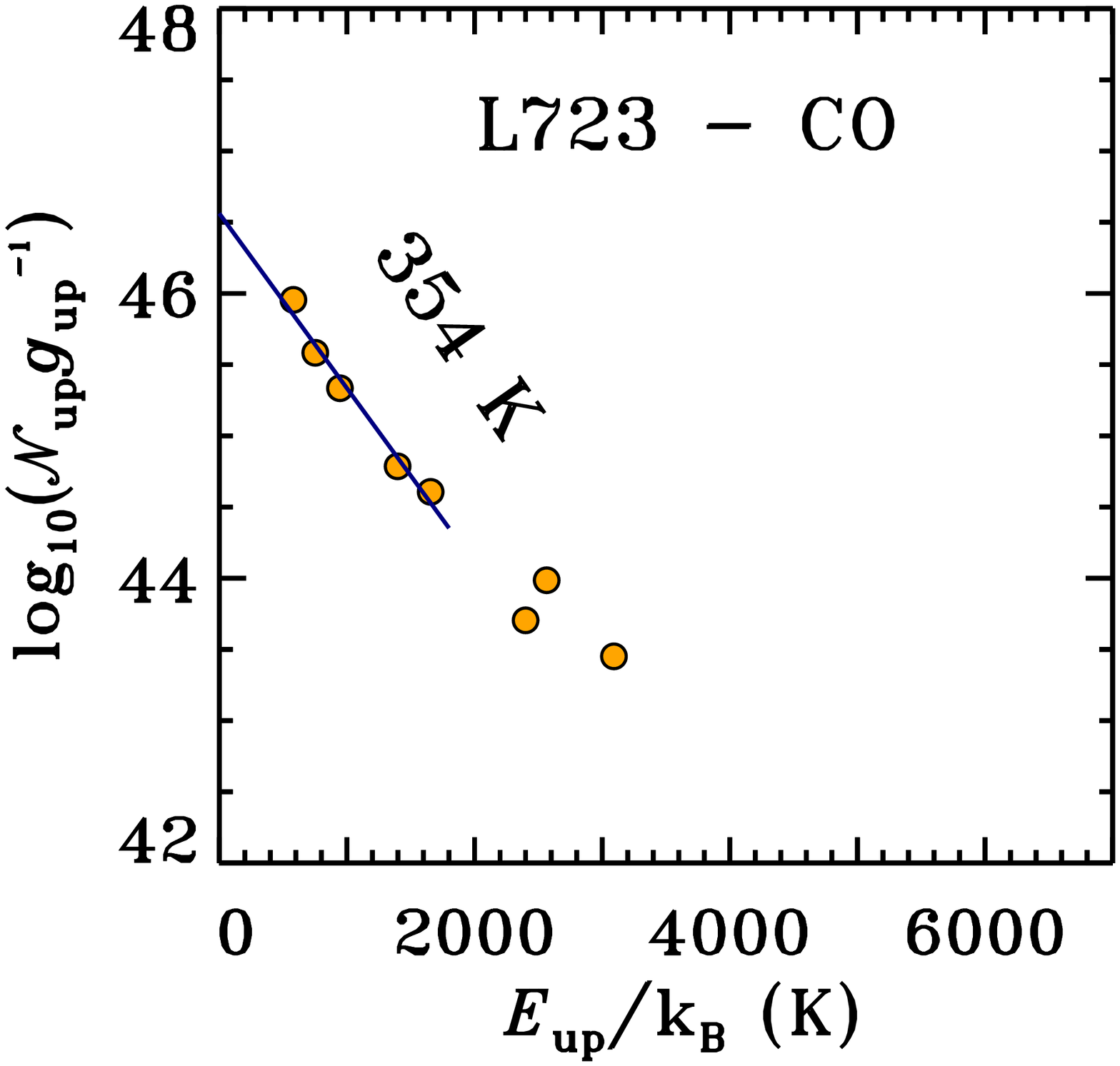} 
  \end{center}
  \end{minipage}
  \hfill
  \begin{minipage}[t]{.3\textwidth}
      \begin{center}
   	   \includegraphics[angle=90,height=4.8cm]{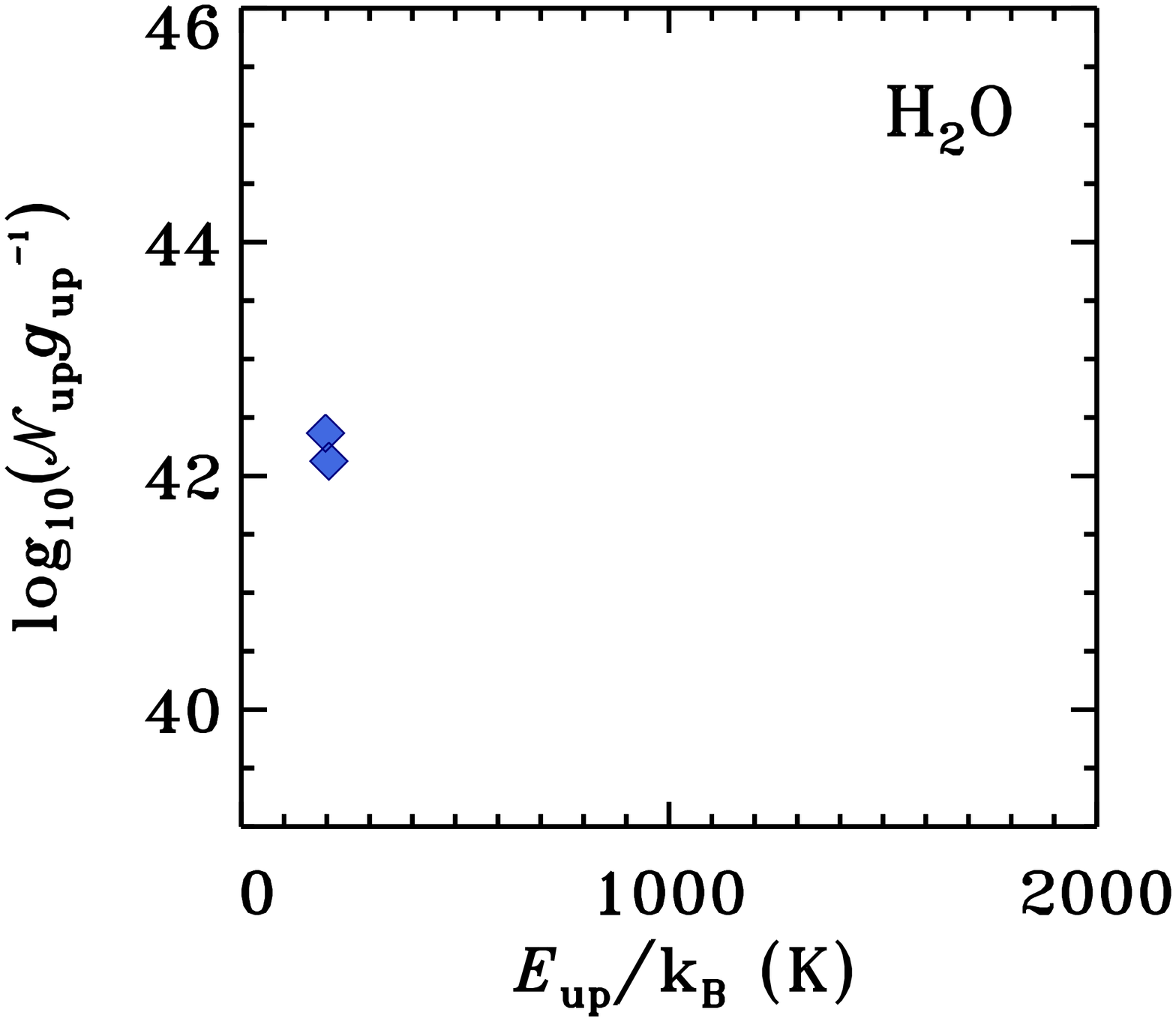} 
       \includegraphics[angle=90,height=4.8cm]{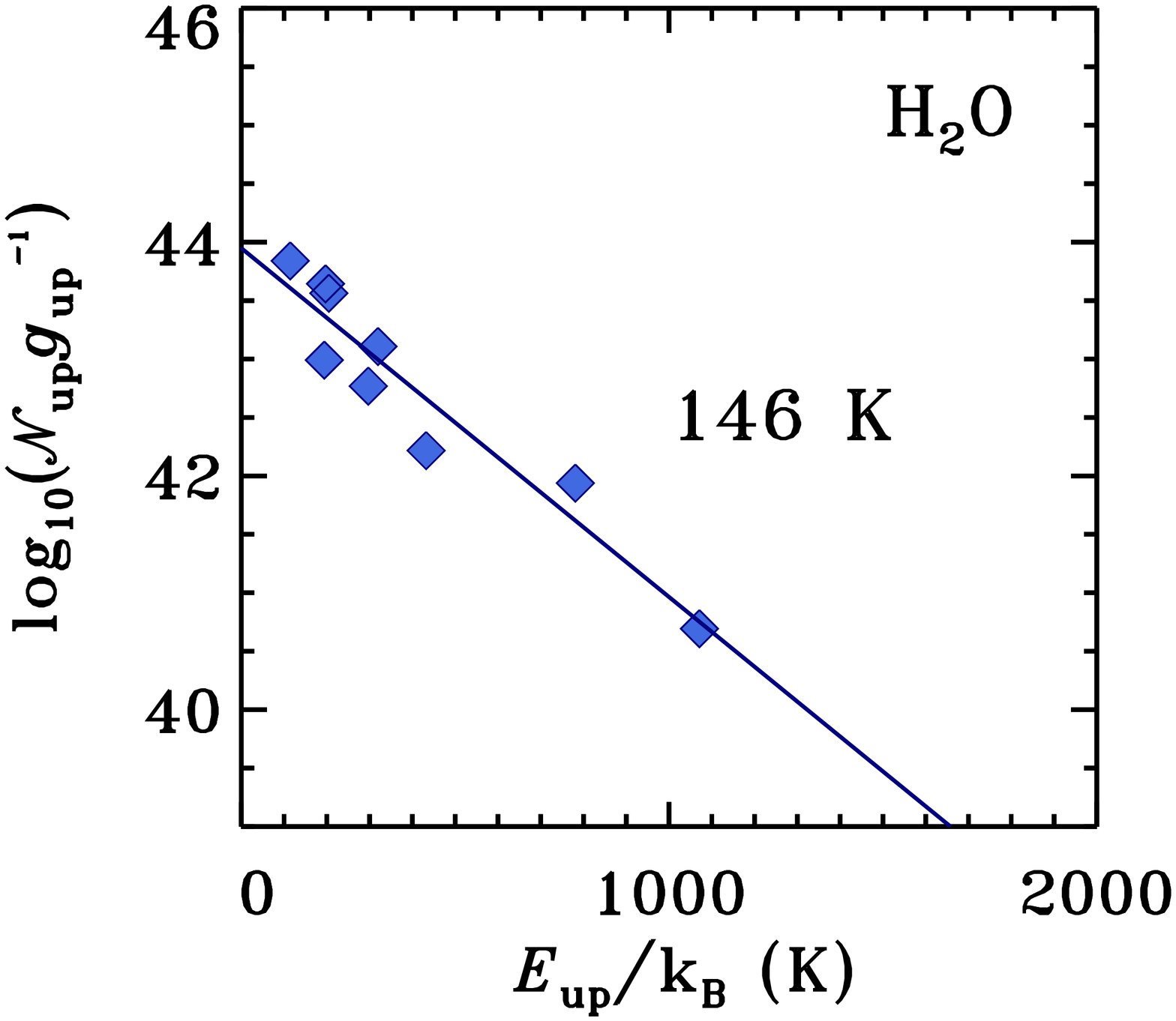} 
       \includegraphics[angle=90,height=4.8cm]{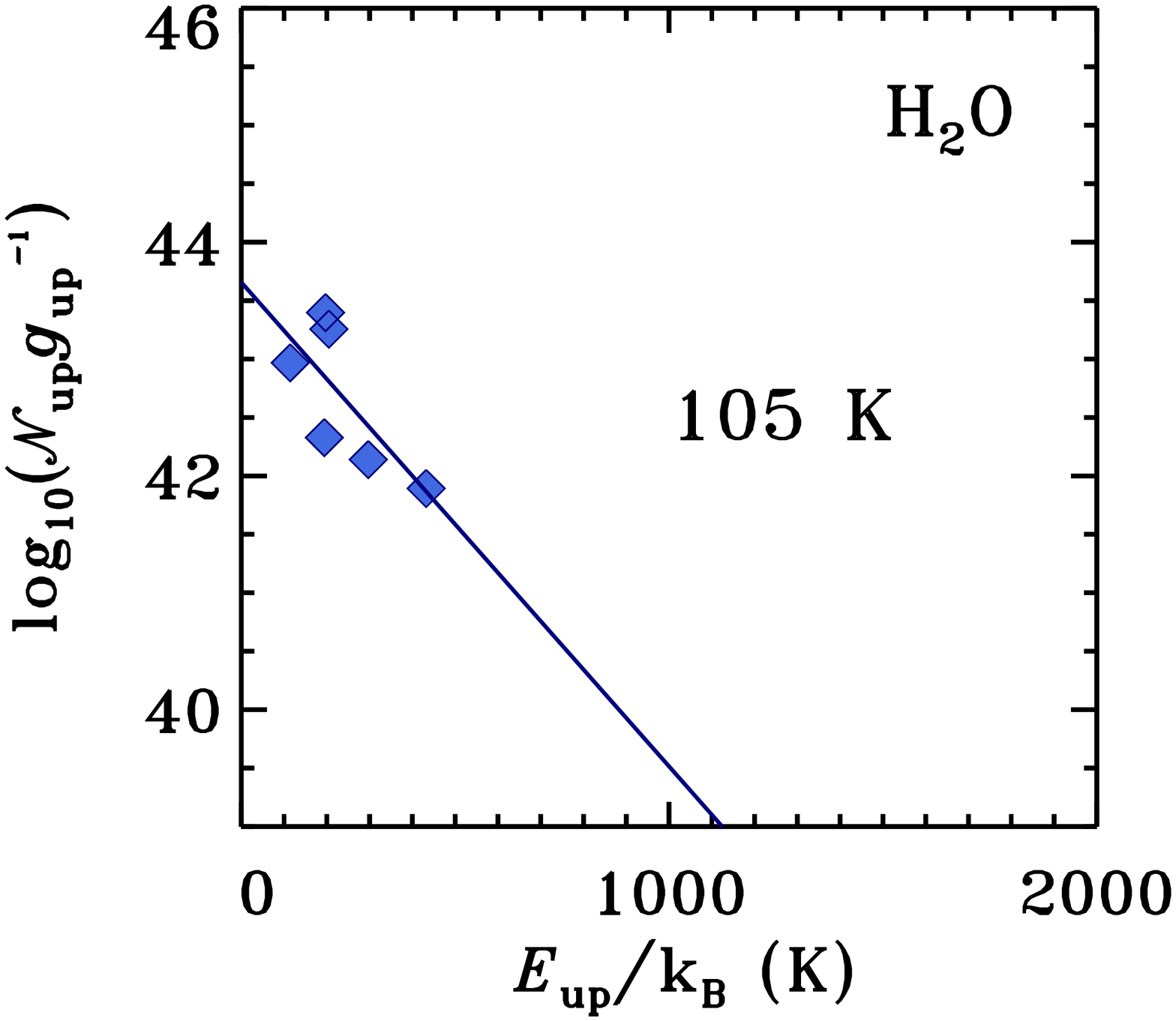}             
      \end{center}
  \end{minipage}
    \hfill
   \begin{minipage}[t]{.3\textwidth}
      \begin{center}
    	\includegraphics[angle=90,height=4.8cm]{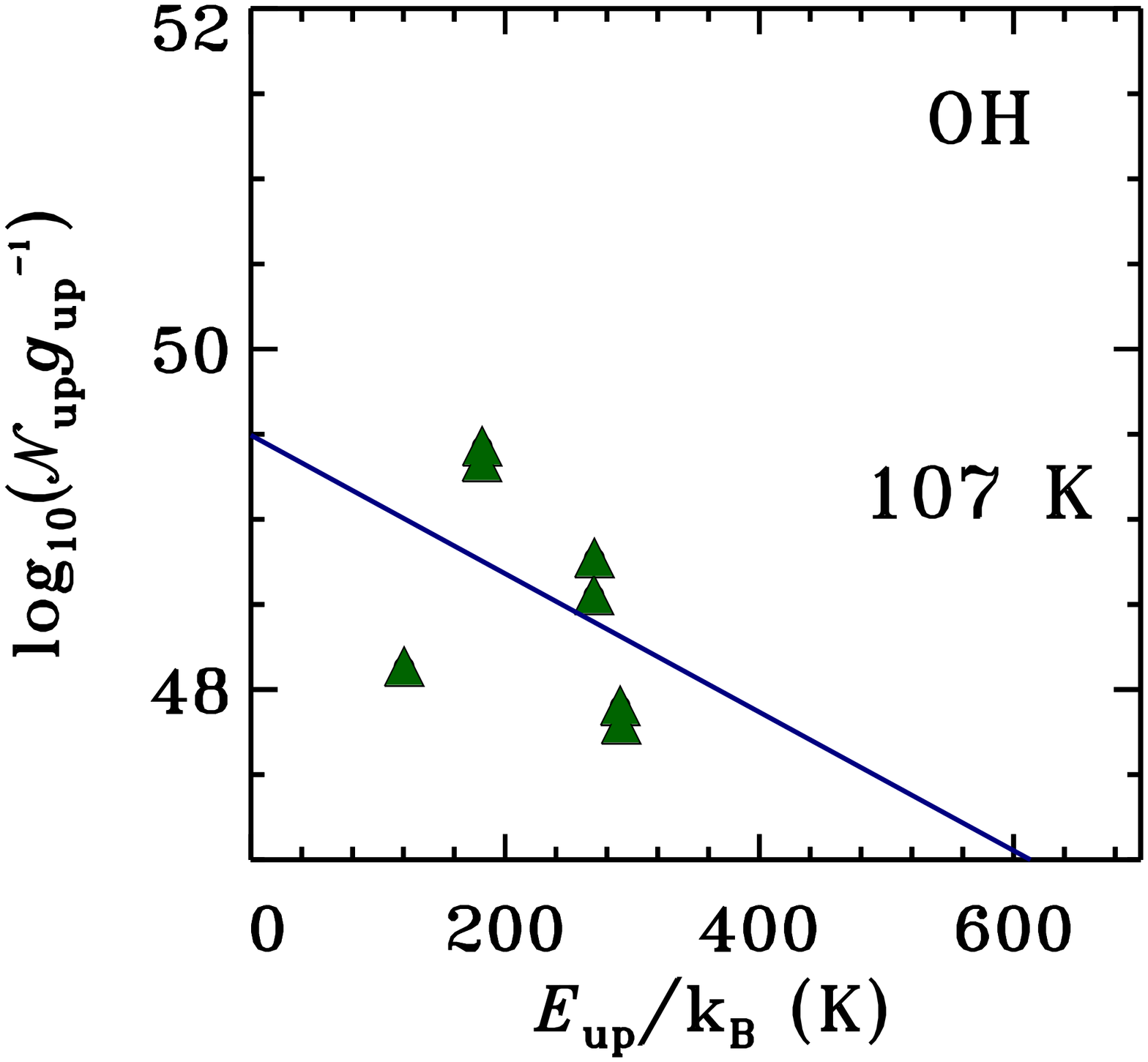} 
        \includegraphics[angle=90,height=4.8cm]{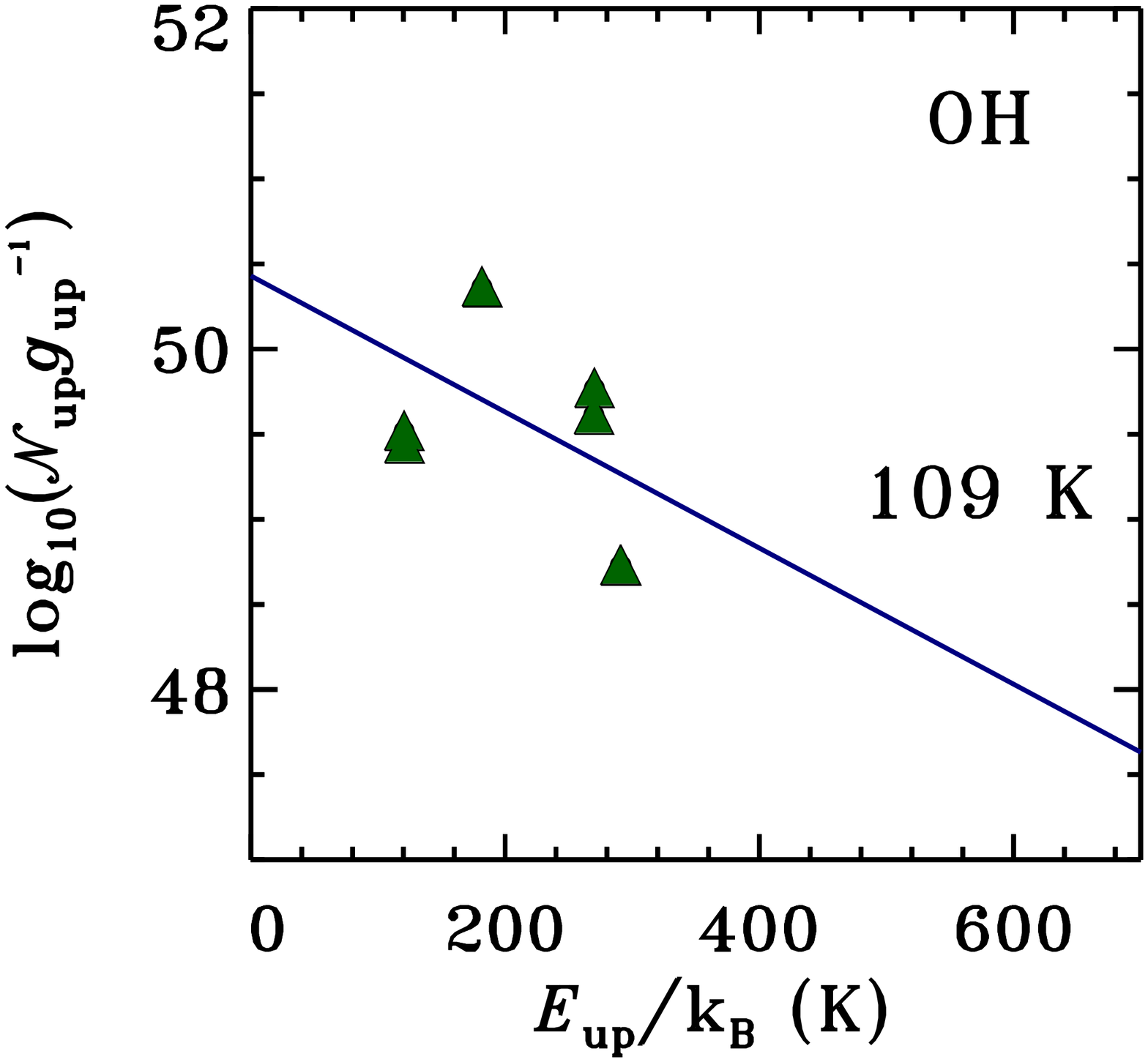} 
        \includegraphics[angle=90,height=4.8cm]{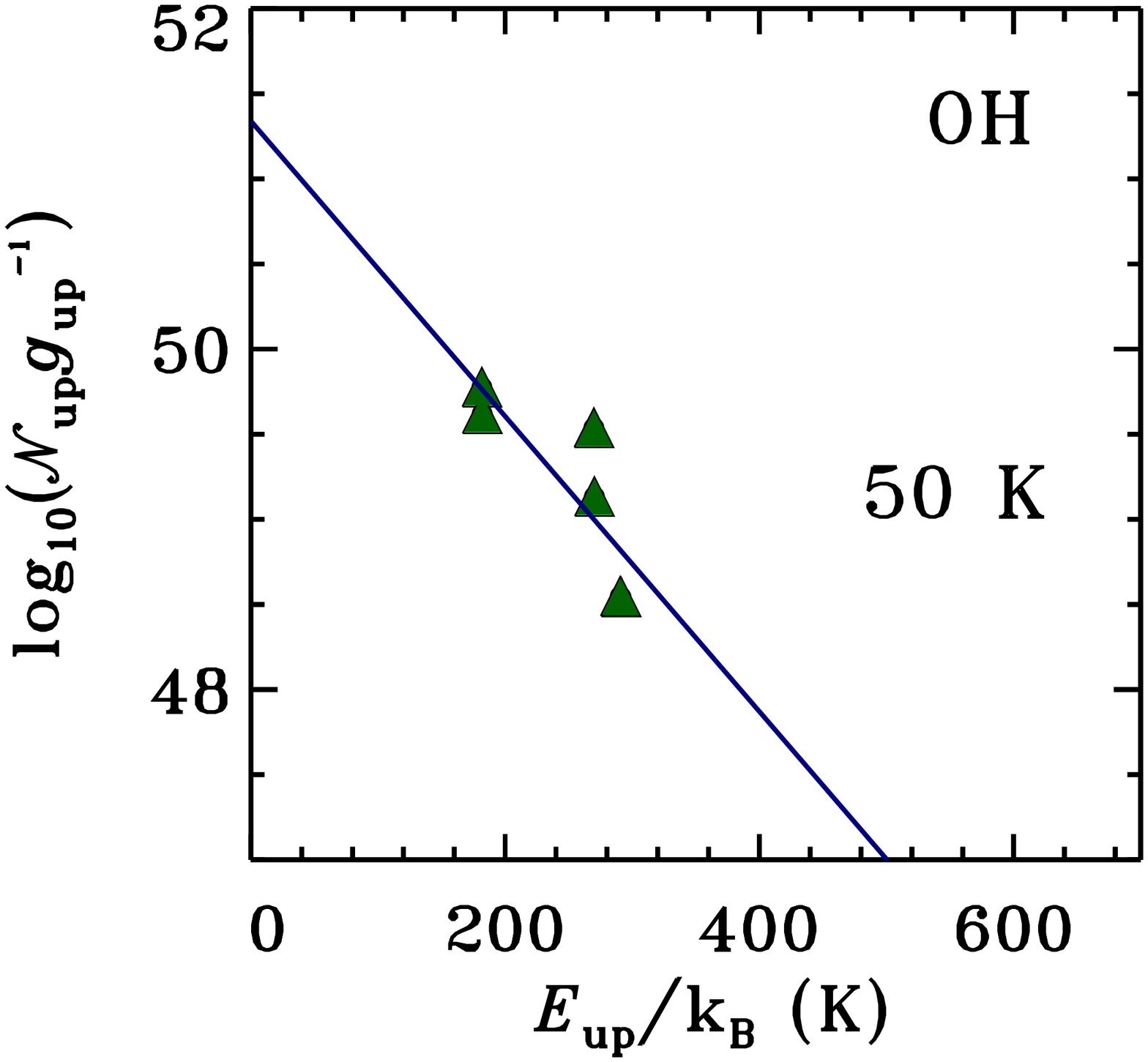} 
                    
      \end{center}
  \end{minipage}
      \hfill
        \caption{\label{wh2} Similar to Figure \ref{molexc}, but for 
        line scan observations of the WISH sources: RNO 91, L483, and L723 (see also \citealt{Ka13}).}
\end{figure*}
\renewcommand{\thefigure}{\thesection.\arabic{figure}}

\begin{deluxetable*}{llccccccccc} 
\tabletypesize{\scriptsize}
\tablecaption{\label{exc_table2a} Comparison of H$_2$O rotational temperatures and 
numbers of emitting molecules for full spectra, and WISH and WILL line scan modes}
\tablehead{
\colhead{ID}           &
\colhead{Source}   &
\multicolumn{2}{c}{Full spectra} &
\multicolumn{2}{c}{WISH line scan} &
\multicolumn{2}{c}{WILL line scan}  \\ \cline{3-4}  \cline{5-6} \cline{7-8} 
\colhead{}   &        
\colhead{}   &
\colhead{$T_\mathrm{rot}$(K)} &
\colhead{$\mathrm{log}_\mathrm{10}\mathcal{N}$}                                          &               
\colhead{$T_\mathrm{rot}$(K)}   &            
\colhead{$\mathrm{log}_\mathrm{10}\mathcal{N}$}  & 
\colhead{$T_\mathrm{rot}$(K)}    & 
\colhead{$\mathrm{log}_\mathrm{10}\mathcal{N}$}    
}
\startdata
3 & L1448 C(N)  &  152(18) &  46.3(0.2) & 151(19) &	46.4(0.2) &	176(24) &		46.2(0.2) \\
4 & L1448 C(S)  &  84(11) & 46.3(0.3) &  89(14) &	46.3(0.3) &	 83(16) &		46.1(0.4) \\		
6 & L1455 IRS1	& 124(30) & 45.6(0.3) &  97(24) &	45.6(0.3) &	132(18) &		45.4(0.2)\\
14 & NGC1333 I4A & 105(18) & 46.4(0.3) & 97(14) &	46.0(0.2) &	97(21) &		45.7(0.4) \\
16 & NGC1333 I4B & 183(15) & 46.5(0.1) & 145(16) &	46.7(0.2) &	153(13) &		46.5(0.2) \\
22 & IRAS03301	&  176(22) & 45.4(0.2) & 162(18) &	45.4(0.2) &	179(25) &		45.4(0.2) \\
25 & B1 a	& 211(54) & 45.3(0.3) & 169(47) &	45.6(0.4) &	186(49) &		45.3(0.3)  \\
31 & L1489  & 191(22) & 45.1(0.2) &  172(22) &	45.2(0.2) &	193(29) &		45.1(0.2)\\
42 & TMR1 & 193(31) & 45.4(0.2) & 211(41) &	45.4(0.2) &	239(66) &		45.4(0.3) \\
44 & TMC1 & 148(24) & 44.9(0.2) & 143(28) &	44.9(0.3) &	174(46) &		44.8(0.3) \\
48 & BHR71 & 125(23) & 45.8(0.3)  & 138(33) &	45.9(0.4) &	146(38) &		45.6(0.4) \\
52 & GSS30 IRS1	&	245(22) & 46.0(0.1) & 217(32) &	46.1(0.2) &	201(23) &		46.0(0.2) \\
53 & VLA 1623	&	165(40) & 45.3(0.3) & 96(15) &	45.5(0.3) &	106(28) &		45.4(0.4)\\
54 & WL12	&	202(25) & 45.0(0.2) & 149(20) &	45.2(0.2) &	187(22) &		45.0(0.2) 	\\
56 & Elias29 & 288(33) & 46.0(0.1) & 236(35) &	46.0(0.2) &	231(33) &		45.9(0.2)  \\
66 & Ser SMM1 &  103(12) & 47.3(0.2) & 107(9) &	47.4(0.2) &	116(12) &		47.2(0.2) \\
82 & RCrA IRS5A	& 219(30) & 45.1(0.2) & 202(33) &	45.2(0.2) &	222(48) &		45.1(0.3)\\
84 & RCrA IRS7A	& 218(24) & 46.0(0.2) & 183(18) &	46.1(0.2) &	209(27) &		45.9(0.2)\\
85 & RCrA IRS7B	 & 208(28) &	45.3(0.2) & 177(23) &	45.5(0.2) &	198(36) &		45.3(0.3) \\
88 & B335 & 140(40) & 44.7(0.4) & 159(51) &	44.8(0.3) &	111(36) &		44.7(0.5) \\
89 & L1157	&  114(20) & 46.3(0.3) & 101(15) &	46.2(0.2) &	99(23) &		46.1(0.4)\\
\enddata
\end{deluxetable*}

\begin{deluxetable*}{llccccccccc} 
\tabletypesize{\scriptsize}
\tablecaption{\label{exc_table2b} Comparison of OH rotational temperatures and 
numbers of emitting molecules for full spectra, and WISH and WILL line scan modes}
\tablehead{
\colhead{ID}           &
\colhead{Source}   &
\multicolumn{2}{c}{Full spectra} &
\multicolumn{2}{c}{WISH line scan} &
\multicolumn{2}{c}{WILL line scan}  \\ \cline{3-4}  \cline{5-6} \cline{7-8} 
\colhead{}   &        
\colhead{}   &
\colhead{$T_\mathrm{rot}$(K)} &
\colhead{$\mathrm{log}_\mathrm{10}\mathcal{N}$}                                          &               
\colhead{$T_\mathrm{rot}$(K)}   &            
\colhead{$\mathrm{log}_\mathrm{10}\mathcal{N}$}  & 
\colhead{$T_\mathrm{rot}$(K)}    & 
\colhead{$\mathrm{log}_\mathrm{10}\mathcal{N}$}    
}
\startdata
3 & L1448 C(N)  &  123(17) & 52.2(0.2) &  113(91) &		52.3(0.7) &	33(6) &	54.4(0.6) \\
4 & L1448 C(S)  &  94(18) & 51.7(0.3) & 54(16) &		52.3(0.6) &	29(4) &	54.1(0.6) \\		
6 & L1455 IRS1	& 37(15) &	53.2(1.2) & 37(15) &		53.4(1.2) &	37(15) &	53.4(1.2)  \\
14 & NGC1333 I4A & 150(214) & 51.6(0.9) & 150(214) &	51.6(0.9) &	35(17) &	53.8(1.5) \\
16 & NGC1333 I4B &  84(14) & 52.8(0.3) & 109(84) &		52.5(0.7) &	35(9) &	54.5(0.8) \\
22 & IRAS03301	&  109(20) & 52.1(0.3) & 181(208) &	51.9(0.6) &	43(13) &	53.5(0.8) \\
25 & B1 a	& 103(20) & 51.8(0.3) & 163(196) &	51.7(0.7) &	44(21) &	53.3(1.2) \\
31 & L1489 &	102(18) & 51.5(0.3) & 102(70) &		51.5(0.7) &	37(12) &	53.3(0.9)  \\
42 & TMR1 & 108(24) & 51.8(0.3) & 132(141) &	51.7(0.8) &	32(8) &	54.1(0.9) \\
44 & TMC1 & 103(19) & 51.5(0.3) & 103(73) &		51.5(0.7) &	35(10) &	53.4(0.9) \\
48 & BHR71 & 87(17) & 52.1(0.4) & 72(41) &		52.2(0.8) &	29(9) &	54.5(1.1) \\
52 & GSS30 IRS1	&	99(18) & 52.2(0.3) & 110(84) &		52.2(0.7) &	34(8) &	54.2(0.7) \\
53 & VLA 1623	&	71(16) & 51.9(0.4) & 126(117) &	51.6(0.7) &	48(30) &	52.8(1.4) \\
54 & WL12	&	94(17) & 51.5(0.3) & 112(94) &		51.4(0.7) &	34(9) &	53.5(0.9) \\
56 & Elias29 &  112(27) & 52.1(0.3) & 149(167) &	52.1(0.7) &	36(10) &	54.2(0.8)  \\
66 & Ser SMM1 &  83(13) & 53.6(0.3) & 109(79) &		53.4(0.7) &	42(17) &	54.9(1.1) \\
82 & RCrA IRS5A	& 113(17) & 51.9(0.3) & 86(70) &		52.0(1.0) &	30(5) &	54.4(0.6) \\
84 & RCrA IRS7A	& 114(17) & 52.9(0.3) & 90(80) &		52.9(1.0) &	31(11) &	55.1(1.2) \\
85 & RCrA IRS7B	 & 119(17) & 52.0(0.3) & 104(100) &	52.0(0.9) &	33(8) &	54.1(0.8)  \\
88 & B335 & 101(63) & 51.1(0.6) & 101(63) &		51.1(0.6) &	60(42) &	51.7(1.3) \\
89 & L1157	&  77(40) & 52.5(0.7) & 77(40) &		52.5(0.7) &	32(9) &	54.4(0.9) \\
\enddata
\end{deluxetable*}
\section{Line cooling}
\begin{figure*}[!tb]
  \begin{minipage}[t]{.5\textwidth}
  \begin{center}
       \includegraphics[angle=90,height=7cm]{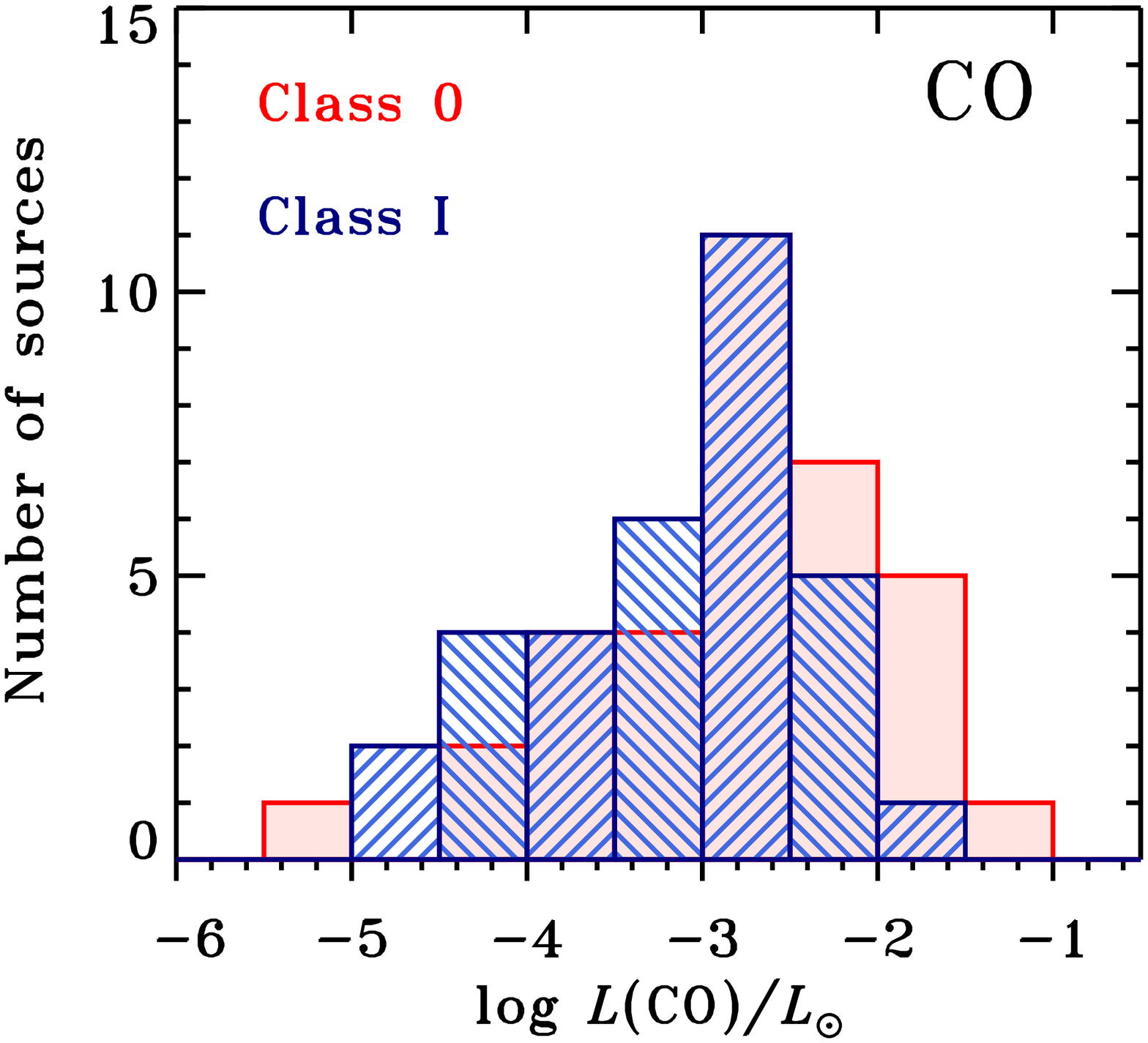}   
       \includegraphics[angle=90,height=7cm]{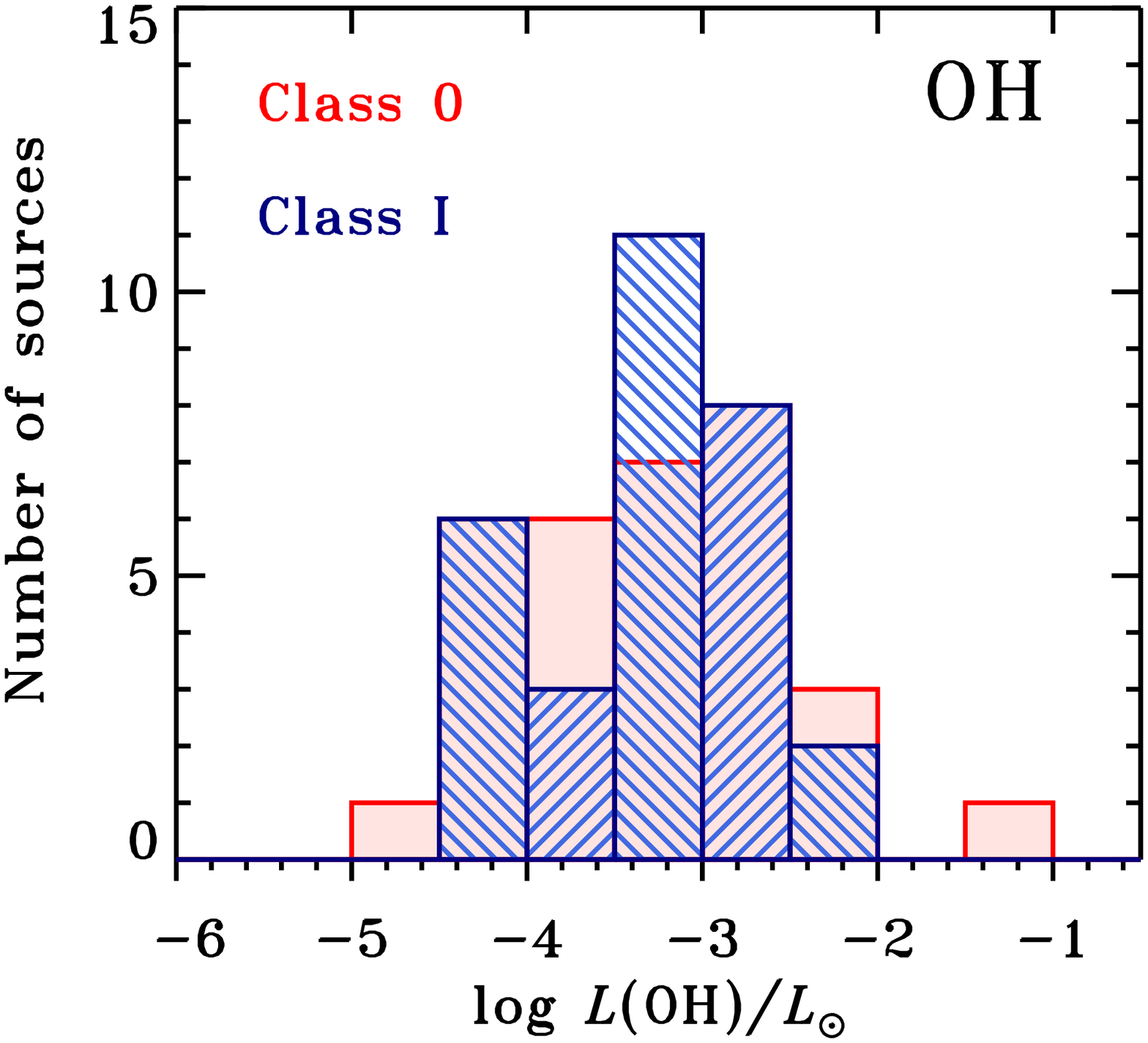}   
  \end{center}
  \end{minipage}
  \hfill
  \begin{minipage}[t]{.5\textwidth}
      \begin{center}
    \includegraphics[angle=90,height=7cm]{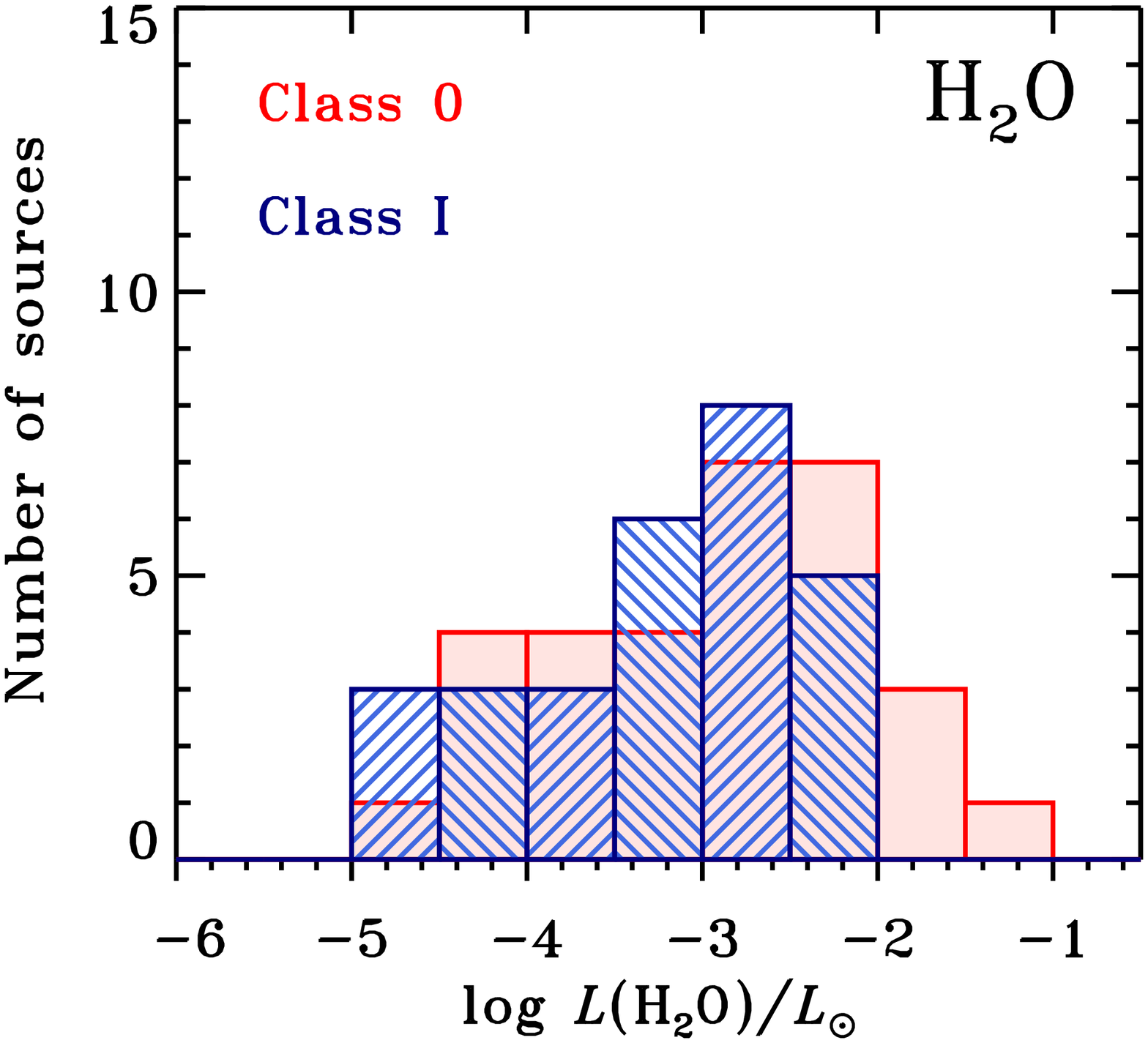} 
    \includegraphics[angle=90,height=7cm]{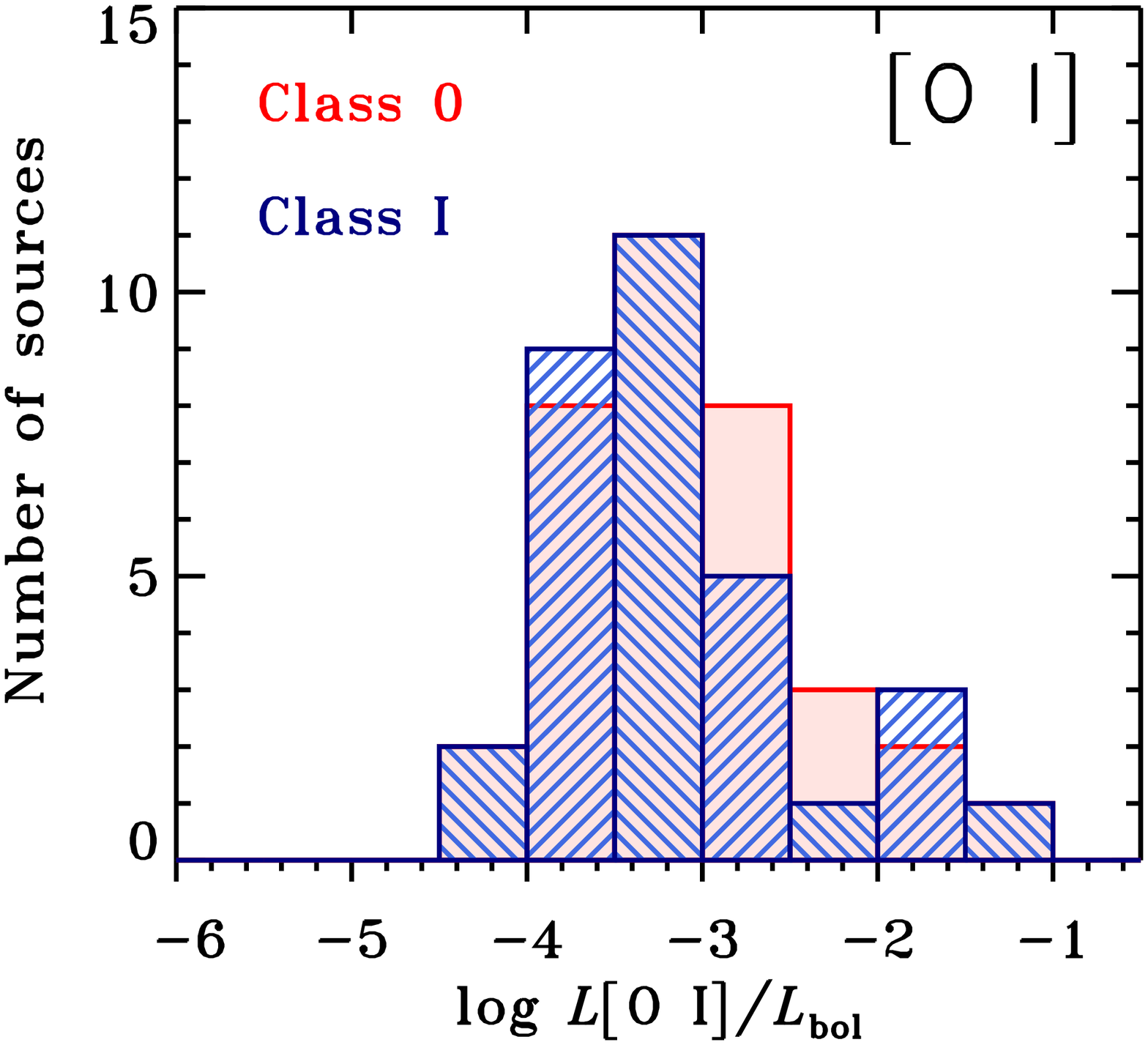} 
      \end{center}
  \end{minipage}
    \hfill
        \caption{\label{hist_new_abs} Histograms of the CO (top left), H$_2$O (top right), 
        OH (bottom left), and [O I] (bottom right) cooling for 
        all sources with at least one detection in a given species.
The red color shows the distributions for Class 0 sources and blue for Class I sources.}
\end{figure*}

The distribution of line luminosities for various species across the entire sample 
is illustrated in Figure \ref{hist_new_abs} (see also Table \ref{lum}). 

\begin{figure*}[!tb]
  \begin{minipage}[t]{.5\textwidth}
      \includegraphics[angle=90,height=6cm]{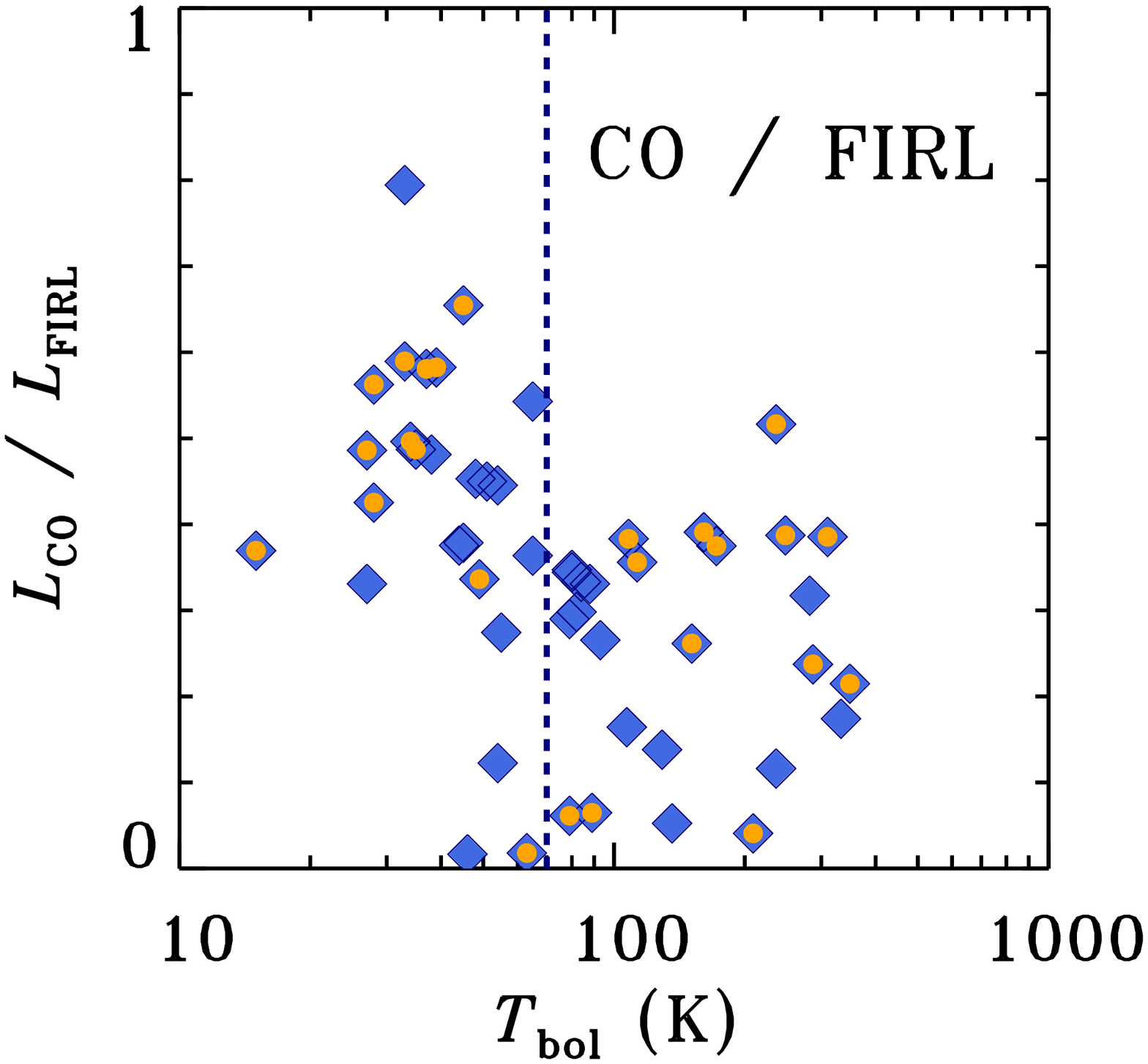}
     \vspace{+0.3cm}
     
        \includegraphics[angle=90,height=6cm]{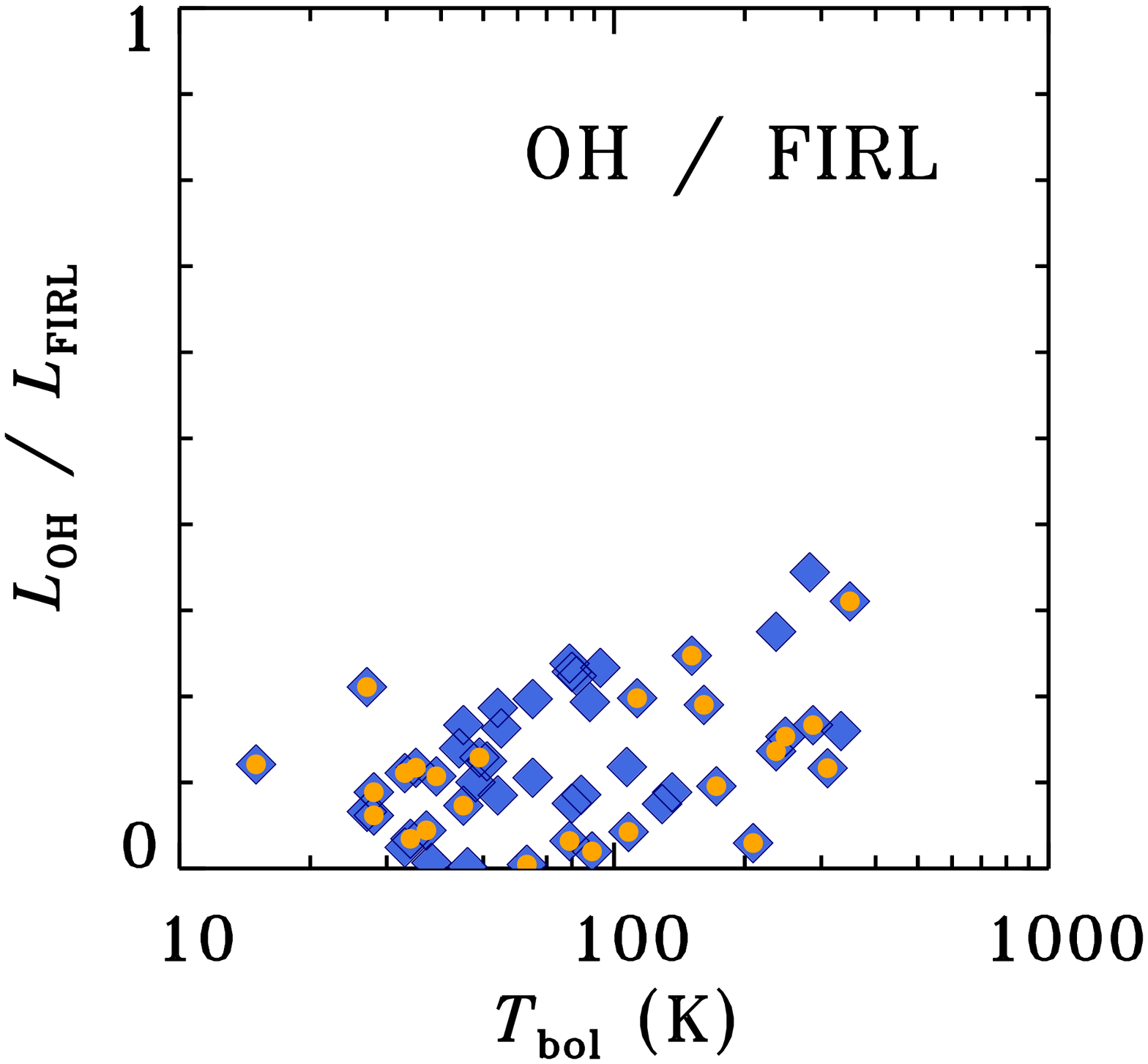}
     \vspace{+0.3cm}
  \end{minipage}
  \hfill
  \begin{minipage}[t]{.5\textwidth}
    \includegraphics[angle=90,height=6cm]{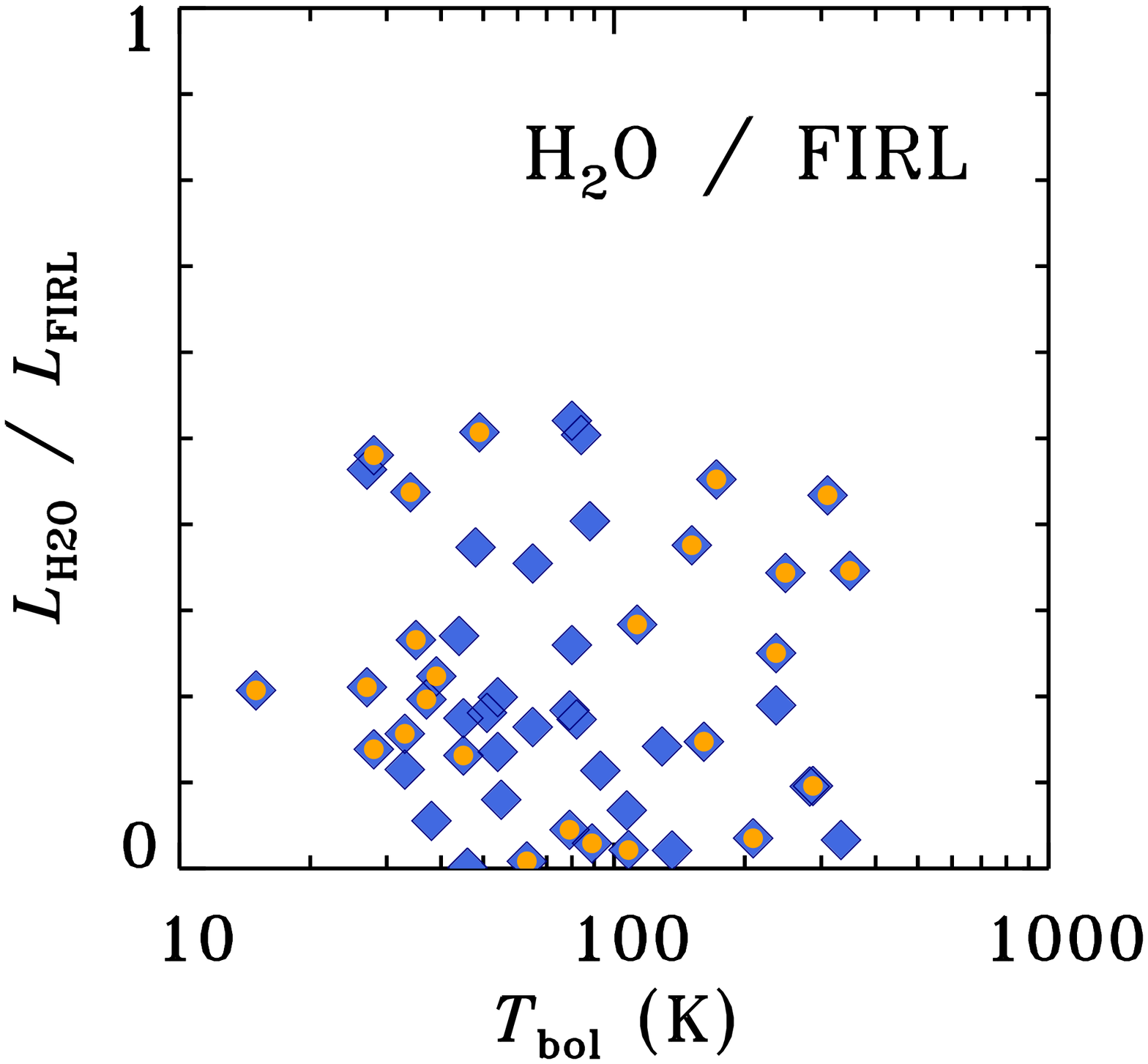} 
         \vspace{0.3cm}
        
    \includegraphics[angle=90,height=6cm]{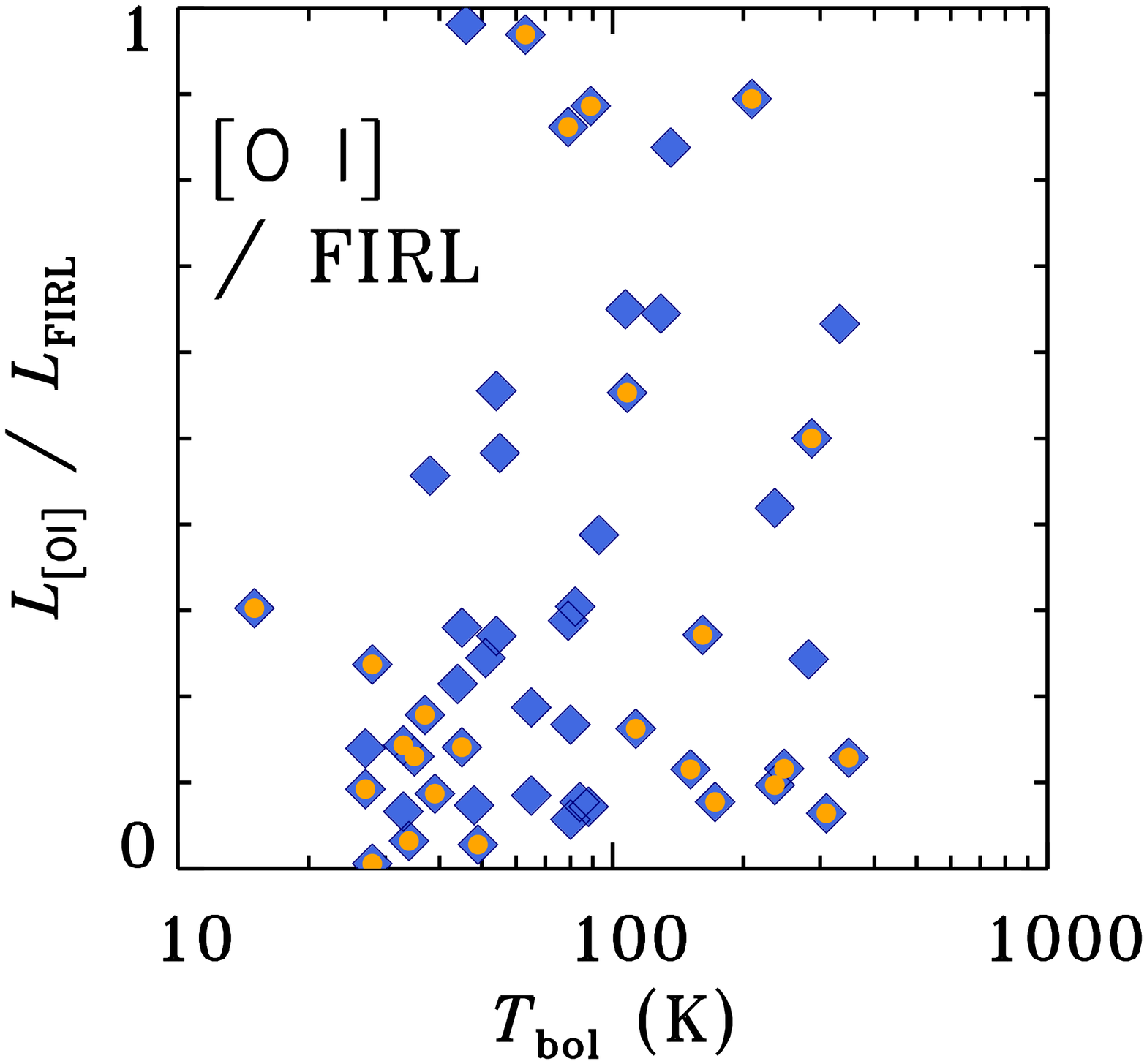} 
         \vspace{0.3cm}
         
  \end{minipage}
        \caption{\label{cool2} Ratio of line luminosities of selected species 
        and the total far-infrared line emission as a function of bolometric temperature.
        Objects observed in the full spectroscopy mode are marked with orange dots.
        The dashed blue vertical line corresponds to $T_\mathrm{bol}=70$ K, with the Class 0 
        sources located to the left and the Class I sources to the right side of the line.}
\end{figure*}

\begin{figure*}[!tb]
  \begin{minipage}[t]{.5\textwidth}
      \includegraphics[angle=90,height=6cm]{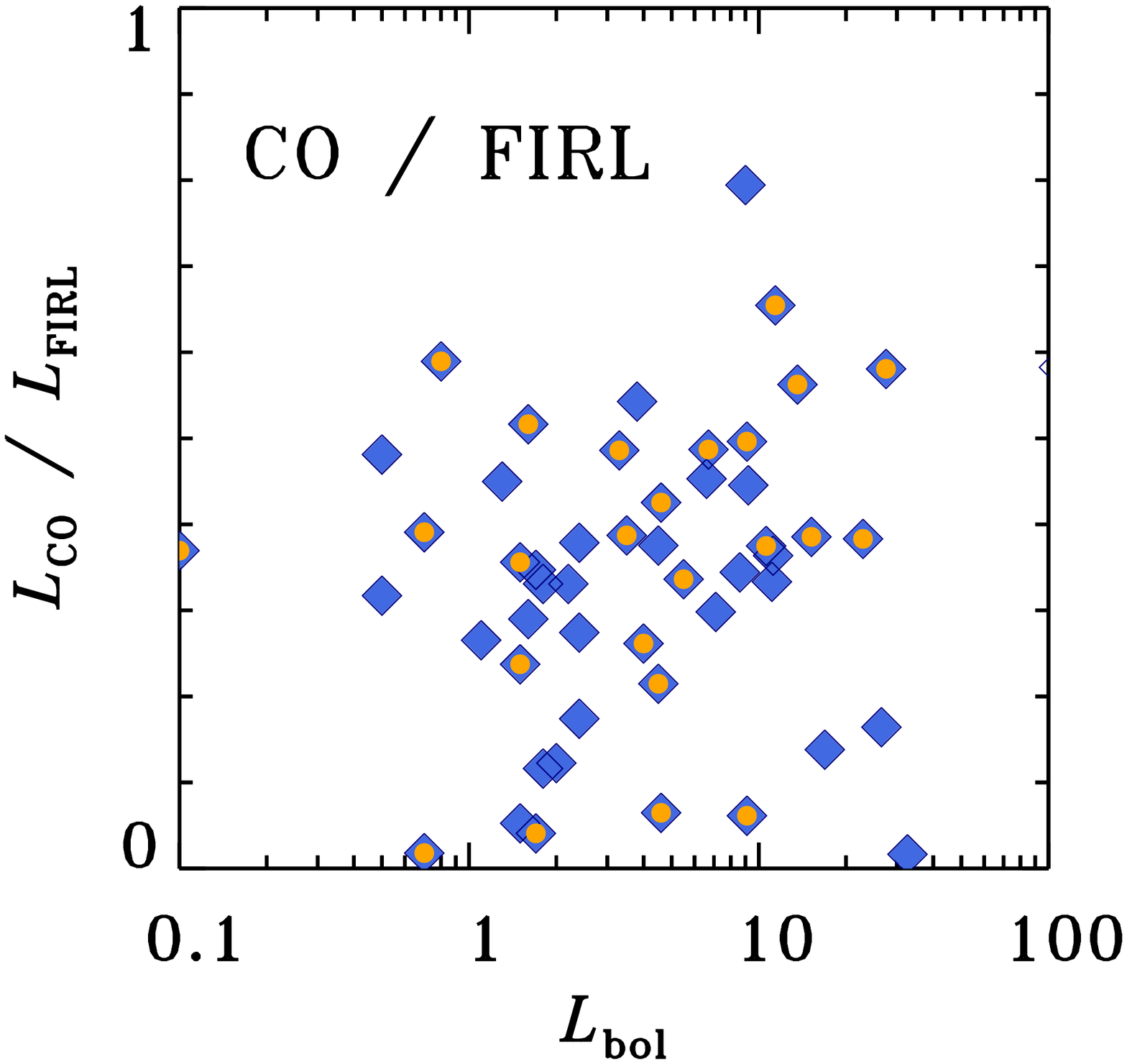}
     \vspace{+0.3cm}
     
        \includegraphics[angle=90,height=6cm]{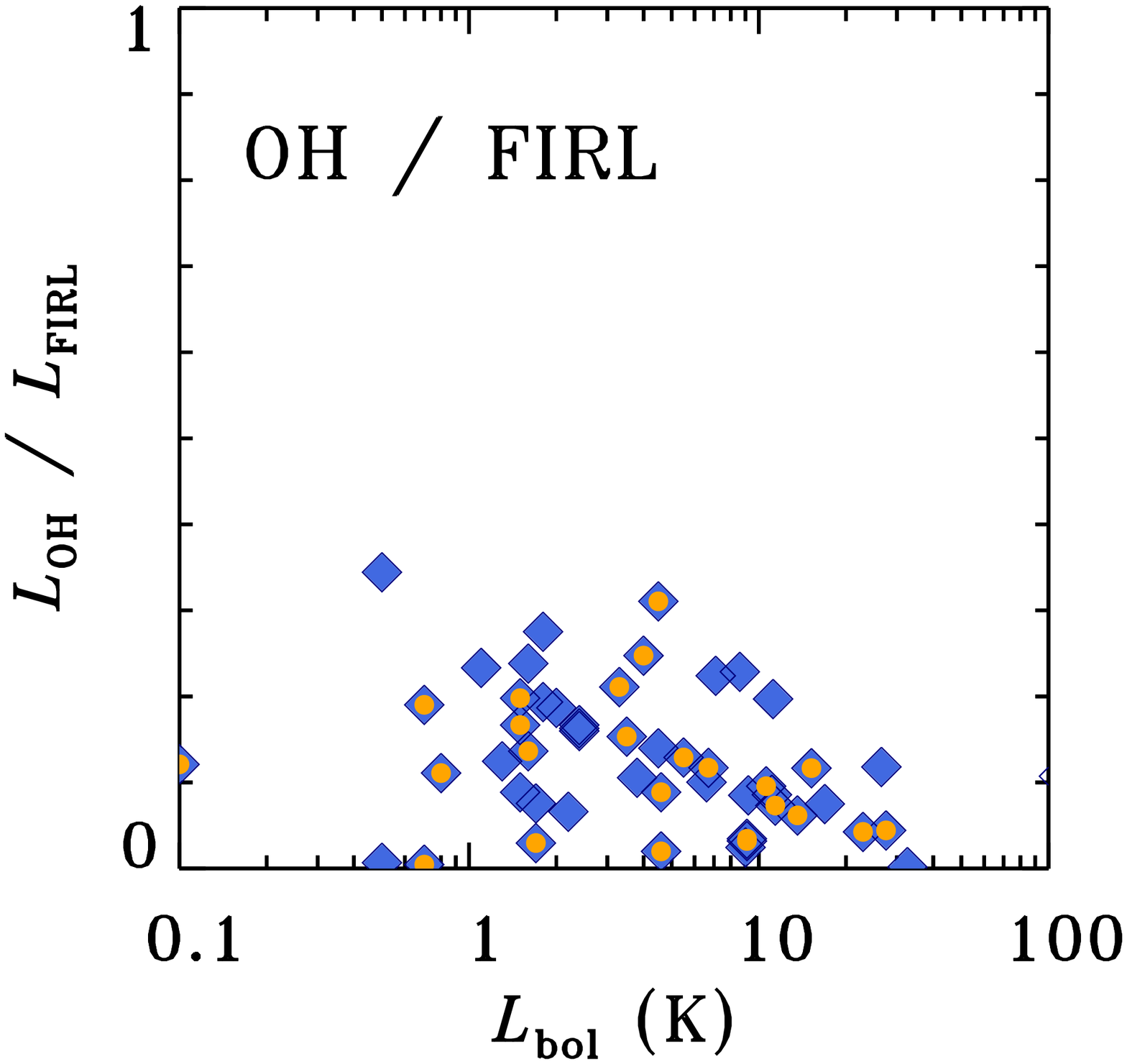}
     \vspace{+0.3cm}
  \end{minipage}
  \hfill
  \begin{minipage}[t]{.5\textwidth}s
    \includegraphics[angle=90,height=6cm]{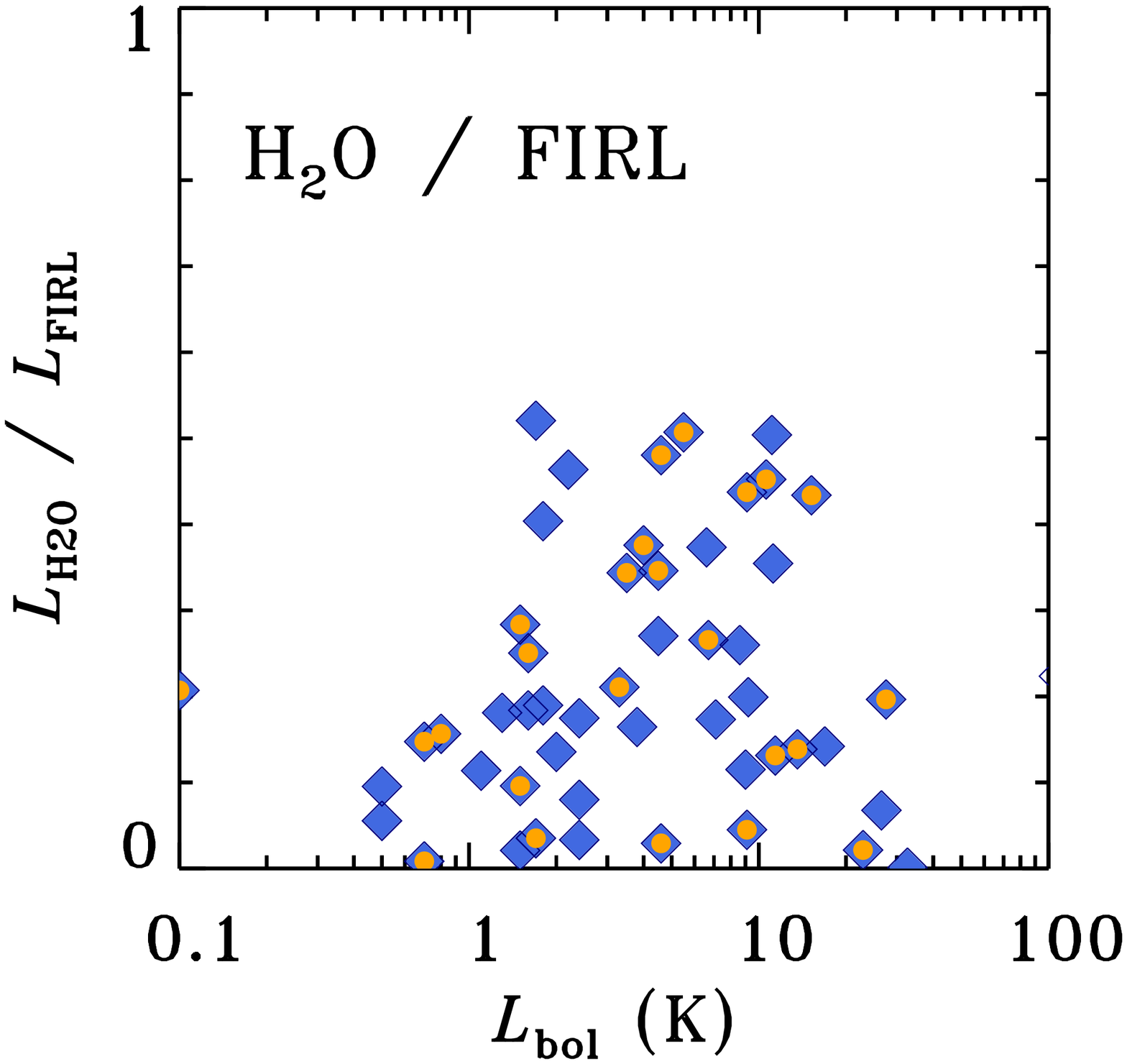} 
         \vspace{0.3cm}
        
    \includegraphics[angle=90,height=6cm]{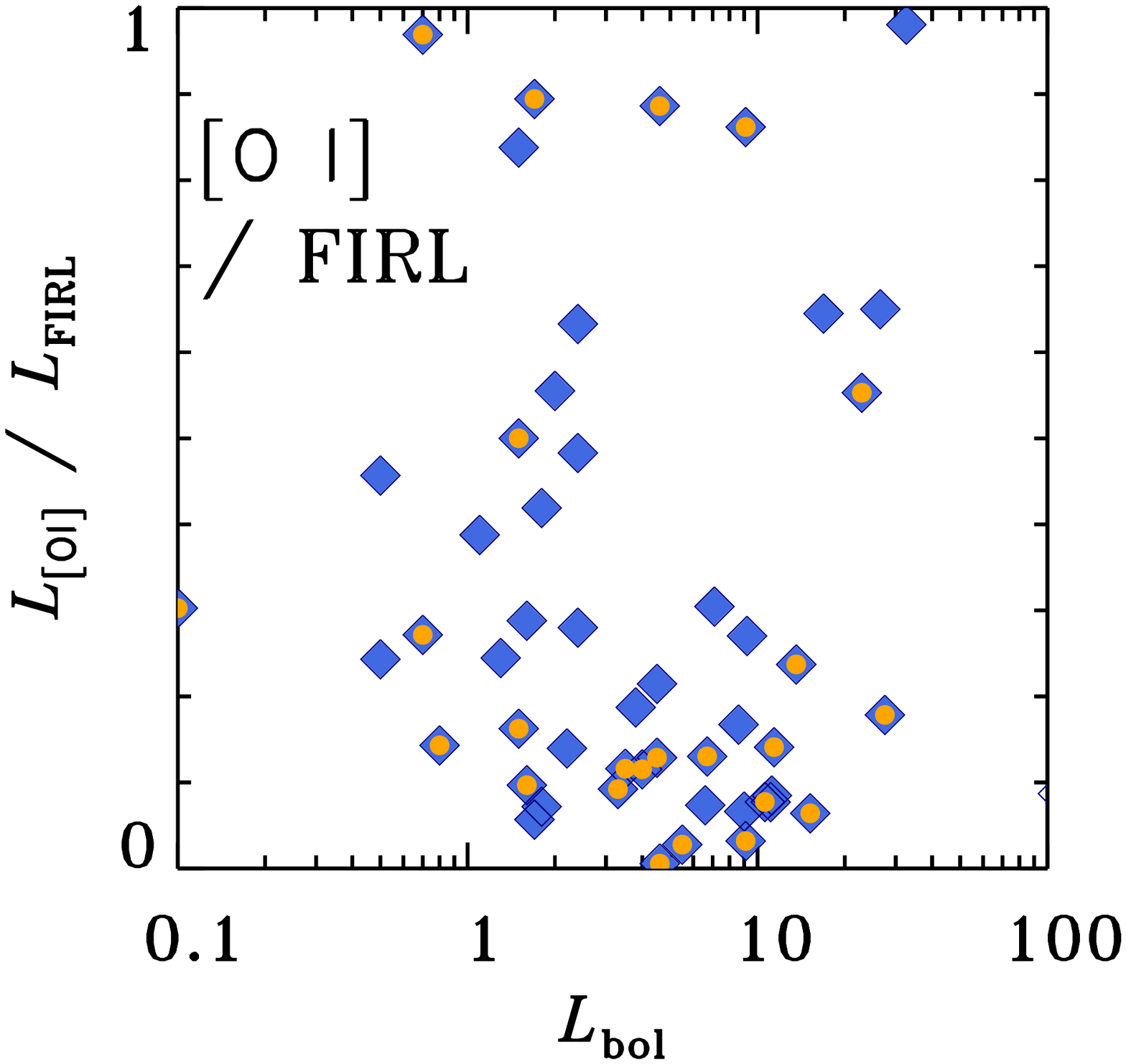} 
         \vspace{0.3cm}
         
  \end{minipage}
        \caption{\label{cool4} Similar to Figure \ref{cool2}, but the ratios of the line luminosities of selected species 
        and the total far-infrared line emission are shown as a function of bolometric luminosity.}
\end{figure*}

Correlations of the far-IR cooling with $L_\mathrm{bol}$ and $T_\mathrm{bol}$ are 
shown in Figure \ref{cool2} and \ref{cool4}.

\begin{figure*}[!tb]
  \begin{minipage}[t]{.5\textwidth}
      \includegraphics[angle=90,height=7cm]{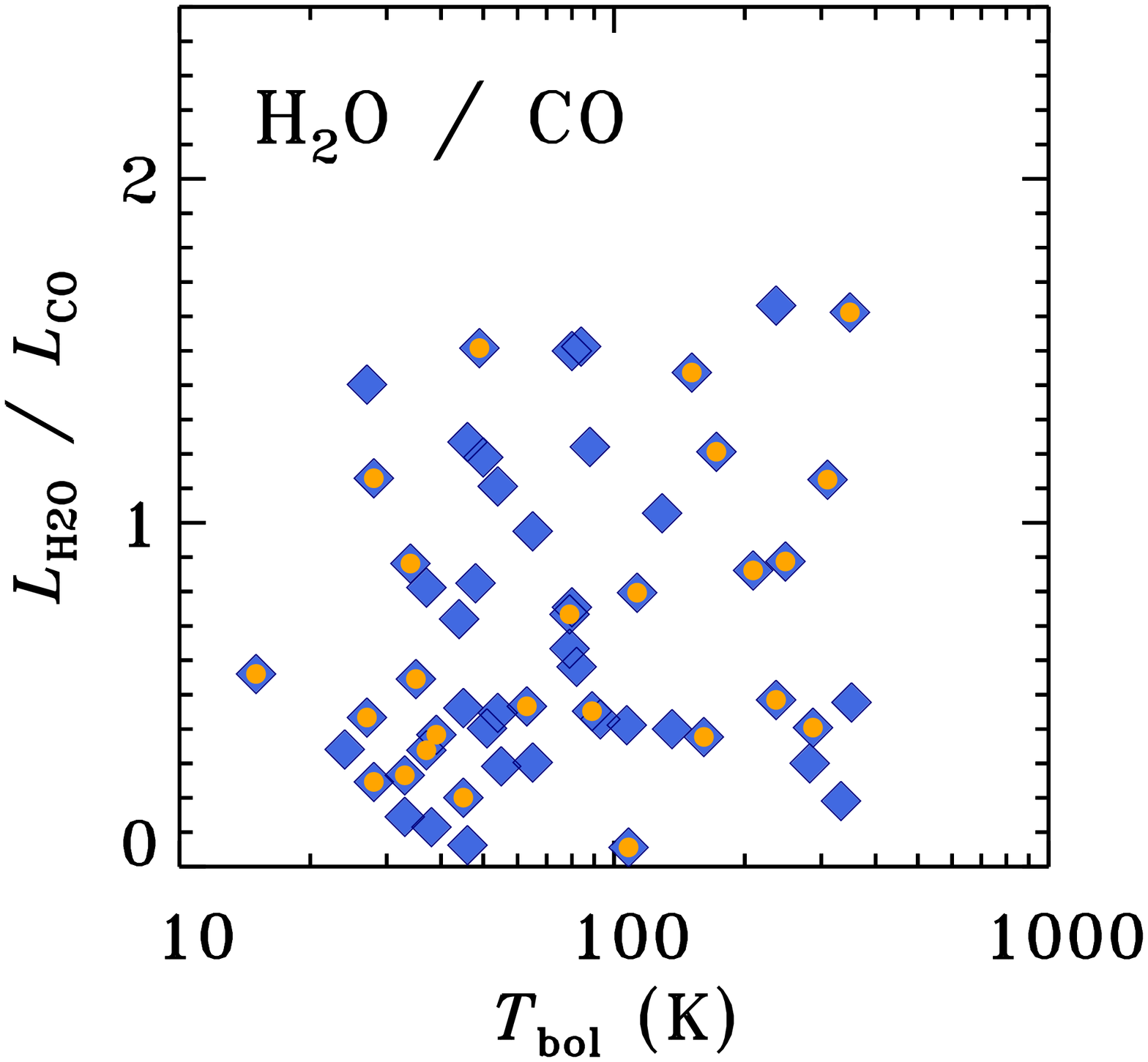}
     \vspace{+0.3cm}
     
        \includegraphics[angle=90,height=7cm]{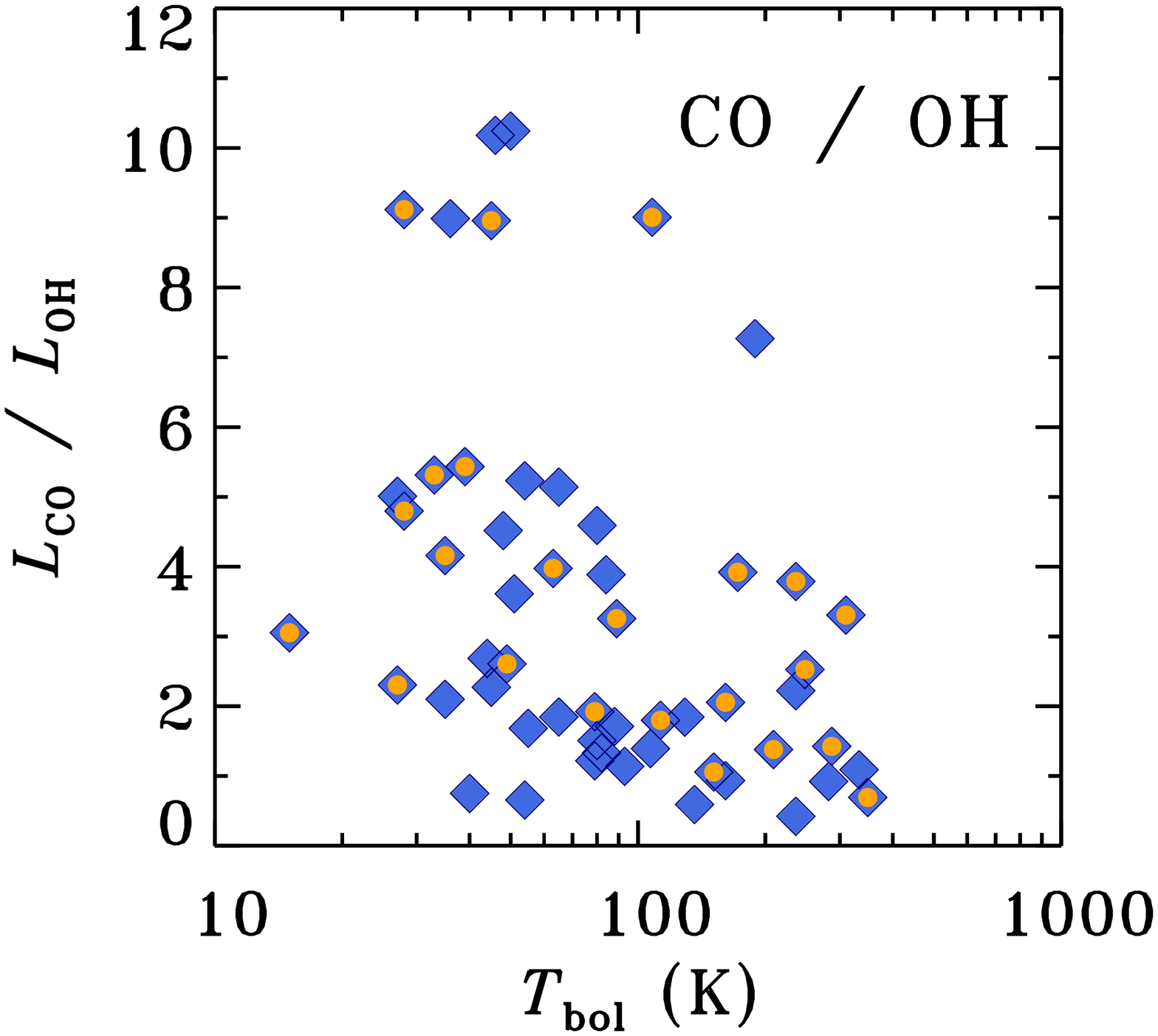}
     \vspace{+0.3cm}
  \end{minipage}
  \hfill
  \begin{minipage}[t]{.5\textwidth}
    \includegraphics[angle=90,height=7cm]{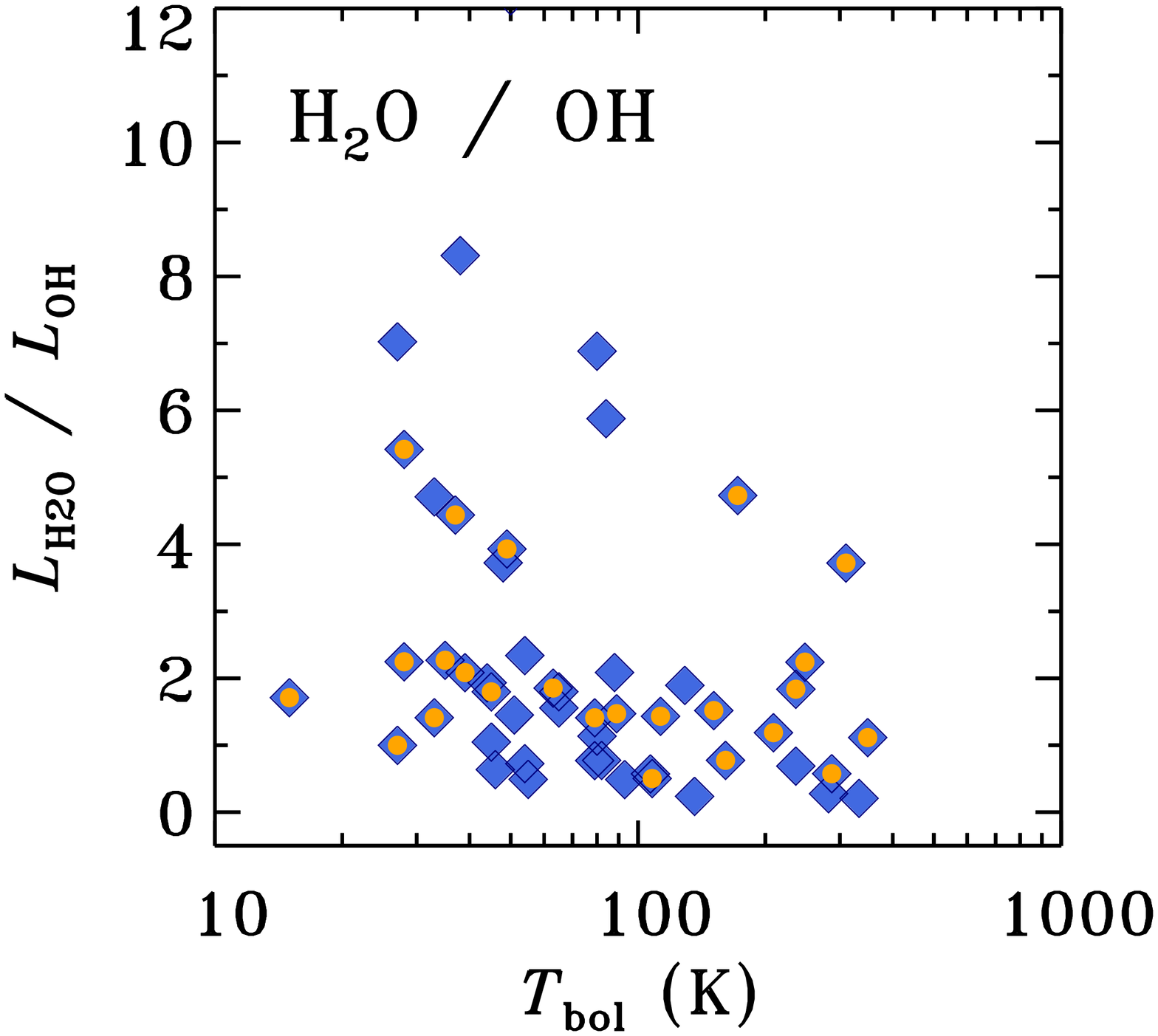} 
         \vspace{0.3cm}
        
    \includegraphics[angle=90,height=7cm]{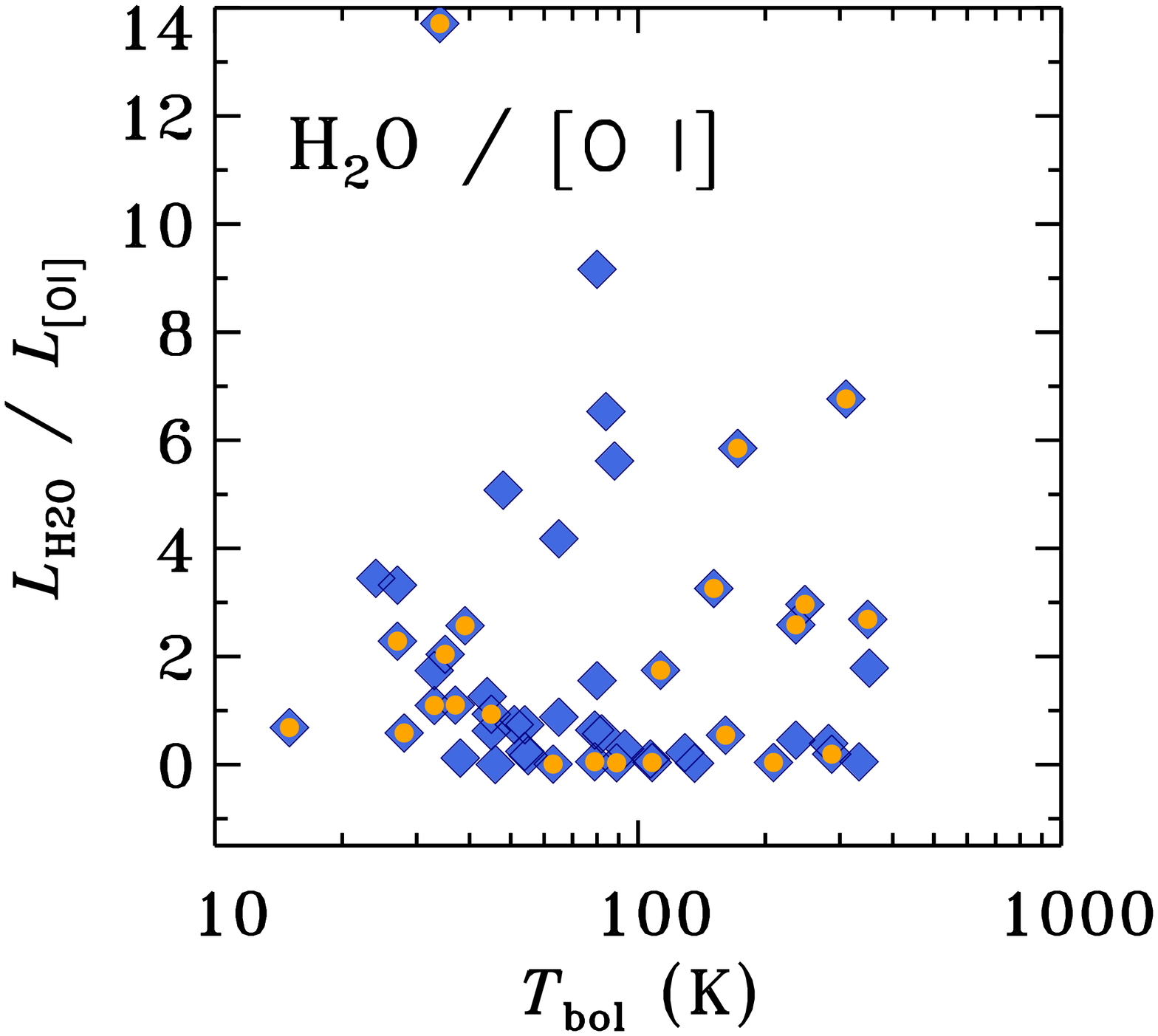} 
         \vspace{0.3cm}
         
  \end{minipage}
        \caption{\label{cool_ratio} Luminosity ratios of selected species 
         as a function of bolometric temperature.
        Objects observed in the full spectroscopy mode are marked with orange dots.}
\end{figure*}

Luminosity ratios of selected species as a function of bolometric temperature 
are shown in Figure \ref{cool_ratio}.


\begin{figure*}[t]
\begin{center}
\vspace{-4cm}
\includegraphics[angle=90,height=10cm]{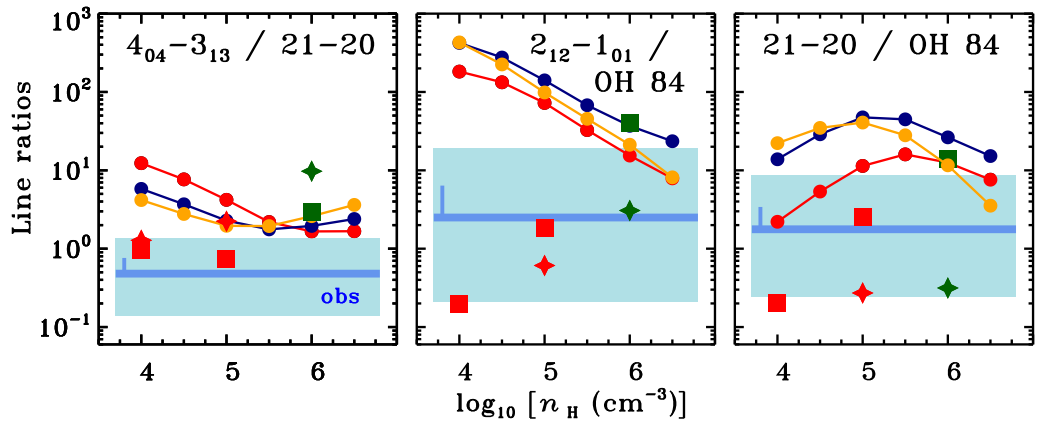}
\caption{\label{obsrat2} Similar to Figure \ref{obsrat}, but for other sets of lines 
(including H$_2$O $2_{12}-1_{01}$ at 179 $\mu$m and CO 21-20 at 124 $\mu$m).}
\end{center}
\end{figure*}

\end{document}